\documentclass{class/NTUASE_MR}
\usepackage[utf8]{inputenc}
\usepackage[english]{babel}
\usepackage{amsmath}
\usepackage{amsfonts}
\usepackage{amssymb}
\usepackage{amsthm}
\usepackage{graphicx}
\usepackage{subfigure}
\usepackage[titletoc, header, page]{appendix}
\usepackage{graphicx}
\usepackage{array}      
\usepackage{booktabs}   
\usepackage{longtable}  
\usepackage{lipsum}
\usepackage{bm}
\usepackage{physics}
\usepackage{mathtools}
\usepackage{MnSymbol}
\usepackage{datetime}   
\usepackage[pdfusetitle,colorlinks=true,linkcolor=blue,urlcolor=blue]{hyperref}
\usepackage{setspace}   
\usepackage[backend=bibtex,bibstyle=ieee,citestyle=numeric-comp,url=false,eprint=false]{biblatex}
\usepackage{multirow}
\usepackage{pdflscape} 
\usepackage{afterpage}
\usepackage{capt-of}
\usepackage{pdfpages} 
\usepackage{stackengine} 
\usepackage{epigraph} 
\usepackage{enumitem}
\usepackage{float}
\usepackage[most]{tcolorbox}
\definecolor{block-gray}{gray}{0.85}
\newtcolorbox{blockquote}{colback=block-gray,grow to right by=-1mm,grow to left by=-1mm,boxrule=0pt,boxsep=0pt,breakable}

\newtcolorbox{Box1}[2][]{
                lower separated=false,
                colback=white,
colframe=white!20!gray,fonttitle=\bfseries,
colbacktitle=white!10!gray,enhanced,
attach boxed title to top left={xshift=1cm,
        yshift=-2mm},
title=#2,#1}

\newcommand{\varket}[1] {\left|{#1}\right\rrangle}
\newcommand{\varbra}[1] {\left\llangle #1\right|}

\newtheorem{prop}{Proposition}[chapter] 

\definecolor{OceanBlue}{rgb}{0,0.35,0.7} 
\usepackage{hyperref}
\hypersetup{colorlinks=true,allcolors=OceanBlue}

\makeatletter
\def\dual#1{\expandafter\dual@aux#1\@nil}
\def\dual@aux#1/#2\@nil{\begin{tabular}{@{}c@{}}#1\\#2\end{tabular}}
\makeatother

\makeatletter
\renewcommand*\env@matrix[1][\arraystretch]{%
  \edef\arraystretch{#1}%
  \hskip -\arraycolsep
  \let\@ifnextchar\new@ifnextchar
  \array{*\c@MaxMatrixCols c}}
\makeatother

\emergencystretch=\maxdimen
\title{QUANTUM-OPTICAL SENSING AND TARGET DETECTION}

\author{THAM GUO YAO}

\affiliation{School of Physical and Mathematical Sciences \\
College of Science}
\affilogo{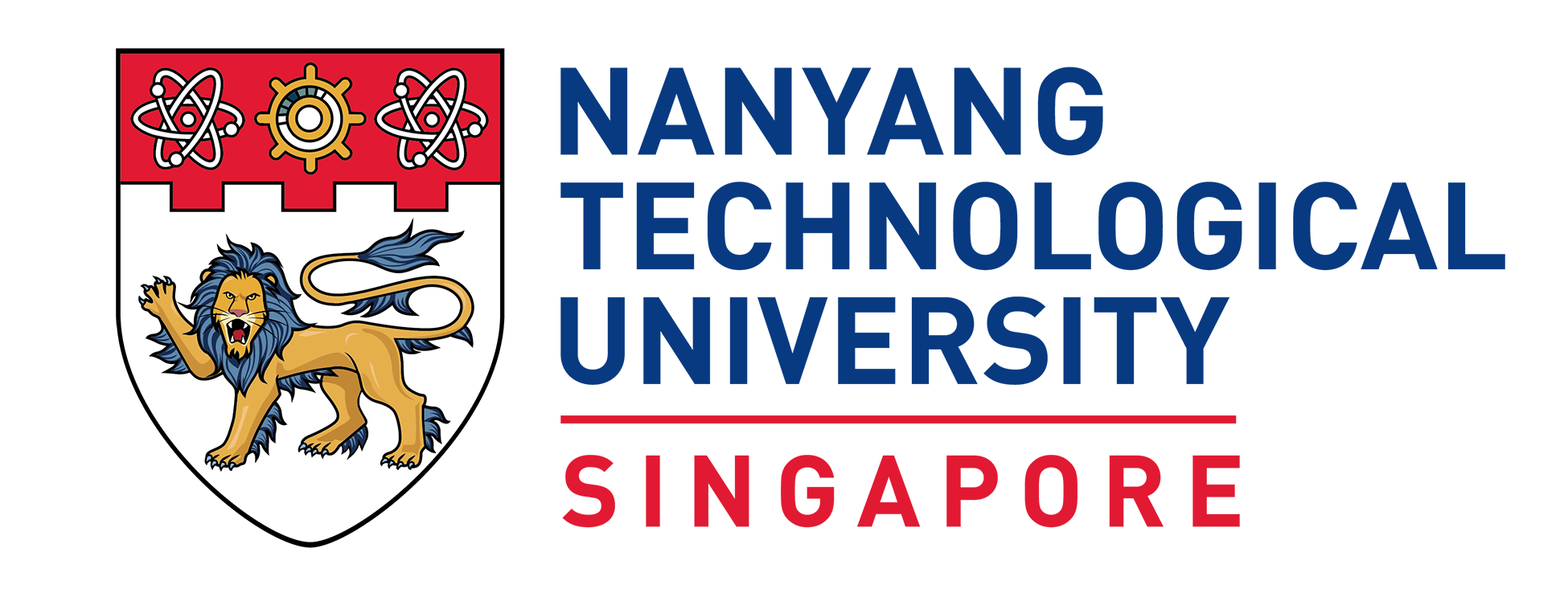}

\TypeOfDocument{A thesis submitted to the Nanyang Technological University in partial fulfilment of the requirement for the degree of
Doctor of Philosophy}

\date{2024}

\begin{document}
    \makefrontmatter
	
        \chapter*{List of Notations}
\setlength{\tabcolsep}{10pt}
\renewcommand{\arraystretch}{1.5}
\begin{longtable}{lp{0.81\textwidth}}
  $\hat{a}_k,\hat{a}^\dag_k$ & Annihilation and creation operators respectively. The subscripts of the operators depends on context of description by notations. \\
  $[\hat{A},\hat{B}]$ & Commutator of quantum operators $\hat{A}$ and $\hat{B}$ defined as $\hat{A}\hat{B}-\hat{B}\hat{A}$.\\
  $\mathbf{A}\oplus \mathbf{B}$ & Direct sum of matrix $\mathbf{A}\oplus \mathbf{B} = \begin{pmatrix}
      \mathbf{A} & \mathbf{0}\\ \mathbf{0} & \mathbf{B}
  \end{pmatrix}$.\\
  $\mathcal{A}_{G}$ &  Quantum-limited amplifier channel with a gain value of $G\geq 1$.\\
  $C_s(\rho,\sigma)$ & s-overlap between the states $\rho$ and $\sigma$.\\
  $D(\rho,\sigma)$ & Trace distance between the states $\rho$ and $\sigma$.\\
  $\det X$ & Determinant of matrix $X$.\\
  $\mathbb{E}[X]$ & Expected value of a random variable $X$. In quantum mechanics, $X$ is an operator, and the expectation value is defined as $\expval{\hat{X}}=\Tr[\rho\hat{X}]$.\\
  $\mathcal{F}_{sq}(\rho,\sigma)$ & Quantum fidelity between the states $\rho$ and $\sigma$.\\
  $\mathcal{F}(\rho,\sigma)$ & Uhlmann square root fidelity between the states $\rho$ and $\sigma$ defined as $\mathcal{F}(\rho,\sigma) = \sqrt{\mathcal{F}_{sq}(\rho,\sigma)}$. In this thesis, unless otherwise stated, any quantification of fidelity will be defined formally as Uhlmann square root fidelity.\\
  $\mathbb{I}$ & Identity matrix with dimension corresponding to the context.\\
  $\inf_{k\in\mathcal{S}}x_k$ & Infimum or greatest lower bound of the set $\{x_k\}_{k\in\mathcal{S}}$.\\
  $\mathcal{J},\mathcal{K}$ & Classical and quantum Fisher information. The dependency on variables can be inferred from the context of description by notations.\\
  $\mathcal{L}_{\tilde{\eta}}$ &  Pure loss channel with transmittance value of $0\leq \tilde{\eta} \leq 1$.\\
  $\mathcal{L}_{\eta,N_B}$ & Noisy attenuator (thermal loss) channel with noise input of strength $N_B$ and transmittance value of $0\leq \eta \leq 1$.\\
  $\mathcal{M}_k$ & Positive operator-valued measure. The subscript of the measure depends on context of description by notations.\\
  $\pi_0,\pi_1$ & Priori probability associated to Bayesian hypothesis testing.\\ 
  $\ket{\psi}$ & Eigenstate representation of a pure quantum state. For Fock state representation, $\ket{\psi}=\ket{n}$ where $n\in\mathbb{N}_0$.\\
  $\rho$ & Density matrix of a quantum state. For the case of a pure state $\ket{\psi}$, $\rho=\ket{\psi}\bra{\psi}$.\\
  $q$ & $2N$-dimensional quadrature operator vector, i.e., $q = (\hat{x}_1,\cdots,\hat{x}_N,\hat{p}_1,\cdots,\hat{p}_N)$.\\
  $\Tr \qty[X]$ & Matrix trace of the matrix $X$.\\
  $\text{Var}[X]$ & Variance of a random variable $X$. In quantum mechanics, $X$ is an operator, and the variance is defined as $\expval{\hat{X}^2}-\expval{\hat{X}}^2 = \Tr[\rho\hat{X}^2] - \qty(\Tr[\rho\hat{X}])^2$. In some context, this quantity is also defined as the mean-squared error (MSE).\\
  $V,\tilde{V}$ & Covariance matrices of a Gaussian state. The different notation represent covariance matrices defined by different quadrature operator vector arrangements.\\
  $\hat{x},\hat{p}$ & Dimensionless position and momentum operator.\\
  $\norm{X}_1$ & Trace norm of the operator $X$ defined as $\Tr\sqrt{X^\dag X}$.\\
  $\chi$ & Error exponent of a given error probability.\\
  $\chi(\xi)$ & Anti-normally ordered characteristic function of a given density state.\\
  $\expval{X}$ & Expectation value of operator $X$ for a given state $\rho$. $\expval{X} = \bra{\psi}X\ket{\psi}$ for pure state $\rho=\ket{\psi}$ and $\expval{X}= \Tr[X\rho]$ for mixed state.
\end{longtable}
\let\cleardoublepage\clearpage
	\listoffigures

    \cleardoublepage

    \pagenumbering{arabic}

    \ifdefined\isdraft
        \doublespacing
    \else
        \onehalfspacing
    \fi

    \chapter[Introduction]{Introduction}\label{chap-1}
\emph{In the introductory chapter of this thesis, we delve into the motivation, objectives, and structural outline of our thesis. Our motivation is firmly rooted in the realm of entanglement-enahanced target detection (quantum illumination), emphasizing how the utilization of quantum entanglement can lead to advantageous performance in target detection tasks. We highlight the unique ability of entanglement to facilitate "better-than-classical" detection outcomes, even in scenarios where an adversary might intercept the signal. This chapter also provides an initial exploration of the fundamental setups for target detection and covert sensing, laying the groundwork for the detailed analyses that follow. By establishing clear objectives and outlining the trajectory of our thesis, we set the stage for a comprehensive examination of quantum illumination and its transformative impact on the field of detection.}
\newpage
	\section{Motivation}

Quantum technologies have emerged as practical applications built upon the foundations of quantum theory. Their growing relevance and research interest stem from the potential they offer to outperform classical systems governed by deterministic mechanics, providing a significant edge in technological capabilities \cite{boixo2018characterizing,arute2019quantum,arvidsson2020quantum,chakrabarti2021threshold,assouly2023quantum}. The potential to revolutionize industries such as computing, communication, and sensing makes quantum technologies a critical area of exploration. Among these, quantum communication stands out, where integrating quantum principles has the potential to radically improve data transmission methods, offering enhanced security, speed, and efficiency over conventional communication systems \cite{gisin2007quantum,cozzolino2019high,sidhu2021advances,kremer1995quantum,chen2021integrated,pirandola2017fundamental}.\\~\\
Communication technologies have long been a topic of extensive research, with a focus on optimizing factors such as energy consumption, range, transmission speed, and signal robustness \cite{min2002framework,pei2008application,izadpanah2001high,young2010robustness}. These parameters remain crucial in the development of modern communication systems. However, the focus of this thesis extends beyond communication to explore the field of sensing technology. Sensing involves detecting and measuring the inherent properties of objects and converting these measurements into signals that can be transmitted and interpreted by the user \cite{davis1978remote,campbell2011introduction}. In the context of modern technology, the advancement of sensing techniques has become a priority. It is essential not only to optimize the energy efficiency of the sensing process but also to enhance the precision of the collected data. This balance of energy conservation and signal accuracy is central to the performance of advanced sensing systems \cite{patton2012efficient, jha2015topology}.\\~\\
One of the key challenges within the domain of sensing is covert sensing, where an adversarial element is introduced. In this scenario, an adversary attempts to intercept and analyse the signals within the environment to discern whether a detection attempt has occurred, effectively determining whether a signal has been sent \cite{tahmasbi2020active,tahmasbi2021signaling,hao2022demonstration,bash2017fundamental}. This adversarial condition complicates the traditional sensing problem, as it introduces a trade-off between minimizing the strength of the transmitted signal (to avoid detection by the adversary) and maintaining the accuracy and reliability of the information being gathered. Achieving an optimal balance requires careful consideration of how much information can be hidden from the adversary without compromising the accuracy of the sensing results. \newpage
\subsection{Target detection}
\begin{figure}[H]
\centering
\includegraphics[width = 0.49\linewidth]{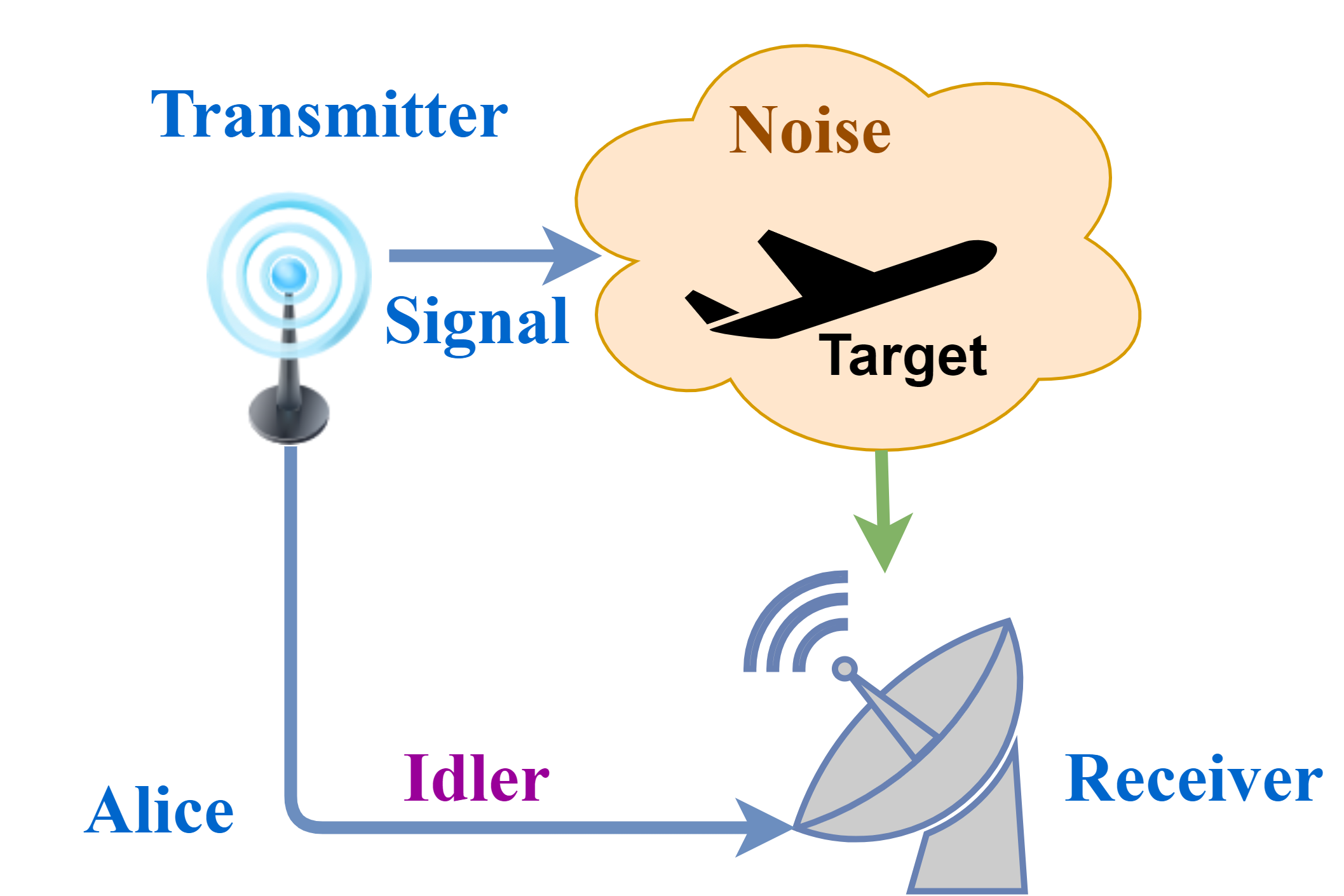}
\captionsetup{width=\linewidth}
\caption{A general setup for quantum target detection: Alice uses entangled ancillas in an attempt to detect the presence of an object. }
\label{fig:setup_nocovert}
\end{figure}
Illumination is a method of target detection involving sending signals (referred to as Alice) to a remote region in space surrounded by some degree of thermal noise, in an attempt to detect the presence or absence of a target in that region \cite{lloyd2008enhanced,nair2020fundamental,tan2008quantum,bradshaw2021optimal,gregory2020imaging,tham2023quantum,shapiro2020quantum}. In classical illumination, strong electromagnetic waves are sent, and the reflected waves from the region are collected and analysed to determine the presence of the target \cite{skolnik1962introduction,richards2010principles,levanon1988radar}. Similarly, in quantum illumination, microwave or optical beams are sent to interrogate the target region. The improvement in quantum illumination over classical illumination lies in the fact that the microwave or optical beam may be entangled with an idler ancilla (hence quantum) held by the receiver \cite{nair2020fundamental,lloyd2008enhanced,tan2008quantum,weedbrook2016discord,bradshaw2017overarching}. A suitable joint quantum measurement is made by the receiver for the reflected signal and the idler ancilla.
\subsection{Covert sensing}
\begin{figure}[h!]
\centering
\includegraphics[width = 0.55\linewidth]{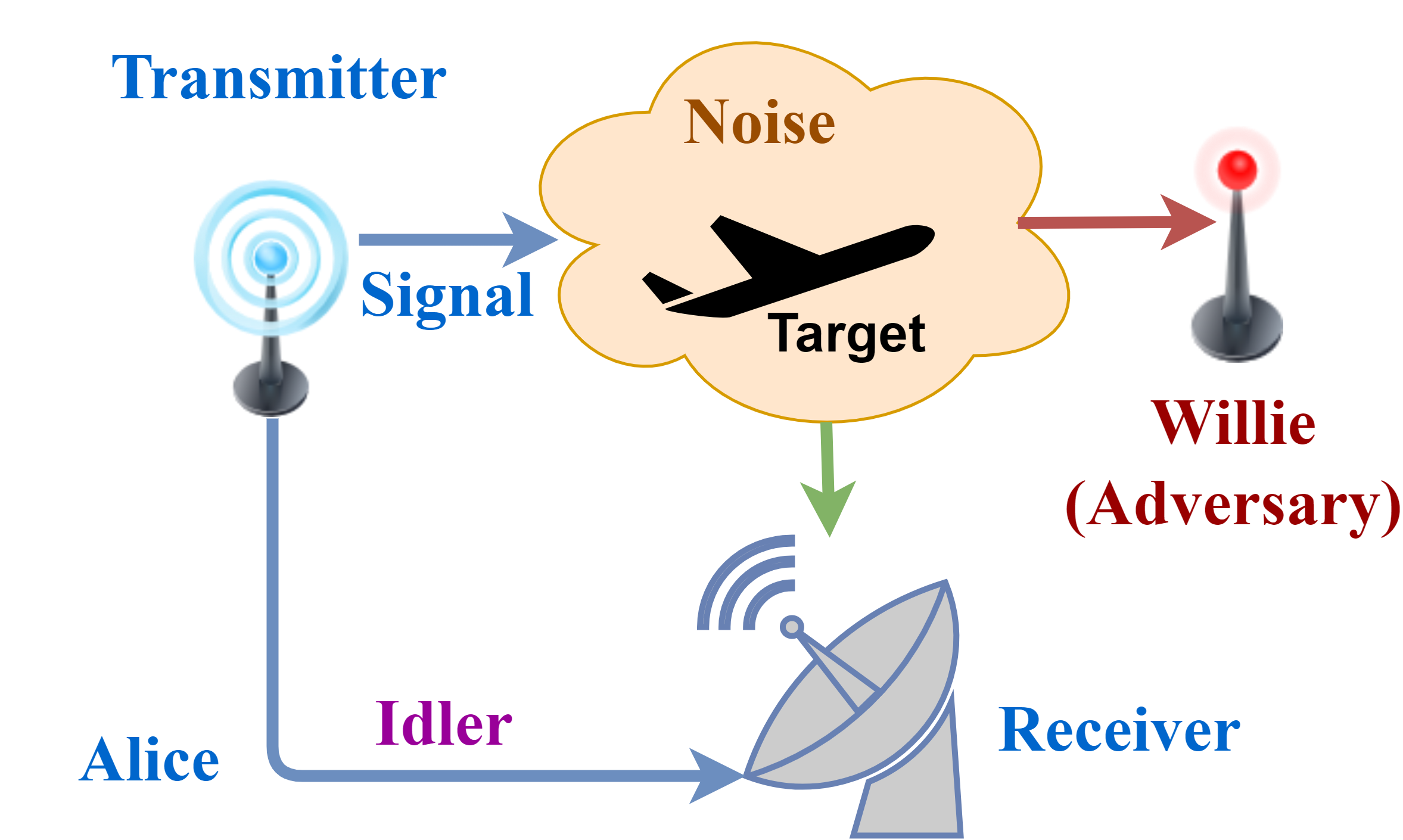}
\captionsetup{width=\linewidth}
\caption{An extension of the setup to include Willie as an adversary: Alice attempts to detect the presence of the adversary Willie while remaining undetected herself. }
\label{fig:setup_covert}
\end{figure}\noindent
In covert illumination, the setup is similar to target detection, with the addition of an adversary (named Willie). Willie is capable of intercepting and collecting all the light that does not return to Alice \cite{tahmasbi2020active,tahmasbi2021signaling,hao2022demonstration,bash2017fundamental,bash2015quantum,bullock2020fundamental,gagatsos2019covert,tham2023quantum}. From all the light intercepted, Willie has to interpret the information and determine if Alice did send a signal to probe the target region. Such covert considerations are particularly important in espionage situations where one wishes to monitor an adversary while ensuring they remain unaware of such surveillance.
\subsection{Gain sensing}
\begin{figure}[h!]
\centering
\includegraphics[width = 0.5\linewidth]{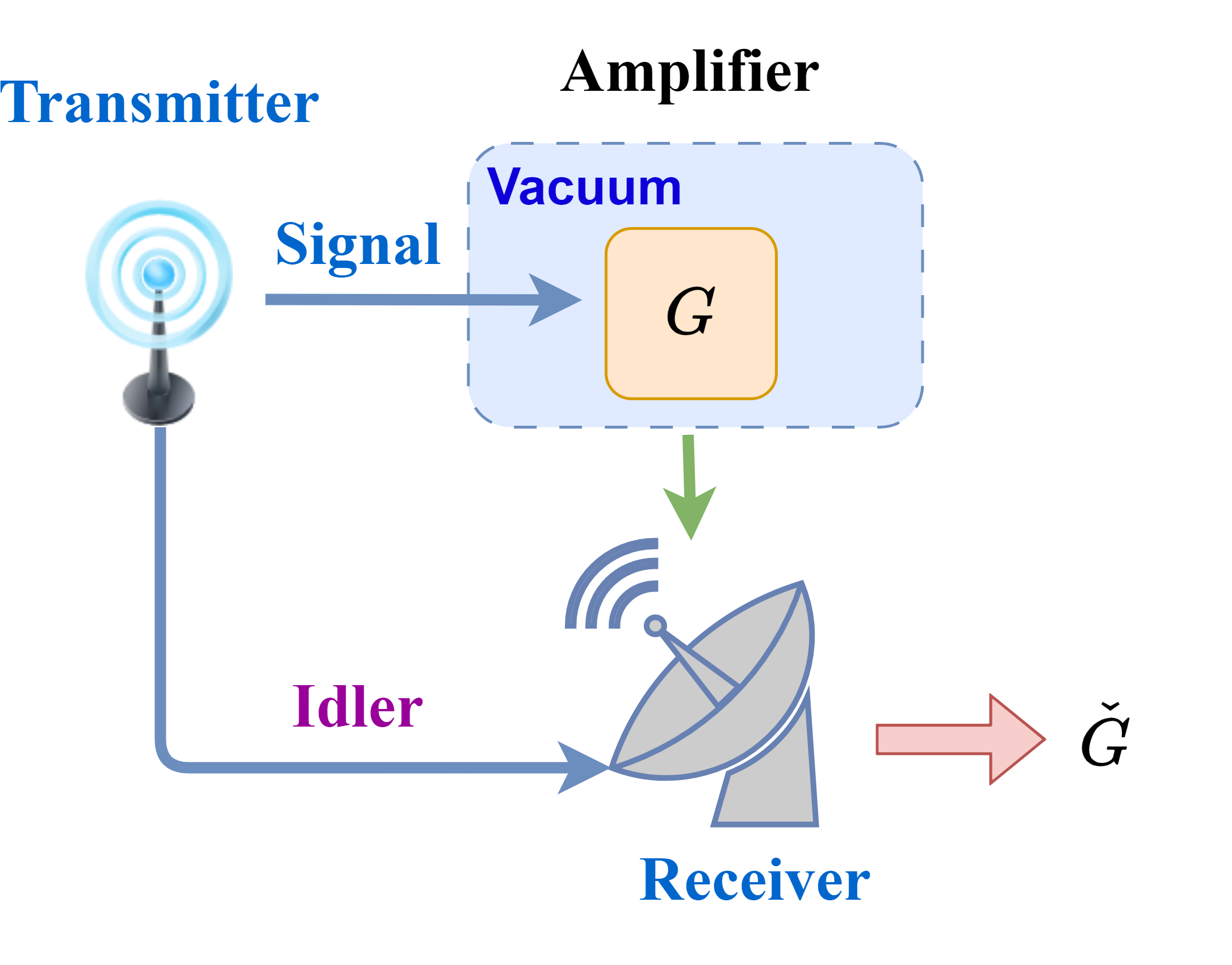}
\captionsetup{width=\linewidth}
\caption{A setup for gain sensing: Using an entangled ancillas pair to determine the gain parameter of an amplifier in a vacuum environment. }
\label{fig:gain_1}
\end{figure}\noindent
Gain sensing involves transmitting signals, which can be either classical or quantum, to estimate the amplification factor (or gain) of a system. In this context, the target area is defined as a vacuum chamber that houses an amplifier characterized by a gain parameter $G$ \cite{nair2022optimal,aspachs2010optimal,gaiba2009squeezed}. The objective here is to send signals into this chamber, allowing them to interact with the amplifier. Upon their return, these signals carry information about the amplifier's properties, specifically the gain parameter $G$. The measured data from these signals are then processed through various computational and statistical methods to accurately estimate the value of the gain parameter $G$.

\section{Outline of thesis}
The outline of the thesis is organised as follows:\\~\\
Chapter~\ref{chap-2} establishes foundational definitions and mathematical tools for understanding metrology, focusing on detection and estimation theory, and the mathematical formulation of Continuous Variable quantum systems. It also explores quantum channels, demonstrating how pure loss channels combined with quantum-limited amplifiers create thermal loss channels, and concludes with a discussion on threshold detector.\\~\\
Chapter~\ref{chap-3} examines the evolution of quantum illumination, detailing its progression from theoretical concepts to experimental milestones. It emphasizes the significant developments that have shaped the field, providing a thorough understanding of both its theoretical foundations and practical applications. The discussion also explores how advancements in quantum technologies have spurred new research questions and goals, especially in the context of this thesis. \\~\\
Chapter~\ref{chap-4} proposes an alternative quantum detection protocol setup whereby a passive signature is present and sending a vacuum state allows target detection to a certain extent. We also address covert target detection, where Alice aims to detect a target without alerting adversary Willie. It establishes quantum limits on covert target detection error probability, compares the performances of various quantum probes, and explores the quantum limit for distinguishing between thermal loss channels under passive signature assumptions.\\~\\
Chapter~\ref{chap-5} examines the precision limits in estimating the gain parameter of phase-insensitive optical amplifiers using quantum methods, potentially entangled probes. We establish that pure-state probes, diagonal in the multimode number basis, achieve optimal quantum precision. The study also compares this precision to classical probes and presents a closed-form expression for the energy-constrained Bures distance between amplifier channels.\\~\\
Chapter~\ref{chap-6} discusses the performance of three probe states - coherent state, two-mode squeezed vacuum (TMSV), and single-photon entangled state (SPES) — in target detection within a No-Passive Signature (NPS) framework. It highlights the unique ability of SPES to maintain non-classical properties post-pure loss channel and uses the error exponent to quantify the target detection performance of the different probes.\\~\\
Chapter~\ref{chap-7} concludes the thesis by summarising all the results, provides applications of the results, and potential future progression.
    \chapter[Theoretical background]{Theoretical background}\label{chap-2}
\emph{This chapter lays the groundwork in providing definitions and mathematical tools essential to comprehend the subsequent research chapter . It starts by delineating the two main branches of metrology, detection and estimation theory, detailing their distinct quantification metrics and mathematical underpinnings. The focus then shifts to the mathematical formulation of Continuous Variable quantum systems, crucial for applying quantum mechanics in metrology.
The chapter also introduces various quantum channels which are essential in quantum information theory. A key highlight is the demonstration of how combining pure loss channels with quantum-limited amplifiers results in thermal loss channels. The chapter concludes with a discussion on the threshold detector, an important tool for hypothesis testing and target detection. It presents a mathematical framework for the threshold detector and provides a quantitative analysis of the error probability limits associated with its use.}
\newpage
\section{Detection theory}
\begin{figure}[h!]
    \centering
    \includegraphics[width = 0.8\linewidth]{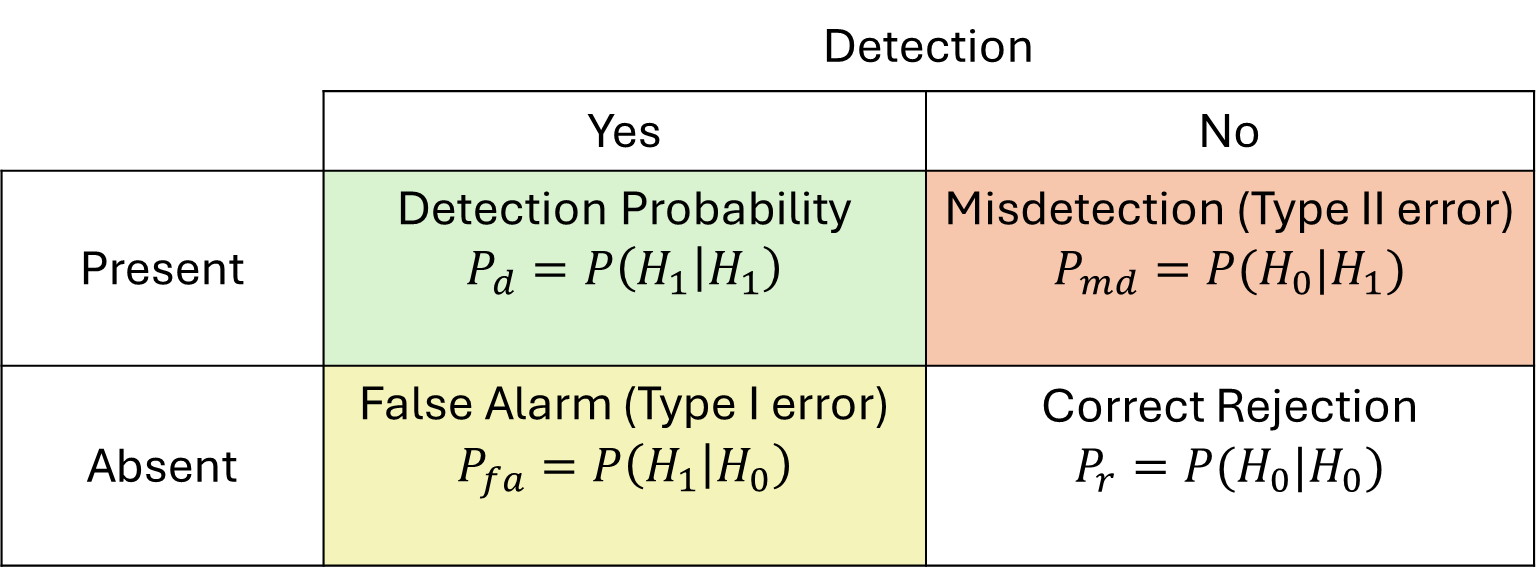}
    \caption{Signal detection theory (SDT) contingency table}
    \label{fig:Detection Theory}
\end{figure}\noindent
Detection theory is a framework for quantifying the ability to discern between the presence and absence of relevant information, also known as the hypothesis-testing problem. The null hypothesis $\textsc{H}_0$ in this context is defined as the target object being absent while the alternative hypothesis $\textsc{H}_1$ indicates that the target object being present. The successful distinction between the two hypotheses is defined by having a detection probability $P_d = P(\textsc{H}_1|\textsc{H}_1)$, i.e., correctly accepting the alternative hypothesis when the target is present. Two types of detection error may occur during the detection process: type-I error (false alarm), which is the error where the null hypothesis is incorrectly rejected, and type-II error (misdetection), which is the error where the alternative hypothesis is incorrectly rejected \cite{helstrom1969quantum}. The minimisation of these probabilities can be executed in various ways, depending on the situation-dependent rules during decision-making.
\subsection{Symmetric hypothesis test}
In the case of a symmetric approach to hypothesis testing, the aim is to minimise the average of both type-I and type-II errors, weighted by their priori probabilities,
\begin{align}
    P_{err} &= P(\textsc{H}_0)P(\textsc{H}_1|\textsc{H}_0) + P(\textsc{H}_1)P(\textsc{H}_0|\textsc{H}_1)\nonumber\\
    &= \pi_0 P(\textsc{H}_1|\textsc{H}_0) + \pi_1 P(\textsc{H}_0|\textsc{H}_1)\label{eq:minerr_classical},
\end{align}
where $\pi_0$ and $\pi_1$ are the priori probabilities associated with the two hypotheses respectively.\\
\noindent 
In quantum information, the Helstrom theorem defines the minimum error probability for symmetric hypothesis testing \cite{helstrom1969quantum}. It sets a fundamental limit on the performance of any quantum measurement used to distinguish between two quantum states. Consider a binary quantum hypothesis testing scenario with two quantum states, represented by density operators $\rho_0$ and $\rho_1$, respectively. The quantum measurement for these states is represented by a corresponding positive operator-valued measure (POVM). A POVM is a generalization of the concept of a measurement in quantum mechanics. Unlike traditional projective measurements, which are described by a set of orthogonal projectors corresponding to the eigenstates of an observable, a POVM consists of a set of positive semi-definite operators $\{\mathcal{M}_i\}$ that sum to the identity operator, $\sum_i\mathcal{M}_i=\mathbb{I}$. Each operator $\mathcal{M}_i$ corresponds to a possible measurement outcome, and the probability of obtaining outcome $i$ when measuring a quantum state $\rho$ is given by $p(i)=\Tr\qty[\rho \mathcal{M}_i]$. For the case of binary quantum hypothesis testing, the POVM is given as $\{\mathcal{M}_0,\mathcal{M}_1\}$ satisfying the completeness relation $\mathcal{M}_0 + \mathcal{M}_1 = \mathbb{I}$, with $\mathbb{I}$ being the identity operator. The error probability of this measurement, analogous to its classical counterpart, is defined as the probability of incorrectly identifying the state $\rho_0$ when it is prepared in $\rho_1$, or vice versa,
\begin{align}
    P_{err} = \pi_0 \Tr\qty[\rho_0\mathcal{M}_1] + \pi_1\Tr\qty[\rho_1\mathcal{M}_0]\label{eq:minerr_quantum},
\end{align}
where $\pi_0=\Tr\qty[\rho_0\mathcal{M}_0]$ and $\pi_1=\Tr\qty[\rho_1\mathcal{M}_1]$ are associated with the probabilities of correctly identifying $\rho_0$ and $\rho_1$ respectively, and $\Tr[X]$ denotes the matrix trace operation. For $\mathcal{M}_1 = \mathbb{I} - \mathcal{M}_0$, we introduce the difference operator defined as $\Gamma \equiv \pi_1\rho_1 - \pi_0\rho_0$ and rewrite Eq.~(\ref{eq:minerr_quantum}) as
\begin{align}
    P_{err} &=\pi_0 + \Tr[(\pi_1\rho_1 - \pi_0\rho_0)\mathcal{M}_0]\nonumber\\
    &=\pi_0 + \Tr\qty[\Gamma\mathcal{M}_0]\label{eq:perr_intermediate}.
\end{align}
From the equation above, the problem can be simplified to finding a set of POVMs $\mathcal{M}_0$ that minimises the error probability,
\begin{align}
    P_{err}^{min} &\coloneqq \min_{0\leq \mathcal{M}_0\leq \mathbb{I}} P_{err}\nonumber\\
    & = \pi_0 + \Tr[\Gamma\mathcal{M}_0].
\end{align}
In order to find the optimum POVM $\mathcal{M}_0$, we first evaluate the difference operator
\begin{align}
    &\Gamma \coloneqq \pi_1\rho_1-\pi_0\rho_0 = \sum_x \lambda_x\ket{x}\bra{x},\\
    &\Gamma_+\coloneqq\sum_{x:\lambda_x\geq 0}\lambda_x\ket{x}\bra{x},\\
    &\Gamma_-\coloneqq\sum_{x:\lambda_x< 0}\abs{\lambda_x}\ket{x}\bra{x},
\end{align}
where we have defined two non-negative terms $\Gamma_+$ and $\Gamma_-$ such that $\Gamma = \Gamma_+-\Gamma_-$, and $\Gamma_+\Gamma_- = 0$.  Introducing two operators $P_\pm$, which project onto the range of $\Gamma_\pm$ and satisfy $P_+P_- = 0$, we can evaluate Eq.~(\ref{eq:perr_intermediate}) as:
\begin{align}
    P_{err}^{min} &= \pi_0 + \Tr\qty[\Gamma\mathcal{M}_0]\nonumber\\
    &=\pi_0 + \Tr\qty[\Gamma_+\mathcal{M}_0] - \Tr\qty[\Gamma_-\mathcal{M}_0]\nonumber\\
    &=\pi_0 + \Tr\qty[\Gamma_+\mathcal{M}_0] - \Tr \qty[P_-\Gamma_-P_-\mathcal{M}_0]\nonumber\\
    &\leq \pi_0 + \Tr\qty[\Gamma_+\mathcal{M}_0] - \Tr\qty[\Gamma_-P_-],
\end{align}
where we have used the property that for $\hat{X}\geq\hat{Y},\hat{A}\geq 0\Rightarrow \Tr\hat{A}\hat{X}\geq\Tr\hat{A}\hat{Y}$.  For the minimisation of the error probability, we can choose the measurement $\mathcal{M}_0 = P_-$, which leads to $\Tr(\Gamma_+\mathcal{M}_0)=\Tr(\Gamma_+P_-)=0$. Hence $\{\mathcal{M}_0=P_-,\mathcal{M}_1=P_+\}$ are the optimum POVMs. The minimum error probability can finally be expressed independently of the measurement choice $\mathcal{M}$
\begin{align}
    P_{err}^{min} &= \pi_0 - \Tr\qty[\Gamma_-]\nonumber\\
    &=\frac{\qty(\pi_0 - \Tr\qty[\Gamma_-] + \pi_0 - \Tr\qty[\Gamma_-])}{2}\nonumber\\
    &=\frac{\qty(\pi_0 - \Tr\qty[\Gamma_-] + \pi_1 - \Tr\qty[\Gamma_+])}{2}\nonumber\\
    &=\frac{1}{2} - \frac{1}{2}\Tr\qty[\Gamma_+ + \Gamma_-]\nonumber\\
    &=\frac{1}{2} - \frac{1}{2}\Tr\abs{\Gamma}\nonumber\\
    &=\frac{1}{2} - \frac{1}{2}\norm{\pi_1\rho_1-\pi_0\rho_0}_1,\label{eq:minerror}
\end{align}
where $\norm{X}_1 = \Tr\sqrt{X^\dag X}$ is the trace norm of matrix $X$. In many settings, the trace norm is often a difficult quantity to calculate analytically. Therefore, in the following section, we will discuss bounds on the minimum error probability $P_{err}^{min}$. For simplicity, in this thesis, $\pi_0$ and $\pi_1$ will be assumed to be $1/2$, representing equal prior probability.
\subsection{Bounds of minimum error probability}
\subsubsection{Quantum fidelity}
Classically, fidelity is a metric for assessing the similarity of two probability distributions. Consider two random variables $X$ and $Y$ producing results $\{1,2,\cdots,N\}$ with probability $\mathbf{p} = \{p_n\}$ and $\mathbf{q} = \{q_n\}$  for $n\in[1,N]$ respectively. The fidelity between $X$ and $Y$ can be represented as 
\begin{align}
    \mathcal{F}_{sq}(X,Y) = \qty(\sum_{n=1}^N \sqrt{p_nq_n})^2,
\end{align}
where $\sum_{n=1}^N \sqrt{p_nq_n}$ is known as the Bhattacharyya coefficient \cite{bhattacharyya1946measure}. 
In quantum theory, the above equation serves as the basis for defining a criterion to differentiate between two quantum states. Consider performing a set of POVMs $\{\mathcal{M}_1,\cdots,\mathcal{M}_N\}$, satisfying the completeness relation $\sum_{n=1}^N\mathcal{M}_n=1$, on two distinct quantum states $\rho_0$ and $\rho_1$. The probability of observing the $n$-th outcome for both the quantum states is defined as $p_n = \Tr\qty[\mathcal{M}_n\rho_0]$ and $q_n = \Tr\qty[\mathcal{M}_n\rho_1]$ respectively. Consequently, the ability to differentiate between quantum states $\rho_0$ and $\rho_1$ is equivalent to distinguishing between their measurement outcome probability distributions $\mathbf{p}$ and $\mathbf{q}$. For all possible POVMs, we are interested in the set of POVMs that maximises the distinguishability of these two probability distributions. Therefore, the quantum fidelity between the two states can be formally written as
\begin{align}
    \mathcal{F}_{sq}(\rho_0,\rho_1) = \min_{\{\mathcal{M}_n\}}\qty(\sum_{n} \sqrt{\qty(\Tr\mathcal{M}_n\rho_0)\qty(\Tr\mathcal{M}_n\rho_1)})^2.
\end{align}
Ref.~\cite{watrous2018theory} demonstrated that the minimum is attained by the projective POVM corresponding to measurement in the eigenbasis of the operator $\rho_0^{-1/2}\abs{\sqrt{\rho_1}\sqrt{\rho_0}}\rho_0^{-1/2}$, where $\abs{X} = \sqrt{X^\dag X}$, allowing quantum fidelity to be evaluated as \cite{jozsa1994fidelity,raginsky2001fidelity,marian2012uhlmann}
\begin{align}
    \mathcal{F}_{sq}(\rho_0,\rho_1) &=\qty(\Tr\abs{\sqrt{\rho_1}\sqrt{\rho_0}})^2\nonumber\\
    &=\qty(\Tr\sqrt{(\sqrt{\rho_1}\sqrt{\rho_0})^\dag(\sqrt{\rho_1}\sqrt{\rho_0})})^2\nonumber\\
    &=\qty(\Tr\sqrt{\sqrt{\rho_0}\rho_1\sqrt{\rho_0}})^2.
\end{align}
In this thesis, we define fidelity as Uhlmann's square root fidelity. which is expressed as \cite{nielsen2010quantum}
\begin{align}
    \mathcal{F}(\rho_0,\rho_1)=\Tr\sqrt{\sqrt{\rho_0}\rho_1\sqrt{\rho_0}}.\label{eq:fidelity}
\end{align}
Some fundamental properties of Uhlmann's square root fidelity are
\begin{enumerate}[itemsep=1pt, topsep=1pt]
    \item \textbf{Bounded values: }$0\leq \mathcal{F}(\rho_0,\rho_1)\leq 1$.
    \item \textbf{Symmetry: }$\mathcal{F}(\rho_0,\rho_1)=\mathcal{F}(\rho_1,\rho_0)$.
    \item \textbf{Monotonicity under CPT maps: }For any completely positive trace preserving (CPT) map $\Phi$, $\mathcal{F}(\Phi(\rho_0),\Phi(\rho_1 ))\geq \mathcal{F}(\rho_0,\rho_1)$.
    \item \textbf{Multiplicativity: }If $\rho_0=\otimes_{m=1}^Mr_{0,m}$ and $\rho_1 = \otimes_{m=1}^Mr_{1,m}$, we have $\mathcal{F}(\rho_0,\rho_1)=\prod_{m=1}^M\mathcal{F}(r_{0,m},r_{1,m})$.
    \item \textbf{Stong concavity: }If $\rho_0$ and $\rho_1$ are distributions of density states such that $\rho_0=\sum_m p_{0,m}r_{0,m}$ and $\rho_1=\sum_m p_{1,m}r_{1,m}$, we have $F\qty(\rho_0,\rho_1)\geq \sum_m p_{0,m}p_{1,m}F\qty(r_{0,m},r_{1,m})$.
\end{enumerate}
The Fuchs-van de Graaf inequalities provide an upper and lower bound to trace distance (half of trace norm, $D(\rho_0,\rho_1)=\frac{1}{2}\norm{\rho_1 - \rho_0}_1$) as a function of fidelity \cite{fuchs1999cryptographic}
\begin{align}
    1 - \mathcal{F}(\rho_0,\rho_1)\leq D(\rho_0,\rho_1)\leq \sqrt{1 - \mathcal{F}(\rho_0,\rho_1)^2}.\label{eq:fidelityerrorbound}
\end{align}
Relating back to Eq.~(\ref{eq:minerror}), the bounds of minimum error probability can thus be derived to be
\begin{align}
    \frac{1}{2}-\frac{1}{2}\sqrt{1-\mathcal{F}(\rho_0,\rho_1)^2}\leq P_{err}^{min}\leq \frac{\mathcal{F}(\rho_0,\rho_1)}{2}\label{eq:QFB}.
\end{align}
\subsubsection{Quantum Chernoff bound}
Chernoff bound provides an upper bound to the minimsation of the average of both type-I and type-II error, as described in Eq.~(\ref{eq:minerr_classical}). For this thesis, we will consider the case where the priori probability is equal to $1/2$ for relevance. Classically, consider two random variables $X$ and $Y$, producing results $\{1,2,\cdots,N\}$ with probabilities $\mathbf{p} = \{p_n\}$ and $\mathbf{q} = \{q_n\}$ for $n\in[1,N]$ respectively. From a single outcome obtained from the outputs of these random variables, the task is to determine the origin of the outcome from either random variable. Intuitively, the optimal minimisation of error probability is achieved by selecting the random variable that maximises the likelihood of the observed outcome, i.e.,
\begin{align}
    P_{err}^{min} &= \frac{1}{2}\sum_{n=0}^N\min\{p_n,q_n\}\nonumber\\
    &\leq \frac{1}{2}\inf_{0\leq s\leq 1}\sum_{n=0}^Np_n^sq_n^{1-s}\label{eq:chernoff_classical},
\end{align}
where the inequality $\min\{p,q\}\leq p^sq^{1-s}$  is used \cite{audenaert2007discriminating,calsamiglia2008quantum}. In the quantum scenario, the objective is to distinguish between two quantum states $\rho_0$ and $\rho_1$ by performing POVM measurements ${\mathcal{M}_0,\mathcal{M}_1}$ on the state. The minimum error probability of incorrectly identifying the states has previously been derived to be
\begin{align}
    P_{err}^{min} = \frac{1}{2} - \frac{1}{2}\norm{\pi_1\rho_1 - \pi_0\rho_0}_1\label{eq:minerr_quantumrepeat}.
\end{align}
Considering the challenge in computing the trace norm of the state differences, we seek a bound for the minimum error probability. Motivated by the classical analogue from Eq.~(\ref{eq:chernoff_classical}), Ref.~\cite{audenaert2007discriminating} \textit{Theorem 1} provided a bound on the trace of two positive operators $A$ and $B$ as follows:
\begin{align}
    \Tr A^sB^{1-s}\geq \Tr[A + B - \abs{A-B}]/2\quad\text{for all }s\in[0,1].
\end{align}
Applying this bound to the definition of the minimum error probability in Eq.~(\ref{eq:minerr_quantumrepeat}), we obtain:
\begin{align}
    P_{err}^{min} &= \frac{1}{2} - \frac{1}{2}\Tr\abs{\pi_1\rho_1 - \pi_0\rho_0}\nonumber\\
    &\leq \frac{1}{2} - \frac{1}{2}\Tr[\pi_0\rho_0 + \pi_1\rho_1] + \Tr[\pi_0^s\rho^s\pi_1^{1-s}\rho_1^{1-s}]\nonumber\\
    &=\Tr[\pi_0^s\pi_1^{1-s}\rho_0^s\rho_1^{1-s}].
\end{align}
Analogous to the classical case, our interest lies in finding $s$ that minimises the minimum error probability. Therefore, the quantum Chernoff bound is formally defined as
\begin{align}
    P_{err}^{min}&\leq \inf_{0\leq s\leq 1}\pi_0^s\pi_1^{1-s}\Tr[\rho_0^s\rho_1^{1-s}]\nonumber\\
    &=\frac{1}{2}\inf_{0\leq s\leq 1}\Tr[\rho_0^s\rho_1^{1-s}]\nonumber\\
    &=\frac{1}{2}\inf_{0\leq s\leq 1} C_s(\rho_0,\rho_1)=\frac{1}{2}Q_s(\rho_0,\rho_1)\label{eq:chernofferrorbound},
\end{align}
where $\pi_0=\pi_1=1/2$ for equal priori probability, $C_s(\rho_0,\rho_1)=\Tr[\rho_0^s\rho_1^{1-s}]$ is the s-overlap between the two states $\rho_0$ and $\rho_1$, and $\inf_{0\leq s \leq 1} C_s$ is the infimum of s-overlap, obtained by minimising s-overlap for all $0\leq s\leq 1$. The Chernoff bound provides an exponentially tighter upper bound on the minimum error probability compared to the fidelity bound \cite{audenaert2007discriminating,calsamiglia2008quantum,kargin2005chernoff,nussbaum2009chernoff}. Similar to fidelity, the Chernoff bound exhibits several basic properties:
\begin{enumerate}[itemsep=1pt, topsep=1pt]
    \item \textbf{Bounded values: }$0\leq Q_s(\rho_0,\rho_1)\leq 1$.
    \item \textbf{Convexity in s: }$Q_s(\rho_0,\rho_1)$ is convex in $s\in[0,1]$ such that the minimisation results in only one local minimum.
    \item \textbf{Monotonicity under CPT maps: }For any CPT map $\Phi$, $Q_s(\Phi(\rho_0),\Phi(\rho_1))\geq Q_s(\rho_0,\rho_1)$.
    \item \textbf{Multiplicativity: }If $\rho_0=\otimes_{m=1}^Mr_{0,m}$ and $\rho_1 = \otimes_{m=1}^Mr_{1,m}$, then we have $Q_s(\rho_0,\rho_1)=\prod_{m=1}^M\inf_{0\leq s\leq 1}\Tr r_{0,m}^sr_{1,m}^{1-s}$
    \item \textbf{Relation to Fidelity: }If either of the state is pure, e.g., $\rho_0=\ket{\psi}\bra{\psi}$, the infimum of the s-overlap is achieved when $s=0$. $Q_s(\rho_0,\rho_1)$ therefore reduces to $\bra{\psi}\rho_1\ket{\psi}=\mathcal{F}(\rho_0,\rho_1)^2$.
\end{enumerate}
A weaker upper bound known as the quantum Bhattacharyya bound \cite{calsamiglia2008quantum,weedbrook2012gaussian} can obtained by equating $s=1/2$ such that
\begin{align}
    P_{err}^{min}\leq \frac{1}{2}\inf_{0\leq s\leq 1} C_s(\rho_0,\rho_1) \leq \frac{1}{2} C_{1/2}(\rho_0,\rho_1)\label{eq:Bhattabound}.
\end{align}
This bound is of interest in this thesis as it is convenient for deriving analytical solutions and sufficiently tight in certain regimes.

\subsection{Error exponent}
In quantum hypothesis testing, the error exponent quantifies the rate of exponential decay of the probabilities of Type I and Type II errors with an increasing number of copies of quantum states involved in the test, providing insights into the asymptotic behaviour of the error probabilities. The mathematical representation of the error exponent is defined as:
\begin{align}
    \chi = \lim_{M\rightarrow \infty}\frac{-\ln P_{err}^{min}}{M},
\end{align}
where $M$ is the number of copies of quantum states \cite{audenaert2008asymptotic}. By applying the bounds of minimum error probability defined in Eq.~(\ref{eq:fidelityerrorbound}) and Eq.~(\ref{eq:chernofferrorbound}), we can similarly bound the error exponent as follows
\begin{align}
    \chi &\approx\lim_{M\rightarrow \infty}\frac{-\ln Q_s(\rho_0,\rho_1)}{M}\geq \lim_{M\rightarrow \infty}\frac{-\ln Q_{1/2}(\rho_0,\rho_1)}{M},\label{eq:exponentlowerbound}\\
    \chi &\leq \lim_{M\rightarrow \infty}\frac{-\ln\qty(1 - \sqrt{1 - \mathcal{F}(\rho_0,\rho_1)^2})}{M}\label{eq:exponentupperbound},
\end{align}
where the Chernoff bound is exponentially tight.

\section{Estimation Theory}
Estimation theory is a framework focusing on the estimation of unknown latent parameters based on observed data (observables). The goal is to find an estimator that can provide the most accurate estimate of the true parameter values based on the observables. \\
Classically, we consider a set of measurement observables $\mathbf{x}= \{x_1,\cdots,x_M\}$ which are influenced by some parameters $\theta$. The probability density function of the distribution that generated the observables must be conditional on the values of the parameters $p(\mathbf{x}|\theta)$. The estimation problem aims to construct an estimator $\hat{\theta}(\mathbf{x})$, which can be interpreted as a function that takes in the observables and outputs an estimation of the parameter $\theta$ \cite{sage1971estimation,kay1993fundamentals}. Good estimators possess the following properties:
\begin{enumerate}[itemsep=1pt, topsep=1pt]
\item \textbf{Unbiasedness: }An estimator is unbiased if its expected value is equal to the true value of the parameter it is estimating $\expval{\hat{\theta}(\mathbf{x})}=\int p(\mathbf{x|\theta})\hat{\theta}(\mathbf{x})\;d\mathbf{x} = \theta$.
\item \textbf{Consistency: }An estimator is consistent if it converges to the true parameter value as the sample size increases: $\lim_{M\rightarrow\infty}\hat{\theta}(\mathbf{x}) = \theta$.
\item \textbf{Efficiency: }For all unbiased estimators, an efficient estimator is one whose variance $\text{Var}[\hat{\theta}]$ is the closest to the lower bound defined by Cram\'er-Rao bound.
\item \textbf{Sufficiency: }An estimator is sufficient if it uses all the information in the sample that is relevant to the estimation of the parameter.
\end{enumerate}
In quantum estimation theory, the aim is to estimate a quantum parameter $\theta$ that characterises a quantum state or evolution with parameter-dependent density matrices $\rho(\theta)$ from the measurement observables \cite{helstrom1969quantum}. Consider a set of POVM measurements $\{\mathcal{M}_1,\cdots,\mathcal{M}_M\}$, obeying the completeness relation $\sum_{m=1}^M\mathcal{M}_m=\mathbb{I}$. The set of measurements produces a corresponding set of observables $\mathbf{y} = \{y_1,\cdots,y_M\}$. The probability distribution of the measurement results follows $p(y_m|\theta) = \Tr \qty[\rho(\theta)\mathcal{M}_m]$. In contrast to the classical scenario, the key distinction in the quantum context lies in the fact that the results of measurements may not be independent due to the phenomenon of entanglement. Quantum estimators $\hat{\theta}(\mathbf{y})$ are constructed to produce an estimated value of the parameter of interest $\theta$ from the set of the measurement outcomes $\mathbf{y}$. Similar to classical estimators, good quantum estimators are unbiased and consistent, with their efficiency evaluated by their closeness to the lower bound defined by the quantum Cram\'er-Rao bound (detailed introduction of Cram\'er-Rao bound can be found in Section~\ref{chap:Cramer-Rao}).

\subsection{Fisher information}
In estimation theory, Fisher information provides a quantification of the amount of information that the measurement observables carry about an unknown parameter. Consider a set of classical observables $\mathbf{x}$ with a probability density function of its distribution conditional on the values of the unknown parameters $p(\mathbf{x}|\theta)$. Some of the relevant functions are defined as follows:
\begin{enumerate}[itemsep=1pt, topsep=1pt]
\item \textbf{Likelihood function: }The likelihood function $\mathcal{L}(\theta|\mathbf{x})$ is defined as the probability of observing the given data $\mathbf{x}$ under different values of $\theta$. It is the same as the probability distribution function but viewed as a function of $\theta$ for fixed $\mathbf{x}$, $\mathcal{L}(\theta|\mathbf{x})\mapsto p(\mathbf{x}|\theta)$.
\item \textbf{Log-likelihood: }The log-likelihood is the natural logarithm of the likelihood function $\ell(\theta)=\log\mathcal{L}(\theta|\mathbf{x}) = \log p(\mathbf{x}|\theta)$.
\item \textbf{Score function: }The score function calculates the gradient of the log-likelihood with respect to the parameter $U(\theta)=\frac{d}{d\theta}\ell(\theta)=\frac{d}{d\theta}\log p(\mathbf{x}|\theta)$.
\end{enumerate}
By applying the regularity condition, which imposes a restriction on the likelihood function such that the order of expectation operator and differentiation can be interchanged, the expected value (first moment) of the score function is shown to be zero following:
\begin{align}
    \mathbb{E}[U(\theta)] &= \int\frac{\partial_{\theta}p(\mathbf{x}|\theta)}{p(\mathbf{x}|\theta)}p(\mathbf{x}|\theta)\;d\mathbf{x}\nonumber\\
    &=\partial_{\theta}\int p(\mathbf{x}|\theta)\; d\mathbf{x}\nonumber\\
    &=\partial_{\theta} 1 = 0.
\end{align}
where $\partial_{\theta} = \frac{\partial}{\partial\theta}$. The Fisher information is defined as the variance (second moment) of the score:
\begin{align}
    \mathcal{J}(\theta) &\coloneqq \text{Var}[U(\theta)]\nonumber\\
    &=\mathbb{E}[U(\theta)^2] - \mathbb{E}[U(\theta)]^2\nonumber\\
    &=\int p(\mathbf{x|\theta})\qty[\partial_{\theta}\ln p(\mathbf{x}|\theta)]^2\;d\mathbf{x}\label{eq:fisherinformation_1}\\
    &=\int \frac{1}{p(\mathbf{x|\theta})}\qty[\partial_{\theta}p(\mathbf{x}|\theta)]^2\;d\mathbf{x}\label{eq:fisherinformation_2}\\
    &=-\int p(\mathbf{x|\theta})\partial^2_{\theta}\ln p(\mathbf{x}|\theta)\;d\mathbf{x}\label{eq:fisherinformation_3},
\end{align}
with the last three being the equivalent representations of Fisher information assuming that $p(\mathbf{x}|\theta)$ is twice differentiable \cite{kay1993fundamentals,ly2017tutorial}.\\
Analogous to its classical counterpart, quantum Fisher information quantifies the sensitivity of a quantum state to changes in a parameter that the state depends on. We consider a quantum state described by a density matrix $\rho(\theta)$ dependent on an unknown parameter $\theta$. A set of POVM measurements $\{\mathcal{M}_y\}_{\mathbf{y}}$ performed on $\rho(\theta)$ is indexed by the measurement results $y\in \mathbf{y}$. To derive the quantum Fisher information, we introduce a Hermitian operator called the Symmetric Logarithmic Derivative $\hat{L}_{\theta}$ associated with the parameter $\theta$. Mathematically, it satisfies the following equation
\begin{align}
    \partial_{\theta}\rho(\theta)= \frac{1}{2}\qty(\hat{L}_\theta\rho(\theta) + \rho(\theta)\hat{L}_{\theta})\label{eq:SLD},
\end{align}
which is a special form of the Lyapunov equation where $\hat{L}_{\theta}$ is a corresponding solution. Taking the derivative of $p(y|\theta)$ with respect to the parameter $\theta$ yields
\begin{align}
    \partial_{\theta} p(y|\theta) &= \partial_{\theta}\Tr\qty[\rho(\theta)\mathcal{M}_y] = \Tr \qty[\partial_{\theta}\rho(\theta)\mathcal{M}_y]\nonumber\\
    &= \Tr\qty[\qty(\frac{\hat{L}_\theta\rho(\theta) + \rho(\theta)\hat{L}_{\theta}}{2})\mathcal{M}_y]\nonumber\\
    &=\frac{1}{2}\Tr\qty[\rho(\theta)\mathcal{M}_y\hat{L}_{\theta}] + \frac{1}{2}\qty[\Tr\qty[\rho(\theta)\mathcal{M}_y\hat{L}_{\theta}]]^*
    \nonumber\\
    &=\text{Re}\qty(\Tr\qty[\rho(\theta)\mathcal{M}_y\hat{L}_{\theta}]),
\end{align}
where the cyclic property of the trace is invoked to arrive at the final expression. From Eq.~(\ref{eq:fisherinformation_2}), the Fisher information obtained from conducting quantum measurements $\{\mathcal{M}_y\}_{\mathbf{y}}$ establishes a classical bound on precision,
\begin{align}
    \mathcal{J}(\theta) = \int dy\; \frac{\text{Re}(\Tr\qty[\rho(\theta)\mathcal{M}_y\hat{L}_{\theta}])^2}{\Tr\qty[\rho(\theta)\mathcal{M}_y]},
\end{align}
which is achievable through proper processing. To evaluate the ultimate bounds of precision, the Fisher information is maximised over all sets of quantum measurements,
\begin{align}
     \mathcal{J}(\theta) &= \int dy\; \frac{\text{Re}(\Tr\qty[\rho(\theta)\mathcal{M}_y\hat{L}_{\theta}])^2}{\Tr\qty[\rho(\theta)\mathcal{M}_y]}\nonumber\\
     &\leq \int dy\; \frac{1}{\Tr\qty[\rho(\theta)\mathcal{M}_y]}\abs{\Tr\qty[\rho(\theta)\mathcal{M}_y\hat{L}_{\theta}]}^2\nonumber\\
     &= \int dy\;\frac{1}{\Tr\qty[\rho(\theta)\mathcal{M}_y]}\abs{\Tr\qty[\qty(\sqrt{\rho(\theta)}\sqrt{\mathcal{M}_y})\qty(\sqrt{\mathcal{M}_y}\hat{L}_{\theta}\sqrt{\rho(\theta)})]}^2\label{eq:fisherinformation_quantummeasure}.
\end{align}
Using the Cauchy-Schwartz inequality $\abs{\Tr \qty(A^\dag B)}^2\leq \Tr \qty(A^\dag A)\Tr \qty(B^\dag B)$, the second term of the integration can be upper bounded as
\begin{align}
    \abs{Tr\qty[\qty(\sqrt{\mathcal{M}_y}\sqrt{\rho(\theta)})^\dag\qty(\sqrt{\mathcal{M}_y}\hat{L}_{\theta}\sqrt{\rho(\theta)})]}^2 &\leq \Tr\qty[\qty(\sqrt{\mathcal{M}_y}\sqrt{\rho(\theta)})^\dag\qty(\sqrt{\mathcal{M}_y}\sqrt{\rho(\theta)})]\nonumber\\
    &\hspace{10pt}\times \Tr\qty[\qty(\sqrt{\mathcal{M}_y}\hat{L}_{\theta}\sqrt{\rho(\theta)})^\dag\qty(\sqrt{\mathcal{M}_y}\hat{L}_{\theta}\sqrt{\rho(\theta)})]\nonumber\\
    &=\Tr\qty[\rho(\theta)\mathcal{M}_y]\Tr\qty[\hat{L}_{\theta}\mathcal{M}_y\hat{L}_{\theta}\rho(\theta)],
\end{align}
which can be substituted back into Eq.~(\ref{eq:fisherinformation_quantummeasure}) to obtain
\begin{align}
    \mathcal{J}(\theta)&\leq \int dy\;\Tr\qty[\hat{L}_{\theta}\mathcal{M}_y\hat{L}_{\theta}\rho(\theta)]\nonumber\\
    &= \Tr\qty[\qty(\int dy\; \mathcal{M}_y)\hat{L}_{\theta}\rho(\theta)\hat{L}_{\theta}]\nonumber\\
    &=\Tr\qty[\rho(\theta)\hat{L}_{\theta}^2].
\end{align}
The series of inequalities presented above demonstrates that the Fisher information $\mathcal{J}(\theta)$  for any quantum measurement is constrained by the quantum Fisher information (QFI) $\mathcal{K}(\theta)$,
\begin{align}
    \mathcal{J}(\theta)\leq \mathcal{K}(\theta)\equiv \Tr\qty[\rho(\theta)\hat{L}_{\theta}^2]=\Tr\qty[\qty(\partial_{\theta}\rho(\theta)\hat{L}_{\theta})]\label{eq:QFI},
\end{align}
where the equivalent representation of the quantum Fisher information is derived using the definition of SLD \cite{paris2009quantum}. If we consider an estimated parameter $\theta'$ that is sufficiently close to the actual parameter such that the state generated from the estimated parameter is close to the actual state, i.e., $\rho(\theta') = \rho(\theta)+\delta\rho$, the quantum Fisher information can be related to the fidelity between the two states as follows \cite{braunstein1994statistical}
\begin{align}
    \mathcal{K}(\theta) = -4\left.\partial^2_{\theta'}\mathcal{F}(\rho(\theta),\rho(\theta'))\right|_{\theta'=\theta}\label{eq:Fisher_fidelity}.
\end{align}

\subsection{Cram\'er-Rao bound}\label{chap:Cramer-Rao}
Cram\'er-Rao bound is a fundamental result in estimation theory, providing a lower bound on the variance of estimators of an unknown parameter \cite{devore1995probability,kay1993fundamentals}. Specifically, an unbiased estimator capable of saturating this bound is classified as fully efficient. For a set of classical biased estimators $\hat{\theta}(\mathbf{x})$ estimating the parameter $\theta$ in a given model with likelihood function $\mathcal{L}(\theta|\mathbf{x})$, the bias of the estimators is defined as the expected value of the difference between the estimators and the actual parameter
\begin{align}
    b\equiv \mathbb{E}[\hat{\theta}(\mathbf{x})-\theta]\equiv\int d\mathbf{x}\;\mathcal{L}(\theta|\mathbf{x})(\hat{\theta}(\mathbf{x})-\theta).
\end{align}
Taking the first derivative of the bias with respect to the parameter gives
\begin{align}
    \partial_{\theta}b &=\int d\mathbf{x}\;(\hat{\theta}(\mathbf{x})-\theta)\partial_{\theta}\mathcal{L}(\theta|\mathbf{x}) - \int d\mathbf{x}\;\mathcal{L}(\theta|\mathbf{x})\nonumber\\
    &=\int d\mathbf{x}\;(\hat{\theta}(\mathbf{x})-\theta)\mathcal{L}(\theta|\mathbf{x})\partial_{\theta}\ln\mathcal{L}(\theta|\mathbf{x}) - 1\\
    \Rightarrow 1 + \partial_{\theta}b &= \int d\mathbf{x}\;(\hat{\theta}(\mathbf{x})-\theta)\mathcal{L}(\theta|\mathbf{x})\partial_{\theta}\ln\mathcal{L}(\theta|\mathbf{x}).
\end{align}
Squaring both sides and using the Cauchy-Schwarz inequality $\abs{\int dx\; \mathcal{F}(x) g(x)}^2\leq\int dx\;\abs{\mathcal{F}(x)}^2\int dx\;\abs{g(x)}^2$, we arrive at the following inequality
\begin{align}
    (1+\partial_{\theta}b)^2 &= \qty[\int d\mathbf{x}\;(\hat{\theta}(\mathbf{x})-\theta)\mathcal{L}(\theta|\mathbf{x})\partial_{\theta}\ln\mathcal{L}(\theta|\mathbf{x})]^2\nonumber\\
    &\leq \qty[\int d\mathbf{x}\;(\hat{\theta}(\mathbf{x})-\theta)^2\mathcal{L}(\theta|\mathbf{x})]\qty[\int d\mathbf{x}\;\mathcal{L}(\theta|\mathbf{x})[\partial_{\theta}\ln\mathcal{L}(\theta|\mathbf{x})]^2]\nonumber\\
    &=\qty[\int d\mathbf{x}\;(\hat{\theta}(\mathbf{x})-\theta)^2p(\mathbf{x}|\theta)]\qty[\int d\mathbf{x}\;p(\mathbf{x}|\theta)[\partial_{\theta}\ln p(\mathbf{x}|\theta)]^2]\nonumber\\
    &=\text{Var}[\hat{\theta}(\mathbf{x}) - \theta]\mathcal{J}(\theta),
\end{align}
where we have used the Fisher information definition from Eq.~(\ref{eq:fisherinformation_1}). Hence, for unbiased estimators with bias value $b=0$, the lower bound of the variance between the estimators and the unknown parameter is
\begin{align}
    \text{Var}[\hat{\theta}(\mathbf{x})-\theta] &\geq \frac{1}{\mathcal{J}(\theta)},
\end{align}
which is the general expression for the Cram\'er-Rao bound. For an unbiased estimator, this variance is also the mean squared error (MSE).\\
Similarly, in the quantum realm, the derivation of the quantum Cramér-Rao bound follows closely to its classical analogue, with the exception that measurements and states are quantum  \cite{pirandola2018advances,polino2020photonic,holevo2011probabilistic}. For a quantum state $\rho=\rho(\theta)$, with $\theta$ being the parameter of interest, a set of POVM measurements $\{\mathcal{M}_y\}_{\mathbf{y}}$ produces the observables $y\in\mathbf{y}$ indexing its corresponding measurement. From these measurements, we deduce an estimator $\hat{\theta}(\mathbf{y})$ that estimates a value of $\theta$. Assuming that the estimators are biased, its bias is similarly quantified as
\begin{align}
    b\equiv \mathbb{E}[\hat{\theta}(\mathbf{y})-\theta] &\equiv\Tr\qty[\mathcal{Q}\rho] - \Tr\qty[\theta\rho]\nonumber\\
    &=\Tr[\rho(\mathcal{Q} - \theta)],
\end{align}
where $\mathcal{Q}$ is a Hermitian operator that produces an observable that approximates the value of the actual parameter when applied to $\rho$. The dimensionality of the parameter $\theta = \theta\mathbb{I}_{n\times n}$ depends on the context. Following the steps of derivation for classical Cram\'er-Rao bound, we take the first derivative of the bias with respect to the parameter $\theta$,
\begin{align}
    \partial_{\theta}b &=\partial_{\theta}\Tr[\rho(\mathcal{Q} - \theta)]\nonumber\\
    &=\Tr\qty[\partial_{\theta}(\rho(\mathcal{Q} - \theta))]\nonumber\\
    &=\Tr[\partial_{\theta}(\rho\mathcal{Q})] -1\label{eq:partialb_QCRB}.
\end{align}
The trace of the density state equals to one implies that $\Tr\rho=1\Rightarrow\Tr\partial_{\theta}\rho=0\Rightarrow\Tr[\partial_{\theta}(\rho\theta)]=0$, hence we can manipulate Eq.~(\ref{eq:partialb_QCRB}) such that
\begin{align}
    1 + \partial_{\theta}b &= \Tr[\partial_{\theta}(\rho\mathcal{Q})] - \Tr[\partial_{\theta}(\rho\theta)]\nonumber\\
    &=\Tr[\partial_{\theta}\rho(\mathcal{Q}-\theta)]\nonumber\\
    &=\Tr[\qty(\frac{1}{2}(\rho\hat{L}_{\theta} + \hat{L}_{\theta}\rho))(\mathcal{Q}-\theta)]\nonumber\\
    &=\frac{1}{2}\Tr\qty[\rho\hat{L}_{\theta}(\mathcal{Q}-\theta)] + \frac{1}{2}\Tr\qty[\rho\hat{L}_{\theta}(\mathcal{Q}-\theta)]^\dag\nonumber\\
    &=\text{Re}\qty{\Tr\qty[\hat{L}_{\theta}\rho(\mathcal{Q}-\theta)]}
\end{align}
where we have used the special form of the Lyapunov equation for SLD from Eq.~(\ref{eq:SLD}). Squaring both sides and using Cauchy-Schwarz inequality for trace as mentioned previously, the inequality can be derived as
\begin{align}
    (1+\partial_{\theta}b)^2 &=\qty(\text{Re}\qty{\Tr\qty[\hat{L}_{\theta}\rho^{1/2}\rho^{1/2}(\mathcal{Q}-\theta)]})^2\nonumber\\
    &\leq \abs{\Tr\qty[\hat{L}_{\theta}\rho^{1/2}\rho^{1/2}(\mathcal{Q}-\theta)]}^2\nonumber\\
    &\leq \Tr\qty[\hat{L}_{\theta}\rho^{1/2}\rho^{1/2}\hat{L}_{\theta}]\Tr\qty[(\mathcal{Q-\theta})\rho^{1/2}\rho^{1/2}(\mathcal{Q}-\theta)]\nonumber\\
    &=\mathcal{K}(\theta)\Tr\qty[\rho(\mathcal{Q}-\theta)^2],
\end{align}
where we have substituted the definition of quantum Fisher information from Eq.~(\ref{eq:QFI}). Evaluating the trace term,
\begin{align}
    \Tr[\rho(\mathcal{Q}-\theta)^2] &=\Tr\qty[\rho\mathcal{Q}^2] - 2\theta\Tr\qty[\rho\mathcal{Q}] + \theta^2\nonumber\\
    &=\mathbb{E}(\hat{\theta}(\mathbf{y})^2) - 2\theta\mathbb{E}(\hat{\theta}(\mathbf{y})) + \theta^2\nonumber\\
    &=\mathbb{E}(\hat{\theta}(\mathbf{y})^2) - 2\mathbb{E}(\hat{\theta}(\mathbf{y}))\mathbb{E}(\theta) + \mathbb{E}(\theta^2)\nonumber\\
    &=\mathbb{E}[(\hat{\theta}(\mathbf{y})-\theta)^2]\\
    &=\text{Var}[(\hat{\theta}(\mathbf{y})-\theta)]
\end{align}
and assuming that the estimator is unbiased such that $b=0$, we recover the general expression of the quantum Cram\'er-Rao bound \cite{paris2009quantum,holevo2011probabilistic}
\begin{align}
    \text{Var}[(\hat{\theta}(\mathbf{y})-\theta)]\geq \frac{1}{\mathcal{K}(\theta)}\label{eq:QCRB}.
\end{align}

\section{Continuous variable quantum systems}
Continuous variable quantum systems rely on infinite-dimensional Hilbert spaces as the foundational mathematical framework \cite{weedbrook2012gaussian,braunstein2005quantum,adesso2014continuous}. However, handling such infinite-dimensional systems presents challenges, particularly in numerical computations where dimensional truncation becomes necessary. To address this issue, the Wigner-Weyl transformation offers a valuable tool by facilitating a bijective mapping of Hilbert space operators from Schrödinger's picture to functions in quantum phase space formulation. Within this context, Gaussian systems represent a noteworthy class of states. These states can be fully characterised by their respective second moments, i.e., $(\hat{x}\hat{p}+\hat{p}\hat{x})/2$, where $\hat{x}$ and $\hat{p}$ correspond to the position and momentum operators. This concise parameterisation of Gaussian systems allows for efficient and compact descriptions of the associated quantum states. 
\subsection{Gaussian states}
For a general $N$-mode Gaussian state, it is characterized by Wigner functions that take the form of normalized Gaussian distribution. The Wigner functions can be described using N pairs of position and momentum, each pair defining the parameters for the corresponding mode 
\begin{align}
    W(\xi) &= \frac{1}{(2\pi)^N\sqrt{\det V}}\exp\qty(-\frac{1}{2}(\mathbf{q}-\bar{q})\qty[V]^{-1}(\mathbf{q}-\bar{q})^T),
\end{align}
where $q$ is the $2N$-dimensional vector having the quadrature operators, $q = (\hat{x}_1,\cdots,\hat{x}_N,\hat{p}_1\cdots,\hat{p}_N)$, $\mathbf{q}\in\mathbb{R}^{2N}$ are the quadrature eigenvalues, and $\bar{q} \coloneq \expval{q}$ is defined as the displacement vector. $V$ is the covariance matrix given by
\begin{align}
    V_{ij} &= \int W(\xi)q_iq_j d^{2N}q\nonumber\\
    &=\expval{\frac{\Delta q_i\Delta q_j + \Delta q_j\Delta q_i}{2}},
\end{align}
where $\Delta q = q-\mathbf{q}$ \cite{weedbrook2012gaussian,braunstein2005quantum,adesso2014continuous,case2008wigner,weyl1950theory}. In this thesis, the quadrature operators will be defined as follows
\begin{align}
    \hat{x} = \frac{\hat{a} + \hat{a}^\dag}{\sqrt{2}},\quad \hat{p} = \frac{\hat{a}-\hat{a}^\dag}{i\sqrt{2}},
\end{align}
where $\hat{a}$ and $\hat{a}^\dag$ are the annihilation and creation operator respectively. The quadrature operators satisfy the following canonical commutation relations
\begin{align}
    [q,q^T] = i\Omega,\quad \Omega = \begin{bmatrix}[0.8]
        0 & 1\\-1 & 0
    \end{bmatrix}\otimes \mathbb{I}_N.
\end{align}
From the Wigner function, we can deduce that the mean vector $\bar{q} = \expval{q}$ (first moment) and the covariance matrix $V$ (second moment) thus provide a full characterisation of a Gaussian state. According to Williamson's theorem, any positive definite matrix with an even dimension can be diagonalised using a symplectic transformation \cite{weedbrook2012gaussian,adesso2014continuous,braunstein2005quantum}. Consequently, for a general $N$-mode Gaussian state, which yields a covariance matrix of size $2N$ by $2N$, it is possible to find a symplectic matrix $S$ such that 
\begin{align}
    V^{(N)}&=S (D\oplus D) S^T,\quad D = \text{diag}(\upsilon_1,\cdots,\upsilon_N),
\end{align}
where $\{\upsilon_1,\cdots,\upsilon_N\}$ are the symplectic eigenvalues of the covariance matrix $V^{(N)}$. Here, the symplectic matrix is a $2N\times 2N$ matrix with real entries and satisfies the condition $S^T\Omega S=\Omega$, ensuring the preservation of symplectic structure. The spectrum of symplectic eigenvalues $\{\upsilon_k\}_{k=1}^N$ are derived from the eigenvalue spectrum of the modified version of the covariance matrix $\abs{i\Omega V^{(N)}}$, where the absolute value notation can be comprehended in operatorial sense \cite{weedbrook2012gaussian}.
\subsection{Distance measures of states}
Given two Gaussian states represented by density matrices $\rho_0$ and $\rho_1$, it is possible to analytically compute the distance measures of states, namely fidelity and s-overlap, based on their corresponding pairs of mean vectors and covariance matrices $(\bar{q}_0,V_0)$ and $(\bar{q}_1,V_1)$.
\subsubsection{Fidelity}\label{section:fidelity_gaussian}
The analytical expression for the fidelity between the two states, defined in Equation (\ref{eq:fidelity}), can be derived specifically for the Gaussian case as \cite{banchi2015quantum}
\begin{align}
    \mathcal{F}(\rho_0,\rho_1) = \mathcal{F}_0(V_0,V_1)\exp\qty[-\frac{1}{4}\delta_u^T(V_0+V_1)^{-1}\delta_u]\label{eq:fidelity_gaussian},
\end{align}
where $\delta_u = \bar{q}_1-\bar{q}_0$ is the difference of mean vector. $\mathcal{F}_0(V_0,V_1)$ is only dependent on the covariance matrices $V_0$ and $V_1$ computed as
\begin{align}
    \mathcal{F}_0(V_0,V_1)&= \frac{F_{tot}}{\sqrt[4]{\det\qty[V_0+V_1]}},\nonumber\\
    F_{tot}&= \sqrt[4]{\det\qty[2\qty(\sqrt{\mathbb{I}+\frac{(V_{aux}\Omega)^{-2}}{4}}+\mathbb{I})V_{aux}]},\label{eq:Ftot}
\end{align}
where $V_{aux}$ is an auxiliary matrix defined to be
\begin{align}
    V_{aux}=\Omega^T(V_0+V_1)^{-1}\qty(\frac{\Omega}{4}+V_1\Omega V_0).
\end{align}
\subsubsection{s-overlap}\label{section:soverlap}
The s-overlap between two Gaussian states can be expressed as follows \cite{pirandola2008computable}
\begin{align}
    C_s = 2^N\frac{\Pi_s}{\sqrt{\det \Sigma_s}}\exp\qty(-\frac{\delta_u^T\Sigma_s^{-1}\delta_u}{2}),\label{eq:s-overlap}
\end{align}
where $\delta_u = \bar{q}_1-\bar{q}_0$ similarly represents the difference of mean vector, $\Pi_s$ and $\Sigma_s$ depends on the covariance matrices $V_0$ and $V_1$.  Introducing two real functions
\begin{align}
    G_s(x) &=\frac{1}{(x+1/2)^s - (x-1/2)^s},\\
    \Lambda_s(x) &=\frac{(x+1/2)^s+(x-1/2)^s}{(x+1/2)^s-(x-1/2)^s},
\end{align}
we can hence define 
\begin{align}
    &\Pi_s \coloneqq \prod _{k=1}^N G_s(\upsilon_k^0) G_{1-s}(\upsilon_k^1),\\
    &\Sigma_s \coloneqq \Tilde{S}_0\qty[\oplus_{k=1}^N\Lambda_s(\upsilon_k^0)\mathbb{I}]\Tilde{S}_0^T + \Tilde{S}_1\qty[\oplus_{k=1}^N\Lambda_{1-s}(\upsilon_k^1)\mathbb{I}]\Tilde{S}_1^T,
\end{align}
where $\{\upsilon_k^0\}$ and $\{\upsilon_k^0\}$ are the symplectic eigenvalues of the matrices $\rho_0$ and $\rho_1$ respectively, with $\tilde{S}_0$ and $\tilde{S}_1$ being the respective symplectic matrices.
\\~\\The following gives a brief description of the methodology for deriving the symplectic eigenvalues and the symplectic matrices for a special class of a two-mode Gaussian state \cite{weedbrook2012gaussian}. We first re-arrange the quadrature operators vector such that quadrature operators of the same mode are grouped together, $q_r = (\hat{x}_1,\hat{p}_1,\cdots,\hat{x}_N,\hat{p}_N)$, the $N$-mode covariance matrix $\Tilde{V}$ follows the symplectic transformation
\begin{align}
    \tilde{V} = \tilde{S}V^\oplus \tilde{S}^T,\quad V^\oplus \coloneqq \bigoplus_{k=1}^N
\upsilon_k\mathbb{I}, \label{eq:symplectic_transformation}
\end{align}
where $V^\oplus$ is a diagonal matrix called the Williamson form of $\tilde{V}$, and $\tilde{S}$ is the symplectic matrix corresponding to $\tilde{V}$. For a two-mode Gaussian state, its covariance matrix can be expressed in a block matrix form
\begin{align}
    \tilde{V}^{(2)} = \begin{bmatrix}[0.8]
        \mathbf{A} & \mathbf{C}\\
        \mathbf{C}^T & \mathbf{B} 
    \end{bmatrix},
\end{align}
where $A=A^T$, $B=B^T$ are $2\times 2$ real matrices describing the individual mode of the Gaussian state, and $C$ is a $2\times 2$ real matrix describing the inter-modal correlations. The distinct category of two-mode Gaussian states, which will be highly discussed in this thesis, possesses a covariance matrix in the standard form
\begin{align}
    \tilde{V}^{(2)} = \begin{bmatrix}[0.8]
        a\mathbb{I} & \mathbf{C}\\
        \mathbf{C} & b\mathbb{I}
    \end{bmatrix},\quad \mathbf{C}=\begin{bmatrix}[0.8]
        c_1 & 0\\ 0 & c_2
    \end{bmatrix},
\end{align}
where $a,b,c_1,c_2\in\mathbb{R}$ are subject to the bonafide condition, which is equivalent to the uncertainty principle
\begin{align}
    \tilde{V}^{(2)}>0,\det \tilde{V}^{(2)}\geq 1,\text{ and }a^2+b^2+2c_1c_2\leq 1 + \det \tilde{V}^{(2)}.
\end{align}
For the case where $c_1=-c_2\coloneqq c\geq 0$, the symplectic eigenvalues can be simply evaluated as
\begin{align}
    \upsilon_\pm = \frac{\sqrt{(a+b)^2-4c^2}\pm (b-a)}{2},
\end{align}
with the corresponding symplectic matrix $\tilde{S}$ given by
\begin{align}
    \Tilde{S} = \begin{bmatrix}[0.8]
        \omega_+\mathbb{I} & \omega_-\mathbf{Z}\\
        \omega_-\mathbf{Z} & \omega_+\mathbb{I}
\end{bmatrix},\quad\omega_\pm \coloneqq \sqrt{\frac{a + b \pm \sqrt{y}}{2\sqrt{y}}},
\end{align}
such that the symplectic transformation in Equation (\ref{eq:symplectic_transformation}) is realised.

\section{Quantum channels}
\subsection{Pure loss}\label{sec:Pure-Loss}
Pure loss channel involves a beam splitter applying a two-mode mixing interaction between the signal (S) mode and an environment (E) mode initialised as a vacuum state $\ket{0}$. The signal mode has the annihilation operator represented by $\hat{a}_{in}$, while the environment has the annihilation operator $\hat{e}_{in}$. In Schrodinger picture, this transformation is defined by the unitary operator $\hat{U}(\phi) = \exp[i\phi(\hat{a}^\dag\hat{e} - \hat{a}\hat{e}^\dag)]$, where $\phi\in[0,\pi/2]$ determines the transmittance of the beam splitter given by $\tilde{\eta} = \cos^2\phi\in[0,1]$ \cite{gerry2023introductory,nair2018quantum}. In the Heisenberg picture, the output annihilation operators of the pure loss channel are obtained by performing the Bogoliubov transformation such that
\begin{align}
    \hat{a}_{out} = \sqrt{\Tilde{\eta}}\hat{a}_{in} + \sqrt{1-\Tilde{\eta}}\hat{e}_{in},\\
    \hat{e}_{out} = \sqrt{1-\Tilde{\eta}}\hat{a}_{in} - \sqrt{\Tilde{\eta}}\hat{e}_{in},
\end{align}
where the negative sign is adopted to ensure that the commutation relations of the above annihilation operators $\qty[\hat{a}_i,\hat{a}_j^\dag]=\delta_{ij}$ and $\qty[\hat{a}_i,\hat{a}_i]=\qty[\hat{a}_i^\dag,\hat{a}_i^\dag]=0$ are satisfied. The average output energy from the beam splitter is
\begin{align}
    \expval{\hat{a}_{out}^\dag\hat{a}_{out}} &= \expval{\qty(\sqrt{\Tilde{\eta}}\hat{a}_{in}^\dag + \sqrt{1-\Tilde{\eta}}\hat{e}_{in}^\dag)\qty(\sqrt{\Tilde{\eta}}\hat{a}_{in} + \sqrt{1-\Tilde{\eta}}\hat{e}_{in})}\nonumber\\
    &=\sqrt{\Tilde{\eta}}\expval{\hat{a}_{in}^\dag\hat{a}_{in}} + (1-\Tilde{\eta})\expval{\hat{e}_{in}^\dag\hat{e}_{in}}\nonumber\\
    &=\sqrt{\Tilde{\eta}}\expval{\hat{a}_{in}^\dag\hat{a}_{in}}.
\end{align}
The state transformation (quantum channel) on the signal mode corresponding to the pure loss channel of transmittance $\tilde{\eta}$ is denoted as $\mathcal{L}_{\tilde{\eta}}$.
\subsection{Quantum-limited amplifier}
Quantum-limited amplifier (QLA) involves an optical parametric amplifier (paramp) applying a two-mode squeezing interaction between the amplified or signal (S) mode and an environment (E) mode initialised as a vacuum state $\ket{0}$. The signal mode has the annihilation operator represented by $\hat{a}_{in}$, while the environment has the annihilation operator $\hat{e}_{in}$. In the interaction picture, the paramp operation is represented by the Hamiltonian $\hat{H}_I = i\hbar\kappa\qty(\hat{a}\hat{e}-\hat{a}^\dag\hat{e}^\dag)$, where $\kappa$ is an effective coupling strength \cite{caves1982quantum}. Evolution for a time $t$ results in the Bogoliubov transformation such that the output of the paramp can be expressed as
\begin{align}
    \hat{a}_{out}&=\sqrt{G}\hat{a}_{in} + \sqrt{G-1}\hat{e}_{in}^\dag,\label{eq:annihilation_QLA}\\
    \hat{e}_{out}&=\sqrt{G}\hat{e}_{in} - \sqrt{G-1}\hat{a}_{in}^\dag,
\end{align}
where $\hat{a}_{in}=\hat{a}(0)$, $\hat{e}_{in}=\hat{e}(0)$ are the input annihilation operators at time zero, and $\hat{a}_{out}=\hat{a}(t)$,$\hat{e}_{out}=\hat{e}(t)$ are the output annihilation operators at time $t$, with the paramp gain $G=\cosh^2\kappa t\equiv \cosh^2\tau\geq 1$ \cite{caves1982quantum}. The average output energy from the paramp is 
\begin{align}
    \expval{\hat{a}_{out}^\dag\hat{a}_{out}} &=\expval{\qty(\sqrt{G}\hat{a}^\dag_{in} - \sqrt{G-1}\hat{e}_{in})\qty(\sqrt{G}\hat{a}_{in} - \sqrt{G-1}\hat{e}_{in}^\dag)}\nonumber\\
    &=G\expval{\hat{a}_{in}^\dag\hat{a}_{in}} + (G-1)\expval{\hat{e}_{in}\hat{e}_{in}^\dag}\nonumber\\
    &=G\expval{\hat{a}_{in}^\dag\hat{a}_{in}} + (G-1),
\end{align}
where the last term represents the added noise of a QLA with gain $G\geq 1$. The state transformation (quantum channel) on the signal mode corresponding to the quantum-limited amplifier of gain $G$ is denoted as $\mathcal{A}_G$.
\subsection{Thermal loss}
A thermal loss channel is a model describes the interaction of a quantum system with a thermal environment. This channel captures the effects of loss and the introduction of thermal noise due to the system's interaction with a background thermal state. Mathematically, it can be represented using a beam splitter model, where the quantum state of interest is partially transmitted while the rest is mixed with the thermal noise. A thermal loss (or noisy attenuator) channel can be expressed as a concatenation of pure loss and quantum-limited amplifier channels \cite{nair2020fundamental}. This can be shown by evaluating the characteristic function of the input and output states of the channels respectively. For any density state $\rho$, it can be fully characterised by its anti-normally ordered characteristic function given as
\begin{align}
    \chi(\xi)=\expval{e^{-\xi^*\hat{a}+\xi\hat{a}^\dag}},
\end{align}
for a complex number $\xi$. The output state annihilation operator for the thermal loss channel $\hat{a}_{TL}^{out}$ with transmittance $0\leq \eta\leq 1$ and excess noise $N_B\geq 0$ can be expressed as \cite{serafini2017quantum}
\begin{align}
    &\hat{a}^{out}_{TL} = \sqrt{\eta}\hat{a}_{in} + \sqrt{1-\eta}\hat{a}_{th},\label{eq:receiver_a}
\end{align}
where $a_{in}$ is the annihilation operator of the input state, $\hat{a}_{th}$ is the annihilation operator of the thermal state $\rho_{\text{th}}(N_B)=\sum_nN_B^n/(N_B+1)^{n+1}\ket{n}\bra{n}$. Hence, the characteristic function of the output state from the thermal loss channel can be expressed as a function of its input state characteristic function
\begin{align}
    \chi_{TL}^{out}(\xi)&=\expval{e^{-\xi^*\hat{a}_{TL}^{out}+\xi\hat{a}_{TL}^{out\dag}}}\nonumber\\
    &=\expval{e^{-\xi^*\sqrt{\eta}\hat{a}_{in}+\xi\sqrt{\eta}\hat{a}_{in}^\dag}}\expval{e^{-\xi^*\sqrt{1-\eta}\hat{a}_{th}+\xi\sqrt{1-\eta}\hat{a}_{th}^\dag}}\nonumber\\\
    &=\chi^{in}(\sqrt{\eta}\xi)e^{-(1-\eta)\abs{\xi}^2(2N_B+1)/2}.\label{eq:characteristicfunction}
\end{align}
The details of evaluating the characteristic function for the thermal state can be found in Appendix~\ref{appendixA}. Similarly, the output state annihilation operator for the pure loss channel ($\hat{a}^{out}_{L}$) of transmittance $0\leq\Tilde{\eta}\leq 1$ and quantum-limited amplifier ($\hat{a}^{out}_{A}$) of gain $G\geq 1$ is respectively
\begin{align}
    \hat{a}^{out}_{L} &= \sqrt{\tilde{\eta}}\hat{a}_{in} + \sqrt{1-\Tilde{\eta}}\hat{e}_{in}\label{eq:pureloss_transform},\\
    \hat{a}^{out}_{A} &=\sqrt{G}\hat{a}_{in} + \sqrt{G - 1}\hat{e}_{in}^\dag\label{eq:quantumlimited_transform},
\end{align}
where $\hat{e}_{in}$ is the annihilation operator for vacuum state.
\begin{figure}[h!]
    \centering
    \includegraphics[width = 0.8\linewidth]{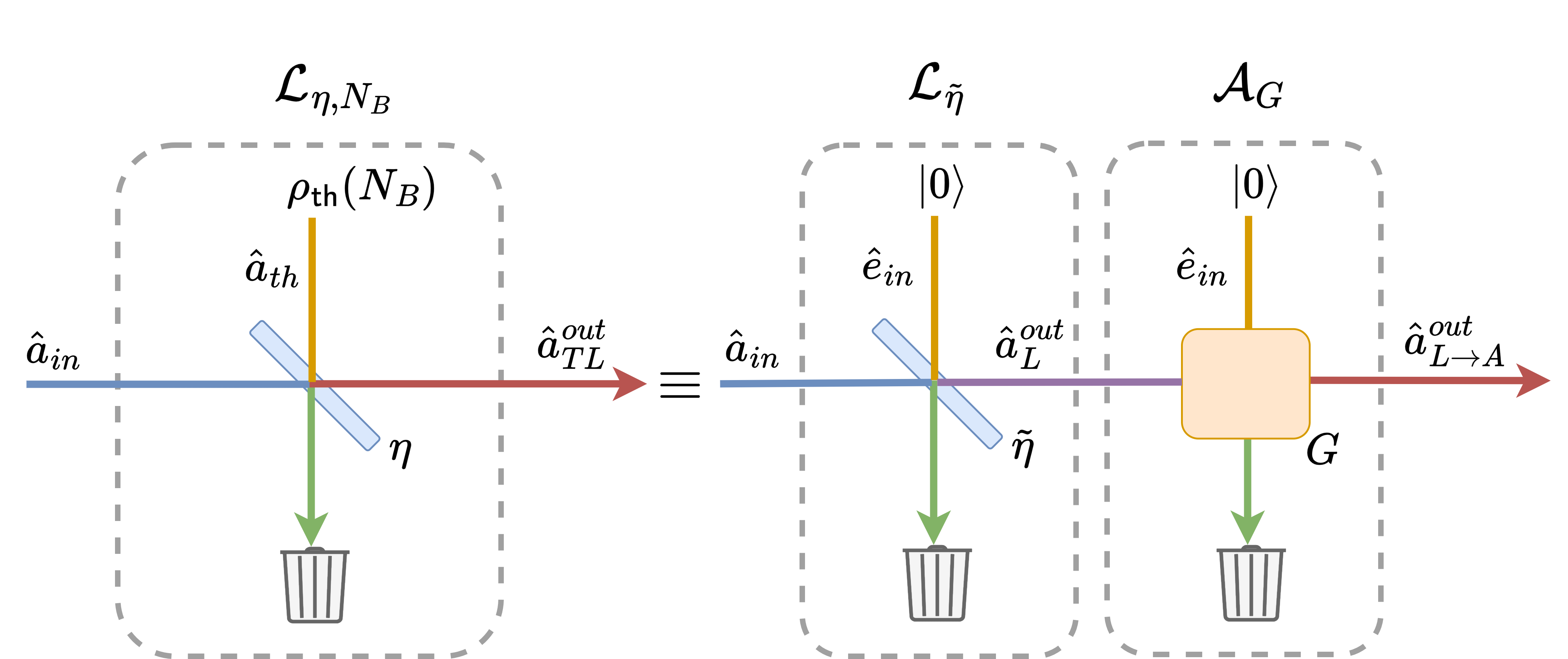}
    \captionsetup{width=\linewidth}
    \caption{Decomposition of a thermal loss channel into a cascade of a pure loss followed by quantum-limited amplifier channels. Left: A thermal loss channel $\mathcal{L}_{\eta,N_B}$ mixes the input state (with annihilation operator $\hat{a}_{in}$) and thermal state of mean brightness $N_B$ at a beam splitter of transmittance $\eta$. Right: Thermal loss channel realised as a pure loss channel $\mathcal{L}_{\tilde{\eta}}$ with transmittance $\tilde{\eta}=\eta/G$ followed by a quantum-limited amplifier channel $\mathcal{A}_G$ of gain $G = (1-\eta)N_B + 1$}. 
    \label{fig:concatenation_setup}
\end{figure}
We now consider a two-stage process where the input state is first passed through the pure loss channel and its output is subsequently fed into a quantum-limited amplifier. The output from the pure loss channel ($\hat{a}^{out}_{L}$) can be related to the output of the entire process ($\hat{a}^{out}_{L\rightarrow A}$)as
\begin{align}
    \chi^{out}_{L\rightarrow A}(\xi) &= \expval{e^{-\xi^*\hat{a}_A^{out} + \xi\hat{a}_A^{out\dag}}}\nonumber\\
    &=\expval{e^{-\xi^*\sqrt{G}\hat{a}_L^{out} + \xi\sqrt{G}\hat{a}_L^{out\dag}}}\expval{e^{-\xi^*\sqrt{G-1}\hat{e}_{in}^\dag + \xi\sqrt{G-1}\hat{e}_{in}}}\nonumber\\
    &=\chi_L^{out}(\sqrt{G}\xi)\bra{\emptyset}\hat{D}_{\sqrt{G-1}\xi}\ket{\emptyset}\nonumber\\
    &=\chi_L^{out}(\sqrt{G}\xi)e^{-\abs{\xi}^2(G-1)/2}\label{eq:twoprocess_gain},
\end{align}
where $\ket{\emptyset}$ represents vacuum state, $D_{\sqrt{G-1}\xi}$ is the displacement operator with $\sqrt{G-1}\xi$ amplitude of displacement. From the above equation, we have determined the characteristic function of the output state from the pure loss channel. Hence, we can relate this characteristic function to that of the input state
\begin{align}
    \chi_L^{out}(\sqrt{G}\xi)&=\expval{e^{-\xi^*\sqrt{G}\hat{a}_L^{out} + \xi\sqrt{G}\hat{a}_L^{out\dag}}}\nonumber\\
    &=\expval{e^{-\xi^*\sqrt{G}\sqrt{\Tilde{\eta}}\hat{a}_{in} + \xi\sqrt{G}\sqrt{\tilde{\eta}}\hat{a}_{in}^\dag}}\expval{e^{-\xi^*\sqrt{G}\sqrt{1-\tilde{\eta}}\hat{e}_{in} + \xi\sqrt{G}\sqrt{1-\tilde{\eta}}\hat{e}_{in}^\dag}}\nonumber\\
    &=\chi^{in}(\sqrt{\tilde{\eta}G})e^{-(1-\tilde{\eta})G\abs{\xi}^2/2}.
\end{align}
By substituting the result above back into Eq. ~(\ref{eq:twoprocess_gain}), we thus form a relation between the characteristic function of the input state and the output state of this two-stage process,
\begin{align}
    \chi_{L\rightarrow A}^{out}(\xi) & = \chi^{in}(\sqrt{\Tilde{\eta}G})e^{-\abs{\xi}^2(2G - \Tilde{\eta}G - 1)/2}.
\end{align}
Comparing with Eq.~(\ref{eq:characteristicfunction}), we can thus equate thermal loss channel to a cascade of pure loss followed by quantum-limited amplifier channel if
\begin{align}
    &G = (1-\eta)N_B + 1,\\
    &\Tilde{\eta} = \eta/G.
\end{align}
Symbolically, we can re-write this concatenation of channels as follows
\begin{align}
\mathcal{L}_{\eta,N_B}=\mathcal{A}_G\circ\mathcal{L}_{\Tilde{\eta}}\label{eq:concatenationofchannel}.
\end{align}

\section{Error exponent of threshold detector}\label{chap-threshold}
Consider a real-valued observation $X$, and the measurements of the two hypotheses yield Gaussian distributions for
\begin{align}
    &\textsc{H}_0: X \sim \mathcal{N}(\mu_0,\sigma_0^2),\\
    &\textsc{H}_1: X \sim \mathcal{N}(\mu_1,\sigma_1^2),
\end{align}
where $\mathcal{N}(\mu,\sigma^2)$ represents a normal distribution with mean $\mu$ and variance $\sigma^2$. The probability mass function of the observation $X$ under the two hypotheses is thus
\begin{align}
    P_{X|\textsc{H}_0}(x)& = \frac{1}{\sqrt{2\pi\sigma_0}}\exp\qty[-\frac{(x-\mu_0)^2}{2\sigma_0^2}],\\
    P_{X|\textsc{H}_1}(x)& = \frac{1}{\sqrt{2\pi\sigma_1}}\exp\qty[-\frac{(x-\mu_1)^2}{2\sigma_1^2}].
\end{align}
Hence, for a given priori probability of $\pi_0$ for Hypothesis $\textsc{H}_0$ and $\pi_1$ for Hypothesis $\textsc{H}_1$, the optimum (minimum error probability) detector when presented with an observation $x$ decides as follows
\begin{align}
    &\pi_0 P_{X|\textsc{H}_0}(x) \overset{\text{say $\textsc{H}_0$}}{\underset{\text{say 
 $\textsc{H}_1$}}\gtreqless}\pi_1 P_{X|\textsc{H}_1}(x).
\end{align}
When $\sigma_0=\sigma_1$, the above takes the form of a decision threshold $t$ \cite{helstrom1969quantum}, such that 
\begin{align}
    &x \overset{\text{say $\textsc{H}_0$}}{\underset{\text{say 
 $\textsc{H}_1$}}\lesseqgtr}t\label{eq:threshold_detector}.
\end{align}
However, when $\sigma_0\neq \sigma_1$, the decision rule is more complicated, and we instead use a sub-optimal threshold detector with threshold $\mu_0\leq t\leq \mu_1$, estimate its error probability and choose an appropriate $t$ to minimise it.
From Eq.~(\ref{eq:threshold_detector}), for a given decision threshold $t$, the probability of a misdetection and false alarm are
\begin{align}
    P_{fa} &= \int_t^\infty P_{X|\textsc{H}_0}(x) dx=\mathcal{Q}\qty(\frac{t-\mu_0}{\sigma_0}),\\
    P_{md} &=\int_{-\infty}^tP_{X|\textsc{H}_1}(x) dx = \mathcal{Q}\qty(\frac{\mu_1-t}{\sigma_1}),
\end{align}
where $\mathcal{Q}(a) = \frac{1}{\sqrt{2\pi}}\int_a^\infty e^{-t^2/2} dt$ is the standard Q-function. The average error probability of the detector is thus
\begin{align}
    P_{err}^{thresh} = \pi_0 \mathcal{Q}\qty(\frac{t-\mu_0}{\sigma_0}) + \pi_1 \mathcal{Q}\qty(\frac{\mu_1-t}{\sigma_1}).
\end{align}
By imposing Chernoff bound on the Q-function $Q(a)\leq e^{-a^2/2}$ for $a > 0$ \cite{chang2011chernoff} , the average error probability can thus be upper bounded as
\begin{align}
    P_{err}^{thresh}\leq \pi_0 e^{-(t-\mu_0)^2/2\sigma_0^2}+\pi_1 e^{-(t-\mu_1)^2/2\sigma_1^2}.\label{eq:chernoff_Qfunction}
\end{align}
Consider now that $X$ is instead the result of the addition of $M\gg 1$ independent and identically distributed (iid) (not necessarily Gaussian distributed) random variables, by central limit theorem, the sampling distribution of the observation $X$ thus approaches Gaussian distribution, with the two hypotheses yielding
\begin{align}
    &\textsc{H}_0: X \sim \mathcal{N}(\mu_0=M\tilde{\mu}_0,\sigma_0=M\tilde{\sigma_0}^2),\\
    &\textsc{H}_1: X \sim \mathcal{N}(\mu_1=M\tilde{\mu}_1,\sigma_1=M\tilde{\sigma}_1^2),
\end{align}
where $\tilde{\mu}_h$ and $\tilde{\sigma}_h^2$ represent the respective mean and variance of each copy of measurement result for hypothesis $h=0,1$. Hence, the above argument is applicable for any large iid random variables of arbitrary distributions. From Eq.~(\ref{eq:chernoff_Qfunction}), in the limit $M\rightarrow \infty$, the exponent approaches infinity and the error exponent is determined by the slower decaying term. To minimise the exponent, we need to set them to equal
\begin{align}
    &\frac{t-\mu_0}{\sigma_0}=\frac{\mu_1 - t}{\sigma_1},
\end{align}
and the detector threshold is thus derived to be
\begin{align}
    \Rightarrow & t_* = \frac{\mu_1\sigma_0 + \mu_0\sigma_1}{\sigma_0+\sigma_1}\label{eq:detector_threshold}.
\end{align}
Considering equal priori probability for hypothesis $\textsc{H}_0$ and $\textsc{H}_1$ ($\pi_0=\pi_1$), the resulting error exponent derived from Eq.~(\ref{eq:chernoff_Qfunction}) is
\begin{align}
    \chi_{thresh} &= \lim_{M\rightarrow\infty}\frac{-\ln P_{err}^{thresh}}{M}\nonumber\\
    &\gtrsim \frac{1}{2}\lim_{M\rightarrow \infty}\frac{1}{M}\qty(\frac{t_*-\mu_0}{\sigma_0})^2 \nonumber\\
    & = \frac{1}{2}\qty[\frac{1}{\tilde{\sigma_0}}\qty(\frac{\tilde{\mu}_1\tilde{\sigma}_0 + \tilde{\mu}_0\tilde{\sigma}_1}{\tilde{\sigma}_0+\tilde{\sigma}_1} - \frac{\tilde{\mu}_0\tilde{\sigma}_0 + \tilde{\mu}_0\tilde{\sigma}_1}{\tilde{\sigma}_0+\tilde{\sigma}_1})]^2\nonumber\\
    &=\frac{1}{2}\qty(\frac{\tilde{\mu}_1-\tilde{\mu}_0}{\tilde{\sigma}_0+\tilde{\sigma}_1})^2.\label{eq:errorexp_threshold}
\end{align}

    \chapter[The story of quantum illumination]{The story of quantum illumination}\label{chap-3}
\emph{The field of quantum illumination has undergone a remarkable transformation, advancing from its theoretical underpinnings to achieving significant experimental milestones. This chapter delves into this journey, highlighting the key developments that have shaped the field.\footnote{This chapter takes its main inspiration from the journal article of Ref.~\cite{shapiro2020quantum}, with additional supporting literatures from various other sources.} It aims to draw a comprehensive picture of how quantum illumination has evolved, shedding light on both its theoretical and practical aspects. As we uncover the potential of quantum technologies, new research questions and objectives emerge. These advancements in quantum illumination have influenced and shaped the research objectives outlined in this thesis.}
\newpage
\section{Development of quantum illumination}
The field of quantum illumination has experienced significant progress, evolving from its initial theoretical foundations to the achievement of encouraging experimental outcomes. This advancement has not only deepened our understanding of quantum protocols but also spurred growing interest in the potential for a substantial technological breakthrough. The journey from conceptual models to tangible experimental demonstrations reflects the rapid development within this domain. As these advancements continue, they pave the way for innovative applications and technologies that harness the unique properties of quantum mechanics, promising a transformative impact on various industries and scientific research areas. This section is dedicated to exploring key developments in quantum illumination, encompassing both theoretical foundations and experimental breakthroughs, to provide a comprehensive understanding of the field's evolution.
\subsection{Target detection using quantum illumination}
\begin{figure}[h!]
    \centering
    \includegraphics[width=0.7\linewidth]{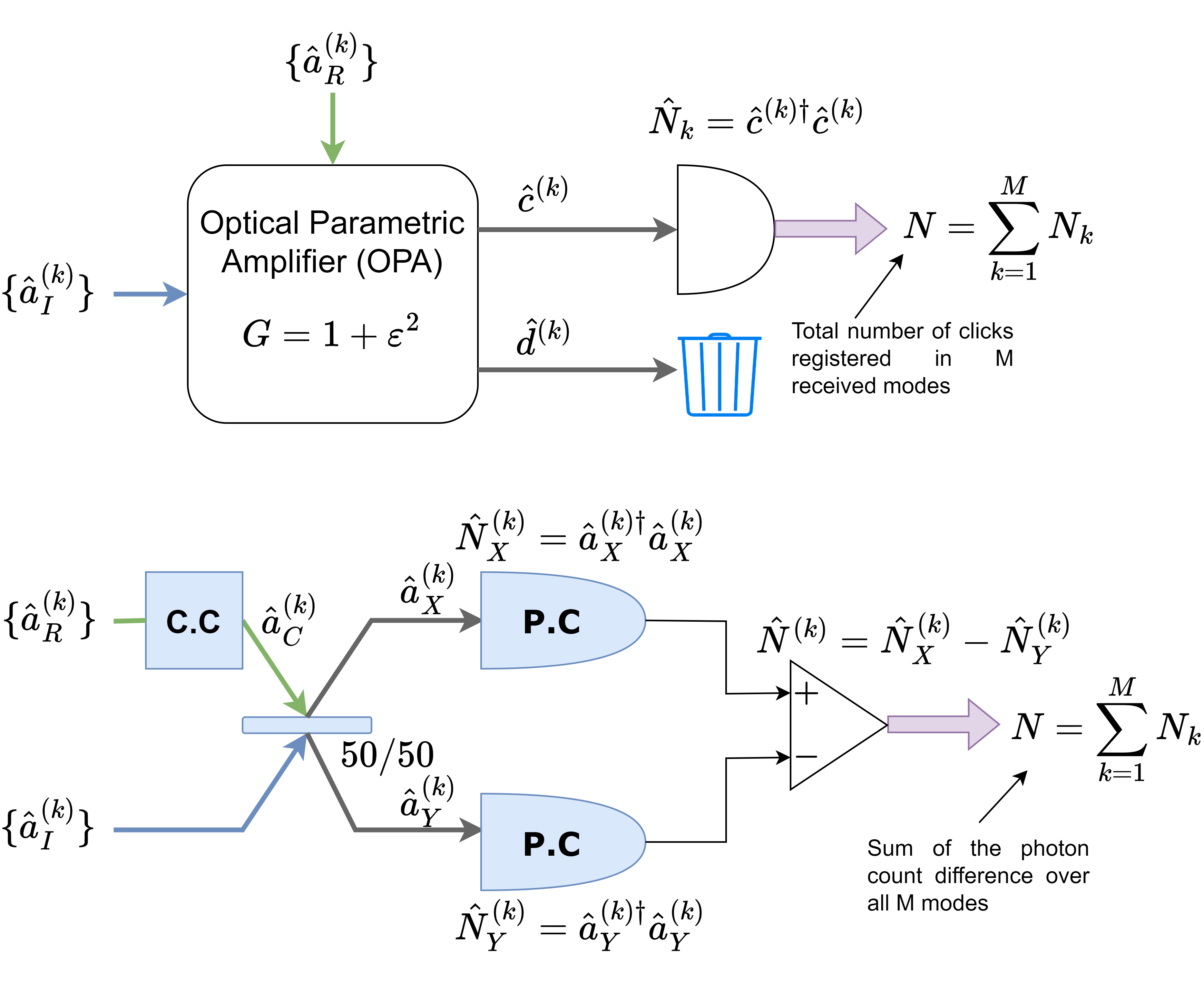}
    \captionsetup{width=\linewidth}
    \caption{Optical parametric amplifier (OPA) receiver (top) and phase-conjugate (PC) receiver (bottom) setup for QI target detection. Return modes and idler modes are inputs to an OPA with gain $G$ in an OPA receiver, with the total number of photons $N$ counted at one output port. For the PC receiver, the return modes undergo phase-conjugation before passing through a balanced difference detector with the idler modes, and the difference in the total number of clicks $N$ is counted. In both setup, if $N$ is less than a threshold count $N_{\text{th}}$, the receiver decides that the target is absent $\textsc{H}_0$, $\textsc{H}_1$ otherwise.}
    \label{fig:guhasetup}
\end{figure}

\begin{figure}[h!]
    \centering
    \includegraphics[width=\linewidth]{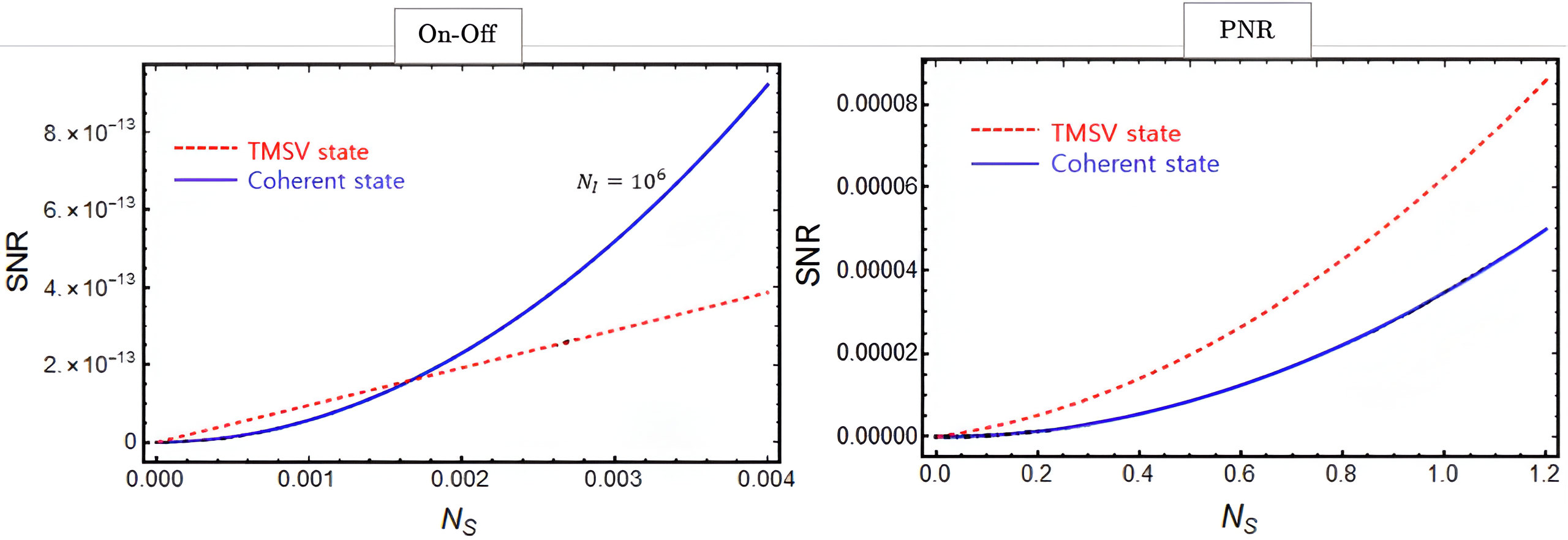}
    \captionsetup{width=\linewidth}
    \caption{Signal-to-noise ratio comparison of coherent state probe (blue solid) and TMSV probe (red dashed) as a function of $N_S$. For the coincident on-off setup (left), QI target detection performance is surpassed by coherent state for $N_S > 0.0016$. Conversely, replacing on-off modulator with a photon number resolving detector (right), QI performance is superior across all values of $N_S$ investigated. In both cases, $\kappa = 0.01,N_B= 600,$ and $M=10^6$. }
    \label{fig:onPNR}
\end{figure}

Lloyd \cite{lloyd2008enhanced} first introduced the concept of "quantum illumination," a method that utilizes quantum entanglement to enhance the ability of an optical radar to identify a weakly reflecting target amid background noise, which can often be stronger than just the signal reflected from the target. This approach is an extension of Sacchi's prior work on quantum operation discrimination \cite{sacchi2005entanglement,sacchi2005optimal}. In his research, Lloyd evaluated the performance of target detection in the following scenario: an optical transmitter illuminating a region of space encapsulated by background light in which a weakly reflecting target has an equal priori probability of being present or absent. 
In the case of single-photon (SP) illumination, the signal sent consists of a sequence of $M$ single-photon pulses, each characterized by a high time-bandwidth product $N=TW\gg 1$, where $T$ is the temporal length and $W$ is the bandwidth. Conversely, for quantum illumination (QI), the area is lit with a sequence of $M$ high time-bandwidth product ($N=TW\gg 1$) SP signal pulses, where each pulse is entangled with a corresponding SP idler pulse. In both scenarios, the detector performs optimal measurement of the reflected signal from the region (joint-measurement with the idler pulse for quantum illumination), making minimum error-probability decision between hypotheses $\textsc{H}_0$ (target absent) and $\textsc{H}_1$ (target present).  Under the following assumptions:
\begin{itemize}
    \item \textbf{Weakly-reflecting object:} The transmissivity for the signal beam when the target is present is $0\leq\kappa\ll 1$. 
    \item \textbf{Low-brightness condition:} The average photon number per temporal mode for background light satisfies $N_B\ll 1$.
    \item \textbf{High signal loss:} For each transmitted signal pulse, at most one photon is returned to the receiver (regardless of either hypothesis) implying that $NN_B\ll 1$.
\end{itemize}
Lloyd distinguished between two operational regimes, termed as the `good' and the `bad', for his SP and QI scenarios and evaluated their performance using quantum Chernoff bound on error probabilities. In the good regime, SP and QI incurred the same bound for their error probabilities,
\begin{align}
    P_e[SP]\leq e^{-M\kappa}/2\quad\text{and}\quad P_e[QI]\leq e^{-M\kappa}/2,
\end{align}
however, QI maintains a significant performance edge over SP illumination because the advantageous 'good' regime for SP is constrained to conditions where $\kappa\gg N_B$, while for QI, this favourable regime expands to a broader range of parameters, specifically where $\kappa \gg N_B/N$. In the bad regime, the performances between QI and SP are radically different
\begin{align}
    P_e[SP]&\leq e^{-M\kappa^2/8N_B}/2, \quad\text{for }\kappa\ll N_B,\\
    P_e[QI]&\leq e^{-M\kappa^2N/8N_B}/2,\quad \text{for }\kappa\ll N_B/N,
\end{align}
and the results revealed double advantage for QI. First, the `bad' regime for QI is relevant over a narrower range of $\kappa$ values compared to SP illumination. Secondly, in scenarios where both systems operate within their respective `bad' regimes, QI benefits from an error probability exponent in the quantum Chernoff bound that is a factor of $N$ higher than that of SP. Remarkably, this advantage is present despite the entanglement between the signal and idler pulse destroyed by the background noise.\\
According to the theory of quantum optics \cite{shapiro2009quantum,shapiro2012physics}, SP states are classified as nonclassical states as they require the use of quantum photodetection theory in the measurement process. As such, Lloyd’s comparison actually involves two quantum radars, where one (QI) employs entanglement, and the other (SP) does not. Hence, Shapiro and Lloyd \cite{shapiro2009quantumcoherent} introduced coherent state in replacement of SP. Measurements involving light beams in coherent states \cite{glauber1963coherent}, or their classical random mixtures, when observed through any of the fundamental photodetection methods — direct, homodyne, or heterodyne detection — do not necessitate the application of quantum photodetection theory for accurate statistical outcomes. Hence, the results that follow provide a comparison between classical radar (coherent state) and quantum radar (entangled signal-idler SP pair). The quantum Chernoff bound for coherent state radar for all $0\leq \kappa\leq 1$ and $N_B\geq 0$ is \cite{tan2008quantum}
\begin{align}
    P_e[CS]\leq e^{-M\kappa\qty(\sqrt{N_B+1}-\sqrt{N_B})^2}/2,
\end{align}
which reduces to 
\begin{align}
    P_e[CS]\leq e^{-M\kappa}/2
\end{align}
under the low background brightness ($N_B\ll 1$) condition. Shapiro and Lloyd demonstrated that a radar using coherent states could equal the performance of Lloyd's QI radar in its optimal (`good') regime and significantly outperform it when the QI radar was in its less effective (`bad') regime. \\
While Shapiro and Lloyd provided a fundamental analysis between classical radar and the QI protocol, they were unable to demonstrate the advantage of QI. In contrast, Tan et al. \cite{tan2008quantum} analysed a Gaussian-state QI system and demonstrated its ability to surpass the performance of all classical radars with equivalent transmitted energy. In her study, the aim is to detect a weakly reflecting target ($0\leq\kappa\ll 1$) that is equally likely to be absent or present in a high thermal background light ($N_B\gg 1$). In the classical case, $M$ identical coherent state (laser) pulse with average photon number $MN_S$, where $N_S\ll 1$ was initialised as probe state described by the annihilation operator $\hat{a}^{(k)}_S$. For the case of QI, $M$ pairs of entangled signal ($\hat{a}_S$) and idler ($\hat{a}_I$) mode pulses generated at the outputs of a continuous-wave spontaneous parametric downconversion (SPDC), with each pair described by the following number-ket representation
\begin{align}
    \ket{\psi}_{SI} = \sum_{n=0}^\infty \sqrt{\frac{N_S^n}{(N_S+1)^{n+1}}}\ket{n}_S\ket{n}_I,
\end{align}
are prepared as probe states, with $N_S$ denoting the average photon number per mode. Such state is termed the two-mode squeezed vacuum (TMSV). The signal state is sent to probe the target region, while the idler state is stored losslessly and subsequently measured jointly with the returned state. As the target is represented by a noisy attenuator with noise input of strength $N_B$, both the coherent state and TMSV retained their Gaussianity property even after undergoing the thermal loss channel transformation. Hence, the quantum Chernoff bound for both states can be computed analytically using the method developed by Pirandola and Lloyd \cite{pirandola2008computable} (previously elaborated in Section~\ref{section:soverlap}). The error probability bound of using $M$ independent and identically distributed (iid) copies of the coherent state as a probe is thus
\begin{align}
    P_e[CS] &\leq e^{-M\kappa N_S\qty(\sqrt{N_B+1}-\sqrt{N_B})^2}/2\\
    &\approx e^{-M\kappa N_S/4N_B}/2,\quad \text{for }N_B\gg 1.
\end{align}
Since the quantum Chernoff bound is equivalent to the Bhattacharyya bound (refer to Eq.~(\ref{eq:Bhattabound})) for the case of coherent state probe, the lower bound of the error probability is derived to be
\begin{align}
    P_e[CS] &\geq \frac{1}{2}\qty[1-\sqrt{1-e^{-2M\kappa N_S\qty(\sqrt{N_B+1}-\sqrt{N_B})^2}}]\\
    &\approx e^{-M\kappa N_S/2N_B}/4,
\end{align}
when $N_B\gg 1$ and $M\kappa N_S/2N_B\gg 1$. For QI when TMSV probe is used, the quantum Bhattacharyya bound when $\kappa \ll 1$, $N_S\ll 1$ and $N_B\gg 1$ can be approximated to as
\begin{align}
    P_e[QI]\leq e^{-M\kappa N_S/N_B}/2.
\end{align}
Comparing this Bhattacharyya error probability bound to the Chernoff error probability bound when using a coherent state probe reveals a factor of four improvement in the error probability exponent. Additionally, the observed twofold enhancement in the error probability exponent, when evaluating the coherent state's lower error exponent bound against the upper bound of the TMSV, conclusively establishes the enhanced efficacy of entanglement in target detection performance. Furthermore, Tan also proved that under strong background light $N_B\gg 1$, any general $M$ mode classical-state (signal and idler) transmitter described by the following density state
\begin{align}
    \rho_{SI} = \iint d^2\bm{\alpha}_S\;d^2\bm{\alpha}_I\; P(\bm{\alpha}_S,\bm{\alpha}_I)\ket{\bm{\alpha_S}}\bra{\bm{\alpha_S}}_S\otimes\ket{\bm{\alpha_I}}\bra{\bm{\alpha_I}}_I,
\end{align}
where $\{\ket{\bm{\alpha}_S}\}$ and $\{\ket{\bm{\alpha}_I}\}$ are the $M$-mode coherent states for the signal and idler system respectively, $P(\bm{\alpha}_S,\bm{\alpha}_I)\geq 0$ is the joint probability distribution, and signal energy constraint $\sum_{k=1}^M\expval{\hat{a}_{S}^{(k)\dag}\hat{a}_{S}^{(k)}}=MN_S$ , incurred error probability exponents that are at least two times inferior to quantum illumination using TMSV with the same signal strength $N_S$.\\
With such promising theoretical results for TMSV, the next question will be whether there exists a detection protocol that has the capability of saturating the theoretical performance of QI. Guha and Erkmen \cite{guha2009gaussian} investigated the designs of two quantum-optical sensor configurations for QI target detection, both of which can be easily implemented in a preliminary experimental demonstration (Fig.~\ref{fig:guhasetup}).
The primary consideration of the designs for both optical parametric amplifier (OPA) and phase-conjugate (PC) receiver is to access the phase-sensitive cross-correlations of the return-idler mode pairs (off-diagonal terms of the receiver-idler covariance matrix $V_{RI}$ which can be seen in Chapter~\ref{chap-6} of this thesis), the distinguishing factor between the two hypotheses. In OPA receiver, this is achieved by passing the return ($\hat{a}_R^{(k)}$) and idler modes ($\hat{a}_I^{(k)}$) through an optical parametric amplifier with gain $G = 1+\epsilon^2,\epsilon\ll 1$, producing the output mode pairs
\begin{align}
    \hat{c}^{(k)} &=\sqrt{G}\hat{a}_I^{(k)} + \sqrt{G-1}\hat{a}_R^{(k)\dag},\\
    \hat{d}^{(k)} &=\sqrt{G}\hat{a}_R^{(k)} + \sqrt{G-1}\hat{a}_I^{(k)\dag}.
\end{align}
The joint quantum measurement required for differentiating between the two hypotheses involves conducting a photon count on each output mode, represented by $N_k=\expval{\hat{c}^{(k)\dag}\hat{c}^{(k)}}$. When $M$ iid copies of TMSV probes are used, the decision between the hypotheses is based on the total photon count $N_k$ across all $M$ detected modes. For sufficiently large $M$, the distribution of $N=\sum_{k=1}^M N_k$ approaches a Gaussian distribution with mean and variance given as $MN_k$ and $M\sigma_k^2$ respectively. From Section~\ref{chap-threshold}, the detection threshold is thus derived as
\begin{align}
N_{th}=\frac{M(\sigma_1 N_0 + \sigma_0 N_1)}{\sigma_0+\sigma_1},
\end{align}
such that hypotheses $H_0$ is chosen when $N<N_{th}$, and $H_1$ otherwise. The error exponent for target detection error probability is hence approximated to be
\begin{align}
    \chi_{OPA} \gtrsim \frac{\epsilon^2\kappa N_S(N_S+1)}{2N_S(N_S+1) + 2\epsilon^2(1+2N_S)(1+N_S+N_B)}\approx \frac{\kappa N_S}{2 N_B},
\end{align}
for a weak transmitter ($N_S\ll 1$) operating in a highly lossy ($\kappa\ll 1$) and noisy ($N_B\gg 1$) regime. Hence, in this regime, the OPA receiver is shown to achieve at least a two-fold improvement in the error exponent over the optimal-receiver classical radar setup. In PC receiver, the receiver phase conjugates all $M$ return mode $\hat{a}_R^{(k)}$, outputting the following mode
\begin{align}
    \hat{a}_C^{(k)} =\sqrt{2}\hat{a}_V^{(k)} + \hat{a}_R^{(k)\dag},
\end{align}
where $\hat{a}_V^{(k)}$ are vacuum-state operators to ensure the operators obey commutation relations. This conjugated mode and the idler modes pass through a $50-50$ beam splitter, giving the following output modes
\begin{align}
    \hat{a}_X^{(k)} &=\qty(\hat{a}_C^{(k)} + \hat{a}_I^{(k)})/\sqrt{2},\\
    \hat{a}_Y^{(k)} &=\qty(\hat{a}_C^{(k)} - \hat{a}_I^{(k)})/\sqrt{2}.
\end{align}
These output modes are detected and fed into a unity-gain difference amplifier such that the final measurement is equivalent to
\begin{align}
    N_k &= \expval{\hat{N}_X^{(k)}}-\expval{\hat{N}_Y^{(k)}}\nonumber\\
    &=\expval{\hat{a}_X^{(k)\dag}\hat{a}_X^{(k)}}-\expval{\hat{a}_Y^{(k)\dag}\hat{a}_Y^{(k)}}.
\end{align}
Similarly, when using $M$ iid copies of TMSV probes, the decision of choosing between the hypotheses is based on the sum of the photon count $N_k$ for all $M$ modes. The decision choosing between the two hypothesis follows the same logic as OPA receiver, and the corresponding error exponent for target detection is given by
\begin{align}
    \chi_{PC} \gtrsim \frac{\kappa N_S(N_S+1)}{2N_B + 4N_SN_B + 6N_S + 4\kappa N_S^2 + 3\kappa N_S + 2}\approx \frac{\kappa N_S}{2N_B},
\end{align}
for the same setup criteria as the OPA detector. The PC receiver achieves the same two-time error exponent performance advantage as the OPA receiver over the optimal classical radar setup. However, the PC receiver's performance is marginally superior in absolute terms, which can be justified by the fact that the balanced dual-detection effectively nullifies the common-mode excess noise in $\hat{a}_X$ and $\hat{a}_Y$. Conversely, the OPA receiver functions at a very low gain, necessitating considerably less pump power compared to unity-gain phase conjugation. \\
Guha and Erkmen underscored the necessity of employing nonlinear optical processes like phase conjugators and optical parametric amplifiers in measurement protocols for QI specifically when a TMSV probe is used. These nonlinear processes are crucial for optimizing the performance of quantum measurement techniques. However, from an experimental standpoint, implementing these nonlinear processes presents significant challenges and inherent complexities. A recent advancement made by Lee et al. \cite{lee2023bound} proposed the use of coincidence counting  (simultaneous photon counting) in measurement protocol when using a TMSV probe. In his research, the target detection performance of different probes is measured by signal-to-noise ratio (SNR) defined as
\begin{align}
    \textsc{SNR} &= \frac{M\qty(\expval{\hat{O}}_\kappa - \expval{\hat{O}}_{\kappa=0})^2}{2\qty(\sqrt{\Delta^2 O_\kappa} + \sqrt{\Delta^2O_{\kappa=0}})^2},
\end{align}
where $M$ is the number of modes, $\expval{\hat{O}}_\kappa$ is the mean value of an observable $\hat{O}$, and $\Delta^2 O_\kappa$ is the corresponding variance. SNR relates to the error probability of target detection following $P_{err}\approx \exp\qty(-\textsc{SNR})$ and can be loosely interpreted as the error exponent. For an on-off modulator coupled with coincidence counting, the observable describing the process is characterised as $\hat{O}_{on} \equiv \bigotimes_{i=S,I}\qty(\mathbb{I}_i - \ket{0}\bra{0}_i)$ for signal (S) and idler (I) modes. 
Using this setup, following the above weak transmitter operating in a highly lossy and noisy paradigm, the SNR for coherent state and TMSV is approximated as follows:
\begin{align}
    &\textsc{SNR}[CS]\approx \frac{M\kappa^2N_S^2}{8N_B^3},\\
    &\textsc{SNR}[TMSV]\approx\frac{M\kappa^2 N_S}{8N_B^4}.
\end{align}
Figure \ref{fig:onPNR} illustrates the Signal-to-Noise Ratios (SNRs) of three Gaussian states as a function of $N_S$ during coincidence counting measurements with on-off detectors. Below $N_S < 0.0016$, the TMSV state shows superior performance compared to the coherent state. However, in the higher mean photon number range of the signal mode, the TMSV state is unable to leverage its quantum correlation to outperform the coherent state under on-off detection conditions.
An alternative measurement protocol is also proposed whereby the on-off detector is replaced by a photon number resolving (PNR) detector. The observable of this coincident PNR detection is given as $\hat{n}_{SI}\equiv \bigotimes_{i=S,I}\hat{a}_i^\dag\hat{a}_i$ such that the signal mode observable is taken for a single-mode probe (coherent state). Similarly, under the same setup criteria, the SNR for both coherent state and TMSV is approximated to be
\begin{align}
    &\textsc{SNR}[CS]\approx \frac{M\kappa^2N_S^2}{8N_B^2},\\
    &\textsc{SNR}[TMSV]\approx \frac{M\kappa^2N_S}{16N_B^2}.
\end{align}
The TMSV state surpasses the performance of the coherent state for the range of signal strength $N_S$ investigated. Specifically, the TMSV state benefits from its quantum correlations, even at high mean photon numbers in the signal mode, when PNR detection is employed (fig.~\ref{fig:onPNR}). This is because PNR detection is more effective at extracting detailed information about the target and the quantum correlations inherent in the TMSV state, in comparison to on-off detection methods. This research has successfully demonstrated the feasibility of developing a measurement protocol that does not require nonlinear optical processes, yet effectively utilizes quantum correlation to attain a performance edge in QI over classical protocol. However, the performance levels achieved in this study are comparatively lower than the theoretical limits previously established by Tan et al. \emph{Therefore, it would be of interest to develop a probe state capable of approaching the theoretical detection performance of TMSV, without necessitating nonlinear processes in its optimal measurement protocols.}
\subsection{Beyond target detection}
Driven by the achievements in quantum illumination, a wide range of applications for entanglement-assisted protocols have been proposed, with the goal of exceeding the performance of the best classical techniques. These applications involve measuring various physical quantities like displacement, time, force, acceleration, temperature, and electric and magnetic fields across different systems such as atoms, ions and spins \cite{giovannetti2014nature,braun2018quantum}. In the realm of optics, the focus has primarily been on measuring interferometric phase shifts, which stands as a classic example where quantum enhancement has been extensively researched \cite{caves1981quantum,demkowicz2015quantum}. In optical systems, a loss is a prevalent phenomenon that extends beyond unitary dynamics, and its accurate measurement is crucial in both scientific and technological contexts. For instance, the loss elements can be simulating actual pixels in an amplitude mask used in image sensing \cite{brida2010experimental}, or representing the absorption coefficients of a sample at different frequencies in an absorption spectroscopy setup \cite{whittaker2017absorption}.  Recognizing the significance of estimating loss parameters, Nair \cite{nair2018quantum} developed a framework for determining multiple loss parameters in an optical system. This approach utilizes the most general entanglement-assisted strategy, factoring in energy constraints. Specifically,  the system is composed of $K$ beam splitters, each characterized by real-valued transmissivities denoted as $\{\sqrt{\eta_k}\}_{k=1}^K$. A general ancilla-assisted entangled probe is prepared, such that the signal modes probing each of the $K$ beam splitters has energy defined as $\{N_k\}_{k=1}^K$. For the $k$-th beam splitter, the unitary operator defining its evolution in Schr\"odinger picture is
\begin{align}
    \hat{U}^{(m)}=\exp\qty[-i\phi_k\qty(\hat{a}^{(m)\dag}\hat{e}^{(m)} + \hat{a}^{(m)}\hat{e}^{(m)\dag})],
\end{align}
where $\hat{a}^{(m)}$ and $\hat{e}^{(m)}$ are the $m$-th signal mode and environmental mode annihilation operator respectively for $1\leq m\leq M\equiv\sum_{k=1}^K M_k$, and $\phi_k\in[0,\pi/2]$ is the ``angle'' parameter defining the transmissivity $\cos\phi_k=\sqrt{\eta_k}$. Extending the approach of Monras and Paris \cite{monras2007optimal} for single-parameter case, Nair derived an upper bound of the quantum Fisher information matrix (QFIM) for estimating $\bm{\phi}=(\phi_1,\cdots,\phi_K)$:
\begin{align}
    \mathcal{K}_{\bm{\phi}}\leq \Tilde{\mathcal{K}}_{\bm{\phi}}=4\;\text{diag}(N_1,\cdots,N_K)\label{eq:MPlimit},
\end{align}
where $N_i$ is the total energy of the signal modes probing the $i$-th beam splitter. The tilde denotes the QFIM assuming access to the environment mode. This bound is referred to as the generalised Monras-Paris (MP) limit and is applicable for any probe state with the given signal energy distribution and independent of the values of $\{M_k\}_{k=1}^K$. For transmittance parametrisation, this optimal QFIM is given by 
\begin{align}
    \mathcal{K}_{\bm{\eta}}\leq \tilde{\mathcal{K}}_{\bm{\eta}}=\text{diag}\qty[\frac{N_1}{\eta_1(1-\eta_1)},\cdots,\frac{N_K}{\eta_K(1-\eta_K)}].
\end{align}
He further shows that for a number-diagonal state (NDS) probe described by the following state
\begin{align}
    \ket{\psi}=\sum_{\mathbf{n\geq 0}}\sqrt{p_\mathbf{n}}\ket{\chi_{\mathbf{n}}}_A\ket{\mathbf{n}}_S.
\end{align}
where $\ket{\mathbf{n}}_S=\ket{n_1}_{S_1}\cdots\ket{n_M}_{S_M}$ is an $M$-mode number state of $S$, $\{\ket{\chi_{\mathbf{n}}}_A\}$ are orthonormal states of $A$, $p_\mathbf{n}$ is the probability distribution of $\mathbf{n}$ with the energy constraints 
\begin{align}
    \sum_{n=0}^\infty np_n=N;\qquad \text{for }p_n=\sum_{\mathbf{n}:n_1+\cdots+n_M=n}p_\mathbf{n},
\end{align}
the QFI for estimating transmissivity value $\phi$ for a singular beam splitter is derived to be $\mathcal{K}_\phi = 4N$. This result demonstrates the capability of NDS in saturating the theoretical performance MP limit in estimating $\phi$ with or without environmental access, for all values of $\phi,N,$ and $M$. Generalising the results for multiparameter case $\bm{\phi}=(\phi_1,\cdots,\phi_K)$, the product probe state $\otimes_{k=1}^K\rho^{(k)} = \otimes_{k=1}^K\ket{\psi^{(k)}}\bra{\psi^{(k)}}$, where $\ket{\psi^{(k)}}$ is any NDS probe of signal energy $N_k$, is considered. The $ij$th element of the QFIM for using this probe state is derived to be $\qty(\mathcal{K}^{NDS}_{\bm{\phi}})_{ij} = 4N_i\delta_{ij}$, such that the QFIM for transmittance parametrisation equates $\mathcal{K}^{NDS}_{\bm{\eta}} = \text{diag}\qty[N_1/\eta_1(1-\eta_1),\cdots,N_K/\eta_K(1-\eta_K)]$, thus showing that the product probe $\otimes_{k=1}^K\rho^{(k)}$ achieves the generalised MP limit from Eq.~(\ref{eq:MPlimit}). Whereas for product coherent-state input of the same energies, the QFIM for estimating the transmittance of the beam splitter is calculated to be $\mathcal{K}^{CS}_{\bm{\eta}}=\text{diag}\qty(N_1/\eta_1,\cdots,N_K/\eta_K)$, such that NDS states have a large advantage over coherent-state if $\eta_k\simeq 1$. Extending on this finding, Nair demonstrated that when NDS probes are utilized, conducting a joint measurement of the output state from the beam splitter in the Schmidt bases constitutes an optimal measurement that is independent of the parameter. \emph{In light of the promising outcomes achieved by quantum illumination setups in accurately estimating loss parameters for corresponding beam splitters, it becomes a compelling avenue for further research to explore the expanded potential of these setups in the realm of parameter estimation.}\\
Since Seth Lloyd's seminal work on quantum illumination \cite{lloyd2008enhanced}, the field has grappled with questions about its effectiveness, particularly compared to simply increasing the brightness of classical light for improved detection. A branch of study on covert illumination offers a pragmatic realization of stealth detection, whereby an adversary is added to the standard quantum illumination setup capable of intercepting all the signals not directed towards the receiver. Under these circumstances, there is a critical trade-off between the strength of the signal (which affects target detection capabilities) and the risk of interception by the adversary. Consequently, enhancing target detection by simply increasing the intensity of the light becomes an unviable strategy. 
Tahmasbi et al.~\cite{tahmasbi2021signaling} developed signalling strategies for discreetly scanning a distant target through a lossy and noisy bosonic channel. In his research, he adopted the no-passive signature model whereby vacuum state $\ket{0}\bra{0}$ is transmitted when no signal is sent to probe for the target. The covertness criteria is defined by the quantum relative entropy between the states intercepted by the adversary
\begin{align}
    \max_{h=0,1} \mathbb{D}\qty(\rho_{W^n}^{(h)}\|\qty(\rho_0^{(h)})^{\otimes n}),
\end{align}
where $\rho_0^{(h)}$ represents the state intercepted by the adversary when no signal is transmitted, while $\rho_{W^n}^{(h)}$ denotes the state intercepted when a signal is sent. Here, $h\in\{0,1\}$ specifies the hypothesis, with $h=0$ indicating the absence of the target and $h=1$ symbolising its presence. Ensuring that the quantum relative entropy remains below a small threshold $\delta>0$ guarantees that the adversary's optimal detection strategy is only marginally more effective than a random guess, thereby effectively maintaining the covertness of the operation. He proposed two signalling protocols: (1) diffused signalling whereby $n$ independent pairs of TMSV probes are prepared and transmitted and (2) sparse signalling where only $n\alpha_n$ ($\alpha_n<1$) independent TMSV probes are selected from $n$ mode and transmitted. The analytical results (error exponents) on the performance of target detection of the two signalling protocols, subjected to the established covertness constraint, were deduced. Additionally, these numerical findings were juxtaposed with the achievable bound on detection error inherent to any covert quantum illumination strategy, a limit which was also derived as part of this research. The findings indicate that the diffuse signalling approach nearly reaches optimal efficiency, whereas the sparse signalling method experiences a performance reduction, exceeding a loss factor of two. \emph{
This study has sparked additional research in the realm of covert sensing, particularly in identifying the essential fundamental limits on target detection efficacy within the parameters of covertness criteria.}
\section{Objectives}
The ongoing progress in quantum illumination is a testament to the dynamic nature of the field and its capacity to drive forward the frontiers of quantum science and technology. Building on the momentum of these advancements, the primary aim of this thesis is to broaden the application of quantum illumination across diverse scenarios and fields. This expansion seeks to capitalize on the promising developments in quantum illumination, exploring its potential in various contexts to fully realize its technological and scientific promise. In particular, one seeks to answer the following:
\begin{enumerate}
    \item Target detection performance under the consideration of covertness.
    \item Precision of quantum states in sensing the gain parameter of a quantum-limited amplifier over classical states.
    \item Assessing the target detection efficacy of a quantum state with a hypothesized measurement protocol that can be feasibly implemented in experimental settings.
\end{enumerate}
In addition to answering these questions, we will also discuss the broader implications and significance of our research findings.
    \chapter[Quantum limits of covert target detection]{Quantum limits of covert target detection}\label{chap-4}

\emph{This chapter explores the realm of covert target detection in quantum metrology, focusing on Alice's challenge to detect a weakly-reflecting target amidst thermal radiation while evading detection by Willie, an adversary co-located with the target. The chapter realistically formulates this problem, deriving quantum-mechanical limits on Alice's error probability in entanglement-assisted detection, balanced against her detectability by Willie.\footnote{Sections of this chapter have been referenced from our published article of Ref.~\cite{tham2023quantum}.} It highlights a key trade-off: Alice must employ a probe with light energy that remains sufficiently close to the thermal background noise to maintain covertness, while still achieving a non-zero probability of target detection, thereby ensuring effective stealth. The performance of two quantum probe types, two-mode squeezed vacuum probes and Gaussian-distributed coherent states, are analysed and compared against the theoretical performance limit. This comparison sheds light on the efficacy of different quantum sensing methods in covert scenarios. Furthermore, the chapter broadens its scope to include quantum limits for distinguishing between thermal loss channels under the passive signature scenario and explores the dynamics of non-adversarial quantum illumination, moving beyond the conventional assumption of no-passive signature.}
\newpage
\section{Motivations and objectives}
The pursuit of understanding and harnessing the capabilities of quantum technologies for covert sensing operations underpins the essential motivation of this research. In the realm of quantum mechanics, the challenge lies in meticulously balancing the act of detection with the imperative of stealth. This dynamic interplay is exemplified in the scenario where an observer, herein referred to as Alice, endeavours to detect a phase shift or identify the presence of a target, all while maintaining her presence undetected by an adversary, known as Willie. This pursuit not only encapsulates the intrigue of covert sensing but also underscores the sophisticated integration of physics and strategic operational execution.\\
Recent advancements in the domain of quantum mechanics have catalysed the development of a multitude of quantum-secure covert protocols, spanning across covert communication \cite{bash2015quantum, bullock2020fundamental}, phase sensing \cite{bash2017fundamental,gagatsos2019covert,hao2022demonstration}, low-probability-of-intercept communication \cite{shapiro2019quantum}, and target detection \cite{tahmasbi2021signaling}, predominantly within the continuous variable framework. These strides have illuminated the potential of quantum technologies to revolutionize sectors such as secure communications and surveillance. However, a comprehensive elucidation of the fundamental limits of covert sensing, particularly in contexts deviating from the no-passive signature paradigm \cite{volkoff2024not}, remains a frontier yet to be fully explored and understood.\\
Therefore, this thesis sets forth the primary objective of investigating these fundamental limits within the quantum mechanical framework. The research aims to delineate the extent to which Alice can accurately sense a phase shift or detect a target while concurrently minimizing her detectability by Willie. This exploration is poised to provide a more profound and nuanced understanding of the potential and constraints inherent in covert quantum operations.\\
Moreover, while existing research predominantly focuses on target detection under the no-passive signature assumption \cite{nair2020fundamental,tan2008quantum}, wherein the transmission of a vacuum state yields an unaltered state irrespective of the target’s presence, this thesis aims to extend beyond this paradigm. A pivotal aspect of this research is to scrutinize scenarios where the mode number $M$ emerges as a critical resource in enhancing performance, even when employing vacuum probes. This area of investigation is inspired by various quantum sensing and discrimination protocols that accentuate the significance of $M$ \cite{nair2022optimal,jonsson2022gaussian,shi2023ultimate,nair2023quantum}, an aspect that has been relatively underemphasized in the extant scholarly discourse.

\section{Covert sensing protocol}
\begin{figure}[h!]
    \centering
    \includegraphics[width=0.6\linewidth]{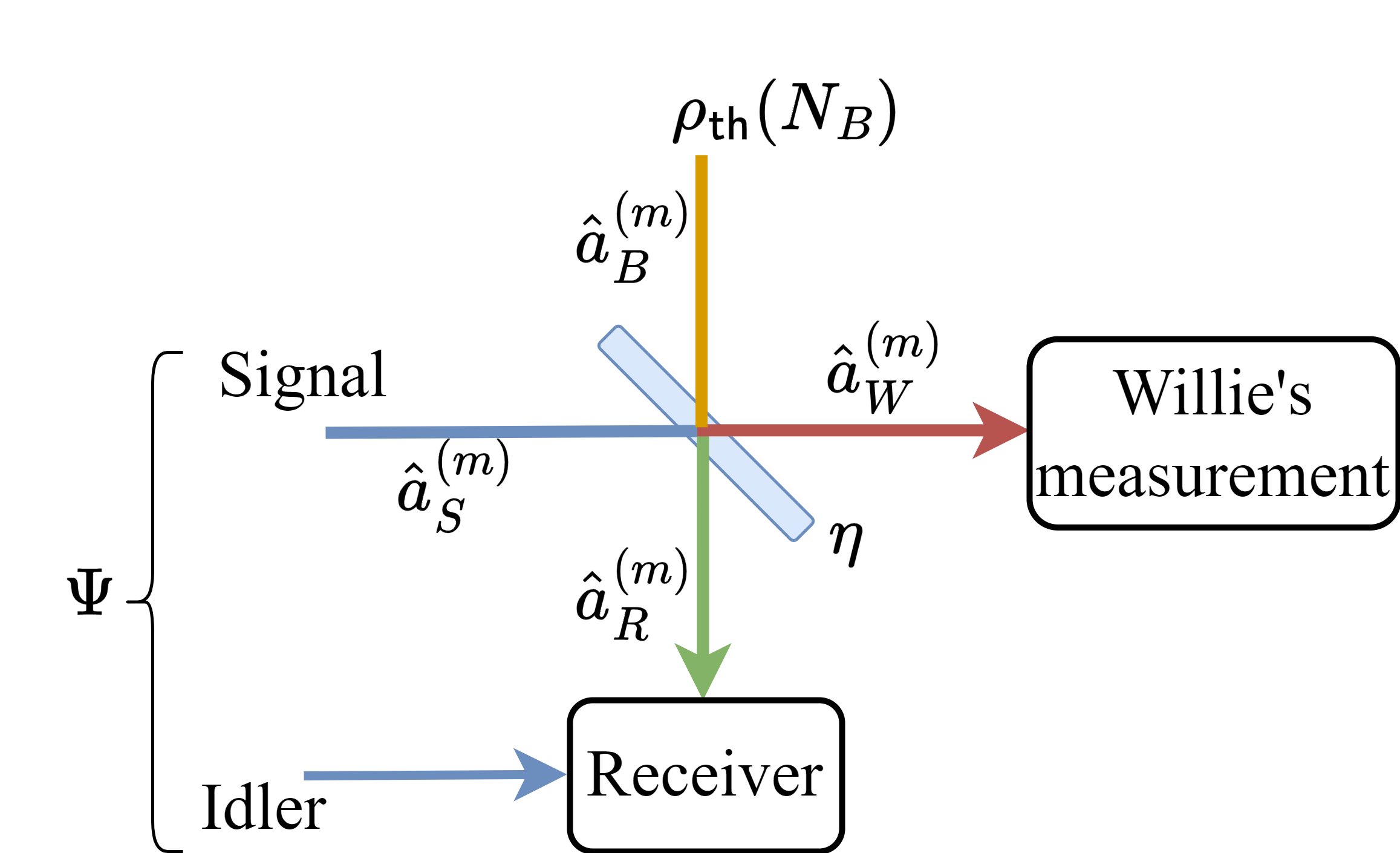}
     \captionsetup{width=\linewidth}
    \caption{A joint state $\Psi$ with $M$ signal and idler modes are prepared, with each signal mode $\hat{a}_S^{(m)}$ sent to probe the target region, which may contain Willie manifesting as a low reflectivity target $\eta\ll 1$. Each signal mode $\hat{a}_S^{(m)}$ is either replaced by a background mode $\hat{a}_B^{(m)}$ when Willie is absent, or mixed with the background at a beam splitter with reflectance  $\eta\ll 1$ representing the target.  Alice makes an optimal joint-measurement of the idler mode $\qty{\hat{a}_I^{(m)}}_{m=1}^M$ and the returned mode $\qty{\hat{a}_R^{(m)}}_{m=1}^M$ while Willie, when present, makes an optimal measurement on all his modes $\qty{\hat{a}_W^{(m)}}_{m=1}^M$.}
    \label{fig:setup}
\end{figure}\noindent
The schematic for the quantum covert illumination setup is illustrated in Figure \ref{fig:setup}. The target of illumination is represented as a noisy beam splitter (thermal loss channel) with reflectivity $\eta$. The transmitter Alice ($A$), who is assumed to be also controlling the receiving apparatus\footnote{If the radar configuration is bistatic, we assume that the idler modes can be transported losslessly to the receiver's location.}, prepares a signal-idler entangled state (called the probe) with $M$ signal and idler modes
\begin{align}
    \ket{\psi}_{IS} = \sum_{\bm{n}}\sqrt{p_{\bm{n}}}\ket{\chi_{\bm{n}}}_I\ket{\bm{n}}_S,\label{eq:purestateprobe}
\end{align}
where $\ket{\bm{n}}=\ket{n_1}_1\ket{n_2}_2\cdots\ket{n_M}_M$ is a $M$-mode number state of the signal ($S$) systems, and $\{\ket{\chi_{\textbf{n}}}\}$ is any normalised (not necessarily orthogonal) set of states of the idler ($I$) system. $p_{\bm{n}}$ is the probability mass function (PMF) of $\bm{n}$, subjected to a constraint on the total signal energy $\sum_{\textbf{n}}p_{\textbf{n}}(n_1+n_2+\cdots+n_M)\equiv \sum_{n=0}^\infty np_n=\mathcal{N}_S$. The signal modes are sent to probe the target region, while the idler modes are held losslessly by Alice. The annihilation operator of the mode $\qty{\hat{a}_R^{(m)}}_{m=1}^M$ returned to Alice obeys the beam splitter transformation given as
\begin{align}
    \hat{a}_R^{(m)}=\sqrt{\eta^{(h)}}\hat{a}_S^{(m)}+\sqrt{1-\eta^{(h)}}\hat{a}_B^{(m)}\label{eq:receivertransform},
\end{align}
where $\hat{a}_S^{(m)}$ and $\hat{a}_B^{(m)}$ are the annihilation operators corresponding to the signal and background modes. The hypotheses for Alice are represented by $h=0\;(1)$, where she must differentiate between the absence (0) or presence (1) of a target in the region, which can be represented by the reflectance of the beam splitter being $\eta^{(0)}=0$ and $\eta^{(1)}=\eta\ll 1$\footnote{$\eta$ represents the residual reflectivity in the entire trip from transmitter to receiver and thus includes diffraction losses as well as the reflectivity of the target object itself.}. Conversely, the adversary Willie ($W$) receives the output mode $\qty{\hat{a}_W^{(m)}}_{m=1}^M$ from the other arm of the beam splitter such that
\begin{align}
    \hat{a}_W^{(m)}=\sqrt{1-\eta}\hat{a}_S^{(m)} - \sqrt{\eta}\hat{a}_B^{(m)}.\label{eq:adversarytransform}
\end{align}
The negative sign is adopted to ensure that the commutation relations $\qty[\hat{a}_i,\hat{a}_j^\dag]=\delta_{ij}$ and $\qty[\hat{a}_i,\hat{a}_i]=\qty[\hat{a}_i^\dag,\hat{a}_i^\dag]=0$ are satisfied. In our study, where the natural passive signature (PS) model is adopted, each background mode $\hat{a}_B^{(m)}$ is in thermal state $\rho_{\text{th}}(N_B)=\sum_{n=0}^{\infty} N_B^n/(N_B+1)^{n+1}\ket{n}\bra{n}$ with background brightness of $N_B$. This model differs from the previous studies which utilised the no-passive signature assumption (NPS). In NPS model, the background brightness is shifted from its nominal value $N_B$ to $N_B/(1-\eta)$ when the target is present such that Alice is unable to detect the target using vacuum probe.\\
\noindent
In our PS model, Alice has to discriminate between the following hypotheses
\begin{align}
    &\text{H$_0$ (Target absent): }\;\rho_0 = \qty(\Tr_S \Psi)\otimes \rho_{\text{th}}(N_B)^{\otimes M},\nonumber\\
    &\text{H$_1$ (Target present): }\rho_1 = \qty(\mathbb{I}_I\otimes \mathcal{L}_{\eta,N_B}^{\otimes M})\Psi\label{eq:Alicehypothesis},
\end{align}
where $\Psi=\ket{\psi}\bra{\psi}_{IS}$ is the density matrix of the probe state, $\mathbb{I}_I$ is the identity channel acting on idler mode, and $\mathcal{L}_{\eta,N_B}$ denotes the thermal loss channel of transmittance $\eta$ and excess noise $N_B$ acting on each signal mode \cite{serafini2017quantum}. Conversely, Willie faces the hypothesis test
\begin{align}
    &\text{H$_0'$ (Signal not sent): }\sigma_0 = \rho_{\text{th}}(N_B)^{\otimes M},\nonumber\\
    &\text{H$_1'$ (Signal  sent): }\hspace{18pt} \sigma_1 = \mathcal{L}_{1-\eta,N_B}^{\otimes M}\qty(\Tr_I \Psi)\label{eq:Williehypothesis}.
\end{align}
In our model, when the signal is not sent by Alice, we assume that the thermal background floods the target region. Consequently, the state intercepted by Willie under the null hypothesis is a thermal state with a background brightness of $N_B$. Both Alice's and Willie's performance, assuming making optimal quantum measurements, are quantified by the minimum-error probability defined by Helstrom limit \cite{helstrom1969quantum}
\begin{align}
    P_e^A &= \frac{1}{2} - \frac{\norm{\rho_0- \rho_1}_1}{4}\leq \inf_{0\leq s\leq 1}\frac{1}{2}\Tr \qty[\rho_0^s\rho_1^{1-s}],\\
    P_e^W &= \frac{1}{2} - \frac{\norm{\sigma_0- \sigma_1}_1}{4}\leq \inf_{0\leq s\leq 1}\frac{1}{2}\Tr \qty[\sigma_0^s\sigma_1^{1-s}],
\end{align}
where the inequality is given by the quantum Chernoff bound discussed earlier in Eq.~(\ref{eq:chernofferrorbound}). It is important to note that within this setup, the achievement of quantum-secure performance relies on two strong assumptions regarding Willie's capabilities. First, it is assumed that Willie can effectively capture all the light that does not return to Alice. Second, Willie is presumed to possess knowledge about the specific probe that Alice intends to use, enabling him to design the optimal detector for conducting his hypothesis test as outlined in Eq.~(\ref{eq:Williehypothesis}).
 
\subsection{Necessary condition for covertness}
Considering that Alice possesses $M$ number of signal modes, she first defines an acceptable level of covertness $0\leq \epsilon \leq 1/2$ at the initialisation of the protocol. Subsequently, she prepares a probe $\Psi$ (namely $\epsilon$-covert probe) that both satisfy the $\epsilon$-covert criteria
\begin{align}
    P_e^W\geq \frac{1}{2} - \epsilon\label{eq:necessarycondition},
\end{align}
and minimises $P_e^A$ over all $\epsilon$-covert probes of $M$ signal modes. To circumvent the difficulty of computing the Helstrom limit for Willie's error probability, a necessary condition is formulated to characterise the $\epsilon$-covert probes. From the upper bound of the Fuchs-van de Graaf inequalities from Eq.~(\ref{eq:QFB}) we have
\begin{align}
    P_e^W=\frac{1}{2} - \frac{\norm{\sigma_0-\sigma_1}_1}{4}\leq \frac{\mathcal{F}(\sigma_0,\sigma_1)}{2},
\end{align}
relating the trace distance $\norm{\sigma_0-\sigma_1}_1$ to the fidelity $\mathcal{F}(\sigma_0,\sigma_1)$ between the hypothesis states of Willie. Combining with the $\epsilon$-covert criteria, it can be deduced that the upper bound of the Fuchs-van de Graaf inequalities must necessarily exceed the $\epsilon$-covert criteria $\mathcal{F}(\sigma_0,\sigma_1)/2\geq 1/2 - \epsilon$. The necessary condition for $\epsilon$-covertness is thus
\begin{align}
    \mathcal{F}(\sigma_0,\sigma_1)\geq 1 - 2\epsilon\label{eq:necessarycondition_fidelity}.
\end{align}
We can evaluate the necessary condition further by exploiting the data processing inequality for fidelity. The null hypothesis state for Willie is 
\begin{align}
    \sigma_0 &=\rho_{\text{th}}(N_B)^{\otimes M}=\prod_{m=1}^M\qty[\sum_{n_m=0}^\infty\frac{N_B^{n_m}}{(N_B+1)^{n_m+1}}\ket{n_m}\bra{n_m}]\nonumber\\
    &=\sum_{\mathbf{n}\geq\mathbf{0}}\frac{N_B^n}{(N_B+1)^{n+M}}\ket{\mathbf{n}}\bra{\mathbf{n}},
\end{align}
where $\ket{\mathbf{n}}=\ket{n_1}\otimes\cdots\otimes\ket{n_m}$ are the $M$-mode number states of Willie's mode, and $n\coloneqq \Tr \mathbf{n}=\sum_{m=1}^M n_m$ is defined to be the total photon number. For the alternative hypothesis state, we can generalise a state
\begin{align}
    \sigma_1 = \sum_{\mathbf{n},\mathbf{n'}\geq\mathbf{0}} q_{\mathbf{n},\mathbf{n'}}\ket{\mathbf{n}}\bra{\mathbf{n'}}.
\end{align}
Introducing a non-destructive measurement $\mathcal{M}$ of the photon number in each of Willie's modes for any input state $\mathcal{M}\sigma=\sum_{\mathbf{n}\geq \mathbf{0}}\bra{\mathbf{n}}\sigma\ket{\mathbf{n}}\ket{\mathbf{n}}\bra{\mathbf{n}}$, the data processing inequality for the fidelity hence bounds the fidelity of Willie's hypothesis states such that
\begin{align}
    \mathcal{F}(\sigma_0,\sigma_1) & \leq \mathcal{F}(\mathcal{M}\sigma_0, \mathcal{M}\sigma_1)\nonumber\\
    &= \mathcal{F}(\sigma_0,\mathcal{M}\sigma_1)\nonumber\\
    &=\mathcal{F}\qty(\sigma_0,\sum_{\mathbf{n}}q_{\mathbf{n}}\ket{\mathbf{n}}\bra{\mathbf{n}})\nonumber\\
    &=\sum_{n=0}^\infty\sqrt{\frac{N_B^n}{(N_B+1)^{n+M}}}\qty(\sum_{\mathbf{n}:\Tr\mathbf{n}=n}\sqrt{q_{\mathbf{n}}})\label{eq:dataprocessing},
\end{align}
where we have simplified the expression $q_{\mathbf{n},\mathbf{n}}=q_{\mathbf{n}}$, and defined $\Tr \mathbf{n} \coloneqq \sum_{m=1}^M n_m$. The second summation remains complicated as it is expressed in terms of Willie's photon number PMF of $\mathbf{n}$. To get a bound in terms of the PMF of  the total photon number seen by Willie under H$_1'$, i.e., $q_n=\sum_{\mathbf{n}:n_1+\cdots+n_M}q_{\mathbf{n}}$, we apply the Cauchy-Schwarz inequality,
\begin{align}
    \sum_{\mathbf{n}:\Tr\mathbf{n}=n}\sqrt{q_{\mathbf{n}}} &\leq \qty(\sum_{\mathbf{n}:\Tr\mathbf{n}=n} 1)^{1/2}\qty(\sum_{\mathbf{n}:\Tr\mathbf{n}=n} q_{\mathbf{n}})^{1/2}\nonumber\\
    &=\sqrt{\binom{n+M-1}{n}}\sqrt{q_n}\label{eq:cauchyschwarz},
\end{align}
where $\binom{n+M-1}{n}=\frac{(n+M-1)!}{n!(M-1)!}$ is the binominal coefficient. This inequality is saturated if $\{q_{\mathbf{n}}\}$ is a product thermal distribution of arbitrary brightness. Substituting the above inequality into Eq.~(\ref{eq:dataprocessing}) and back into Eq.~(\ref{eq:necessarycondition_fidelity}), we get the necessary condition for $\epsilon$-covertness in terms of $q_n$
\begin{align}
    \sum_{n=0}^\infty\sqrt{q_n}\sqrt{\binom{n+M-1}{n}\frac{N_B^n}{(N_B+1)^{n+M}}}\geq 1 - 2\epsilon\label{eq:ecovert_constraint}.
\end{align}
In the remainder of the chapter, the necessary condition for $\epsilon$-covertness refers to the above inequality unless otherwise defined.
\section{Fundamental limit of target detection}\label{section:limitsofQI_PS}
Before discussing the covert sensing protocol and the $\epsilon$-covertness for finite covertness $\epsilon > 0$, we are interested in Alice's target detection performance in the absence of the adversary Willie. Previous research by Nair \textit{et al.} \cite{nair2020fundamental} sets a bound on the best performance achievable by Alice in the NPS model. Consider Alice sending a general quantum pure-state probe  described in Eq.~(\ref{eq:purestateprobe}), her detection error probability is lower bounded as
\begin{align}
    P_e^A\geq \frac{1}{4}\qty{\sum_{n=0}^\infty p_n\qty[1-\frac{\eta}{N_B+1}]^{n/2}}^2\geq \frac{\exp(-\beta \mathcal{N}_S)}{4}\label{eq:minerror_NPS},
\end{align}
where the last inequality is obtained by applying Jensen's inequality. $p_n=\sum_{\mathbf{n}:n_1+\cdots+n_M}p_{\mathbf{n}}$ is the PMF of the total photon number for the state sent by Alice, $\mathcal{N}_S$, $\beta=-\ln\qty[1 - \eta/(N_B+1)]$ is defined as the error exponent and $\mathcal{N}_S=\sum_{n=0}^\infty n p_n$ is the total signal energy of the probe. However, in our model where we invoke the PS criteria, the above bound on Alice's detection performance does not apply. Hence, we introduce a fundamental target detection error probability limit applicable to all quantum illumination systems under the PS model. \\
\noindent
For a general result, we consider two output states of any two arbitrary thermal loss channels with the input state $\Psi$,
\begin{align}
    \tau_0 &= \qty(\mathbb{I}_I\otimes \mathcal{L}_{\kappa_0,N_0}^{\otimes M})\Psi,\nonumber\\
    \tau_1 &= \qty(\mathbb{I}_I\otimes \mathcal{L}_{\kappa_1,N_1}^{\otimes M})\Psi,
\end{align}
where $\mathcal{L}_{\kappa_b,N_b}$ represents the thermal loss channel with transmittance $\kappa_b$ and excess noise $N_b$ for $b\in\{0,1\}$. Using the strong concavity property of fidelity, the fidelity of the output states $\mathcal{F}(\tau_0,\tau_1)=\Tr\sqrt{\sqrt{\tau_0}\tau_1\sqrt{\tau_0}}$ satisfies 
\begin{align}
    \mathcal{F}(\tau_0,\tau_1)\geq \nu^M\sum_{n=0}^\infty p_n\qty[\nu\sqrt{\Tilde{\kappa}_0\Tilde{\kappa}_1}+\sqrt{(1 - \Tilde{\kappa}_0)(1 - \Tilde{\kappa}_1)}]^n\label{eq:fidelity_thermalloss},
\end{align}
where $\nu=\qty(\sqrt{G_0G_1}-\sqrt{(G_0-1)(G_1-1)})^{-1}$ and $\Tilde{\kappa}_b = \kappa_b/G_b$ for $G_b = (1-\kappa_b)N_B + 1$. For our current target detection setup in PS model, we define $\kappa_0 = 0$, $\kappa_1 = \eta$, and $N_0 = N_1 = N_B$ to get the fidelity lower bound of Alice's target detection
\begin{align}
    \mathcal{F}(\rho_0,\rho_1)\geq \nu^M\sum_{n=0}^\infty p_n\qty[1- \frac{\eta}{(1-\eta)N_B + 1}]^{n/2}\label{eq:fidelitybound_Alice}.
\end{align}
Applying the lower bound of the Fuchs-van de Graaf inequalities from Eq.~(\ref{eq:QFB}), Alice's target detection performance is hence quantified by error probability lower bound
\begin{align}
    P_e^A &\geq \frac{1}{2}\qty{1 - \sqrt{1- \nu^{2M}\qty[\sum_{n=0}^\infty p_n\qty(1 - \frac{\eta}{(1-\eta)N_B+1})^{n/2}]^2}}\label{errorprobability_Alice}\\
    &\geq \frac{1}{2}\qty[1-\sqrt{1-\nu^{2M}\qty(1 - \frac{\eta}{(1-\eta)N_B+1})^{\mathcal{N}_S}}]\label{eq:errorprobabilityJensen_Alice},
\end{align}
where $\mathcal{N}_S=\sum_{n=0}^\infty np_n$ is the total signal energy of the probe sent by Alice, and the last inequality is obtained by invoking Jensen's inequality. The detailed derivation for the above results can be found in Appendix~\ref{appendixB}. This result shows that no quantum illumination system in the PS model can achieve a target detection (Alice) error probability smaller than Eq.~(\ref{eq:errorprobabilityJensen_Alice}). A comparison with the NPS results in Eq.~(\ref{eq:minerror_NPS}) reveals the additional $\nu^{2M}$ term in our result characterises the passive signature. For PS model, the number of mode M becomes an important
resource for increasing the target detection performance. The detection error probability result for NPS model can be recovered from our result by equating $N_1 = N_B/(1-\eta)$ instead.
\section{Perfect covertness}
Consider the scenario whereby Alice sends a $M$-mode probe yet Willie is unable to obtain any useful information from his intercepted signals, i.e. $\epsilon = 0$, achieving perfect covertness. This implies that the state received by Willie in either of the hypotheses is identical to each other $\sigma_1 = \sigma_0 = \rho_{\text{th}}(N_B)^{\otimes M}$. For quantum probes, achieving perfect covertness is possible only with $M$-mode TMSV state of per-mode energy $N_B$,
\begin{align}
    \Psi = \ket{\psi}\bra{\psi}_{TMSV},\quad \ket{\psi}_{TMSV} = \prod_{m=1}^M\qty[\sum_{n_m=0}^\infty\qty(\frac{N_B^{n_m}}{(N_B+1)^{n_m + 1}})^{1/2}\ket{n_m}_S\ket{n_m}_I].
\end{align}
For classical probe, one possible way of achieving perfect covertness is Alice generating coherent states in each signal mode with amplitude $\alpha\in\mathbb{C}$ chosen according to the circular Gaussian distribution $P(\alpha) = \exp\qty(-\abs{\alpha}^2/N_B)/\pi N_B$. This state is known as the Gaussian-distributed coherent state (GCS) probe,
\begin{align}   \Psi=\prod_{m=1}^M\int_{\mathbb{C}}d^{2}\alpha_m\;\frac{e^{-\abs{\alpha_m}^2/N_B}}{\pi N_B}\ket{\alpha_m}\bra{\alpha_m}.
\end{align}
Alice's measurement can be contingent on her knowledge of the transmitted amplitude in each of the M shots.
\subsection{Two-mode squeezed vacuum}
Considering multiple identical copies of the TMSV states of per-mode brightness $N_S$, $\ket{\psi} = \sum_{n=0}^\infty\sqrt{N_S^n/(N_S+1)^{n+1}}\ket{n}_S\ket{n}_I$. This state has a zero mean vector and the covariance matrix is
\begin{align}
    V_{SI} = \begin{bmatrix}[0.8]
        S & C\\ C & S
    \end{bmatrix}\oplus\begin{bmatrix}[0.8]
        S & -C\\ -C & S
    \end{bmatrix},
\end{align}
where $S=N_S+1/2$ corresponds to the local mode, and $C=\sqrt{N_S(N_S+1)}$ quantifies the inter-modal correlations. Using the annihilation operator of the mode returned to Alice from Eq.~(\ref{eq:receivertransform}), we can compute the mean vectors and the covariance matrices corresponding to the states received by Alice for both hypotheses. Both states remain zero-mean with covariance matrices
\begin{align}
    &\text{H}_0:\;V_0 = \begin{bmatrix}[0.8]
        B & 0\\ 0 & S
    \end{bmatrix}\oplus\begin{bmatrix}[0.8]
        B & 0\\ 0 & S
    \end{bmatrix}\label{eq:covariancematrix_H0},\\
    &\text{H}_1:\;V_1 = \begin{bmatrix}[0.8]
        A & \sqrt{\eta}C\\ \sqrt{\eta}C & A
    \end{bmatrix}\oplus\begin{bmatrix}[0.8]A & -\sqrt{\eta}C\\ -\sqrt{\eta}C & A
    \end{bmatrix}\label{eq:covariancematrix_H1},
\end{align}
where
\begin{align}
    A &= \eta N_S + (1-\eta)N_B + 1/2,\\
    B &= N_B + 1/2.
\end{align}
Hence, if Alice sends multiple identical copies of the TMSV probes, her target detection error probability is bounded by quantum Chernoff bound from Eq.~(\ref{eq:chernofferrorbound}):
\begin{align}
    P_e^A[TMSV] &\leq \frac{1}{2}\prod_{m=1}^M \inf_{0\leq s\leq 1} C_s(\rho_{0,m},\rho_{1,m})\nonumber\\
    & = \frac{1}{2}\qty[\inf_{0\leq s\leq 1}C_s(\tilde{\rho}_0,\tilde{\rho}_1)]^M,
\end{align}
where the multiplicative property of the quantum Chernoff bound is invoked. $\tilde{\rho}_0$ and $\tilde{\rho}_1$ correspond to the states received by Alice for the respective hypotheses when a single copy of TMSV probe is used. Since the returned states to Alice remains Gaussian, $C_s(\tilde{\rho}_0,\tilde{\rho}_1)$ can be computed using the covariance matrices from Eq.~(\ref{eq:covariancematrix_H0}) and Eq.~(\ref{eq:covariancematrix_H1}), with the method discussed in Section~\ref{section:soverlap}. The corresponding quantum Chernoff exponent is hence
\begin{align}
    \chi_{TMSV} & = \lim_{M\rightarrow \infty}\frac{-\ln P_e^A[TMSV]}{M}\nonumber\\
    &\simeq\lim_{M\rightarrow\infty}\frac{-\ln\qty[\inf_{0\leq s\leq 1}C_s(\tilde{\rho}_0,\tilde{\rho}_1)]^M}{M}\nonumber\\
    &=-\ln\qty[\inf_{0\leq s\leq 1}C_s\qty(\tilde{\rho}_0,\tilde{\rho}_1)]\label{eq:Perfect_TMSV}.
\end{align}
 By setting $N_S = N_B$, we obtain the quantum Chernoff bound of Alice for perfect covertness when using TMSV probe.

\subsection{Gaussian-distributed coherent state}
For GCS, Alice transmits a string of $M$ coherent states $\ket{\bm{\alpha}}\equiv\ket{\alpha_1}\otimes\cdots\otimes\ket{\alpha_M}$ with amplitudes drawn independently from a circular Gaussian distribution
\begin{align}
    P(\alpha) = \frac{1}{\pi N_T}\exp\qty(-\frac{\abs{\alpha}^2}{N_T}),
\end{align}
where $N_T$ is the average per-mode probe energy. The reduced state intercepted by Willie in each signal mode is a thermal state $\rho_{\text{th}}(N_T)$. As the probe is generated by Alice, she knows the vector $\bm{\alpha}$ and is hence able to incorporate this knowledge in constructing her receiver. Her error probability is therefore 
\begin{align}
    P_e^A[GCS] = \int_{\mathbb{C}^M}\;d^{2M}\bm{\alpha} P(\bm{\alpha}) P_e^A[\bm{\alpha}],
\end{align}
where $P_e^A[\bm{\alpha}]$ represents Alice's optimal error probability of discriminating the states received for the two hypotheses when using $\ket{\bm{\alpha}}$ probe. When a single mode coherent state $\ket{\alpha} = \ket{\alpha_r + i\alpha_i}$ is used as a probe, the states received by Alice have the respective mean vectors and covariance matrices
\begin{align}
    &\text{H}_0: \;q_0 = \begin{bmatrix}[0.8]
        0\\0
    \end{bmatrix};\quad V_0 = \begin{bmatrix}[0.8]
        N_B+\frac{1}{2} & 0\\ 0 & N_B+\frac{1}{2}
    \end{bmatrix},\\
    &\text{H}_1: \;q_1 = \begin{bmatrix}[0.8]
        \sqrt{2\eta\alpha_r}\\ \sqrt{2\eta\alpha_i}
    \end{bmatrix};\quad V_1 = \begin{bmatrix}[0.8]
        (1-\eta)N_B + \frac{1}{2} & 0 \\0 & (1-\eta)N_B+\frac{1}{2}
    \end{bmatrix},
\end{align}
which correspond to a thermal state and a displaced thermal state respectively. Similarly, using quantum Chernoff bound, Alice's target detection error probability is upper bounded as
\begin{align}
    P_e^A[GCS] &= \int_{\mathbb{C}^M} d^{2M}\bm{\alpha}\;\qty(\prod_{m=1}^M P(\alpha_m))P_e^A[\bm{\alpha}]\nonumber\\
    &\leq \frac{1}{2}\int_{\mathbb{C}^M}d^{2M}\bm{\alpha}\; \prod_{m=1}^M\qty[P(\alpha_m)\qty(\inf_{0\leq s\leq 1} C_s[\alpha_m])]\nonumber\\
    &=\frac{1}{2}\qty[\int_{\mathbb{C}}d^{2}\alpha\; P(\alpha)\qty(\inf_{0\leq s \leq 1}C_s[\alpha])]^M,
\end{align}
where we have similarly used the property of multiplicativity of quantum Chernoff bound. $C_s[\alpha]$ here represents the s-overlap between the states received by Alice corresponding to the two hypotheses when a single mode coherent probe state $\ket{\psi}=\ket{\alpha}$ is used. Her performance quantified by quantum Chernoff exponent is hence
\begin{align}
    \chi_{GCS} &= \lim_{M\rightarrow\infty}\frac{-\ln P_e^A[GCS]}{M}\nonumber\\
    &\simeq \lim_{M\rightarrow\infty}\frac{-\ln\qty[\int_{\mathbb{C}}d^{2}\alpha\; P(\alpha)\qty(\inf_{0\leq s \leq 1}C_s[\alpha])]^M}{M}\nonumber\\
    &=-\ln\int_{\mathbb{C}}d^{2}\alpha\; P(\alpha)\qty(\inf_{0\leq s \leq 1}C_s[\alpha])\label{eq:Perfect_GCS}.
\end{align}
In the scenario of perfect covertness, the average per-mode probe energy of the coherent state $N_T$ is set to be equal to $N_B$.
\subsection{Performance comparison}
\begin{figure}
    \centering
    \begin{subfigure}
    \centering
        \includegraphics[width = 0.49\linewidth]{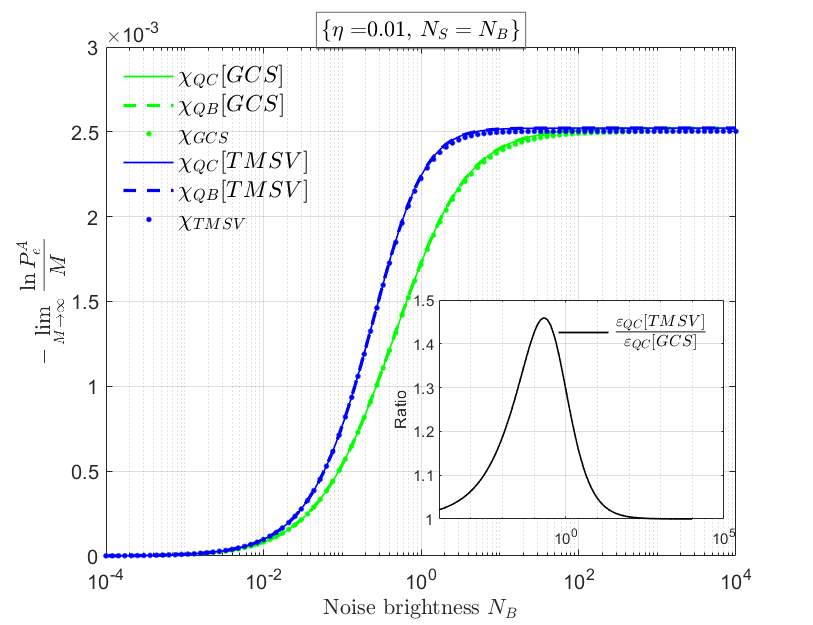}
    \end{subfigure}
        \begin{subfigure}
    \centering
        \includegraphics[width = 0.49\linewidth]{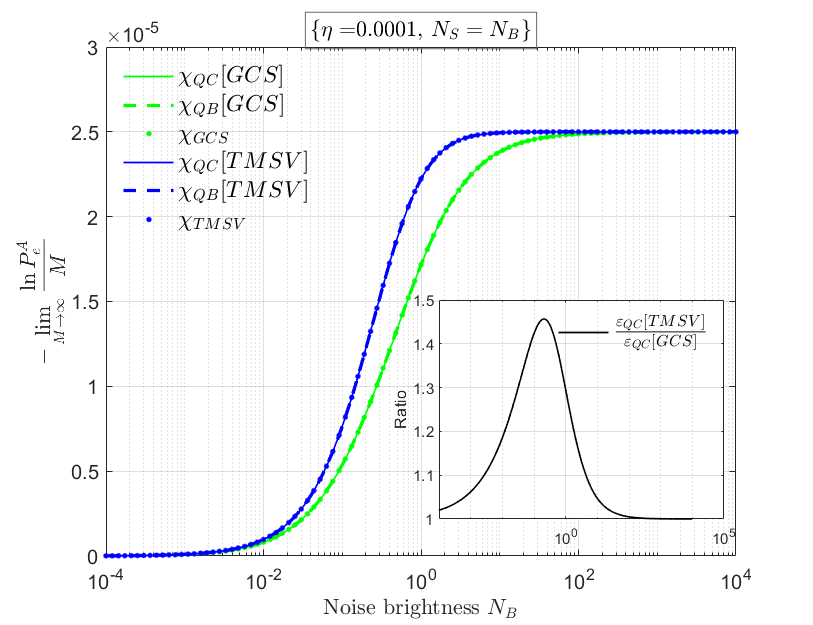}
    \end{subfigure}
    \captionsetup{width = \linewidth}
    \caption{Alice's quantum Chernoff (QC) (solid), quantum Bhattacharyya (QB) (dashed) and analytical quantum Bhattacharyya (following Eq.~(\ref{eq:perfcov_analytic_TMSV}) and Eq.~(\ref{eq:perfcov_analytic_GCS})) (dotted) error probability exponent as a function of noise brightness $N_B$ with target reflectivity of $\eta=10^{-2}$ (left) and $\eta=10^{-4}$ (right). Perfectly covert TMSV (blue) and GCS (green) probes are used. The inset shows the ratio of Chernoff exponent between TMSV and GCS probes, with a maximum value of $\simeq 1.45$ for $N_B\simeq 0.2$. }
    \label{fig:perfect_covert}
\end{figure}\noindent
For TMSV probe, Alice's target detection Chernoff error exponent can be computed by numerically optimising Eq.~(\ref{eq:Perfect_TMSV}) over all $0\leq s\leq 1$. A comparison with the Bhattacharyya exponent ($s=1/2$) shows that the Bhattacharyya exponent is sufficiently close to the Chernoff exponent (refer to Fig.~\ref{fig:perfect_covert}). Using the Bhattacharyya exponent, we thus approximate an analytical result in the scenario of a weakly reflecting target ($\eta\ll 1$),
\begin{align}
    \chi_{TMSV}\simeq -\ln\qty[1 - \frac{\eta}{4}\qty[1 - \frac{1}{(2N_B+1)^2}]]\label{eq:perfcov_analytic_TMSV}.
\end{align}
This analytical result was plotted against the Chernoff exponent and Bhattacharyya exponent as well, showing that it is sufficiently accurate in representing Alice's target detection error exponent when using TMSV probe across the range of noise brightness $N_B$ investigated.\\
For GCS, the optimisation of $s\in [0,1]$ for Alice's quantum Chernoff exponent inside the integral of Eq.~(\ref{eq:Perfect_GCS}) is dependent on the variable of integration $\abs{\alpha}$. However, using numerical approximation, we can obtain a highly accurate value of the Chernoff exponent. Comparing with its Bhattacharyya exponent, we conclude that $s=1/2$ is nearly optimal for all values of $\abs{\alpha}$ and $N_B$ for a weakly reflecting target $\eta \ll 1$ (refer to Fig.~\ref{fig:perfect_covert}). Hence, an analytical result of Alice's target detection error exponent when using GCS probe can be approximated as
\begin{align}
    \chi_{GCS}\simeq -\ln\qty[1 - 2\eta N_B\qty(N_B - \sqrt{N_B(N_B+1)} + \frac{1}{2})]\label{eq:perfcov_analytic_GCS}.
\end{align}
A detailed calculation of the process of deriving this result can be found in Appendix~\ref{appendixC}. Similarly, a comparison of the analytical result with the Chernoff and Bhattacharyya exponent was performed across the range of $N_B$, thereby proving its accuracy in representing Alice's error exponent when using GCS probe.\\
The ratio between the two analytical error exponents can be further approximated by performing Puiseux series expansion for $\eta\ll 1$, thereby giving
\begin{align}
    \frac{\chi_{TMSV}}{\chi_{GCS}}\simeq (N_B+1)\frac{2\qty(N_B + \sqrt{N_B(N_B+1)}) + 1}{(2N_B+1)^2}\geq 1, \quad \text{for $N_B\in\mathbb{R}_{\geq 0}$}\label{eq:perfectcovratio}.
\end{align}
This result indicates that TMSV probe is superior to the GCS probe in perfect covertness setup, with the maximum advantage occurring when the noise brightness $N_B\simeq 1/\sqrt{2} - 1/2$, while the exponents converge in the regime of small and large $N_B$. This result is consistent with the numerical outcome observed when evaluating the Chernoff exponents, as depicted in the inset of Fig.~(\ref{fig:perfect_covert}). In the regime where $N_B$ is small, the condition $N_S\ll 1$ necessary for achieving quantum advantage in target detection (similar to NPS QI \cite{tan2008quantum,nair2020fundamental}) coincides with the requirement for covertness ($N_S\simeq N_B$). However, this advantage is limited as GCS probes with the same $N_S$ exhibit similar levels of covertness and near-identical detection performance. When $N_B\gg 1$, the condition for optimal detection advantage ($N_S\ll 1$) conflicts with the requirement for covertness ($N_S\approx N_B$). Therefore, we expect the maximum quantum advantage to occur at some intermediate value of $N_B$. In this study, two different curves were plotted corresponding to two different reflectance values $\eta$ of the beam splitter. It is observed that both the quantum Chernoff and quantum Bhattacharyya metrics for both TMSV and GCS decrease proportionally with the decrease in $\eta$. However, the maximum advantage provided by the TMSV probe compared to the GCS probe remains consistent. Additionally, the value of $N_B$ at which this maximum advantage is achieved also remains constant, regardless of the changes in $\eta$. These can be explained from the fact that the ratio of the TMSV and GCS error exponents, as expressed in Eq.~(\ref{eq:perfectcovratio}), is independent of $\eta$.
\section{Generating functions}
In this research, we also established the multimode generalisation mathematical formulation, demonstrating that both the falling-factorial moment generating function (falling-factorial MGF) and rising-factorial moment generating function (rising-factorial MGF) can be expressed as a function of the photon-number probability generating function (PGF). Furthermore, we relate the output PGF and the input PGF of both pure loss and quantum-limited amplifier channels. Note that all the multimode generalisations derived as follows are based on the single-mode definitions \cite{haus2000electromagnetic}. 
\subsection{Three kinds of generating functions}
Suppose $\mathcal{H}$ is the Hilbert space of a $M$-mode quantum system, the PGF is expressed as a function of the $M$-vector $\xi=(\xi_1,\cdots,\xi_M)\in\mathbb{R}^M$,
\begin{align}
    \mathcal{P}(\xi)\coloneqq \sum_{\mathbf{n\geq 0}}p_{\mathbf{n}}\prod_{m=1}^M\xi_m^{n_m},
\end{align}
where $p_\mathbf{n}=\bra{\mathbf{n}}\rho\ket{\mathbf{n}}$ is the photon number PMF on the $M$ modes of the system. As the photon number PMF is normalised such that $\sum_\mathbf{n}p_\mathbf{n}=1$, the probability generating function converges absolutely (therefore exist) for all $\xi$ such that $\abs{\xi_m}\leq 1$ for all $m$ between $1$ and $M$. From the PGF, the photon-number PMF can be recovered as follows
\begin{align}
    p_\mathbf{n}=\frac{\left.\partial_1^{n_1}\cdots\partial_M^{n_m}\mathcal{P}(\xi)\right|_{\bm{\xi}\mathbf{=0}}}{n_1!\cdots n_M!},
\end{align}
where $\partial_m\equiv \partial/\partial\xi_m$. For a vector of non-negative integer $\mathbf{r}=(r_1,\cdots,r_M)$, the $\mathbf{r}$-th falling-factorial moment of the state $\rho$ is defined as
\begin{align}
    F_\mathbf{r}\coloneqq \Tr\qty[\rho\qty(\prod_{m=1}^M\hat{a}_m^{r_m\dag}\hat{a}_m^{r_m})],
\end{align}
where $\{\hat{a}_m\}_{m=1}^M$ are the annihilation operators of the $M$ modes of the system. The falling-factorial MGF, containing full information of all the falling-factorial moments, is thus 
\begin{align}
    \mathcal{F}(\xi)\coloneqq \sum_{\mathbf{r\geq 0}}F_{\mathbf{r}}\frac{\xi_1^{r_1}}{r_1!}\cdots\frac{\xi_M^{r_M}}{r_M!}\label{eq:falling_factorial_mgf},
\end{align}
which can be evaluated to obtain its relationship with the PGF
\begin{align}
    \mathcal{F}(\xi)&=\sum_{\mathbf{r\geq0}}\Tr\qty[\rho\qty(\prod_{m=1}^M\hat{a}_m^{r_m\dag}\hat{a}_m^{r_m})]\frac{\xi_1^{r_1}}{r_1!}\cdots\frac{\xi_M^{r_M}}{r_M!}\nonumber\\
    &=\sum_{\mathbf{r\geq0}}\qty[\sum_{\mathbf{n\geq 0}}p_\mathbf{n}\prod_{m=1}^Mn_m(n_m-1)\cdots(n_m-r_m+1)]\frac{\xi_1^{r_1}}{r_1!}\cdots\frac{\xi_M^{r_M}}{r_M!}\nonumber\\
    &=\sum_{\mathbf{n\geq 0}}p_{\mathbf{n}}\prod_{m=1}^M\qty[\sum_{r_m=0}^{n_m}\frac{\xi_m^{r_m}}{r_m!}n_m(n_m-1)\cdots(n_m-r_m+1)]\nonumber\\
    &=\sum_{\mathbf{n\geq 0}}p_{\mathbf{n}}\prod_{m=1}^M(1+\xi_m)^{n_m}=\mathcal{P}(1+\xi)\label{eq:ffmgf_pgf}.
\end{align}
Similarly, the $\mathbf{r}$-th rising factorial moment of $\rho$ for any $\mathbf{r}=(r_1,\cdots,r_M)$ is defined to be
\begin{align}
R_{\mathbf{r}}\coloneqq\Tr\qty[\rho\qty(\prod_{m=1}^M\hat{a}_m^{r_m}\hat{a}_m^{r_m\dag})].
\end{align}
Also, since the rising-factorial MGF contains all the information of the rising-factorial moment,
\begin{align}
    \mathcal{R}(\xi)\coloneqq\sum_{\mathbf{r\geq 0}}R_{\mathbf{r}}\frac{\xi_1^{r_1}}{r_1!}\cdots\frac{\xi_M^{r_M}}{r_M!}\label{eq:rising_factorial_mgf},
\end{align}
the relationship of rising-factorial MGF with PGF is thus
\begin{align}
     \mathcal{R}(\xi)&=\sum_{\mathbf{r\geq 0}}\Tr\qty[\rho\qty(\prod_{m=1}^M\hat{a}_m^{r_m}\hat{a}_m^{r_m \dag})]\frac{\xi_1^{r_1}}{r_1!}\cdots\frac{\xi_M^{r_M}}{r_M!}\nonumber\\
     &=\sum_{\mathbf{r\geq 0}}\qty[\sum_{\mathbf{n\geq 0}}p_{\mathbf{n}}\prod_{m-1}^M (n_m+1)(n_m+2)\cdots(n_m+r_m)]\frac{\xi_1^{r_1}}{r_1!}\cdots\frac{\xi_M^{r_M}}{r_M!}\nonumber\\
     &=\sum_{\mathbf{n\geq 0}}p_{\mathbf{n}}\prod_{m=1}^M\qty[\sum_{r_m=0}^\infty \frac{\xi_m^{r_m}}{r_m!}(n_m+1)(n_m+2)\cdots(n_m+r_m)]\nonumber\\
     &=\sum_{\mathbf{n\geq 0}}p_{\mathbf{n}}\prod_{m=1}^M\qty(\frac{1}{1-\xi_m})^{n_m+1}=\qty[\prod_{m=1}^M\qty(\frac{1}{1-\xi_m})]\mathcal{P}\qty(\frac{1}{1-\xi})\label{eq:rfmgf_pgf}.
\end{align}
Hence, any of these generating functions can be derived from any other.
\subsection{PGF of pure loss and quantum-limited channels output}\label{section:PGF_pl_qla}
Generalising Eq.~(\ref{eq:pureloss_transform}), for any arbitrary $M$-mode state sent as an input to a pure loss channel, the transformation follows
\begin{align}
    \hat{B}_i = \sqrt{\Tilde{\eta}}\hat{A}_i + \hat{N}_i,
\end{align}
where $\hat{A}_i$ and $\hat{B}_i$ refers to the input and output annihilation operators of mode $i\in[1,M]$ respectively, and $\hat{N}_i$ is the annihilation operator of noise. The commutator of the noise source is chosen so that the commutators are preserved in the passage through the pure-loss channel: $[\hat{N}_i,\hat{N}_i^\dag]= 1-\tilde{\eta}$. For a pure loss channel, since the noise source is a vacuum state, the falling-factorial moment of the output does not include contributions from the noise source. Therefore, it is practical to compute the falling-factorial moment for assessing the probability-generating function. The falling-factorial moment of the pure-loss channel output can be described in terms of its input falling-factorial moment
\begin{align}
    F_{\mathbf{r}}^{out} &=\expval{\qty(\hat{B}_1^\dag\hat{B}_1)^{r_1}\qty(\hat{B}_2^\dag\hat{B}_2)^{r_2}\cdots\qty(\hat{B}_M^\dag\hat{B}_M)^{r_M}}\nonumber\\
    &=\expval{\qty(\sqrt{\Tilde{\eta}}\hat{A}_1^\dag+\hat{N}_1^\dag)^{r_1}\qty(\sqrt{\Tilde{\eta}}\hat{A}_1+\hat{N}_1)^{r_1}\cdots\qty(\sqrt{\Tilde{\eta}}\hat{A}_M^\dag+\hat{N}_M^\dag)^{r_M}\qty(\sqrt{\Tilde{\eta}}\hat{A}_M+\hat{N}_M)^{r_M}}\nonumber\\
    &=\Tilde{\eta}^{r_1}\Tilde{\eta}^{r_2}\cdots\Tilde{\eta}^{r_M}\expval{\hat{A}_1^{r_1\dag}\hat{A}_1^{r_1}\hat{A}_2^{r_2\dag}\hat{A}_2^{r_2}\cdots\hat{A}_M^{r_M\dag}\hat{A}_M^{r_M}}\nonumber\\
    &=\Tilde{\eta}^{r_1}\Tilde{\eta}^{r_2}\cdots\Tilde{\eta}^{r_M}F_{\mathbf{r}}^{in}.
\end{align}
Using the definition of falling-factorial MGF from Eq.~(\ref{eq:falling_factorial_mgf}), we can relate the output falling-factorial MGF as a function of the input falling-factorial MGF
\begin{align}
    \mathcal{F}^{out}(\xi) &\coloneqq \sum_{\mathbf{r\geq 0}} F_{\mathbf{r}}^{out}\frac{\xi_1^{r_1}}{r_1!}\cdots\frac{\xi_M^{r_M}}{r_M!}\nonumber\\
    &=\sum_{\mathbf{r\geq 0}}F_{\mathbf{r}}^{in}\frac{(\Tilde{\eta}_1\xi_1)^{r_1}}{r_1!}\cdots\frac{(\Tilde{\eta}_M\xi_M)^{r_M}}{r_M!}\nonumber\\
    &=\mathcal{F}^{in}(\tilde{\eta}\xi).
\end{align}
Equating PGF to the falling-factorial MGF using Eq.~(\ref{eq:ffmgf_pgf}), we can thus relate the output PGF of a pure loss channel to its input PGF
\begin{align}
    \mathcal{F}^{out}(\xi) &= \mathcal{P}^{out}(1 + \xi)\\
    \mathcal{F}^{in}(\Tilde{\eta}\xi) &=\mathcal{P}^{in}(1 + \Tilde{\eta}\xi)\\
    \Rightarrow \mathcal{P}^{out}(\xi) &=\mathcal{P}^{in}(1+\Tilde{\eta}(\xi-1)).
\end{align}
Similarly, for a given quantum-limited amplifier channel, the $M$-mode input state undergoes the following transformation generalised from Eq.~(\ref{eq:quantumlimited_transform})
\begin{align}
    \hat{B}_i = \sqrt{G}\hat{A}_i + \hat{N}_i^\dag,
\end{align}
where the commutator of the noise source required for the conservation of the commutator from input to output is $[\hat{N}_i,\hat{N}_i^\dag]=1-G$. In the case of a quantum-limited amplifier channel, the rising-factorial moment of the output does not contain the noise source contribution. Therefore, it is practical to calculate the rising-factorial moment for the evaluation of the probability-generating function. Relating the output rising-factorial moment to the input rising-factorial moment of the quantum-limited amplifier channel,
\begin{align}
    R_{\mathbf{r}}^{out}&=\expval{\qty(\hat{B}_1\hat{B}_1^\dag)^{r_1}\qty(\hat{B}_2\hat{B}_2^\dag)^{r_2}\cdots\qty(\hat{B}_M\hat{B}_M^\dag)^{r_M}}\nonumber\\
    &=\expval{\qty(\sqrt{G}\hat{A}_1 + \hat{N}_1^\dag)^{r_1}\qty(\sqrt{G}\hat{A}^\dag_1 + \hat{N}_1)^{r_1}\cdots\qty(\sqrt{G}\hat{A}_M + \hat{N}_M^\dag)^{r_M}\qty(\sqrt{G}\hat{A}^\dag_M + \hat{N}_M)^{r_M}}\nonumber\\
    &=G^{r_1}G^{r_2}\cdots G^{r_M}\expval{\hat{A}_1^{r_1}\hat{A}_1^{r_1 \dag}\hat{A}_2^{r_2}\hat{A}_2^{r_2 \dag}\cdots\hat{A}_M^{r_M}\hat{A}_M^{r_M \dag}}\nonumber\\
    &=G^{r_1}G^{r_2}\cdots G^{r_M}R_{\mathbf{r}}^{in}.
\end{align}
Using Eq.~(\ref{eq:rising_factorial_mgf}), we obtain a relationship between the output rising-factorial MGF and the input PGF of the amplifier channel
\begin{align}
    \mathcal{R}^{out}(\xi) &\coloneqq \sum_{\mathbf{r\geq 0}} R_{\mathbf{r}}^{out}\frac{\xi_1^{r_1}}{r_1!}\cdots\frac{\xi_M^{r_M}}{r_M!}\nonumber\\
    &=\sum_{\mathbf{r\geq 0}}R_{\mathbf{r}}^{in}\frac{(G_1\xi_1)^{r_1}}{r_1!}\cdots\frac{(G_M\xi_M)^{r_M}}{r_M!}\nonumber\\
    &=\mathcal{R}^{in}(G\xi).
\end{align}
Hence, by relating the PGF to the rising-factorial MGF using Eq.~(\ref{eq:rfmgf_pgf}), we find the output PGF of a quantum-limited amplifier channel as a function of its input PGF
\begin{align}
    \mathcal{R}^{out}(\xi) &=\qty[\prod_{m=1}^M\qty(\frac{1}{1-\xi_m})]\mathcal{P}\qty(\frac{1}{1-\xi}) = \qty(\frac{1}{1-\xi})^M\mathcal{P}^{out}\qty(\frac{1}{1-\xi})\\
    \mathcal{R}^{in}(G\xi)&=\qty[\prod_{m=1}^M\qty(\frac{1}{1-G\xi_m})]\mathcal{P}^{in}\qty(\frac{1}{1-G\xi}) = \qty(\frac{1}{1-G\xi})^M\mathcal{P}\qty(\frac{1}{1-G\xi})\\
    \Rightarrow \mathcal{P}^{out}(\xi) &= \qty[\frac{1}{G - \xi(G - 1)}]^M\mathcal{P}^{in}\qty(\frac{\xi}{G - \xi(G - 1)}).
\end{align}

\section{\texorpdfstring{$\epsilon$}{epsilon}-covertness}
By relaxing the condition of covert operation to permit a non-zero $\epsilon$ for the error probability of Willie's detection, our focus shifts to understanding the spectrum of permissible probe energies. Additionally, we aim to explore the optimal levels of target detection performance that Alice can attain under these conditions. In essence, this involves investigating the trade-off between probe energy levels, Alice's target detection performance and the effectiveness of maintaining covert operations, considering a margin of detection tolerance represented by the parameter $\epsilon$.
\subsection{Bounds on transmitter's energy}
In the previous section, we examined the utilization of $M$-mode quantum and classical probes for maintaining perfect covertness against the adversary Willie. This requires an energy expenditure of $MN_B$ photons. However, by introducing a relaxation to the covertness condition, allowing for $\epsilon>0$, there exists a spectrum of probe energies that satisfies the $\epsilon$-covertness criterion. By using the necessary condition of $\epsilon$-covertness, we obtain an upper and lower limit on Alice probe energy beyond which covert operations are compromised. Mathematically, we can extremise the average energy $\sum_{n=0}^\infty nq_n$ of the state intercepted by Willie for alternate hypothesis $\sigma_1$ under the constraint of Eq.~(\ref{eq:ecovert_constraint}). The corresponding range of probe energies $\mathcal{N}_S = \qty[\sum_{n=0}^\infty n q_n - \eta N_B]/(1-\eta)$ computed using Eq.~(\ref{eq:adversarytransform}) can thus be derived.
\begin{align}
&\textbf{Extremise }f(q_1,q_2,\cdots)=\sum_{n=0}^\infty nq_n\nonumber\\
&\textbf{subject to }\nonumber\\
&\quad g(q_1,q_2,\cdots)=1-2\epsilon-\sum_{n=0}^\infty\sqrt{\binom{n+M-1}{n}\frac{N_B^n}{(N_B+1)^{n+M}}q_n}\leq 0\nonumber\\
&\quad h(q_1,q_2,\cdots)=\sum_{n=0}^\infty q_n-1=0\nonumber
\end{align}
where $g(q_1,q_2,\cdots)$ is the covertness constraint and  $\{q_n\}$ is the PMF of Willie's total photon number under $\mathrm{H}_1'$. Applying the Karush–Kuhn–Tucker (KKT) approach to non-linear programming, we obtain two equations
\begin{align}
    &\sum_{n=0}^d\mathcal{N}_1\frac{\lambda_1^2}{4(n-\lambda_2)^2}\qty[\binom{n+M-1}{n}\frac{N_B^n}{(N_B+1)^{n+M}}]=1\label{eq:KKT_1}\\
&\sum_{n=0}^d \mathcal{N}_2\qty[\binom{n+M-1}{n}\frac{N_B^n}{(N_B+1)^{n+M}}]\abs{\frac{\lambda_1}{2(n-\lambda_2)}}=1-2\epsilon\label{eq:KKT_2},
\end{align}
where $d$ is the finite upper summation limit, $\mathcal{N}_1$ and $\mathcal{N}_2$ is the re-normalisation factor for finite summation (details of KKT condition analysis can be found in Appendix~\ref{appendixD}). The simultaneous equations can be solved numerically by selecting a sufficiently large $d$, and appropriate initial values for the KKT multipliers $\lambda_1$ and $\lambda_2$ using \textsc{Matlab}.
\begin{figure}[h!]
    \centering
    \includegraphics[width = 0.8\linewidth]{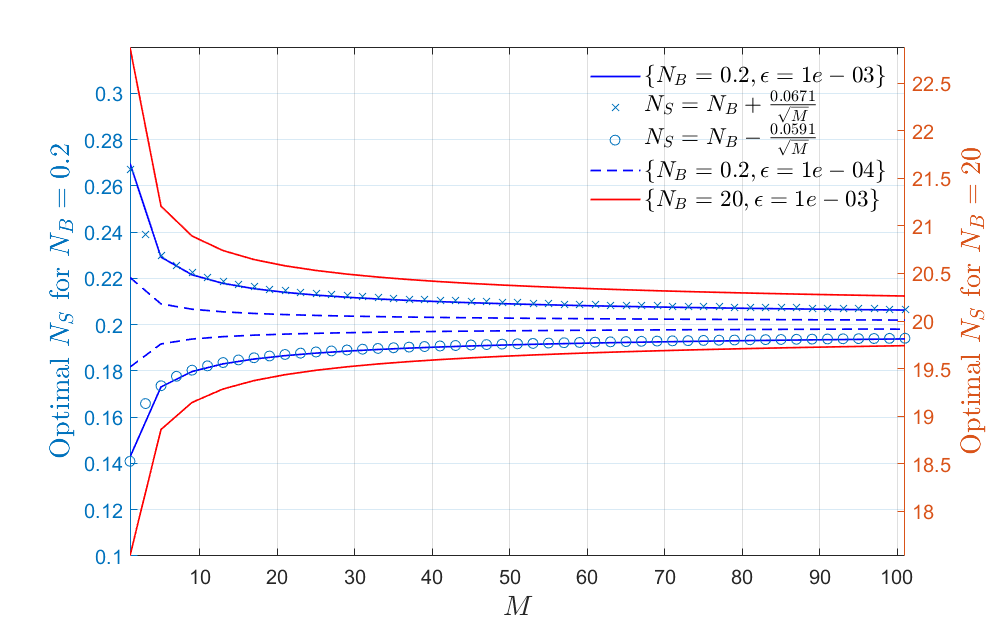}
    \captionsetup{width = \linewidth}
    \caption{Upper and lower bounds of allowable per-mode energy $N_S$ for an $\epsilon$-covert probe according to Eq.~(\ref{eq:ecovert_constraint}) as a function of $M$ for $\{N_B=0.2,\epsilon = 10^{-3}\}$ (solid blue). Curve-fitting the numerical data generate estimated functions $N_S = N_B + 0.0671/\sqrt{M}$ (maximum energy) and $N_S = N_B - 0.0591/\sqrt{M}$ (minimum energy). The allowable ranges of per-mode probe energies $N_S$ for $\{N_B=0.2,\epsilon = 10^{-4}\}$ (dashed blue) and $\{N_B=20,\epsilon = 10^{-3}\}$ (solid red) are also shown. The reflectivity of target $\eta = 0.01$ for all curves.}
    \label{fig:signalenergy}
\end{figure}
Due to the nature of the simultaneous equations, we are unable to provide a viable analytical solution to the energy bounds of a $\epsilon$-covert probe $\mathcal{N}_S$ as a function of the system parameters $\{\epsilon,\eta,M,N_B\}$. However, numerical results (details on the derivation of the allowable per-mode energy $N_S$ from the numerical optimisation can be found in Appendix~\ref{appendixD} ) suggest that the per-mode allowable energies for $\epsilon$-covert probe follow the relation $N_S\sim N_B\pm A_{\pm}/\sqrt{M}$, where the fitting parameter $A_{\pm}$ depend only on $\eta,N_B$ and $\epsilon$ (See Fig.~\ref{fig:signalenergy}). Therefore, considering the specific parameters of the covert sensing setup, we can derive a numerical solution. Within this solution, any energy values that lie outside the boundaries defined by the generated curve are conclusively determined not to meet the $\epsilon$-covert criteria. In other words, this numerical analysis allows us to identify and exclude probe energies that do not align with the specified constraints, providing a clear distinction for what can be considered non-covert in the context of the given $\epsilon$ threshold.

\subsection{Bounds on Alice's error probability}
Previously, we found the numerical solution to the bounds of Alice's probe energies for which beyond the limits, the $\epsilon$-covert constraint is violated. In this section, we find the lower bound on Alice's target detection error probability $P_e^A$ that must be satisfied under $\epsilon$-covertness criteria. The thermal loss channel $\mathcal{L}_{1-\eta,N_B}^{\otimes M}$ of Willie's alternate hypothesis $\textsc{H}'_1$ can be expressed as a concatenation of a pure-loss channel followed by a quantum-limited amplifier channel based to Eq.~({\ref{eq:concatenationofchannel}})
\begin{align}
    \mathcal{L}_{1-\eta,N_B}^{\otimes M}=\mathcal{A}^{\otimes M}_{G_W}\circ\mathcal{L}^{\otimes M}_{\eta_W},
\end{align}
where $G_W = \eta N_B+1$ is the gain parameter of the quantum-limited amplifier, and $\eta_W = (1-\eta)/G$ is the transmittance of the pure loss channel \cite{nair2020fundamental}. From Section~\ref{section:PGF_pl_qla}, we have derived the relationship between the PGF of the input and output state of both pure loss and quantum-limited amplifier channels. Hence, by evaluating the results from the two channels, the PGF at the output of a thermal loss channel can be expressed as a function of its input PGF following:
\begin{align}
    \mathcal{P}^{out}(\xi) = \qty[\frac{1}{G_W - \xi(G_W - 1)}]^M\mathcal{P}^{in}\qty(1 -\eta_W + \frac{\eta_W\xi}{G_W - \xi(G_W - 1)})\label{eq:thermalloss_pgf}.
\end{align}
Re-writing the $\epsilon$-covertness criteria of Eq.~(\ref{eq:ecovert_constraint}),
\begin{align}
    \sum_{n=0}^\infty \sqrt{\binom{n+M-1}{n}\frac{N_B^n}{(N_B+1)^{n+M}}x^{-n}x^nq_n}\geq 1 - 2\epsilon\label{eq:ecovertness_pgf},
\end{align}
where $x$ will be evaluated later, we can invoke the Cauchy-Schwartz inequality to factorise the summation
\begin{align}
     &\qty(\sum_{n=0}^\infty\binom{n+M-1}{n}\frac{N_B^n}{(N_B+1)^{n+M}}x^{-n})^{1/2}\qty(\sum_{n=0}^\infty q_nx^n)^{1/2} \nonumber\\
     &\hspace{150pt}\geq\sum_{n=0}^\infty \sqrt{\binom{n+M-1}{n}\frac{N_B^n}{(N_B+1)^{n+M}}x^{-n}x^nq_n}\nonumber\\
     &\qty(\sum_{n=0}^\infty\binom{n+M-1}{n}\frac{N_B^n}{(N_B+1)^{n+M}}x^{-n})^{1/2}\qty(\sum_{n=0}^\infty q_nx^n)^{1/2}\geq 1- 2\epsilon\nonumber\\
     &\qty(\frac{1}{N_B+1-\frac{N_B}{x}})^M\mathcal{P}_W(x)\geq (1-2\epsilon)^2\label{eq:ecovertness_converg},
\end{align}
where $\mathcal{P}_W(x)$ is the PGF of the intercepted state for Wille's alternate hypothesis $\sigma_1$. To ensure the absolute convergence of the first summation, we must have $N_B/\abs{x}\leq N_B+1$. Using Eq.~(\ref{eq:thermalloss_pgf}), the above inequality is expressed in terms of the PGF $\mathcal{P}_S$ of the probe sent by Alice
\begin{align}
    \mathcal{P}_S\qty(1+\frac{(x-1)(1-\eta)}{\eta N_B(1-x) + 1})\geq (1-2\epsilon)^2\qty(N_B+1-\frac{N_B}{x})^M[\eta N_B(1-x)+1]^M\label{eq:probe_pgf1}.
\end{align}
In Section~\ref{section:limitsofQI_PS}, we have derived the lower bound of Alice's target detection error probability as a function of the fidelity between the states returned to Alice for respective hypothesis $P_e^A\geq 1/2 - \sqrt{1-\mathcal{F}^2(\rho_0,\rho_1)}/2$, where the fidelity can be expressed as a lower bound of the probe PGF following: 
\begin{align}
    \mathcal{F}(\rho_0,\rho_1) & \geq \nu^M\sum_{n=0}^\infty p_n\qty[1-\frac{\eta}{(1-\eta)N_B+1}]^{n/2}\nonumber\\
    &= \nu^M \mathcal{P}_S\qty(\qty(1 - \frac{\eta}{(1-\eta)N_B+1})^{1/2})\label{eq:probe_pgf2}.
\end{align}
Equating Eq.~(\ref{eq:probe_pgf1}) and Eq.~(\ref{eq:probe_pgf2}), we can hence solve for the independent variable $x$ 
\begin{align}
    & 1 + \frac{(x-1)(1-\eta)}{\eta N_B(1-x) + 1} = \qty[1 - \frac{\eta}{(1-\eta)N_B + 1}]^{1/2}\\
    \Rightarrow &x = 1 - \frac{\Theta}{\eta[1 + N_B(1-\Theta)]},\quad \text{for }\Theta = \frac{\sqrt{(1-\eta)(N_B+1)}}{\sqrt{1 + (1-\eta)N_B}},
\end{align}
and applying the condition to ensure absolute convergence of the first summation in Eq.~(\ref{eq:ecovertness_converg}) yields
\begin{align}
    \frac{N_B}{N_B+1}\leq 1 - \frac{\Theta}{\eta[1 + N_B(1 - \Theta)]} \leq 1.
\end{align}
In the limit of target with small reflectivity such that $\eta \ll 1$, we can apply Puiseux series expansion on $x$ about the expansion point $\eta = 0$ to obtain
\begin{align}
    x = 1 - \frac{\eta}{2(N_B+1)} + \mathcal{O}(\eta^2),
\end{align}
which can be deduced that in this regime, the convergence condition is always true for finite values of $N_B$. Numerically, the convergence condition is satisfied for $\eta\lesssim 0.4$. Hence, the lower bound of the fidelity between Alice's received states can be expressed as
\begin{align}
    \mathcal{F}(\rho_0,\rho_1) \geq (1 - 2\epsilon)^2\nu^M\qty(N_B+1-\frac{N_B}{x})^M[\eta N_B(1-x) + 1]^M\label{eq:Alice_fidelitybound_PGF}
\end{align}
where $x=1 - \Theta/[\eta[1 + N_B(1-\Theta)]]$. In Appendix~\ref{appendixE}, we show that this bound is sufficiently tight in the regime of interest. Applying the Fuchs-van de Graaf inequality, we thus derive a probe-independent analytical lower bound on Alice's target detection error probability under $\epsilon$-covertness criteria
\begin{align}
    P_e^A\geq \frac{1 - \sqrt{1 - (1-2\epsilon)^4\mathcal{F}(\eta,N_B,M)}}{2}\label{eq:ecovert_performance_bound},
\end{align}
where $\mathcal{F}(\eta, N_B,M)$ is a function dependent on the parameter of quantum illumination setup
\begin{align}
    \mathcal{F}(\eta,N_B,M)= \nu^{2M}\qty(N_B+1-\frac{N_B}{x})^{2M}[\eta N_B(1-x)+1]^{2M}.
\end{align}
As previously stated, TMSV and GCS are the two states representing quantum and classical probes respectively that allow for perfect covertness detection. In this context, our focus is directed towards assessing and contrasting their target detection performance with the analytical bound derived in Equation (\ref{eq:ecovert_performance_bound}) under $\epsilon$-covert constraint. This comparison aims to gauge the proximity of these states to optimality, providing insights into the effectiveness of TMSV and GCS in achieving the best possible target detection performance within the given $\epsilon$-covert framework. When either TMSV or GCS states are used as probes by Alice, the states intercepted by Wille will be thermal states. To ensure the compliance with $\epsilon$-covertness condition, i.e., $P_e^W\geq 1/2 - \epsilon$, we evaluate the Helstrom bound on Willie's error probability and require that:
\begin{align}
    \frac{1}{2} - \frac{1}{4}\norm{\sigma_0-\sigma_1}_1&\geq \frac{1}{2}-\epsilon\nonumber\\
    \norm{\sigma_0-\sigma_1}_1&\leq 4\epsilon.
\end{align}
The trace norm between the two states intercepted by Willie can be calculated as
\begin{align}
    \norm{\sigma_0-\sigma_1}_1 &= \Tr\qty[\sum_{\mathbf{n}\geq\mathbf{0}}\abs{\frac{N_B^n}{(N_B+1)^{n+M}}-\frac{N^n}{(N+1)^{n+M}}}\ket{\mathbf{n}}\bra{\mathbf{n}}]\nonumber\\
    &=\sum_{n=0}^\infty\binom{n+M-1}{n}\abs{\frac{N_B^n}{(N_B+1)^{n+M}}-\frac{N^n}{(N+1)^{n+M}}}\nonumber\\
    &=\sum_{n=0}^{n_t}\binom{n+M-1}{n}\frac{N_B^n}{(N_B+1)^{n+M}} -\sum_{n=0}^{n_t}\binom{n+M-1}{n}\frac{N^n}{(N+1)^{n+M}}\nonumber\\
    &\hspace{40pt}+\sum_{n=n_t+1}^\infty\binom{n+M-1}{n}\frac{N^n}{(N+1)^{n+M}}-\sum_{n=n_t+1}^\infty\binom{n+M-1}{n}\frac{N_B^n}{(N_B+1)^{n+M}},
\end{align}
where $n\coloneqq\tr\mathbf{n}=\sum_{m=1}^Mn_m$, $N = (1-\eta)N_S + \eta N_B$, and $n_t = \left\lfloor \left.M\ln\qty[\frac{N+1}{N_B+1}]\right/\ln\qty[\frac{N(N_B+1)}{(N+1)N_B}]\right\rfloor$. Solving it numerically would require the evaluation of the binomial coefficient up to a very large value of $M$ and $n$, usually resulting in numerical overflow. As such, a trick is deployed where we instead calculate the logarithm of the binomial coefficient. A brief explanation of the method used to compute the first summation of the trace norm is discussed as follows. The expression is first re-expressed in the following form
\begin{align}
    \sum_{n=0}^{n_t}\binom{n+M-1}{n}\frac{N_B^n}{(N_B+1)^{n+M}}&=\sum_{n=0}^{n_t}e^{\ln\qty[\binom{n+M-1}{n}\frac{N_B^n}{(N_B+1)^{n+M}}]}.
\end{align}
The exponent is evaluated as
\begin{align}
    \ln\qty[\binom{n+M-1}{n}\frac{N_B^n}{(N_B+1)^{n+M}}]&=\ln\qty[\frac{\Gamma(n+M)}{\Gamma(n+1)\Gamma(M)}\frac{N_B^n}{(N_B+1)^{n+M}}]\nonumber\\
    &=\ln\Gamma(n+M) - \ln\Gamma(n+1) - \ln\Gamma(M) \nonumber\\
    &\hspace{70pt}+ n\ln N_B - (n+M)\ln(N_B+1),
\end{align}
where the gamma function $\Gamma(x) = (x-1)!$ can be approximated to a high order of accuracy using Lanczos approximation. Deploying the same method to evaluate the summations for the other three terms and selecting a sufficiently large value for the upper limit of summation, we can thus compute the error probability for given values of $\eta, M, N_B$, and $N_S$. For fixed $\eta, M$ and $N_B$, we can hence vary $N_S$ to obtain the optimal strength of the probe state $N_S$. This implies that TMSV or GCS probes with brightness up to this maximum value are assured to satisfy $\epsilon$-covert criteria. Subsequently, using the quantum Chernoff bound, we can perform numerical calculations to determine Alice's target detection performance when employing TMSV and GCS probes. 
\begin{figure}[t!]
\centering
\begin{subfigure}
        \centering
    \includegraphics[width=\linewidth]{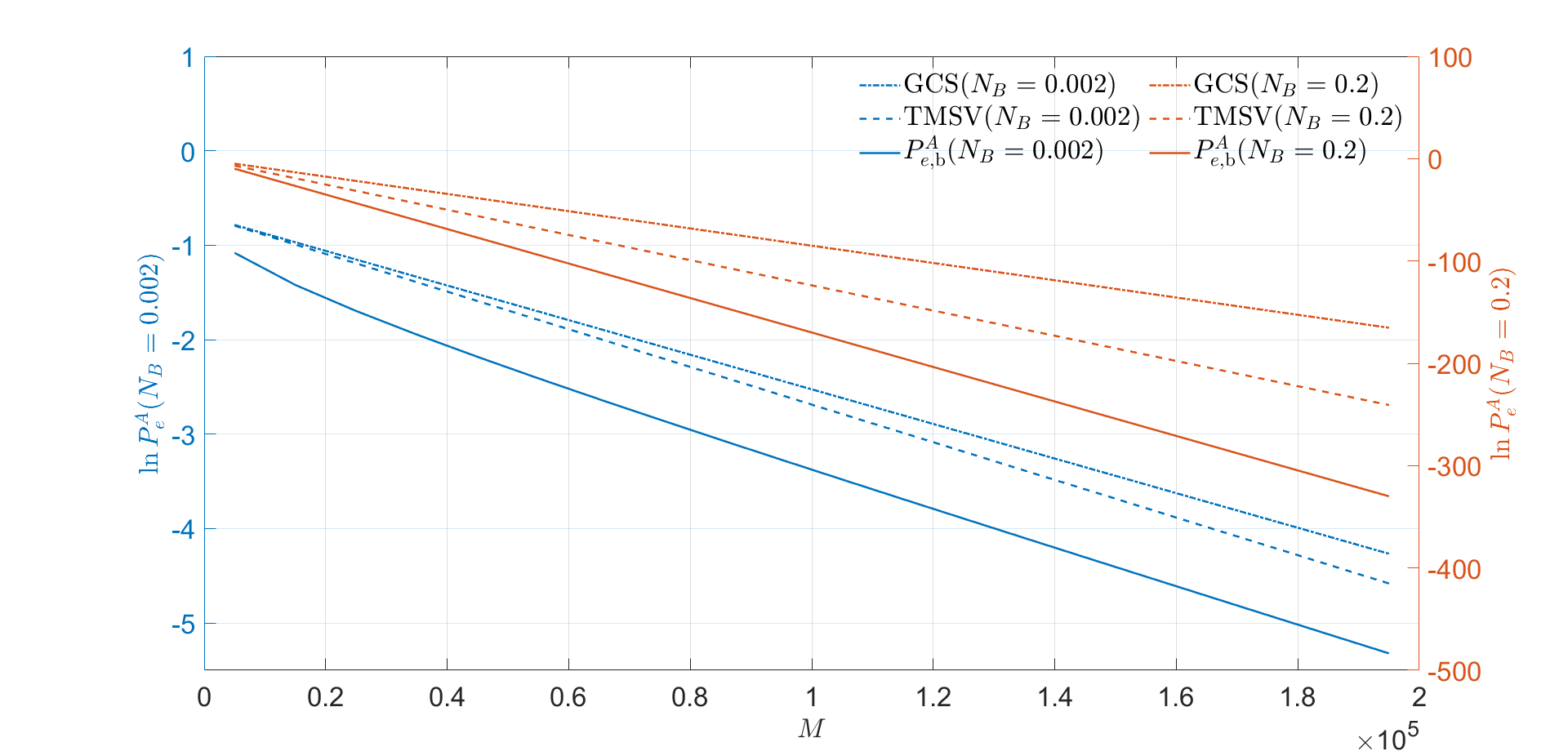}
\end{subfigure}
    \captionsetup{width=\linewidth}
    \caption{The lower bound  Eq.~\eqref{eq:ecovert_performance_bound} (solid) on Alice's error probability is compared to that of $\epsilon$-covert TMSV (dashed) and GCS probes ( dash-dotted line) for $N_B =0.2$ (blue) and $N_B = 0.002$ (red). $\epsilon=10^{-3}$ for both. For large $M$, the ratio between the error exponents predicted by the bound \eqref{eq:ecovert_performance_bound} and using TMSV probes are $1.37$ (for  $N_B = 0.2$)  and $1.16$ (for $N_B=0.002$) respectively.}
    \label{fig:TMSVvsTheoretical}
\end{figure}
Figure \ref{fig:TMSVvsTheoretical} compares the theoretical lower limit of Alice's target detection error probability, as defined by Eq.~(\ref{eq:ecovert_performance_bound}), with her performance when utilizing independent and identically distributed (iid) $\epsilon$-covert TMSV and GCS probes. Across various values of $N_B$ investigated, a discernible trend emerged: for smaller values of $N_B$, TMSV tends to closely approach the theoretical bound. This is demonstrated by the comparison of the error exponent ratios predicted by the analytical bound with those of the TMSV probe, yielding values of $1.37$ for  $N_B = 0.2$ and $1.16$ for $N_B=0.002$. Comparing the performance between classical (GCS) and quantum (TMSV) probes, we observed the greatest quantum advantage occurring at $N_B=0.2$, with their performance converging for large and small $N_B$. This result is consistent with the result obtained in the perfect covertness setup as evident from Fig.~\ref{fig:perfect_covert}. From the result, we have shown the capability of TMSV probes to operate close to the theoretically defined optimal performance in $\epsilon$-covert scenario. Furthermore, while the quantum advantage is most pronounced at $N_B=0.2$, it remains a robust feature across varying levels of background noise, contributing to the versatility and effectiveness of quantum probes in covert communication setups.

\section{Summary}
In this study, we introduced the natural passive signature model of illumination, providing a framework to understand and analyse covert sensing and detection protocols. We derived fidelity bound for the outputs of thermal loss channels, offering possible valuable insights into the characterisation of scheme such as quantum reading \cite{pirandola2011quantum}, pattern recognition \cite{banchi2020quantum} and channel position finding \cite{zhuang2020entanglement,pereira2021idler}. Additionally, we established a lower bound for the performance of Alice's target detection under $\epsilon$-covertness assumptions. This bound serves as a fundamental limit across all parameter ranges, providing a crucial benchmark for evaluating the effectiveness of covert communication.\\
Our findings revealed that, under $\epsilon$-covertness assumptions, the quantum advantage observed for quantum probes is modest when compared to classical probes, with the advantage observed in a different range than the $N_S\ll 1$, $N_B\gg 1$ regime of NPS model. The nuances in the quantum advantage underscore the complexity of covert sensing scenarios and highlight the importance of considering specific parameter regimes for optimal probe performance.\\
We hope that the novel technique introduced in deriving performance bounds under $\epsilon$-covertness will stimulate further research in the field of covert sensing protocols such as phase and transmittance sensing \cite{bash2017fundamental,gagatsos2019covert,hao2022demonstration}. The fundamental limits established in this study provide a solid foundation for exploring and refining covert communication strategies, opening avenues for advancements in the design and implementation of secure and efficient sensing systems.

    \chapter[Optimal gain sensing of quantum-limited phase-insensitive amplifiers]{Optimal gain sensing of quantum-limited phase-insensitive amplifiers}\label{chap-5}
\emph{This chapter focuses on understanding the precision boundaries in estimating the gain of phase-insensitive optical amplifiers using quantum approaches.\footnote{Sections of this chapter have been referenced from our published article of Ref.~ \cite{nair2020fundamental}.} These amplifiers uniformly amplify each quadrature of an input field, holding significant value for both foundational research and practical technological applications. The research investigates the quantum limit for gain estimation precision using a multimode, possibly entangled, probe and ancilla system. The chapter further identifies that all pure-state probes, whose reduced state on the input modes is diagonal in the multimode number basis, achieve quantum optimality under the same gain-independent measurement. A comparison is made between the precision attainable with classical probes to the performance of a photon-counting-based estimator in quantum probes. A closed-form expression for the energy-constrained Bures distance between two product amplifier channels is also presented.}
\newpage
\section{Motivations and objectives}
The impetus for this research is anchored in the critical examination of phase-insensitive amplifiers, quintessentially exemplified by laser gain media with population inversion, a cornerstone in both laser technology and contemporary optical communication networks. Such amplifiers, notably including erbium-doped fibre amplifiers, are indispensable in today's optical communications for their role in signal amplitude restoration and offset detection noise \cite{ramaswami2009optical}. Despite the diversity of mechanisms inducing phase-insensitive amplification across varied platforms \cite{clerk2010introduction,caves2012quantum,lahteenmaki2014advanced,chia2020phase}, a commonality persists – they are all invariably subject to the constraints imposed by the unitarity of quantum dynamics, necessitating the incorporation of gain-dependent excess noise \cite{clerk2010introduction,haus1962quantum,caves1982quantum}. This noise is minimised under conditions of complete inversion of the effective population in the active levels of the gain medium \cite{caves2012quantum,agarwal2012quantum}. Such minimum-noise phase-insensitive amplifiers are known as quantum-limited amplifiers (QLAs), and the significance of QLAs transcends their practical utility, underpinning the foundational aspects of continuous-variable quantum information. They are instrumental in constructing phase-covariant Gaussian channels by concatenation with pure-loss channels \cite{caruso2006one,serafini2017quantum}.\\
Despite the extensive research landscape encompassing the sensing of loss channels (see, e.g., \cite{nair2018quantum,pirandola2018advances,braun2018quantum,polino2020photonic} and references therein), a comprehensive understanding and optimisation of gain sensing precision within QLAs have not been adequately addressed. Previous work on sensing gain of a QLA is limited to the context of detecting Unruh-Hawking radiation using single-mode probes \cite{aspachs2010optimal} or assumes access to the internal degrees of freedom of the amplifier \cite{gaiba2009squeezed}. This research is dedicated to filling this research gap by optimizing the precision of gain sensing in QLAs. This entails an exploration encompassing all multimode ancilla-entangled probes and joint quantum measurements, bound only by the constraints of energy and the number of input modes of the probe. \\
The overarching objective of this research is to propose and delineate practical probes, measurements, and estimators that are not only theoretically robust but also amenable to empirical demonstration using present-day technology, specifically considering the prevalent limitations of nonunity-efficiency photodetection in contemporary experimental frameworks. This endeavour is expected to elucidate a quantum advantage in gain sensing with QLAs, contributing to the theoretical advancement for a vast suite of detection and estimation problems involving Gaussian channels with excess noise - see e.g., Ref.~\cite{tan2008quantum,nair2020fundamental,bradshaw2021optimal,gregory2020imaging,pirandola2011quantum,ortolano2021experimental,monras2011measurement,sharma2018bounding,sharma2022optimal,wang2020quantum,jonsson2022gaussian,zhuang2020entanglement,harney2021ultimate,zhuang2021quantum,bash2017fundamental,tahmasbi2021signaling}. The ambition of this work is to enhance the efficacy and precision of quantum information processing and communication systems, leveraging the unique capabilities of quantum technologies in the realm of optical amplification.

\section{Gain sensing protocol}
\begin{figure}[h!]
    \centering
    \includegraphics[width = 0.9\linewidth]{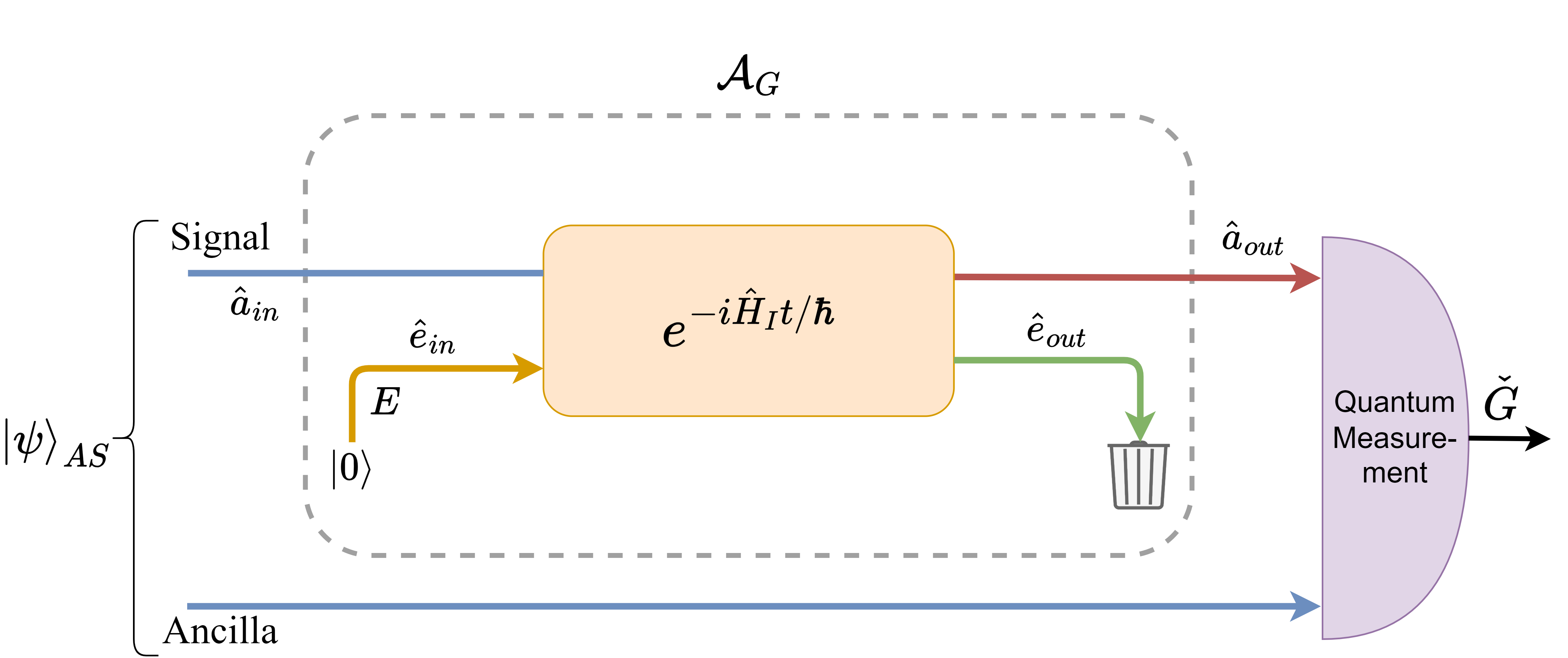}
    \captionsetup{width=\linewidth}
    \caption{General setup of an ancilla-assisted parallel strategy for sensing of gain $G$ of a quantum-limited amplifier (QLA) channel $\mathcal{A}_G$ (dashed box). A joint state $\ket{\psi}_{AS}$ with $M$ signal and idler mode prepared, with each signal (S) mode $\hat{a}_{in}$ subjected to a two-mode squeezing interaction $\hat{H}_I = i\hbar\kappa(\hat{a}\hat{e}-\hat{a}^\dag\hat{e}^\dag)$ with an environment (E) mode $\hat{e}_{in}$ initialised in the vacuum state.  From the optimal joint measurement on the output $\hat{a}_{out}$ and the ancilla, an estimate $\check{G}$ of $G = \cosh^2\kappa t$ is obtained}
    \label{fig:setup_gain}
\end{figure}\noindent
Figure~\ref{fig:setup_gain} depicts a general setup for an ancilla-assisted parallel estimation strategy for gain sensing. A pure state $\ket{\psi}_{AS}$ (called the probe) of $M$ pairs of entangled signal (S) and arbitrary ancilla (A) modes is prepared, with each signal mode passing through the QLA. The output from the QLA is jointly measured with the ancilla modes using an optimal (possibly probe-dependent) measurement for estimating $G$. The probe prepared has the general form
\begin{align}
    \ket{\psi}_{AS}=\sum_{\mathbf{n\geq 0}}\sqrt{p_{\mathbf{n}}}\ket{\chi_{\mathbf{n}}}_A\ket{\mathbf{n}}_S\label{eq:pure_NDS},
\end{align}
where $\ket{\mathbf{n}}=\ket{n_1}_{S_1}\ket{n_2}_{S_2}\cdots\ket{n_M}_{S_M}$ is a $M$-mode number state of the signal system, $\{\ket{\chi_{\mathbf{n}}}_A\}$ are normalised (not necessarily orthogonal) states of $A$, and $\{p_{\mathbf{n}}\geq 0\}$ is the probability distribution of $\mathbf{n}$. The number of available signal modes $M$ depends on operational constraints such as measurement time and bandwidth and will be later proven to be fundamental in determining the sensing precision. A constraint on the average photon number in the signal modes is imposed such that $\bra{\psi}\mathbb{I}_A\otimes\qty(\sum_{m=1}^M\hat{N}_m)\ket{\psi}=N$, where $\hat{N}_m=\hat{a}_m^\dag\hat{a}_m$ is the number operator of the $m$-th signal mode and $\mathbb{I}_A$ is the identity operation on the ancilla system. Rephrasing the constraint to be dependent on the probability distribution of $\mathbf{n}$: $\sum_{n=0}^\infty n p_n = N$, where $p_n=\sum_{\mathbf{n }\; :\;n_1+\cdots+n_M=n}p_{\mathbf{n}}$ is the probability mass function of the total photon number in the signal modes. Since a mixed-state probe can be purified using an additional ancilla, leading to the purification once again taking the form Eq.~(\ref{eq:pure_NDS}) with the same value of $N$ and $M$, the optimisation over probes of the form of Eq.~(\ref{eq:pure_NDS}) is sufficient. \\
We are interested in comparing the performance of quantum probe of the form of Eq~(\ref{eq:pure_NDS}) to the best performance achievable using classical probe under the same resource constraints, i.e. probes that consist of mixtures of $M$-mode coherent states, possibly correlated with an arbitrary number $M'$ of ancilla modes. Such classical probes can be prepared using a laser source, and mathematically presented as
\begin{align}
    \rho_{AS} &=\iint d^{2M'}\bm{\alpha}d^{2M}\bm{\beta}\;P(\bm{\alpha},\bm{\beta})\ket{\bm{\alpha}}\bra{\bm{\alpha}}_A\otimes\ket{\bm{\beta}}\bra{\bm{\beta}}_S\label{eq:classical_probe},
\end{align}
where $\bm{\alpha} = \qty(\alpha^{(1)},\cdots,\alpha^{(M')})\in\mathbb{C}^{M'}$ and $\bm{\beta} = \qty(\beta^{(1)},\cdots,\beta^{(M)})\in\mathbb{C}^{M}$ index $M'-$ and $M-$mode coherent states of ancilla and signal system respectively, and $P(\bm{\alpha},\bm{\beta})\geq 0$ is the joint probability distribution. The signal energy constraint then takes the form $\int_{\mathbb{C}^{M'}}d^{2M'}\bm{\alpha}\;\int_{\mathbb{C}^{M}}d^{2M}\bm{\beta}\;P(\bm{\alpha},\bm{\beta})\qty(\sum_{m=1}^M\abs{\bm{\beta}^{(m)}}^2)=N$. Hence, for both quantum and classical probe, the output states from the QLA are respectively:
\begin{align}
    &\text{Quantum: }\rho_G = \qty(\mathbb{I}_A\otimes\mathcal{A}_G^{\otimes M})\ket{\psi}\bra{\psi}_{AS},\\
    &\text{Classical: }\rho_G = \qty(\mathbb{I}_A\otimes\mathcal{A}_G^{\otimes M})\rho_{AS},
\end{align}
where $\mathbb{I}_A$ is the identity channel on ancilla system, $G = \cosh^2\kappa t\equiv\cosh^2\tau$ is the gain parameter of the QLA. Estimation of the gain parameter $G$ is subjected to the quantum Cram\'er-Rao bound (QCRB) \cite{barnett2002methods}, defined previously in Eq.~(\ref{eq:QCRB}).
\section{Optimal gain sensing}\label{sec:optimalsensing}
A canonical realisation of a QLA utilises an optical parametric amplifier (or paramp), which induces a two-mode squeezing interaction between the amplified or signal ($S$) mode (annihilation operator $\hat{a}$) and an environmental mode ($E$) (annihilation operator $\hat{e}$), after which the $E$ mode is discarded \cite{caves2012quantum,agarwal2012quantum,serafini2017quantum}. In the interaction picture, the paramp Hamiltonian can be expressed as $\hat{H}_I=i\hbar\kappa\qty(\hat{a}\hat{e}-\hat{a}^\dag\hat{e}^\dag)$, where the gain parameter $G$ is related to the effective coupling strength $\kappa$ as $G=\cosh^2\kappa t$.  We first compute the action of the paramps unitary $\hat{U}(\tau)\coloneqq\exp\qty(-it\hat{H}_I/\hbar)=\exp[\tau(\hat{a}\hat{e}-\hat{a}^\dag\hat{e}^\dag)]$ on the state $\ket{n}_S\ket{0}_E$, which will enable us to compute the output states of the paramps for arbitrary probes $\ket{\psi}_{AS}$. Since $\hat{U}(\tau)$ is a two-mode squeezing operation, we can rewrite it in the alternate form (Cf. Eq. (3.7.50) of \cite{barnett2002methods}) following
\begin{align}
    \hat{U}(\tau) = \exp[(\tanh\tau)\hat{a}^\dag\hat{e}^\dag]\exp[(\ln\sech\tau)(\hat{a}^\dag\hat{a} + \hat{e}\hat{e}^\dag)]\exp[-(\tanh\tau)\hat{a}\hat{e}].
\end{align}
Expanding the exponentials into power series and applying them onto the product state $\ket{n}_S\ket{0}_E$ yields
\begin{align}
    \hat{U}(\tau)\ket{n}_S\ket{0}_E = \sech^{(n+1)}\tau\sum_{a=0}^\infty\sqrt{\binom{n+a}{a}}\tanh^a\tau\ket{n+a}_S\ket{a}_E,
\end{align}
which shows that paramp coherently adds a random number $a$ of photons to both $S$ and $E$ system according to negative binomial distribution $\textsc{NB}(n+1,\sech^2\tau)$ \cite{rohatgi2015introduction}. Hence for a given probe $\ket{\psi}_{AS}$, we hold the state family $\{\Psi_\tau=\ket{\psi_\tau}\bra{\psi_\tau}\}$ defined by 
\begin{align}   \ket{\psi_\tau}_{ASE}&=\mathbb{I}_I\otimes\qty(\otimes_{m=1}^M\hat{U}_m(\tau))\ket{\psi}_{AS}\ket{\mathbf{0}}\nonumber\\
&=\sum_{\mathbf{a\geq 0}}\sum_{\mathbf{n\geq 0}}\sqrt{p_{\mathbf{n}}A_\tau(\mathbf{n,a})}\ket{\chi_\mathbf{n}}_A\ket{\mathbf{n+a}}_S\ket{\mathbf{a}}_E\nonumber\\
&=\sum_{\mathbf{a\geq 0}}\varket{\psi_{\mathbf{a};\tau}}_{AS}\ket{\mathbf{a}}_E\label{eq:NDS_output},
\end{align}
where $\varket{\psi_{\mathbf{a};\tau}}_{AS}$ are non-normalised states of $AS$ and $A_\tau(\mathbf{n,a})=\prod_{m=1}^M\binom{n_m+a_m}{a_m}\sech^{2(n_m+1)}\tau\tanh^{2a_m}\tau$ is the product of negative binomial probabilities. Returning to our setup described in Fig.~\ref{fig:setup_gain}, the environment system is inaccessible \cite{caves2012quantum,agarwal2012quantum,serafini2017quantum}, hence the state that is accessible for detection is obtained by tracing out the environment mode 
\begin{align} \rho_\tau&=\Tr_E\Psi_\tau=\Tr_E(\ket{\psi_\tau}\bra{\psi_\tau}_{ASE})\nonumber\\
&=\sum_{\mathbf{a,a'\geq 0}}\varket{\psi_{\mathbf{a};\tau}}_{AS}\varbra{\psi_{\mathbf{a'};\tau}}_{AS}\bra{\mathbf{a}}\ket{\mathbf{a'}}_E\nonumber\\
&=\sum_{\mathbf{a\geq 0}}\varket{\psi_{\mathbf{a};\tau}}_{AS}\varbra{\psi_{\mathbf{a};\tau}}_{AS}.
\end{align}
For a given probability distribution $p_\mathbf{n}$ in Eq.~(\ref{eq:pure_NDS}), we consider probes for which $\{\ket{\chi_\mathbf{n}}_A\}$ is an orthonormal set. This class of probes is known as number-diagonal signal probes and have been shown in previous literatures to be optimal probes for diverse sensing problems \cite{nair2018quantum,sharma2018bounding,nair2011optimal}. Hence, using NDS probes, the fidelity between two QLA outputs of different gains can be computed to be (see Appendix.~\ref{appendixB} \textbf{Proposition B.1.}):
\begin{align} \mathcal{F}(\rho_\tau,\rho_{\tau'})=\sum_{\mathbf{a\geq 0}}\llangle{\psi_{\mathbf{a};\tau}}|{\psi_{\mathbf{a};\tau'}}\rrangle=\sum_{n=0}^\infty p_n\nu^{n+M}\label{eq:fidelity_amplifier},
\end{align}
where $\nu=\sech(\tau'-\tau)=\qty[\sqrt{GG'}-\sqrt{(G-1)(G'-1)}]^{-1}\in(0,1]$. Assuming $\rho_\tau$ and $\rho_{\tau'}$ to be sufficiently close to each other, we can thus calculate the quantum Fisher information (QFI) of the output states from the QLA on $\tau$ and $G$ following Eq.~(\ref{eq:Fisher_fidelity})
\begin{align}
    \mathcal{K}_\tau &= -4\left.\frac{\partial^2 \mathcal{F}(\Psi_\tau,\Psi_{\tau'})}{\partial\tau'^2}\right|_{\tau'=\tau}\nonumber\\
    &= -4\left.\qty[\sum_{n=0}^\infty p_n(n+M)\qty[(M+n+1)\tanh^2(\tau'-\tau)-1]\sech^{M+n}(\tau'-\tau)]\right|_{\tau'=\tau}\nonumber\\
    &=4(N+M)\label{eq:QFI_gain_tau},\\
    \mathcal{K}_G &=-4\left.\frac{\partial^2 \mathcal{F}(\Psi_\tau,\Psi_{\tau'})}{\partial G'^2}\right|_{G'=G}=\mathcal{K}_\tau\qty(\frac{\partial\tau}{\partial G})^2\nonumber\\
    &=4(N+M)[4(\cosh^2\tau-1)(\cosh^2\tau)]^{-1}\nonumber\\
    &=\frac{N+M}{G(G-1)}\label{eq:QFI_gain_G}.
\end{align}
These results show the non-dependency of the quantum Fisher information on the exact signal photon number distribution $\{p_{\mathbf{n}}\}$, implying that any NDS probe with a given signal energy $N$ and number of signal modes $M$ is always quantum optimal. This finding extends the scope of the optimality concept previously established for single-mode Fock states, as referenced in \cite{aspachs2010optimal}. The extension is not merely confined to multimode Fock states but encompasses an expansive array of multimode NDS probes that are entangled with ancillary systems, which includes the widely-utilized two-mode squeezed vacuum (TMSV) state, a cornerstone in the field of optical quantum information. \\
Furthermore, the results show the explicit dependency of gain sensing efficiency on $M$, which distinctly sets it apart from the dynamics of loss sensing, where the QFI does not vary with changes in $M$, as detailed in \cite{nair2018quantum}. The physical basis of the observed difference between gain and loss sensing is rooted in the nature of the gain-dependent quantum noise introduced by the QLAs. Due to the gain-dependent nature of the quantum noise introduced by the QLAs, the output states produced by two QLAs of different gains are distinguishable from one another even when the input to these amplifiers is a vacuum state. Increasing the number of signal modes serves to further enhance the distinguishability between these output states. In contrast, vacuum probes of any $M$ are invariant states of loss channels and are therefore useless for sensing them. \\
\begin{figure}[h!]
\centering
\begin{subfigure}
    \centering
    \includegraphics[width=\linewidth]{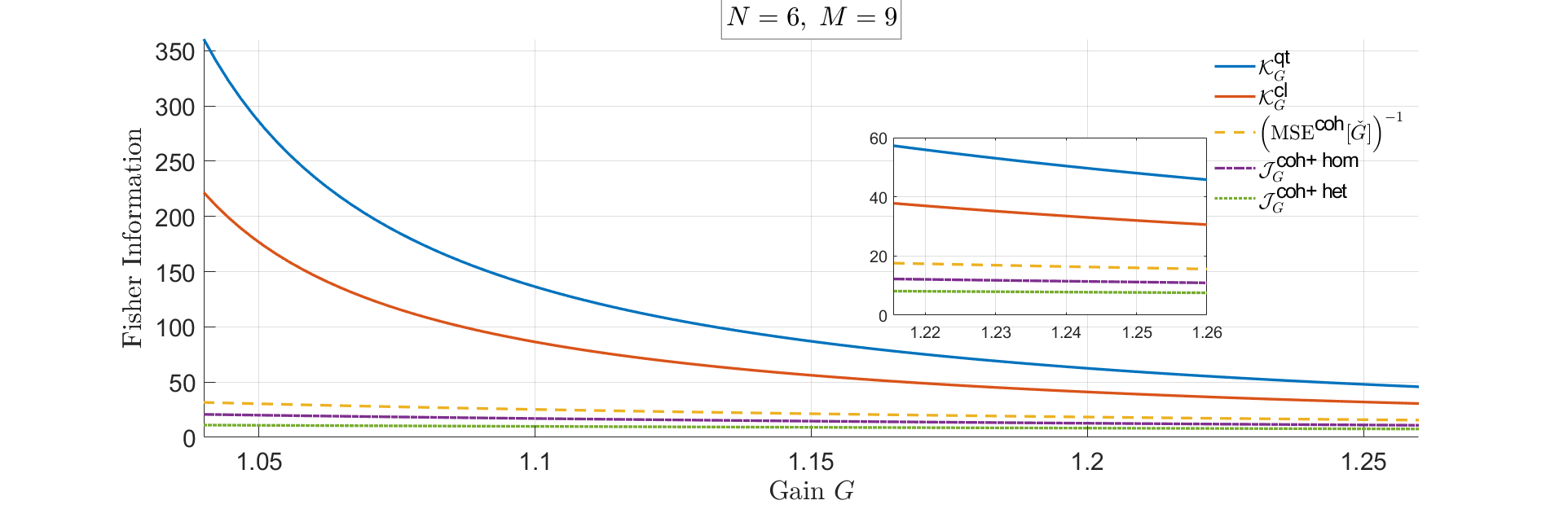}
    \end{subfigure}\\
    \begin{subfigure}
    \centering
        \includegraphics[width=\linewidth]{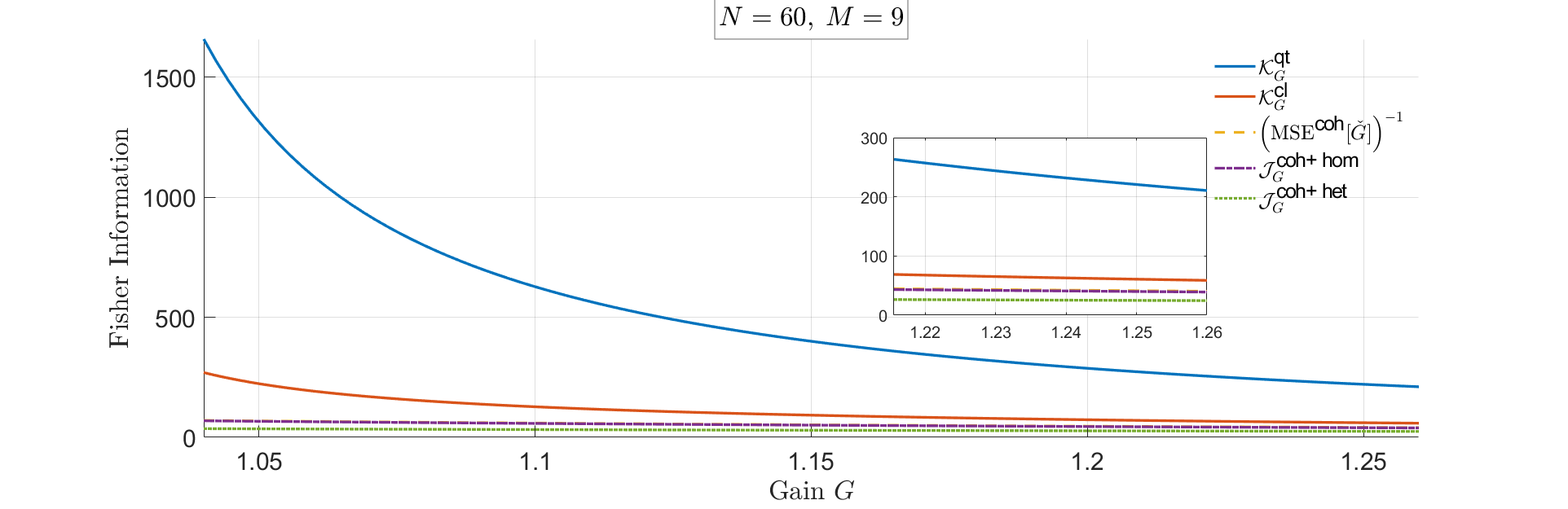}
    \end{subfigure} \\
    \begin{subfigure}
    \centering
        \includegraphics[width=\linewidth]{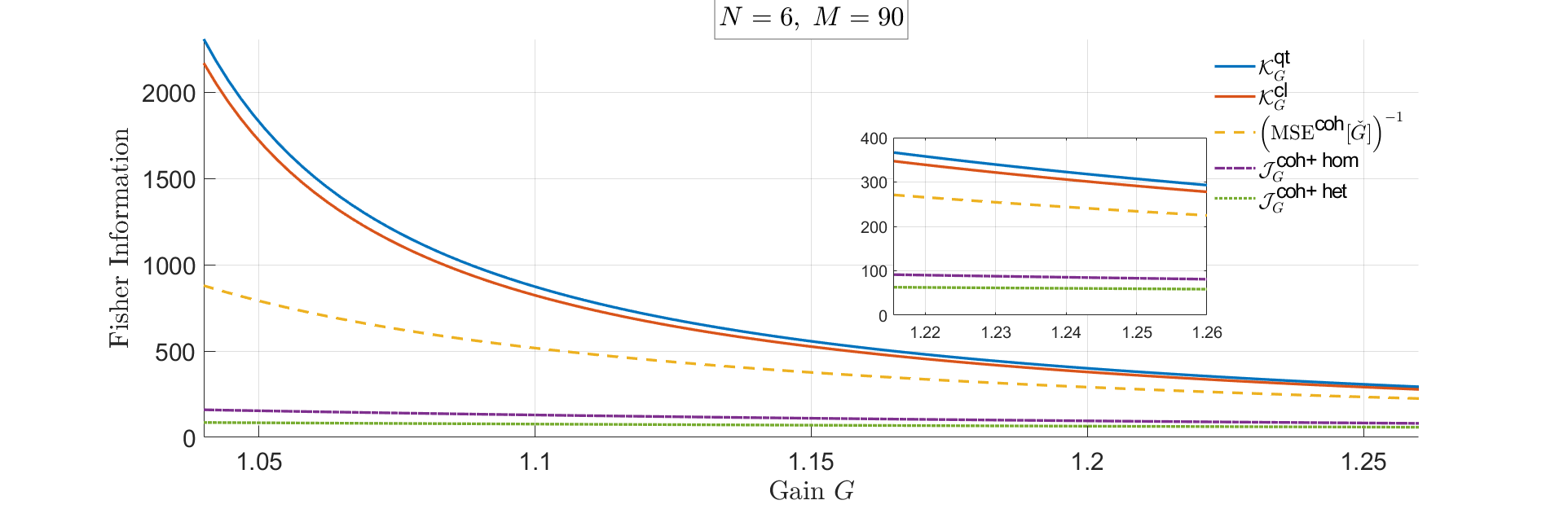}
    \end{subfigure}
    \captionsetup{width=\linewidth}
    \caption{Comparison of the Fisher information between using quantum probe (blue) (Eq.~(\ref{eq:QFI_gain_G})) and classical probe (red) (Eq.~(\ref{eq:QFI_gain_G_classical})) for $\{N=6$,\;$M=9\}$ (top), $\{N=60$,\;$M=9\}$ (centre) and $\{N=6$,\;$M=90\}$ (bottom). Quantum advantage for the range $G$ investigated, with the most prominent differences observed at $G\sim 1$. Classical Fisher information for homodyne (purple dashed-dotted), heterodyne detection (green dotted), and the inverse MSE of the photodetection-based estimator (Eq.~(\ref{eq:G_estimator})) (yellow dashed) when using a coherent state probe are shown as well.}
    \label{fig:clvsqt}
\end{figure}\noindent
From Eq.~(\ref{eq:QFI_gain_tau}) and Eq.~(\ref{eq:QFI_gain_G}), $N$ and $M$ are observed to be interchangeable, indicating that one resource can be substituted for the other. This interchangeability offers enhanced versatility in selecting optimal probes. Essentially, this means that adjustments in the values of $N$ can be compensated by corresponding changes in $M$, and vice versa, without affecting the overall efficacy of the system. Such a feature allows for greater flexibility in tailoring the system to specific requirements or constraints, as it provides multiple avenues to achieve optimal performance through the strategic allocation of these two resources.\\
For an $M$-mode signal-only coherent state probe $\ket{\sqrt{N_1}}_{S_1}\cdots\ket{\sqrt{N_M}}_{S_M}$ with total signal energy $\sum_{m=1}^MN_m = N$, the output state $\rho_\tau$ is a product of single-mode Gaussian states. The QFI on $G$ can be calculated to be (refer to Appendix~\ref{appendixF} for detailed calculation)
\begin{align}
    \mathcal{K}_G^{coh} = \frac{N}{G(2G-1)} + \frac{M}{G(G-1)}\label{eq:QFI_gain_G_classical}.
\end{align}
The convexity of the QFI in the state, as referenced in \cite{fujiwara2001quantum}, along with the linear relationship of the first term in the above equations with respect to $N$, suggests that classical probes outlined in Eq.~(\ref{eq:classical_probe}) with $M$ signal modes cannot surpass the QFI as defined in Eq.~(\ref{eq:QFI_gain_G_classical}). Both Eq.~(\ref{eq:QFI_gain_G}) and (\ref{eq:QFI_gain_G_classical}) feature two distinct components: one that is directly proportional to $N$, which can be understood as the photon contribution, and another that is proportional to $M$, representing the modal contribution. The modal contribution remains consistent between the optimal quantum and classical QFI calculations. However, a significant distinction arises when considering the photon contribution. In the context of quantum-optimal scenarios, the photon contribution is at least double that of the classical photon contribution. This difference becomes even more pronounced in the regime where G is approximately 1, as depicted in Fig.~\ref{fig:clvsqt}. In practical terms, this implies that in specific scenarios, especially those close to the $G \sim 1$ regime, quantum-optimal probes offer significantly higher efficiency in terms of their photon contribution compared to classical probes, leading to enhanced performance and potentially more accurate sensing capabilities. This contribution extend towards enhancing the quantum advantages of using quantum probes over classical probes for higher signal strength $N$ (comparing the top and centre graph in Fig.~(\ref{fig:clvsqt})), suggesting that for larger $N$, quantum probes become increasingly superior, highlighting the potential scalability of quantum advantage with respect to signal energy. 

\section{Performance of standard measurements}
Building upon the earlier derivation of the theoretical quantum-optimal Fisher information in Eq.~(\ref{eq:QFI_gain_tau}) and (\ref{eq:QFI_gain_G}), we want to investigate and implement measurements that can fully realise or saturate this theoretical boundary. Consider the NDS probe $\ket{\psi}=\sum_{\mathbf{n\geq 0}}\sqrt{p_{\mathbf{n}}}\ket{\chi_{\mathbf{n}}}_A\ket{\mathbf{n}}_S$ is deployed and we measure the output of the amplifier in Schmidt bases, i.e., the basis $\{\ket{\chi_{\mathbf{n}}}_A\}$ in the ancilla system and also the photon number in each of the $M$ output signal modes. The measurement result is denoted by $(\mathbf{X},\mathbf{Y})$, where $\mathbf{X}=(X_1,X_2,\cdots,X_M)$ is the measurement results on the ancilla system ($\mathbf{X}=\mathbf{x}$ if the result $\ket{\chi_{\mathbf{x}}}$ is obtained) and $\mathbf{Y}=(Y_1,Y_2,\cdots,Y_M)$ denotes the outcome of the photon number measurement in the $M$ signal modes. From Eq.~(\ref{eq:NDS_output}), the joint distribution of this observation is given as
\begin{align}
    P(\mathbf{x},\mathbf{y};\tau) &= p_{\mathbf{x}}A(\mathbf{x},\mathbf{y-x})\nonumber\\
    &=p_{\mathbf{x}}\qty[\prod_{m=1}^M\binom{y_m}{x_m}\sech^{2(x_m+1)}\tau\tanh^{2(y_m-x_m)}\tau].
\end{align}
From Eq.~(\ref{eq:fisherinformation_3}), we have the general expression for Fisher information
\begin{align}
    \mathcal{J}_{\theta}[\mathbf{x}]&=-\int p(\mathbf{x|\theta})\partial^2_{\theta}\ln p(\mathbf{x}|\theta)\;d\mathbf{x}\nonumber\\
    &=-\mathbb{E}[\partial_\theta\ln p(\mathbf{x}|\theta)],
\end{align}
and applying it for the measurements of $[\mathbf{X},\mathbf{Y}]$ on the parameter $\tau$ for the case of our setup, we obtain
\begin{align}
    \mathcal{J}_{\tau}[\mathbf{X},\mathbf{Y}] &=-\mathbb{E}[\partial_\tau\ln P(\mathbf{x},\mathbf{y};\tau)]\nonumber\\
    &=2\sum_{m=1}^M\qty[(\mathbb{E}[X_m]+1)\sech^2\tau + \mathbb{E}[Y_m-X_m]\qty(\sech^2\tau+\csch^2\tau)]\nonumber\\
    &=4\sum_{m=1}^M\qty[\mathbb{E}[X_m]+1]=4(N+M)\label{eq:QFI_schmidt},
\end{align}
where we have used the fact that, conditioned on $X_m=x_m$, $Y_m\sim \text{NB}(x_m+1,\sech^2\tau)$ such that $\mathbb{E}[Y_m-X_m]=\qty(\mathbb{E}[X_m]+1)\sinh^2\tau$, and the energy constraint is used in the last step. This statement indicates that Eq.~(\ref{eq:QFI_schmidt}) aligns with the theoretical QFI for an NDS probe, as detailed in Eq.~(\ref{eq:QFI_gain_tau}). The alignment of the QFI values in the two equations suggests that the theoretical maximum of information extraction from using the NDS probe in gain sensing protocol, as conceived in quantum estimation theory, can indeed be achieved in practical scenarios when using NDS probes and measuring in the Schmidt bases. This implies that using a maximum likelihood estimator based on $(\mathbf{X}, \mathbf{Y})$, one can reach the quantum limit of precision for parameter estimation, provided there is a sufficiently large $M$ as referenced in \cite{kay1993fundamentals,rohatgi2015introduction}. However, in the case of a finite sample size, a quantum-optimal estimator that achieves this level of precision may not be feasible or might not exist \cite{kay1993fundamentals}. For a multimode number-state probe $\otimes_{m=1}^M\ket{n_m}_{s_m}$ with $\sum_{m=1}^Mn_m=N$, we consider the estimator
\begin{align}
    \check{G}\coloneqq \frac{Y+M}{N+M}\label{eq:G_estimator},
\end{align}
where $Y=\sum_{m=1}^MY_m$ is the total photon number measured in the signal modes. Using the fact that $Y-N\sim \text{NB}(N+M,\sech^2\tau)$, $\check{G}$ can be shown to be unbiased following
\begin{align}
    \mathbb{E}[Y]&=\expval{\sum_{m=1}^M Y_m}=\sum_{m=1}^M\expval{\hat{a}_{out}^{(m)\dag}\hat{a}_{out}^{(m)}}\nonumber\\
    &=\sum_{m=1}^M\qty[G\expval{\hat{a}_{in}^{(m)\dag}\hat{a}_{in}^{(m)}} + (G-1)\expval{\hat{e}_{in}^{(m)}\hat{e}_{in}^{(m)\dag}}]\nonumber\\
    &=GN + M(G-1),
\end{align}
which results in the mean of the estimator $\mathbb{E}[\check{G}]=\frac{GN+M(G-1)+M}{N+M}=G$ fulfilling the property of an unbiased estimator. The variance, also the MSE, of this estimator is thus (refer to Appendix~\ref{appendixG}.\ref{appendixG_3} for detailed calculation by setting $\eta_d = 1$)
\begin{align}
    \text{Var}[\check{G}] &=\text{Var}\qty[\frac{Y+M}{N+M}]=\frac{\text{Var}[Y]}{(N+M)^2}\nonumber\\
    &=\frac{\sum_{m=1}^M\qty[\expval{Y_m^2}-\expval
    {Y_m}^2]}{(N+M)^2}\nonumber\\
    &=\frac{G(G-1)(N+M)}{(N+M)^2}\nonumber\\
    &=\frac{G(G-1)}{N+M}.
\end{align}
Here, the mean and second moments of the photon count operator for the output states $\expval{Y_m}=\expval{\hat{N}_{out}}=\expval{\hat{a}_{out}^{(m)\dag}\hat{a}_{out}^{(m)}}$ follows from the Heisenberg-evolution equations for a QLA given in Eq.~(\ref{eq:annihilation_QLA}). In this case, from the quantum Cram\'er-Rao bound, we have a saturation on the quantum Fisher information 
\begin{align}
    \text{Var}[\check{G}] = \frac{G(G-1)}{N+M} = \frac{1}{\mathcal{K}_G},
\end{align}
where the last equality follows the result from Eq.~(\ref{eq:QFI_gain_G}). Hence, QCRB is achieved even on a finite sample for any multimode number-state probe. For the classical case, a $G$-independent measurement that achieves the coherent-state QFI in Eq.~(\ref{eq:QFI_gain_G_classical}) remains unknown. For the same estimator $\check{G}$ as above, it remains unbiased when the probe is changed to $M$-mode coherent state $\otimes_{m=1}^M\ket{\sqrt{N_m}}$. However, the resulting variance (obtained from Appendix~\ref{appendixG}.\ref{appendixG_3} by setting $\eta_d=1$) is suboptimal
\begin{align*}
\text{MSE}^{\text{coh}}[\check{G}]=\frac{G(G-1)}{N+M}+\frac{G^2N}{(N+M)^2},
\end{align*}
where the second term represents an additional error due to the suboptimality of the coherent state probe. Applying a homodyne or a heterodyne detection in each output mode produces the respective Fisher information which too are suboptimal:
\begin{align}
    \mathcal{J}_G^{\text{coh+hom}}=\frac{N}{G(2G-1)};\quad\mathcal{J}_G^{\text{coh+het}}=\frac{N/2+M}{G^2}.
\end{align}
These Fisher information quantities are compared in Fig.~\ref{fig:clvsqt}, where we observe a large deviation between the theoretical classical Fisher information, and the MSE of the estimator from Eq.~(\ref{eq:G_estimator}) when using multimode coherent state probe, especially in the region where $G\sim 1$. Homodyne detection in general offers a modest performance improvement over heterodyne detection. As the signal energy $N$ increases, the performance of homodyne detection converges towards the MSE of photodetection-based estimator $\check{G}$.
\section{Practical quantum advantage}
\begin{figure}[h!]
    \centering
    \includegraphics[width=0.7\linewidth]{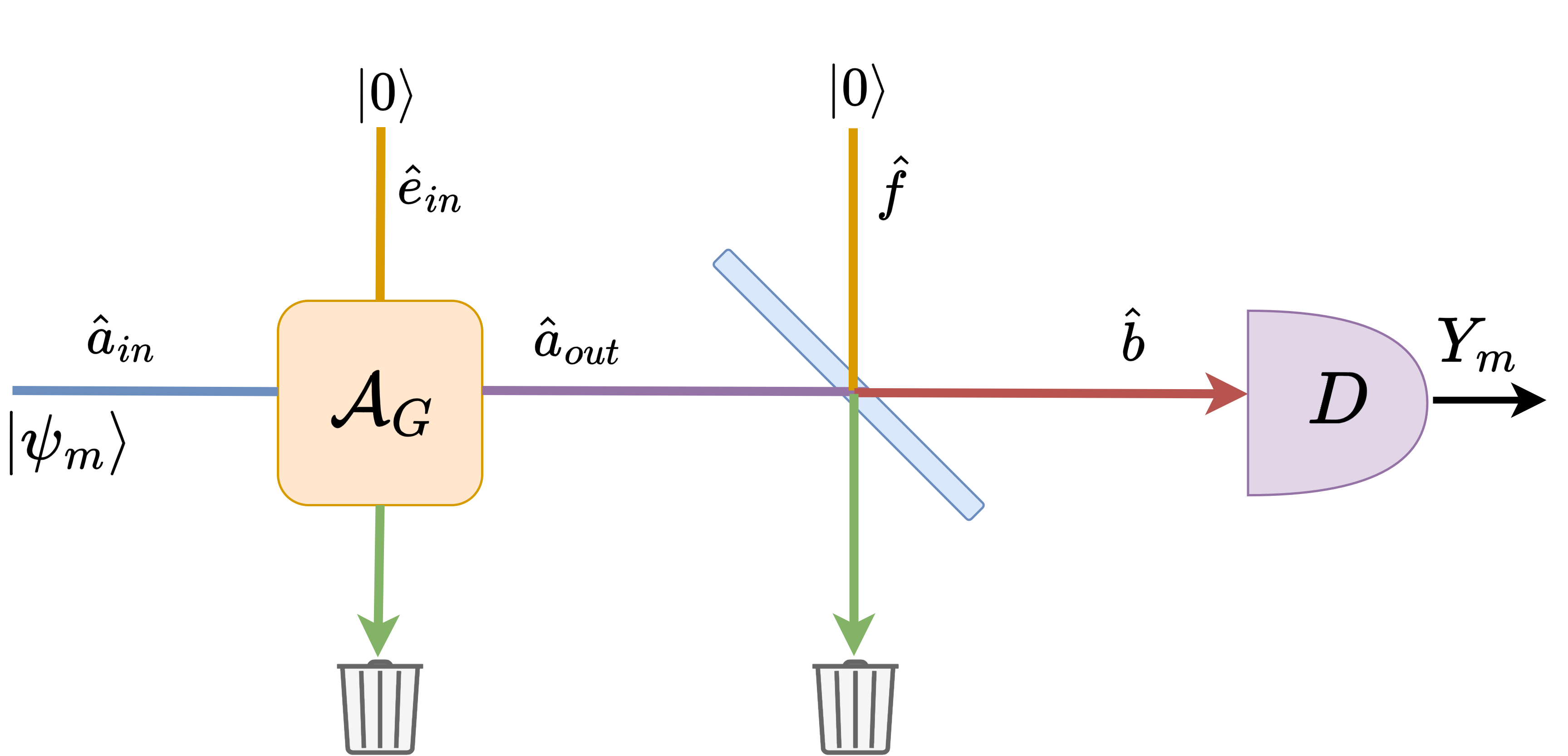}
    \captionsetup{width=\linewidth}
    \caption{Gain estimation with an inefficient detection incurring loss. Each mode of a product signal-only probe $\otimes_{m=1}^M\ket{\psi_m}$ passes through a QLA $\mathcal{A}_G$ with gain parameter $G$. Detection with quantum efficiency $\eta_d$ is modelled by a beam splitter with vacuum background input $\ket{0}$ with mode $\hat{f}$. The output mode $\hat{b}$ is measured using an ideal photodetector $D$, giving photon count $Y_m$.}
    \label{fig:nonunity}
\end{figure}
\begin{figure}[h!]
    \centering
    \begin{subfigure}
    \centering
        \includegraphics[width = 0.49\linewidth]{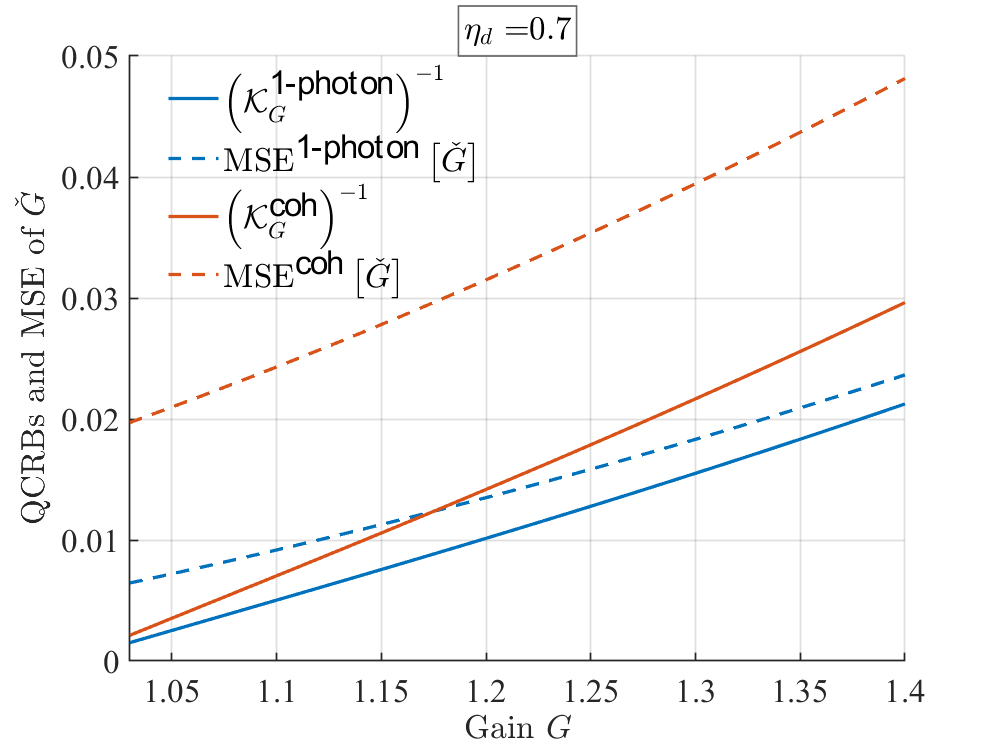}
    \end{subfigure}
        \begin{subfigure}
    \centering
        \includegraphics[width = 0.49\linewidth]{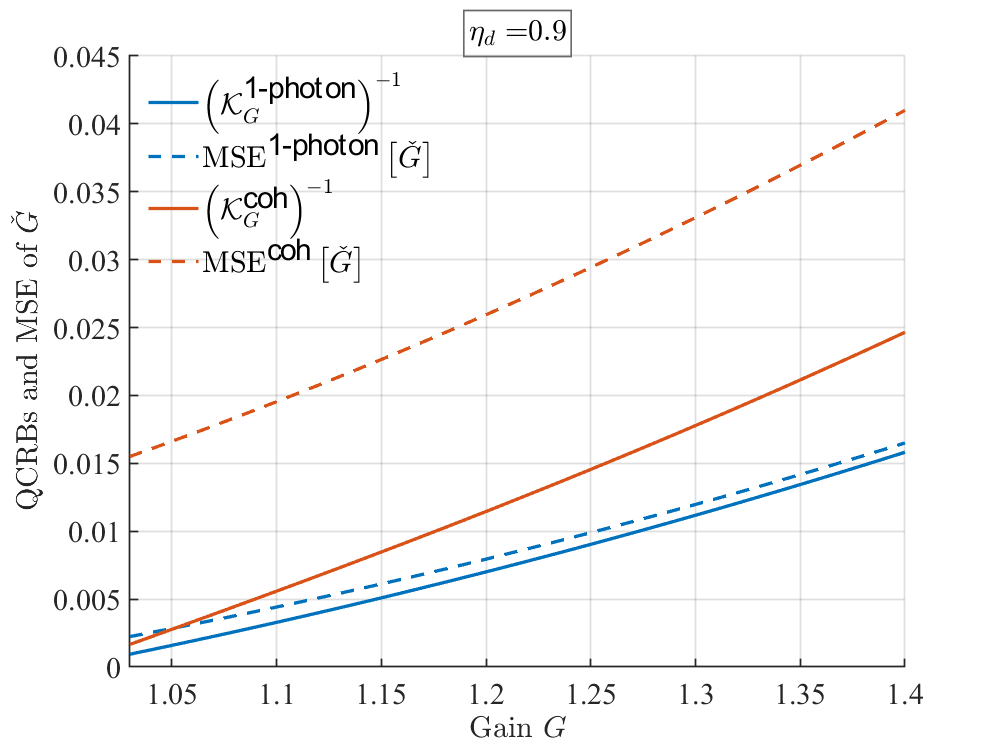}
    \end{subfigure}
    \captionsetup{width = \linewidth}
    \caption{Performance of single-photon probes with inefficient detection: QCRBs of multimode single-photon (Eq.~(\ref{eq:MSE_num}) for $M=N$) (blue solid) and coherent state (Eq.~(\ref{eq:MSE_coh})) (red solid) probes along with the MSE of $\check{G}$ of Eq.~(\ref{eq:G_estimator_loss}) for single-photon (blue dashed) and coherent state (red dashed) probes for $\eta_d=0.7$ (left) and $\eta_d=0.9$ (right) with $M=N=20$.}
    \label{fig:loss_sensing}
\end{figure}

\noindent
Realistically, in an experimental setup, detectors have a nonunity efficiency. Hence, to investigate the possibility of demonstrating quantum advantage in a laboratory setting,  we study the estimation of the gain parameter $G$ involving single-photon probes and photodetectors with efficiency $\eta_d < 1$ (see Fig.~(\ref{fig:nonunity})). Consider any multimode number-state probe $\otimes_{m=1}^M\ket{n_m}$, performing photon counting in each output mode $\hat{b}$ remains the optimal method for QFI-achieving measurement. The resulting QFI can be computed by first calculating the classical Fisher information using the measured state $\rho_\tau' = \mathcal{L}_{\eta_d}^{\otimes M}\circ\mathcal{A}_{G}^{\otimes M}\rho_{in}$, and saturating its inequality to QFI by performing photodetection measurement (Appendix~\ref{appendixG}.\ref{appendixG_2}). The numerical outcome of the QFI has been calculated and is depicted graphically in Fig.~\ref{fig:loss_sensing}. This approach was adopted because the analytical expression of the QFI turned out to be exceedingly intricate, making it challenging to extract any useful or practical insights directly from the formula. We also consider using multimode coherent state probe $\otimes_{m=1}^M\ket{\sqrt{N_m}}$ of the same value of $N$ and $M$. The resulting QFI from this setup is 
\begin{align}
    \mathcal{K}_G^{\text{coh}} = \frac{\eta_d N}{G[2\eta_d(G-1)+1]} + \frac{\eta_d M}{(G-1)[\eta_d(G-1)+1]}\label{eq:nonunity_coherent},
\end{align}
where we have generalised the result obtained in Appendix~\ref{appendixG}.\ref{appendixG_1} for multimode coherent state probe. Similar to the case of perfect detection, we consider an estimator 
\begin{align}
    \check{G} = \frac{\eta_d^{-1}Y + M}{N+M},\label{eq:G_estimator_loss}
\end{align}
for $Y = \sum_{m=1}^M Y_m$. We show in Appendix~\ref{appendixG}.\ref{appendixG_3} that this estimator is unbiased and derived the MSE achieved using multimode number-state probe $\otimes_{m=1}^M\ket{n_m}$ of total signal energy $N = \sum_{m=1}^M n_m$ to be
\begin{align}
    &\text{MSE}^{\text{num}}[\check{G}] = \frac{G(G-1)}{N+M} + \frac{1-\eta_d}{\eta_d(N+M)}\qty[G-\frac{M}{N+M}]\label{eq:MSE_num},
\end{align}
 where the first term represents the QCRB under perfect photodetection, and the second term quantifies the additional error introduced as a result of inefficiency. Additionally, we also show that for coherent state probe $\otimes_{m=1}^M\ket{\sqrt{N_m}}$ with total signal energy $N=\sum_{m=1}^MN_m$, the MSE of the estimator $\check{G}$ is computed to be
\begin{align}
    &\text{MSE}^{\text{coh}}[\check{G}] = \frac{G(G-1)}{N+M} + \frac{G^2N}{(N+M)^2} + \frac{1-\eta_d}{\eta_d(N+M)}\qty[G-\frac{M}{N+M}]\label{eq:MSE_coh}.
\end{align}
Similarly, the first term corresponds to the quantum-optimal error while the middle term accounts for the extra error that arises due to the suboptimality of the coherent state probe. In fact, the sum of the first two terms is strictly greater than the QCRB $\mathcal{K}_G^{\text{coh}^{-1}}$ (from Eq.~(\ref{eq:QFI_gain_G_classical})), corresponding to the fact that photon counting is not a QFI-achieving measurement for coherent state. The last term characterises the excess error arising from inefficient detection, which is identical to the corresponding term in Eq.~(\ref{eq:MSE_num}). 
The practicality of preparing single-photon states over multiphoton Fock states \cite{meyer2020single} is a significant consideration in an experimental setup. As such, we performed a comparative analysis of single-photon state probe performance against coherent states in Fig.~\ref{fig:loss_sensing}.  The MSE of estimator $\check{G}$ when using single-photon probes (for which $M=N$), denoted as $\text{MSE}^{\text{1-photon}}[\check{G}]$, is consistently lower than the MSE observed with coherent states probe.
\begin{figure}[h!]
    \centering
    \includegraphics[width = 0.7\linewidth]{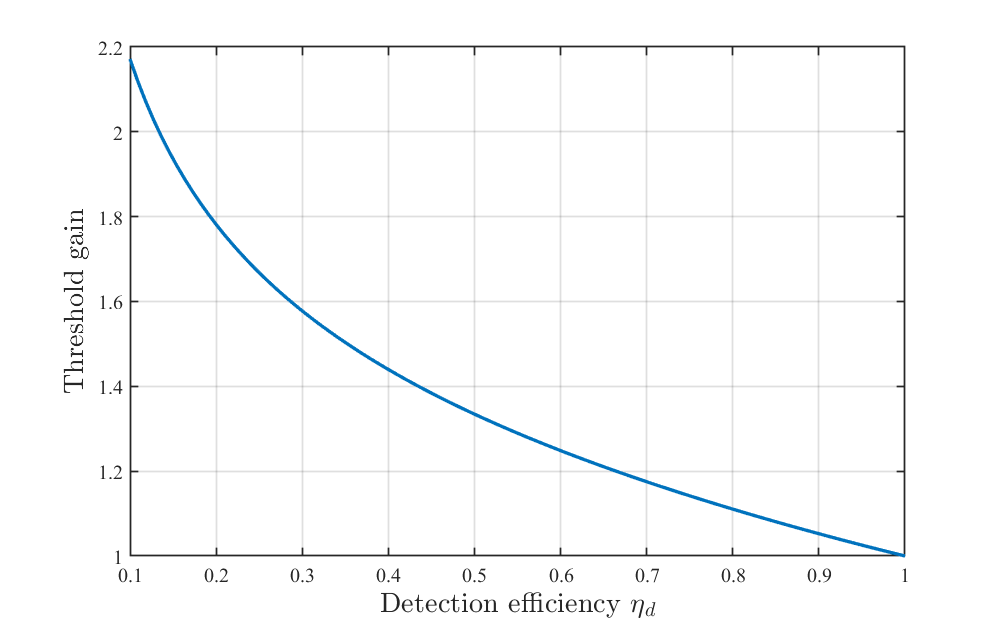}
    \captionsetup{width=\linewidth}
    \caption{The threshold gain beyond which single-photon probes and photon counting surpass the Quantum Cram\'er-Rao Bound (QCRB) of coherent state.}
    \label{fig:threshold_G}
\end{figure}\noindent
Furthermore, for each value of efficiency $\eta_d$, there exists a critical gain value, independent of $M$, above which $\text{MSE}^{\text{1-photon}}[\check{G}]$ falls below the QCRB for coherent states (Fig.~\ref{fig:threshold_G}). This implies that for gain parameters known to exceed this threshold, employing single-photon modes ensures a quantum advantage in sensing. The existence of a threshold gain value suggests that once the gain exceeds a certain level, the precision of gain measurement with single-photon modes not only surpasses that of coherent states but also breaches the fundamental limit set by the QCRB for coherent states. 

\section{Energy-constrained Bures distance}\label{sec:5-6}
In recent research, several measures for distinguishing between channels under an energy limitation, referred to as energy-constrained channel divergences, have been introduced. Such measures include the energy-constrained diamond distance and the energy-constrained Bures (ECB) distance \cite{shirokov2018uniform}, among others, which have found various applications in the field of quantum information and sensing \cite{sharma2018bounding,sharma2022optimal,shirokov2018uniform,pirandola2017fundamental,pirandola2017ultimate,winter2017energy,shirokov2018energy,shirokov2019energy,becker2020convergence,becker2021energy,takeoka2016optimal}. In our research, we present the result of the energy-constrained Bures distance between two product amplifier channels $\mathcal{A}_G^{\otimes M}$ and $\mathcal{A}_{G'}^{\otimes M}$ acting on $M$ modes. To define this quantity, we consider the ancilla-assisted channel discrimination problem whereby each signal mode of a probe consisting of $M$ signal modes and total energy $N$ entangled with an arbitrary ancilla interrogating a black box containing one of the quantum-limited amplifiers $\mathcal{A}_G$ or $\mathcal{A}_{G'}$. Since a constraint is imposed on the total signal energy, one can ask what probe state maximises a chosen state distinguishability measure at the output. The ECB distance between $\mathcal{A}_\tau^{\otimes M}$ and $\mathcal{A}_{\tau'}^{\otimes M}$ is given by the expression
\begin{align}
    B_N(\mathcal{A}_\tau^{\otimes M},\mathcal{A}_{\tau'}^{\otimes M})\coloneqq \sup_{\rho_{AS}:\Tr\rho_{AS}\mathbb{I}_A\otimes\hat{N}_S=N}\sqrt{1-\mathcal{F}\qty(\mathbb{I}_A\otimes\mathcal{A}_\tau^{\otimes M}(\rho_{AS}),\mathbb{I}_A\otimes\mathcal{A}_{\tau'}^{\otimes M}(\rho_{AS}))}\label{eq:bures_gain},
\end{align}
where $\mathcal{F}$ is the fidelity, $\tau = \cosh^{-1}\sqrt{G}$, $\mathbb{I}_A$ is the identity channel acting on $A$, $\hat{N}_S$ is the total photon number operator on $S$, and the optimisation is over all states $\rho_{AS}$ of $AS$ with signal energy $N$. The formulation of the ECB distance in Eq.~(\ref{eq:bures_gain}) differs from the definition in Ref.~\cite{shirokov2018uniform} in using an equality energy constraint, and is normalised to lie between $0$ and $1$ rather than $0$ and $\sqrt{2}$. Any arbitrary $\rho_{AS}$ satisfying the constraint can be purified using an additional ancilla to give a probe of the form from Eq.~(\ref{eq:pure_NDS}). From the expression of the ECB distance given above, maximising the output Bures distance implies the minimisation of the output fidelity. From Section~\ref{sec:optimalsensing}, we claimed that the optimal probe $\ket{\psi}_{AS}$ must be of the NDS form, i.e. the ancilla state$\{\ket{\chi_\mathbf{n}}_A\}$ appearing in Eq.~(\ref{eq:pure_NDS}) must be orthonormal. To see this, we first recall that if the environment modes are accessible, the output fidelity of the purified states takes the value given in Eq.~(\ref{eq:fidelity_amplifier}). Since the fidelity between the accessible outputs $\rho_\tau$ and $\rho_{\tau'}$ on the $AS$ system cannot be less than between their purification, using the results of fidelity from Eq.~(\ref{eq:fidelity_amplifier}), we can lower bound the fidelity as
\begin{align}
    \mathcal{F}(\rho_\tau,\rho_{\tau'})\geq\nu^M\sum_{n=0}^\infty p_n\nu^n,
\end{align}
where $\nu=\sech(\tau'-\tau)$. Since we have shown in \text{Section}~\ref{sec:optimalsensing} that the right-hand side can be achieved by any NDS probe with the same photon number distribution $\{p_n\}$, hence the optimal fidelity is achieved using an NDS probe. Alternatively, the same conclusion can be directly drawn from the discussion in Section 12 of Ref.~\cite{sharma2018bounding}, which argues that NDS probes are optimal for differentiating between any two phase-covariant channels, which amplifier channels are.\\
As $M$ is fixed, we need to minimise $\sum_np_n\nu^n$ subjecting to the energy constraint $\sum_n np_n=N$. Since the function $x\mapsto\nu^x$ is convex, we can invoke Jensen's inequality to show that $\sum_n p_n\nu^n\geq \nu^{\sum_n np_n} = \nu^N$ for any probe. If $N$ is an integer, this inequality is saturated if and only if the probe satisfies $p_N=1$, e.g., the probe could be a multimode number state of total photon number $N$. For general $N$, we can reprise an argument from Ref.~\cite{nair2018quantum} following:
\begin{Box1}{Theorem}
For any $\{p_n\}$ satisfying the energy constraint, let $A_\downarrow=\sum_{n\leq \lfloor N \rfloor} p_n$, and $A_\uparrow = 1 - A_\downarrow$. For $N_\downarrow = A_\downarrow^{-1}\sum_{n\leq\lfloor N\rfloor}n p_n\leq\lfloor N\rfloor$ and $N_\uparrow =A_\uparrow^{-1}\sum_{n\geq\lfloor N\rfloor}n p_n\geq\lfloor N\rfloor$, we have $A_\downarrow N_\downarrow + A_\uparrow N_\uparrow = N$. Convexity of $x\mapsto\nu^x$ implies that $\sum_np_n\nu^n\geq A_\downarrow\nu^{N_\downarrow} + A_\uparrow\nu^{N_\uparrow}$. Convexity also implies that the chord joining $(N_\downarrow,\nu^{N_\downarrow})$ and $(N_\uparrow,\nu^{N_\uparrow})$ lies above that joining $(\lfloor N\rfloor,\nu^{\lfloor N\rfloor})$ and $(\lceil N \rceil,\nu^{\lceil N \rceil})$ in the interval $\lfloor N\rfloor \leq x\leq \lceil N \rceil\;-\;$this follows from the fact that the intersection of the epigraph of $x\mapsto\nu^x$ and the region above the line joining $(\lfloor N\rfloor,\nu^{\lfloor N\rfloor})$ and $(\lceil N \rceil,\nu^{\lceil N \rceil})$ is the intersection of two convex sets and hence is also convex. Denote the fractional part of $N$ by $\{N\} = N - \lfloor N\rfloor$. Since the energy constrain can be satisfied by taking $N_\downarrow = \lfloor N\rfloor,N_\uparrow=\lceil N\rceil,p_{\lfloor N\rfloor} = 1-\{N\}$, and $p_{\lceil N\rceil}=\{N\}$, the energy-constrained minimum fidelity equals:
\begin{align}
    \mathcal{F}^{\min}(\rho_\tau,\rho_{\tau'}) = \nu^M\qty[(1-\{N\})\nu^{\lfloor N\rfloor} + \{N\}\nu^{\lceil N\rceil}]\label{eq:fidelity_gain_NS}.
\end{align}
\end{Box1}\noindent
The ECB distance $B_N(\mathcal{A}_\tau^{\otimes M},\mathcal{A}_{\tau'}^{\otimes M})$ then follows from Eq.~(\ref{eq:bures_gain}). Since the ECB distance is an increasing function of $N$, it equals (up to normalisation) the ECB distance defined using an inequality constraint in Ref.~\cite{shirokov2018uniform}. These results add QLAs to the short list of channels for which exact values of energy-constrained channel divergences are known and also give bounds on other divergences between QLAs \cite{audenaert2012comparisons}.\\
For quantifying the advantage of using quantum probes, we can define a classical energy-constrained Bures distance (CECB distance) analogous to Eq.~(\ref{eq:bures_gain}) by restricting the usage of the probes $\rho_{AS}$ to be in the set of classical states of the form as defined previously in Eq.~(\ref{eq:classical_probe})
\begin{align}
    \rho_{AS}=\int_{\mathbb{C}^{M'}} d^{2M'}\bm{\alpha}\int_{\mathbb{C}^{M}} d^{2M}\bm{\beta}\;P(\bm{\alpha},\bm{\beta})\ket{\bm{\alpha}}\bra{\bm{\alpha}}_A\otimes\ket{\bm{\beta}}\bra{\bm{\beta}}_S,
\end{align}
where $\bm{\alpha} = \qty(\alpha^{(1)},\cdots,\alpha^{(M')})\in\mathbb{C}^{M'}$ and $\bm{\beta} = \qty(\beta^{(1)},\cdots,\beta^{(M)})\in\mathbb{C}^{M}$ index $M'-$ and $M-$mode coherent states of ancilla and signal system respectively, and $P(\bm{\alpha},\bm{\beta})\geq 0$ is the joint probability distribution. The signal energy constraint implies that this joint probability distribution $P(\bm{\alpha})$ should satisfy
\begin{align}
    \int_{\mathbb{C}^{M'}}d^{2M'}\bm{\alpha}\int_{\mathbb{C}^M}d^{2M}\bm{\beta}\;P(\bm{\alpha},\bm{\beta})\qty(\sum_{m=1}^M\abs{\alpha_S^{(m)}}^2)=N.
\end{align}
Denoting $\mathcal{S}_N^{\text{cl}}$ as the set of classical states satisfying the energy constraint defined above, the CECB distance is then defined as
\begin{align}
    B_N^{\text{cl}}(\mathcal{A}_\tau^{\otimes M},\mathcal{A}_{\tau'}^{\otimes M})\coloneqq \sup_{\rho_{AS}\in\mathcal{S}_N^{\text{cl}}}\sqrt{1-\mathcal{F}\qty(\mathbb{I}_A\otimes\mathcal{A}_\tau^{\otimes M}(\rho_{AS}),\mathbb{I}_A\otimes\mathcal{A}_{\tau'}^{\otimes M}(\rho_{AS}))}\label{eq:bures_classical}.
\end{align}
We now consider a coherent state probe $\ket{\sqrt{N}}_S$ of a single signal mode. Since, the QLA channel is Gaussian-preserving, the corresponding output state is a Gaussian state $\rho_\tau=\Tr_E\qty[\hat{U}(\tau)\ket{\sqrt{N}}\bra{\sqrt{N}}_S\otimes\ket{0}\bra{0}_E\hat{U}^\dag(\tau)]$, specifically a displaced thermal state. We can thus use the method detailed in Appendix~\ref{appendixF} to compute the results of the fidelity between the two output Gaussian states $\mathcal{F}(\rho_\tau,\rho_{\tau'})$:
\begin{align}
    \mathcal{F}(\rho_\tau,\rho_{\tau'})&=\sech(\tau'-\tau)\exp\qty[-\frac{(\cosh\tau'-\cosh\tau)^2}{2(\sinh^2\tau'+\sinh^2\tau+1)}N]\nonumber\\
    &=\nu\exp\qty[-\frac{(\cosh\tau'-\cosh\tau)^2}{2(\sinh^2\tau'+\sinh^2\tau+1)}N]\label{eq:fidelity_QLA_SC}.
\end{align}
For a multimode coherent state $\ket{\sqrt{N_1}}\otimes\cdots\otimes\ket{\sqrt{N_M}}\in\mathcal{S}_N^{\text{cl}}$ for $\sum_{m=1}^MN_m = N$, it follows from Eq.~(\ref{eq:fidelity_QLA_SC}) that the output fidelity is
\begin{align}
    \mathcal{F}^{\text{coh}}&=\nu^M\exp\qty[-\frac{(\cosh\tau'-\cosh\tau)^2}{2(\sinh^2\tau'+\sinh^2\tau+1)}N]\label{eq:fidelity_QLA_MC}.
\end{align}
The strong concavity of the fidelity \cite{nielsen2010quantum} over mixtures and the convexity with respect to $N$ of the exponential appearing in Eq.~(\ref{eq:fidelity_QLA_MC}) imply that the above expression is the minimum fidelity over all probes in $\mathcal{S}_N^{\text{cl}}$. The minimum output Bures distance over classical probes then follows from Eq.~(\ref{eq:bures_classical}). Note that the dependence on $M$ of the minimum fidelity appears in both the quantum (Eq.~(\ref{eq:fidelity_gain_NS})) and classical (Eq.~(\ref{eq:fidelity_QLA_MC})) expressions as the factor $\nu^M$.
\begin{figure}
    \centering
    \includegraphics[width = \linewidth]{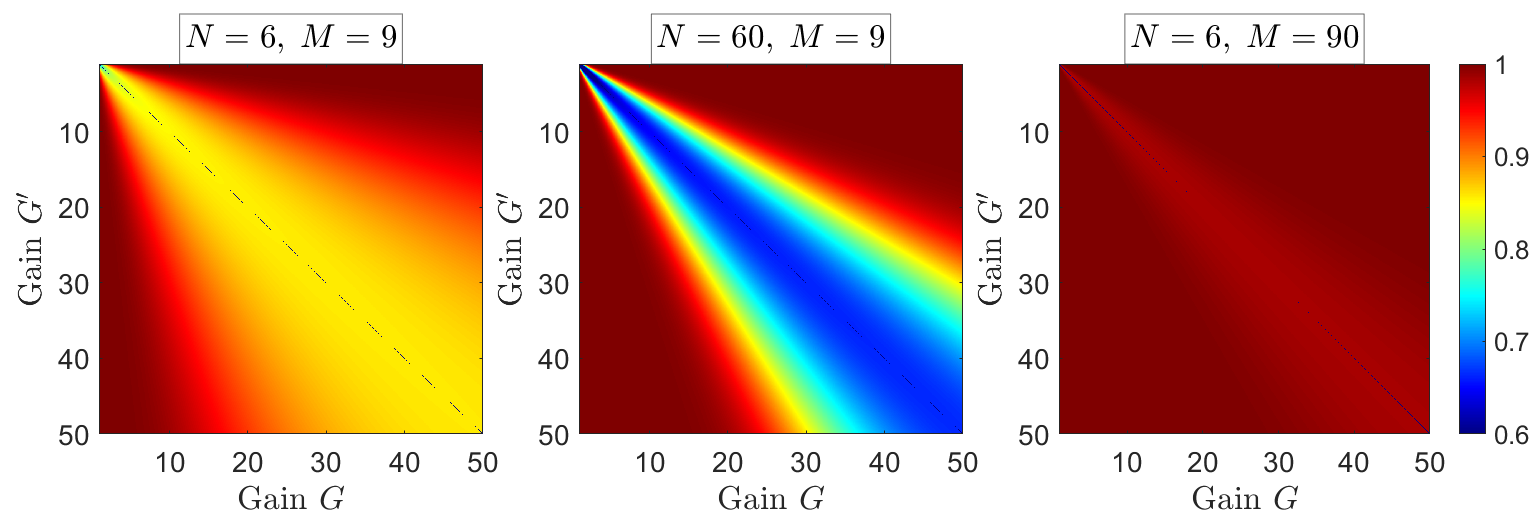}
    \captionsetup{width = \linewidth}
    \caption{Comparison of energy-constrained Bures distance (ECB) between quantum-optimal and classical probes with the colour mapping the ratio of the quantity $B_N^{\text{cl}}$ (Eq.~(\ref{eq:bures_classical})) and $B_N^{\min}$ (Eq.~(\ref{eq:bures_gain})) for $\{N=6,\;M=9\}$ (left), $\{N=60,\;M=9\}$ (centre), and $\{N=6,\;M=90\}$ (right) respectively.  }
    \label{fig:bures_distance}
\end{figure}\noindent
To compare the advantage of using quantum probes against classical probes, we plot the ratio of the CECB (Eq.~(\ref{eq:bures_classical})) against that of the ECB (Eq.~(\ref{eq:bures_gain})) for different values of $N$ and $M$ in Fig.~\ref{fig:bures_distance}. Across all values of the gain parameter for the two QLAs, the ratio $B_N^{\text{cl}}/B_N^{\min}$ never exceeds unity, thereby implying that the quantum-optimal probe is superior to any classical probe in correctly identifying the gain parameter between two distinct QLAs. An imaginary diagonal axis spanning from the top left to the bottom right of the graph lies the region $G=G'$, where the quantum advantage is most prominent, beyond which the quantum advantage depletes. This observation suggests that a balance between the number of modes and the signal strength is a critical factor for maximizing quantum advantage. Increasing the number of mode $M$ and signal strength $N$ significantly decreases the region for which quantum advantage is observable. However, for the case of increasing $N$, a larger quantum advantage is observed neighbouring the region of $G=G'$.

\section{Summary}
In this study, we have outlined the optimal precision of sensing the gain of QLAs, independent of their implementation platform and explicit physical realisation. In our approach, we limited the average energy of the signal to equal $N$. However, given that the optimal QFI grows with increasing $N$, it follows that NDS states, with an average energy of $N$ are the most effective among all probe states with an average energy that is less than or equal to $N$.\\
For multimode number-state probes, we have established a quantum-optimal estimator and demonstrated the practical possibility of enhancing gain sensing of QLA using standard single-photon sources \cite{meyer2020single} and photon counting, even with suboptimal detection efficiency. Moreover, our analytical methods can also incorporate additional losses in the signal path upstream of the QLA. The use of brighter TMSV sources could more effectively harness the photon contribution to the QFI of Eq.~(\ref{eq:QFI_gain_G}), and investigating suitable measurements and estimators for TMSV probes under imperfect detection conditions stands as a promising avenue for future research. Our study can be generalized to the estimation of multiple \cite{liu2020quantum} and distributed \cite{guo2020distributed} gain parameters. The implications of our results for relativistic metrology problems \cite{aspachs2010optimal,martin2011using,ahmadi2014quantum,ahmadi2014relativistic} is yet another intriguing prospect that warrants further exploration.\\
Noisy attenuator channels, which are pertinent to applications like quantum illumination, noisy imaging, and quantum reading \cite{tan2008quantum,gregory2020imaging,ortolano2021experimental,pirandola2011quantum}, along with noisy amplifier channels that simulate laser amplifiers with incomplete inversion \cite{agarwal2012quantum,serafini2017quantum}, and additive noise channels relevant for noisy continuous-variable teleportation \cite{braunstein1998teleportation}, are compositions of pure-loss channels with QLA channels. Our current research, in conjunction with existing studies on loss sensing \cite{nair2018quantum}, is anticipated to lay the groundwork for a comprehensive theory on the fundamental limits of sensing for such noisy phase-covariant Gaussian channels, while highlighting the role of M as an important resource therein.
    \chapter[Single-photon entangled states for quantum-optimal target detection]{Single-photon entangled states for quantum-optimal target detection}\label{chap-6}
\emph{This chapter compares the target detection performance of three different probe states: coherent state, two-mode squeezed vacuum (TMSV), and single-photon entangled state (SPES) in a No-Passive Signature (NPS) regime.\footnote{Sections of this chapter have been referenced from poster presented at International Conference on Quantum Communication, Measurement and Computing (QCMC), Lisbon.} SPES introduces a novel perspective on target detection, as it retains non-classical characteristics even after undergoing a lossy attenuator operation. Performance assessment of these states is carried out using the error exponent bound, derived from the Bhattacharyya bound between the receiver's state for two hypotheses ($\textsc{H}_0:$ No target, $\textsc{H}_1:$ Target present). Two mode-by-mode mixing operations are also investigated as viable detection protocols.}

\newpage
\section{Motivations and objectives}
Previous research in quantum illumination using Gaussian probe state in NPS regime \cite{tan2008quantum,nair2022optimal} has shown that for target of weak reflectivity ($\eta\ll 1$) immersed in a strong thermal background ($N_B\gg 1$), using coherent state and TMSV of weak per-mode signal strength ($N_S \ll 1$) results in the following Bhattacharyya error exponent
\begin{align}
    \chi_{BB}^{coh} &\gtrsim \frac{\eta N_S}{4N_B},\\
    \chi_{BB}^{TMSV} &\gtrsim \frac{\eta N_S}{N_B},
\end{align}
which showed that theoretically, TMSV has the capability of achieving a factor of four exponential target detection performance over coherent state (best classical state). While much research are focused on TMSV, an alternative state known as the NOON state remains elusive. A NOON state can be described as a generalised form
\begin{align}
    \ket{\psi}_{NOON} &=\frac{\ket{N}_I\ket{0}_S + e^{iN\theta}\ket{0}_I\ket{N}_S}{\sqrt{2}},
\end{align}
representing a superposition of $N$ particles in mode $S$ with zero particle in mode $I$ and vice versa \cite{sanders1989quantum,boto2000quantum,lee2002quantum,dowling2008quantum}. While NOON state has been investigated in phase sensing \cite{joo2011quantum,slussarenko2017unconditional,mitchell2004super} and parameter estimation \cite{hosler2013parameter}, few research have evaluated its target detection performance in a quantum illumination setup \cite{lee2021quantum}.  In this chapter, we consider a state inspired by the NOON state, which only has one photon in either mode
\begin{align}
    \ket{\psi}_{SPES} & = \sqrt{N_S}\ket{0}_I\ket{1}_S + \sqrt{1-N_S}\ket{1}_I\ket{0}_S.\label{eq:SPESstate}
\end{align}
By harnessing the advantage of SPES in retaining its non-classical characteristic prior to passing through a lossy attenuator \cite{gerry2023introductory,dowling2008quantum}, it remains a viable state for quantum target detection. A primary inspection of its returned state covariance matrix theorised that the detection protocol when using SPES can be optimised without necessarily using non-linear operation. The covariance matrix $V_{RI} = \expval{\qty[\hat{a}_R,\hat{a}_I,\hat{a}_R^\dag,\hat{a}_I^\dag]^T\qty[\hat{a}_R^\dag,\hat{a}_I^\dag,\hat{a}_R,\hat{a}_I]}$ of the return-idler state for using TMSV probe is
\begin{align}
    V^{TMSV}_{RI} &= \begin{bmatrix}
        \eta N_S + N_B  + 1 & 0 & 0 & \color{red}\sqrt{\eta N_S(N_S+1)}\\
        0 & N_S+1 & \color{red}\sqrt{\eta N_S(N_S+1)} & 0\\
        0 & \color{red}\sqrt{\eta N_S(N_S+1)} & \eta N_S + N_B & 0\\
        \color{red}\sqrt{\eta N_S(N_S+1)} & 0 & 0 & N_S
    \end{bmatrix},
    \end{align}
    where the target signature is in the phase-sensitive correlation term, therefore the detection protocol require non-linear quantum operation $\hat{c}^{(m)}=\sqrt{G}\hat{a}_I^{(m)} + \sqrt{G - 1}\hat{a}_R^{(m)\dag}$ ($G\geq 1$) for mode-mixing in order to access the target signature. However, when SPES probe is used, the covariance matrix of the return-idler state follows
    \begin{align}
    V^{SPES}_{RI} &= \begin{bmatrix}
        \eta N_S + N_B + 1 & \color{red}\sqrt{\eta N_S(N_S+1)} & 0 & 0 \\
        \color{red}\sqrt{\eta N_S(N_S+1)} & 2 - N_S & 0 & 0\\
        0 & 0 & \eta N_S + N_B & \color{red}\sqrt{\eta N_S(N_S+1)}\\
        0 & 0 & \color{red}\sqrt{\eta N_S(N_S+1)} & 1-N_S
    \end{bmatrix},
\end{align}
where the target signature is in the phase-insensitive correlation term, therefore linear quantum operation $\hat{c}^{(m)}=\sqrt{\kappa}\hat{a}_I^{(m)} + \sqrt{1-\kappa}\hat{a}_R^{(m)}$ ($\kappa \leq 1$) for mode-mixing is sufficient to access the target signature. Experimentally,  linear operations are conceptually simpler and easier to implement than non-linear operations as non-linear operations involve complex interactions between photons which require specialised nonlinear crystals and are more challenging to control and optimize. \\
In this chapter, we derive the Bhattacharyya error exponent of target detection when using SPES as probe state and show that in the regime of interest ($\eta,N_S\ll 1$, $N_B\gg 1$), its performance is comparable to that of TMSV and approaches quantum optimal. We also explore two designs of quantum-optical sensors for target detection using SPES probe, both of which can be easily implemented for proof-of-concept experiments. 

\section{Generation of SPES state}
\begin{figure}[h!]
    \centering
    \includegraphics[width = 0.7\linewidth]{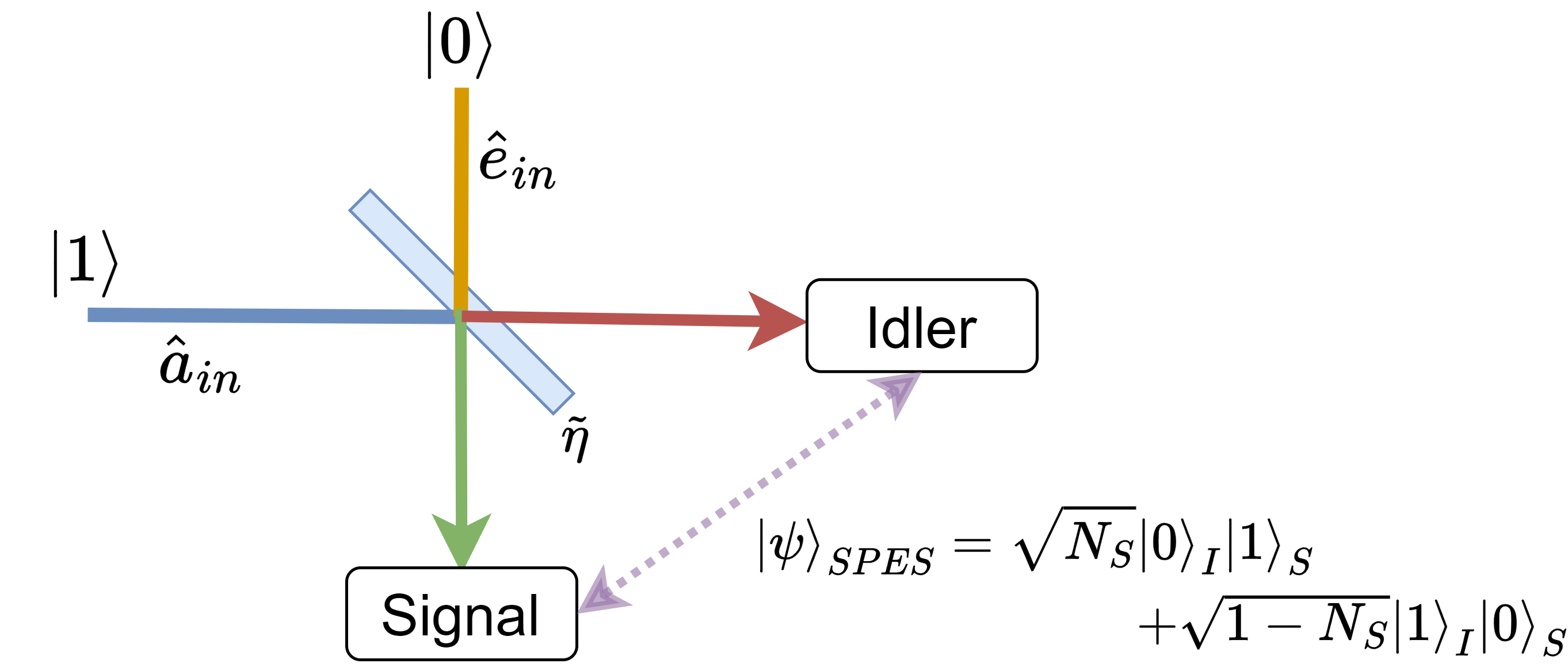}
    \captionsetup{width=\linewidth}
    \caption{A simplified schematic of SPES state generation. A single photon $\ket{1}$ and a vacuum state $\ket{0}$ is subjected to a two-mode mixing unitary operation $\hat{U}(\phi) = \exp[i\phi(\hat{a}^\dag\hat{e}-\hat{a}\hat{e}^\dag)]$, where the transmittance of the beam splitter is given by $\tilde{\eta}=\cos^2\phi\in[0,1]$. The resulting output is the SPES state (Eq.~\ref{eq:SPESstate}) where the $N_S$ is determined by the transmittance of the beam splitter.}
    \label{fig:SPES_gen}
\end{figure}\noindent
Figure~\ref{fig:SPES_gen} illustrates a straightforward theoretical approach for preparing the single-photon entangled state (SPES). The SPES state can be generated by performing a two-mode mixing operation between a single photon and a vacuum state using a beam splitter (also known as a pure loss channel from Section \ref{sec:Pure-Loss}).  In the interaction picture, the unitary of the beam splitter operation on the two mode can be expressed as $\hat{U}(\phi)=\exp[i\phi(\hat{a}^\dag\hat{e}-\hat{a}\hat{e}^\dag)]$, where $\phi\in[0,2\pi]$ is related to the transmittance of the beam splitter as $\tilde{\eta}=\cos^2\phi\in[0,1]$ \cite{gerry2023introductory,nair2018quantum}. By expanding the exponentials into power series and applying them onto the product state with $n$ number of photons in one mode, we can generalise the operation as
\begin{align}
\hat{U}(\phi)\ket{0}\ket{n}=\sum_{l=0}^n[\cos\phi]^{n-l}[\sin\phi]^l\sqrt{\binom{n}{l}}\ket{l}\ket{n-l}.
\end{align}
For the specific case of generating the SPES state, we set $n = 1$, corresponding to a single photon initially present. This leads to the output state:
\begin{align}
\ket{\psi}_{SPES}&=\sum_{l=0}^1[\cos\phi]^{1-l}[\sin\phi]^l\sqrt{\binom{1}{l}}\ket{l}\ket{1-l}\nonumber\\
&=\sqrt{\tilde{\eta}}\ket{0}\ket{1} + \sqrt{1-\tilde{\eta}}\ket{1}\ket{0}.
\end{align}
Thus, the resulting SPES state is a superposition of the photon being in either mode, with the coefficients determined by the beam splitter transmittance. Specifically, the signal strength $N_S$ of the signal mode is directly proportional to the transmittance $\tilde{\eta}$, i.e., $N_S\equiv\tilde{\eta}$.

\section{Quantum illumination protocol}
\begin{figure}[h!]
    \centering
    \includegraphics[width = 0.5\linewidth]{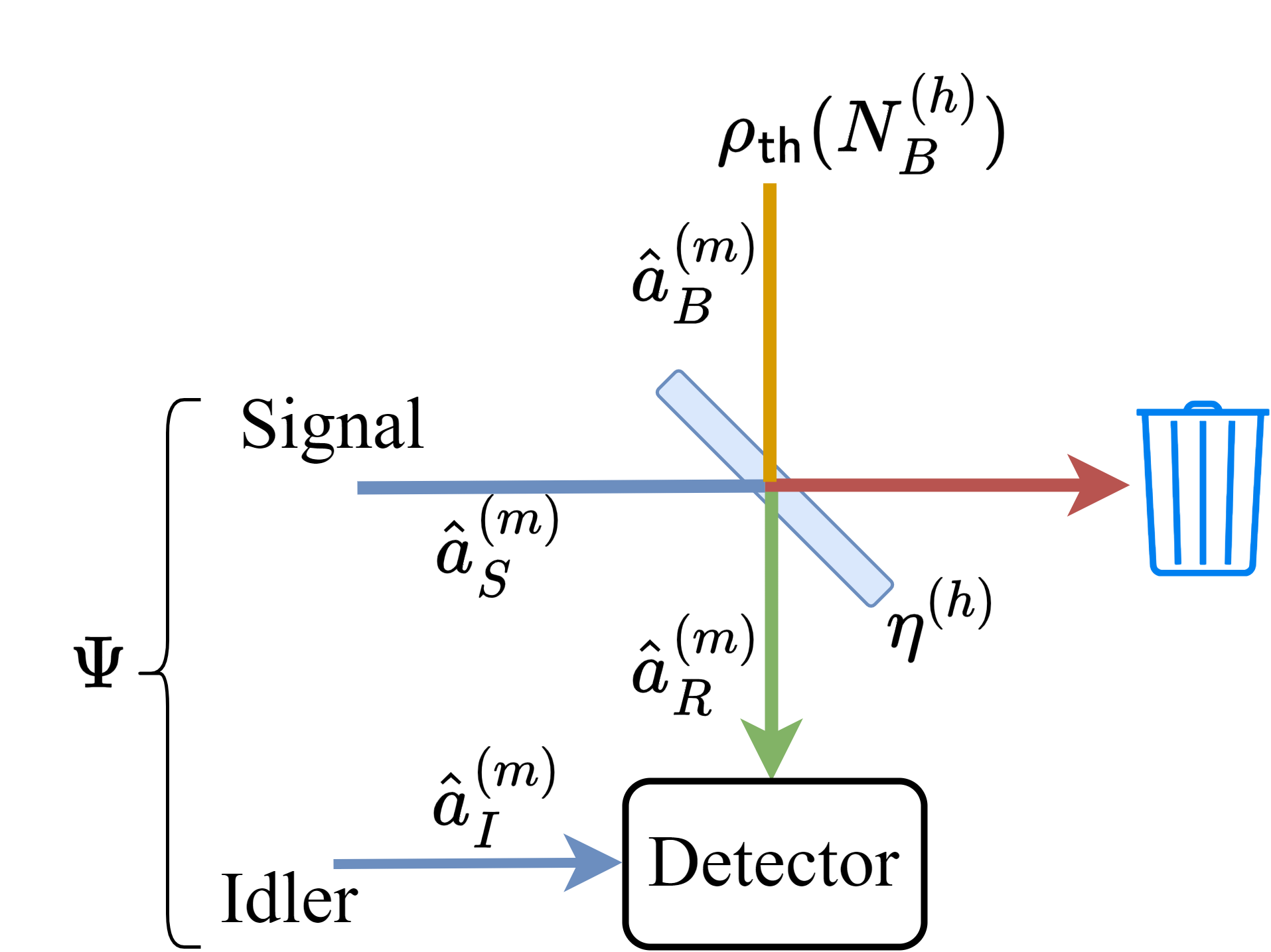}
    \captionsetup{width=\linewidth}
    \caption{A joint state $\Psi$ with $M$ signal and idler modes prepared. Each signal mode $\hat{a}_S^{(m)}$ is sent to probe the target region encapsulated by thermal radiation $\hat{a}_B^{(m)}$ which may contain a target of low effective reflectance $\eta\ll 1$. The modes returned from the region $\qty{\hat{a}_R^{(m)}}_{m=1}^M$ are jointly measured with the unperturbed idler mode $\qty{\hat{a}_I^{(m)}}_{m=1}^M$ to decide the existence of the target in the target region.}
    \label{fig:setup_NPS}
\end{figure}\noindent
The schematic for quantum illumination is illustrated in Fig.~\ref{fig:setup_NPS}. The target of illumination is represented as a beam splitter with reflectivity $\eta^{(h)}$. A signal-idler entangled state with $M$ signal and idler modes is prepared, with the signal mode being sent to probe the target region. The annihilation operator of the returned mode $\qty{\hat{a}_R^{(m)}}_{m=1}^M$ obeys the beam splitter transformation given as
\begin{align}
    \hat{a}_R^{(m)} = \sqrt{\eta^{(h)}}\hat{a}_S^{(m)} + \sqrt{1 - \eta^{(h)}}\hat{a}_B^{(m)},
\end{align}
where $h = 0(1)$ represents the absence (presence) of a target which can be emulated by the reflectance of the beam splitter $\eta^{(0)}=0$ and $\eta^{(1)}\ll 1$. Differing from the Passive signature model mentioned in Chapter~\ref{chap-3}, here each background mode is assumed to be in a thermal state $\rho_{\text{th}}(N_B^{(h)}) = \sum_{n=0}^\infty N_B^{(h)^n}/(N_B^{(h)}+1)^{n+1}\ket{n}\bra{n}_B$ ($\{\ket{n}_B\}$ are number states of mode $B$) of brightness
\begin{align}
    \expval{\hat{a}_B^{(m)^\dag}\hat{a}_B^{(m)}}_h\equiv N_B^{(h)} = N_B/(1-\eta^{(h)}),
\end{align}
whereby the background brightness is adjusted from its nominal value $N_B$ to the value $N_B/(1-\eta)$ under the hypothesis when target is present. This adjustment ensures that when a vacuum probe (hence "passive") is transmitted, the returned state is the same regardless of the absence or presence of the target (hence "no signature"). The states received by the detector for the respective hypotheses can hence be described as follows:
\begin{align}
    \textsc{H}_0: \;  \rho_0 &= \qty(\Tr_S \Psi) \otimes \rho_{\text{th}}\qty(N_B)^{\otimes M},\\ 
    \textsc{H}_1: \; \rho_1 &= \qty(\mathbb{I}_I\otimes\mathcal{L}^{\otimes M}_{\eta,N_B/(1-\eta)})\Psi\label{eq:SPES_H1},
\end{align}
where $\Psi=\ket{\psi}\bra{\psi}$ is the density state of the probe, $\rho_{\text{th}}(N_B)$ is a thermal state with a mean photon number $N_B$, $\mathcal{L}_{\tilde{\eta},N}$ denotes a thermal loss channel of transmittance $\tilde{\eta}$ and excess noise of $N$.
\section{Theoretical error exponent bound}
\begin{figure}[H]
    \centering
    \begin{subfigure}
    \centering
        \includegraphics[width = 0.49\linewidth]{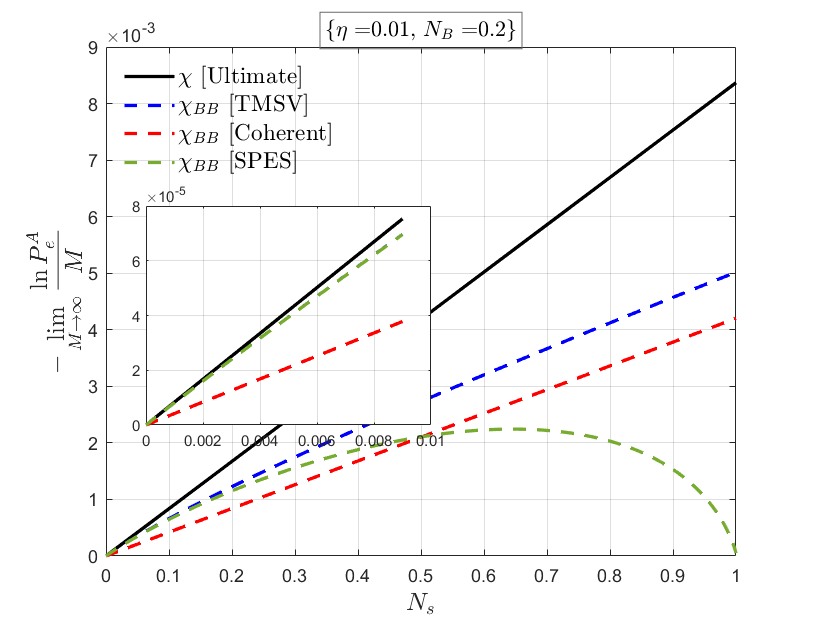}
    \end{subfigure}
        \begin{subfigure}
    \centering
        \includegraphics[width = 0.49\linewidth]{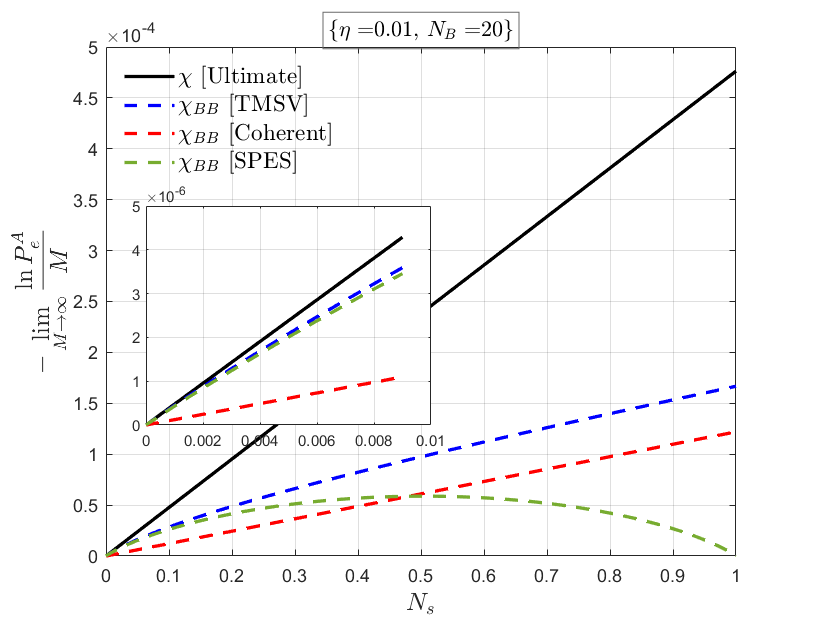}
    \end{subfigure}
    \captionsetup{width = \linewidth}
    \caption{Quantum Bhattacharyya error probability exponent as a function of per-mode signal strength $N_S$ with noise brightness $N_B=0.2$ (left) and $N_B=20$ (right). $\eta =0.01$ for both the graphs. The inset shows the exponents for $N_S\ll 1$.}
    \label{fig:BBexponent_SPES}
\end{figure}

As SPES is not a Gaussian state, we are unable to exploit the method from Eq.~(\ref{eq:s-overlap}) to derive an analytical result for the Bhattacharyya error exponent. We can however numerically compute the exponent from the returned state of both hypotheses. By decomposing the thermal loss channel into a concatenation of pure-loss and quantum-limited amplifier channels according to Eq.~(\ref{eq:concatenationofchannel}) 
\begin{align}
    \mathcal{L}_{\eta,N_B/(1-\eta)} = \mathcal{A}_G\circ\mathcal{L}_{\Tilde{\eta}},
\end{align}
with $G = N_B+1$ and $\tilde{\eta} = \eta/G$, and assuming independent and identically distributed (iid) SPES probe sent, the evaluation of the respective state for the two hypotheses gives
\begin{align}
    \rho_0 &= \qty(\Tr_S\ket{\psi}\bra{\psi}_{SPES})\otimes\rho_{\text{th}}(N_B)^{\otimes M}\nonumber\\
    &=\qty[\frac{1}{N_B+1}\sum_{n=0}^\infty\qty(\frac{N_B}{N_B+1})^n\bigg[N_S\ket{0}\bra{0}\otimes\ket{n}\bra{n} + (1-N_S)\ket{1}\bra{1}\otimes\ket{n}\bra{n}\bigg]]^{\otimes M},\\
    \rho_1 &= \qty(\mathbb{I}_I\otimes\mathcal{A}_G^{\otimes M}\mathcal{L}^{\otimes M}_{\tilde{\eta}})\ket{\psi}\bra{\psi}_{SPES}\nonumber\\
    &=\left[ \frac{1-N_S}{N_B+1}\sum_{n=0}^\infty \qty(\frac{N_B}{N_B+1})^n\ket{1}\bra{1}\otimes\ket{n}\bra{n}\right.\nonumber\\
    &\hspace{20pt}+ N_S\qty(\frac{1- \eta + N_B}{(N_B+1)^2})\sum_{n=0}^\infty \qty(\frac{N_B}{N_B+1})^n\ket{0}\bra{0}\otimes\ket{n}\bra{n}\nonumber\\
    &\hspace{30pt} + \frac{\eta N_S}{(N_B+1)^3}\sum_{n=0}^\infty \qty(\frac{N_B}{N_B+1})^n(n+1)\ket{0}\bra{0}\otimes\ket{n+1}\bra{n+1}\nonumber\\
    &\hspace{40pt}+\frac{\sqrt{\eta N_S(1-N_S)}}{(N_B+1)^2}\sum_{n=0}^\infty\qty(\frac{N_B}{N_B+1})^n\sqrt{n+1}\ket{1}\bra{0}\otimes\ket{n}\bra{n+1}\nonumber\\
    &\hspace{50pt}\left.+\frac{\sqrt{\eta N_S(1-N_S)}}{(N_B+1)^2}\sum_{n=0}^\infty\qty(\frac{N_B}{N_B+1})^n\sqrt{n+1}\ket{0}\bra{1}\otimes\ket{n+1}\bra{n}\right]^{\otimes M}\label{eq:SPES_rho1}.
\end{align}
The detailed steps taken for the evaluation of the states can be found in Appendix~\ref{appendixH}. The target detection error probability can hence be computed using the Bhattacharyya bound
\begin{align}
    P_e^A\leq \frac{1}{2}\Tr\rho_0^{1/2}\rho_1^{1/2}.
\end{align}
By confining the upper summation limit to a sufficiently large finite value, the Bhattacharyya error exponent of target detection when using SPES probe can thus be numerically computed and compared with the known error exponents of using TMSV and the coherent states as probes. The results are shown in Fig.~\ref{fig:BBexponent_SPES}.
\noindent 
The ultimate upper bound of target detection error exponent ($\chi[\text{Ultimate}]$) is referenced from \cite{nair2020fundamental}, which is derived from the lower bound of the error exponent and describes the quantum-optimal performance for target detection in NPS protocol. In the regime of weak per-mode signal strength ($N_S\ll 1$), the lower error exponent (Bhattacharyya error exponent) of SPES probe approaches the TMSV, implying that the performance of target detection using SPES is just as good as TMSV. This is true in both low ($N_B=0.2$) and high ($N_B=20$) background noise regime. Furthermore, both of these error exponents are sufficiently near the ultimate upper error exponent bound, therefore numerically proving that in the regime of $N_S\ll 1$, TMSV and SPES probes approach quantum-optimal performance. For $N_S \leq 0.47$, SPES provides
a more accurate measurement as compared to using the best classical state, the coherent state.
\section{Detection protocols}
In this section, the performances of the detectors when using SPES probe in target detection setup are evaluated based on the threshold detector error exponent derived in Eq.~(\ref{eq:errorexp_threshold}).
\subsection{Mode-mixing photon counting}
\begin{figure}[h!]
    \centering
    \includegraphics[width = 0.6\linewidth]{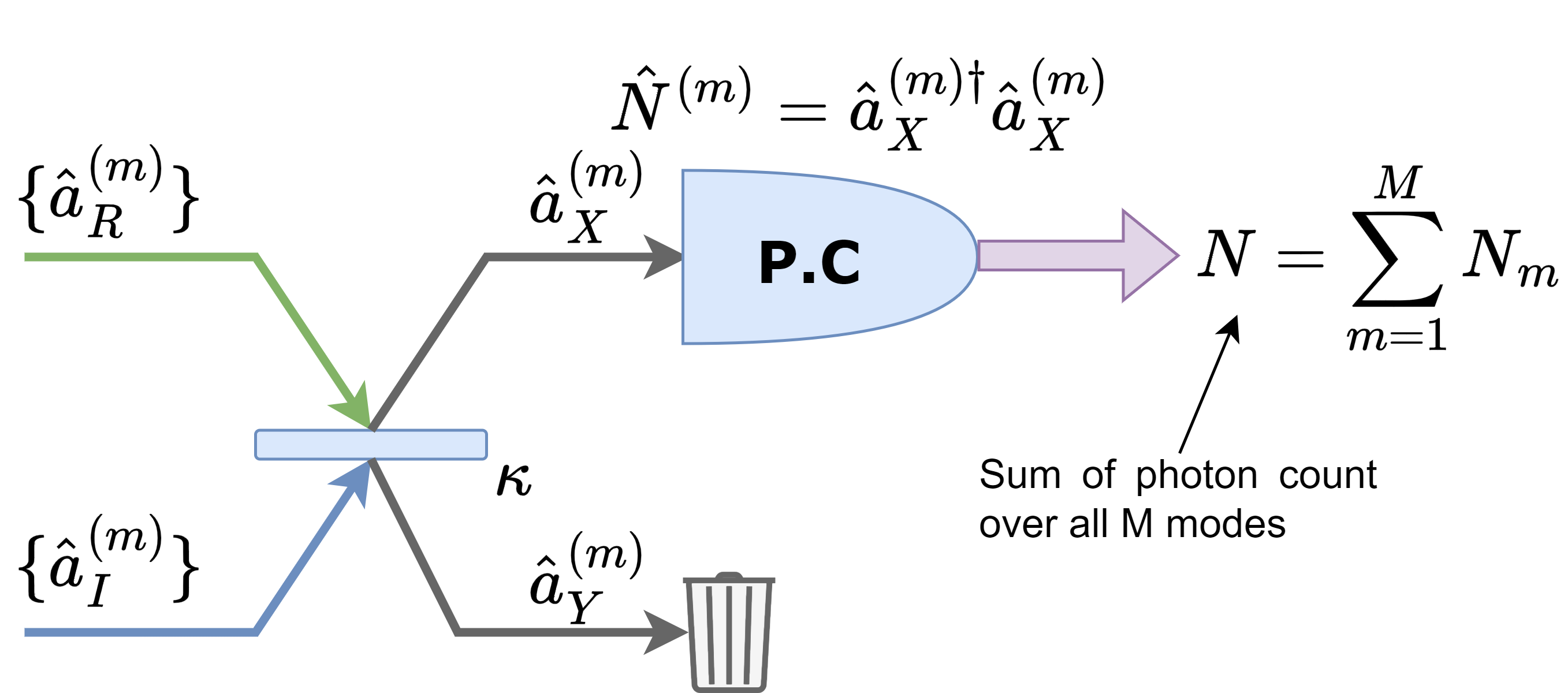}
    \captionsetup{width=\linewidth}
    \caption{Mode-mixing photon counter with tunable beam splitter for user-defined reflectivity $\kappa$. Returned modes and idler modes are mixed at the beam splitter, with one arm of the beam splitter output fed to a photon counter (P.C). If the sum of the photon counts for all $M$ modes is less than the threshold defined in Eq.~(\ref{eq:detector_threshold}), $\textsc{H}_0$ is decided, otherwise $\textsc{H}_1$ will be chosen.}
    \label{fig:photoncount}
\end{figure}
In a mode-mixing photon counting detection protocol setup, a tunable beam splitter is implemented which provides a mode-by-mode mixing of the returned signal state and the idler state. The output annihilation operators of this mode-mixing beam splitter is 
\begin{align}
    \hat{a}_X^{(m)} &= \sqrt{\kappa}\hat{a}_R^{(m)}\pm \sqrt{1-\kappa}\hat{a}_I^{(m)}\label{eq:X_MMPC},\\
    \hat{a}_Y^{(m)} &=\sqrt{1-\kappa}\hat{a}_R^{(m)}\mp \sqrt{\kappa}\hat{a}_I^{(m)}\label{eq:Y_MMPC}.
\end{align}
Photons are collected from only one output arm of the mode-mixing beam splitter (photons output from the `$X$' arm is considered) before passing through a photon counter which performs the following measurement
\begin{align}
    \hat{N}^{(m)} = \hat{a}_X^{(m)\dag}\hat{a}_X^{(m)}.
\end{align}
For Hypothesis $\textsc{H}_0$, whereby the target is absent from the target region, the parameters of the photon distributions are
\begin{align}
    \bar{N}^{\textsc{H}_0}_m &= \expval{\hat{N}_{\textsc{H}_0}^{(m)}}=\kappa N_B + (1-\kappa)(1-N_S),\\
    \qty(\sigma^{\textsc{H}_0}_m)^2 &= \expval{\qty(\hat{N}_{\textsc{H}_0}^{(m)})^2} - \expval{\hat{N}_{\textsc{H}_0}^{(m)}}^2\nonumber\\
    &= \kappa^2N_B(2N_B+1) +(1-\kappa)^2(1-N_S)\nonumber\\
    &\hspace{20pt} +\kappa(1-\kappa)[(1-N_S)(4N_B+1)+N_B]- \qty(\bar{N}_m^{\textsc{H}_0})^2.
\end{align}
For Hypothesis $\textsc{H}_1$, the target is present in the target region, and the parameters of the photon distributions are
\begin{align}
    \bar{N}^{\textsc{H}_1}_m &= \expval{\hat{N}_{\textsc{H}_1}^{(m)}}=\kappa (\eta N_S + N_B) + 2\sqrt{\kappa\eta (1-\kappa)N_S(1-N_S)} + (1-\kappa)(1-N_S),\\
    \qty(\sigma^{\textsc{H}_1}_m)^2 &= \expval{\qty(\hat{N}_{\textsc{H}_1}^{(m)})^2} - \expval{\hat{N}_{\textsc{H}_1}^{(m)}}^2\nonumber\\
    &= \kappa(1-\kappa)[1 + N_B(5-4N_S) - N_S(1-\eta)]+ (1-\kappa)^2(1-N_S)  \nonumber\\
    &\hspace{20pt} +2(1-\kappa)\sqrt{\kappa (1-\kappa)\eta N_S (1-N_S)}+ \kappa^2[\eta N_S(4N_B+1) + N_B(2N_B+1)]\nonumber\\
    &\hspace{30pt}+ 2\kappa(4N_B+1)\sqrt{\kappa (1-\kappa)\eta N_S (1-N_S)} - \qty(\bar{N}^{\textsc{H}_1}_m)^2.
\end{align}
Assuming that a large number of iid states are sent for target detection ($M\gg 1$), the photon distribution of $N=\sum_{m=1}^MN_m$ approaches Gaussian distribution. Hence, substituting the appropriate values of means and variances into Eq.~(\ref{eq:errorexp_threshold}), and taking the limit of $N_S\ll 1$, the error exponent of using MMPC threshold detector with SPES probe is
\begin{align}
    \chi_{MMPC}^{SPES}&\gtrsim \frac{1}{2}\qty(\frac{\Bar{N}_m^{\textsc{H}_1}-\Bar{N}_m^{\textsc{H}_0}}{\sigma_m^{\textsc{H}_0} + \sigma_m^{\textsc{H}_1}})^2\nonumber\\
    &=\frac{(1-\kappa)\eta N_S}{2[3N_B -\kappa(-N_B^2 + 2N_B + 1) + 1]}.\label{eq:errorexp_MMPC}
\end{align}
When the target is immersed in a strong thermal background $N_B\gg 1$, the optimum reflectivity of the mode-mixing beam splitter should be adjusted to $\kappa=0$ to maximise the error exponent such that
\begin{align}
    \chi_{MMPC}^{SPES} \gtrsim \frac{\eta N_S}{6N_B + 2},
\end{align}
which unfortunately shows a worse target detection performance compared to the Bhattacharyya bound error exponent of a coherent state.
\subsection{Mode-mixing photon difference counting}
\begin{figure}[h!]
    \centering
    \includegraphics[width = 0.7\linewidth]{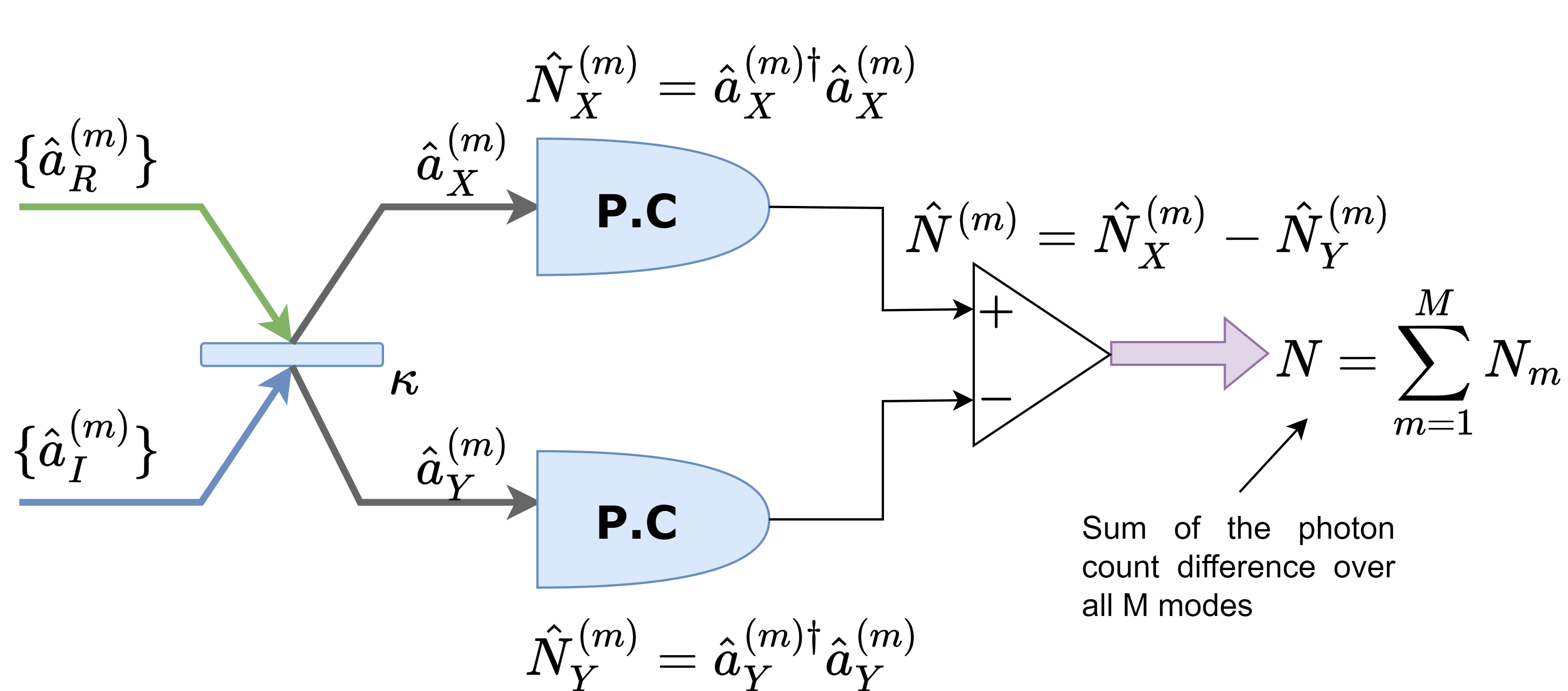}
     \captionsetup{width=\linewidth}
    \caption{Mode-mixing photon difference counting, whereby instead of performing photon counting on one arm of the tunable beam splitter output, photon counting is performed on both the output arm. The outputs of the photon counters are fed into a balanced difference detector, and the sum of the differences for all $M$ are calculated. Either hypothesis is decided by comparing these sum of differences against the threshold. }
    \label{fig:photondifference}
\end{figure}
\noindent
The setup for mode-mixing photon difference counting (MMPDC) detection protocol setup is similar to the MMDC. The difference is that both outputs from the mode-mixing beam splitter are fed into a balanced difference detector before passing through the photon counter. The resulting measurement taken is
\begin{align}
    \hat{N}^{(m)} &=\hat{N}_X^{(m)} - \hat{N}_Y^{(m)}\nonumber\\
    &=\hat{a}_X^{(m)\dag}\hat{a}_X^{(m)} - \hat{a}_Y^{(m)\dag}\hat{a}_Y^{(m)},
\end{align}
where the annihilation operators $\qty{\hat{a}_{X}^{(m)},\hat{a}_{Y}^{(m)}}$ follows from Eq.~(\ref{eq:X_MMPC}) and Eq.~(\ref{eq:Y_MMPC}) respectively, and the creation operators are obtained by taking the Hermitian conjugate of their corresponding annihilation operators.\\
For Hypothesis $\textsc{H}_0$, the parameters of the resulting photon distributions are
\begin{align}
    \bar{N}^{\textsc{H}_0}_m &= \expval{\hat{N}_{\textsc{H}_0}^{(m)}}=(2\kappa - 1)(N_S + N_B - 1),\\
    \qty(\sigma^{\textsc{H}_1}_m)^2 &= \expval{\qty(\hat{N}_{\textsc{H}_0}^{(m)})^2} - \expval{\hat{N}_{\textsc{H}_0}^{(m)}}^2\nonumber\\
    &= (2\kappa - 1)^2[(2N_B - 1)(N_B + N_S) + 1]\nonumber\\
    &\hspace{20pt} +4\kappa(1 - \kappa)[3N_B - N_S(2N_B + 1) + 1]- \qty(\bar{N}^{\textsc{H}_0}_m)^2.
\end{align}
For Hypothesis $\textsc{H}_1$, we have the following
\begin{align}
    \bar{N}^{\textsc{H}_1}_m &= \expval{\hat{N}_{\textsc{H}_0}^{(m)}}=4\sqrt{\kappa(1-\kappa)}\sqrt{\eta N_S(1-N_S)} + (2\kappa-1)[\eta(N_S+1) + N_B - 1],\\
    \qty(\sigma^{\textsc{H}_1}_m)^2 &= \expval{\qty(\hat{N}_{\textsc{H}_1}^{(m)})^2} - \expval{\hat{N}_{\textsc{H}_1}^{(m)}}^2\nonumber\\
    &= \kappa(1-\kappa)[2N_B(7 - 6N_B) - 2N_S(1-\eta) - 4N_SN_B(2\eta + 1) + 2]\nonumber\\
    &\hspace{20pt} + [\kappa^2 + (1-\kappa)^2][\eta N_S(4N_B + 1)+ N_B(2N_B+1) + (1-N_S)(1-2N_B)]  \nonumber\\
    &\hspace{30pt}+ 16N_B(2\kappa - 1)\sqrt{\eta\kappa(1-\kappa)N_S(1-N_S)} - \qty(\bar{N}^{\textsc{H}_1}_m)^2.
\end{align}
Similarly, we assume a large number of iid states are sent for target detection such that the photon distribution of $N=\sum_{m=1}^M N_m$ approaches Gaussian distribution. Substituting the appropriate values of means and variances into Eq.~(\ref{eq:errorexp_threshold}), and taking the limit of $N_s\ll 1$ and $N_B\gg 1$ we arrived at the error exponent
\begin{align}
    \chi_{MMPDC}^{SPES} \gtrsim \frac{\eta N_S}{6N_B + 2},
\end{align}
where the optimal reflectivity of the mode-mixing beam splitter is determined to be $\kappa=0.5$. It is noted that both MMPC and MMPDC threshold detector setup fail to exceed the Bhattacharyya bound error exponent of coherent state (\textbf{$\mathbf{\approx\frac{2}{3}}$ times worse}). A similar setup was proposed in ~\cite{guha2009gaussian} as a threshold detector for TMSV probe, with an additional phase conjugation (non-linear operation) of all $M$ return mode $\hat{a}_R^{(m)}$ following
\begin{align}
    \hat{a}_C^{(m)} = \sqrt{2}\hat{a}_{vac}^{(m)} + \hat{a}_R^{(m)\dag},
\end{align}
where $\hat{a}_{vac}^{(m)}$ is the annihilation operator for the vacuum state to preserve the commutation relationship. Their result showed that when TMSV state is used as the probe state, the resulting error exponent from the detector is
\begin{align}
    \chi^{TMSV}\approx \frac{\eta N_S}{2N_B}
\end{align}
which was two times better than the theoretical Bhattacharyya bound error exponent of a coherent state.
\section{Summary}
In this study, we introduced SPES as a novel probe for target detection in quantum illumination. The study demonstrated that within the NPS model, the theoretical performance of SPES is comparable to that of the TMSV state and approaches the quantum optimal limit in the low per-mode signal regime, specifically when analysing a weakly reflecting target embedded in strong thermal noise. The current findings highlight that mode-by-mode mixing operations between the signal and idler modes, achieved through the use of a tunable beam splitter followed by photon counting, result in a detection performance characterized by the error exponent $\chi_{MMPC/MMPDC}^{SPES}\gtrsim \eta N_S/(6N_B+2)$. Further optimization of the detection protocol for SPES is possible by investigating multi-mode mixing strategies. Such optimisation could enhance the performance of SPES, potentially surpassing the capabilities of all classical detection schemes. With an optimized detection protocol, the SPES state has the potential to serve as a quantum-enhanced probe, delivering performance comparable to the TMSV state in the low signal strength regime ($N_S\ll 1$), making it a promising candidate for quantum-enhanced target detection in practical applications.
    \chapter[Conclusions and future works]{Conclusions and future works}\label{chap-7}
\emph{This concluding chapter summarises the extensive work presented throughout the thesis, highlighting the substantial advancements made in the field of quantum metrology. It provides a comprehensive overview of the key findings and contributions of the research, while also offering a foundation for future exploration. A central focus of this chapter is the discussion of open questions and unresolved challenges that have arisen from the core research. These issues point to areas requiring further investigation, indicating pathways for continued progress and development in quantum metrology.}
\newpage
\section{Conclusions}
This thesis encompasses a comprehensive exploration of quantum-enhanced sensing and detection, delving into covert communication, gain sensing of quantum-limited amplifiers, and quantum illumination. The studies presented offer groundbreaking insights and establish fundamental benchmarks in the field of quantum information and communication.

Chapter~\ref{chap-4} introduces a passive signature model of illumination, providing a framework for covert sensing and detection protocols. It establishes a fidelity bound for the outputs of thermal loss channels, crucial for understanding quantum reading, pattern recognition, and channel position finding. Importantly, a lower performance bound under $\epsilon$- covertness assumptions is derived, serving as a fundamental limit for covert communication performance. The findings reveal that the quantum advantage of quantum probes over classical ones is modest and depends significantly on specific parameter regimes, emphasising the complexity of covert sensing. 

Chapter~\ref{chap-5} focuses on the precision of sensing the gain parameter for Quantum-Limited Amplifiers (QLAs), crucial in various quantum technologies. The study finds that Number-Diagonal States (NDS) with average energy $N$ are optimally effective for all probes with energy less than or equal to $N$. It also establishes a quantum-optimal estimator for multimode number-state probes and demonstrates enhanced gain sensing using standard single-photon sources. The chapter extends to investigate the performance of brighter Twin Mode Squeezed Vacuum (TMSV) sources under imperfect detection and lays the foundation for a general theory of sensing in noisy phase-covariant Gaussian channels.

Finally, Chapter~\ref{chap-6} introduces the Single-Photon Entangled State (SPES) as a novel probe for quantum illumination. It shows that SPES, in the Non-Passive Signature (NPS) model, performs comparably to the Two-Mode Squeezed Vacuum (TMSV) in low signal regimes and approaches quantum optimal performance for weakly reflecting targets. 
The study demonstrates that, while mode-by-mode mixing operations followed by photon counting do not yield significant performance improvements when utilizing the SPES probe, it remains a potentially viable probe for quantum illumination schemes. The possibility of optimising SPES detection protocols is emphasized, offering the prospect of performance advantages over all classical states.

In summary, this thesis delivers a detailed exploration of quantum sensing and detection, presenting both theoretical insights and practical applications. These findings hold considerable potential for advancing the field of quantum technologies and metrology in the future. The work paves the way for advancing secure communications, efficient quantum amplification, and enhanced target detection under challenging conditions, marking a substantial contribution to the field of quantum information science.

\section{Future works}
\subsubsection{Extending covert sensing model}
This thesis has established a baseline for understanding and implementing covert sensing in the context of quantum mechanics. Looking ahead, there is significant potential for expanding this research to encompass environments where the properties of targets and the surrounding radiation are not static but change over time. Such an advancement would involve developing and testing adaptive strategies for deploying probes, where adjustments are made in real-time based on immediate feedback. This approach would make covert sensing more applicable and effective in real-world situations, where conditions are often unpredictable and dynamic \cite{pirandola2018advances}. In addition, a critical area for future exploration is identifying and optimising the most effective measurement strategies and their practical physical setups in covert sensing operations. This inquiry would encompass both quantum and classical probes, comparing and contrasting their effectiveness in various scenarios. The goal would be to not only understand the theoretical underpinnings of these detection methods but also to explore their practical implementation. 

\subsubsection{Multi-parameter and relativistic gain sensing}
This research lays a foundational groundwork that can be expanded in several promising directions. One notable extension is the application of our study's methodologies to the estimation of multiple gain parameters. This expansion would involve adapting our current models and techniques to handle a more complex scenario where multiple parameters are simultaneously estimated, potentially offering a more comprehensive understanding and broader utility in practical applications. Additionally, our approach can be extended to distributed gain parameters. In such a scenario, the focus would be on estimating parameters that are not localised but distributed across different spatial or temporal scales. This advancement could lead to significant improvements in the field of distributed quantum sensing, where the ability to accurately estimate parameters over a distributed network is crucial. Furthermore, the implications of our findings present a fascinating avenue for further research in the realm of relativistic metrology \cite{aspachs2010optimal,martin2011using,ahmadi2014quantum,ahmadi2014relativistic}. Relativistic metrology involves the application of metrological techniques in contexts where relativistic effects are significant, such as in high-velocity or strong gravitational fields. Investigating how our results can contribute to or be adapted for such scenarios could not only enhance the understanding of relativistic systems but also open up new possibilities in high-precision measurements in extreme physical conditions.

\subsubsection{Optimal measurement strategy for SPES target detection}
This thesis has successfully laid the groundwork for understanding the theoretical performance of Single-Photon Entangled States (SPES) in target detection, demonstrating their potential to reach quantum-optimal levels. However, the development of a practical and effective measurement protocol for SPES is still an unresolved challenge. A promising direction for future research is the exploration of multi-mode mixing, which involves manipulating various modes of SPES in different configurations to optimize measurement results. Multi-mode mixing with SPES could open new avenues for enhancing the effectiveness of quantum sensing. By experimenting with various combinations and sequences of entangled modes, we can explore how different configurations affect the measurement outcomes. The objective is to determine the optimal setup that maximises the advantages offered by quantum entanglement in SPES, thereby improving the precision and accuracy of target detection.

    \printbibliography

@article{assouly2023quantum,
  title={Quantum advantage in microwave quantum radar},
  author={Assouly, R{\'e}ouven and Dassonneville, R{\'e}my and Peronnin, Th{\'e}au and Bienfait, Audrey and Huard, Benjamin},
  journal={Nature Physics},
  pages={1--5},
  year={2023},
  publisher={Nature Publishing Group UK London},
  doi={10.1038/s41567-023-02113-4},
  URL={https://doi.org/10.1038/s41567-023-02113-4}
}

@article{lloyd2008enhanced,
  title={Enhanced sensitivity of photodetection via quantum illumination},
  author={Lloyd, Seth},
  journal={Science},
  volume={321},
  number={5895},
  pages={1463--1465},
  year={2008},
  publisher={American Association for the Advancement of Science},
  doi = {10.1126/science.1160627},
URL = {https://www.science.org/doi/abs/10.1126/science.1160627},
}

@article{arvidsson2020quantum,
  title={Quantum advantage in postselected metrology},
  author={Arvidsson-Shukur, David RM and Yunger Halpern, Nicole and Lepage, Hugo V and Lasek, Aleksander A and Barnes, Crispin HW and Lloyd, Seth},
  journal={Nature communications},
  volume={11},
  number={1},
  pages={3775},
  year={2020},
  publisher={Nature Publishing Group UK London},
  doi = {10.1038/s41467-020-17559-w}
}

@article{chakrabarti2021threshold,
  title={A threshold for quantum advantage in derivative pricing},
  author={Chakrabarti, Shouvanik and Krishnakumar, Rajiv and Mazzola, Guglielmo and Stamatopoulos, Nikitas and Woerner, Stefan and Zeng, William J},
  journal={Quantum},
  volume={5},
  pages={463},
  year={2021},
  publisher={Verein zur F{\"o}rderung des Open Access Publizierens in den Quantenwissenschaften},
  doi = {10.22331/q-2021-06-01-463}
}

@article{arute2019quantum,
  title={Quantum supremacy using a programmable superconducting processor},
  author={Arute, Frank and Arya, Kunal and Babbush, Ryan and Bacon, Dave and Bardin, Joseph C and Barends, Rami and Biswas, Rupak and Boixo, Sergio and Brandao, Fernando GSL and Buell, David A and others},
  journal={Nature},
  volume={574},
  number={7779},
  pages={505--510},
  year={2019},
  publisher={Nature Publishing Group},
  doi = {10.1038/s41586-019-1666-5}
}

@article{boixo2018characterizing,
  title={Characterizing quantum supremacy in near-term devices},
  author={Boixo, Sergio and Isakov, Sergei V and Smelyanskiy, Vadim N and Babbush, Ryan and Ding, Nan and Jiang, Zhang and Bremner, Michael J and Martinis, John M and Neven, Hartmut},
  journal={Nature Physics},
  volume={14},
  number={6},
  pages={595--600},
  year={2018},
  publisher={Nature Publishing Group UK London},
  doi = {10.1038/s41567-018-0124-x}
}

@inproceedings{min2002framework,
  author={Min, R. and Chandrakasan, A.},
  booktitle={Proceedings of the International Symposium on Low Power Electronics and Design}, 
  title={A framework for energy-scalable communication in high-density wireless networks}, 
  year={2002},
  volume={},
  number={},
  pages={36-41},
  doi={10.1109/LPE.2002.146705} 
}

@inproceedings{pei2008application,
    author={Zhongmin Pei and Zhidong Deng and Bo Yang and Xiaoliang Cheng},
  booktitle={2008 IEEE International Conference on Industrial Technology}, 
  title={Application-oriented wireless sensor network communication protocols and hardware platforms: A survey}, 
  year={2008},
  volume={},
  number={},
  pages={1-6},
  doi={10.1109/ICIT.2008.4608532}
}

@misc{izadpanah2001high,
  title={High-speed broadband wireless communication system architecture},
  author={Izadpanah, Hossein},
  year={2001},
  month=may # "~29",
  publisher={Google Patents},
  note={US Patent 6,240,274}
}

@inproceedings{young2010robustness,
   author={Young, George Forrest and Scardovi, Luca and Leonard, Naomi Ehrich},
  booktitle={Proceedings of the 2010 American Control Conference}, 
  title={Robustness of noisy consensus dynamics with directed communication}, 
  year={2010},
  volume={},
  number={},
  pages={6312-6317},
  doi={10.1109/ACC.2010.5531506}
}

@article{gisin2007quantum,
  title={Quantum communication},
  author={Gisin, Nicolas and Thew, Rob},
  journal={Nature photonics},
  volume={1},
  number={3},
  pages={165--171},
  year={2007},
  publisher={Nature Publishing Group UK London},
  doi = {10.1038/nphoton.2007.22}
}

@article{cozzolino2019high,
  author = {Cozzolino, Daniele and Da Lio, Beatrice and Bacco, Davide and Oxenløwe, Leif Katsuo},
title = {High-Dimensional Quantum Communication: Benefits, Progress, and Future Challenges},
journal = {Advanced Quantum Technologies},
volume = {2},
number = {12},
pages = {1900038},
doi = {10.1002/qute.201900038},
url = {https://onlinelibrary.wiley.com/doi/abs/10.1002/qute.201900038},
eprint = {https://onlinelibrary.wiley.com/doi/pdf/10.1002/qute.201900038},
year = {2019}
}

@article{sidhu2021advances,
  author = {Sidhu, Jasminder S. and Joshi, Siddarth K. and Gündoğan, Mustafa and Brougham, Thomas and Lowndes, David and Mazzarella, Luca and Krutzik, Markus and Mohapatra, Sonali and Dequal, Daniele and Vallone, Giuseppe and Villoresi, Paolo and Ling, Alexander and Jennewein, Thomas and Mohageg, Makan and Rarity, John G. and Fuentes, Ivette and Pirandola, Stefano and Oi, Daniel K. L.},
title = {Advances in space quantum communications},
journal = {IET Quantum Communication},
volume = {2},
number = {4},
pages = {182-217},
doi = {10.1049/qtc2.12015},
url = {https://ietresearch.onlinelibrary.wiley.com/doi/abs/10.1049/qtc2.12015},
eprint = {https://ietresearch.onlinelibrary.wiley.com/doi/pdf/10.1049/qtc2.12015},
year = {2021}
}

@book{kremer1995quantum,
  title={Quantum communication},
  author={Kremer, Ilan},
  year={1995},
  publisher={Citeseer}
}

@article{chen2021integrated,
  title={An integrated space-to-ground quantum communication network over 4,600 kilometres},
  author={Chen, Yu-Ao and Zhang, Qiang and Chen, Teng-Yun and Cai, Wen-Qi and Liao, Sheng-Kai and Zhang, Jun and Chen, Kai and Yin, Juan and Ren, Ji-Gang and Chen, Zhu and others},
  journal={Nature},
  volume={589},
  number={7841},
  pages={214--219},
  year={2021},
  publisher={Nature Publishing Group UK London},
  doi = {10.1038/s41586-020-03093-8}
}

@article{pirandola2017fundamental,
  title={Fundamental limits of repeaterless quantum communications},
  author={Pirandola, Stefano and Laurenza, Riccardo and Ottaviani, Carlo and Banchi, Leonardo},
  journal={Nature communications},
  volume={8},
  number={1},
  pages={15043},
  year={2017},
  publisher={Nature Publishing Group UK London},
  doi = {10.1038/ncomms15043}
}

@book{campbell2011introduction,
  title={Introduction to remote sensing},
  author={Campbell, James B and Wynne, Randolph H},
  year={2011},
  publisher={Guilford Press}
}

@article{davis1978remote,
 author={Swain, Philip H. and Davis, Shirley M.},
  journal={IEEE Transactions on Pattern Analysis and Machine Intelligence}, 
  title={Remote Sensing: The Quantitative Approach}, 
  year={1981},
  volume={PAMI-3},
  number={6},
  pages={713-714},
  doi={10.1109/TPAMI.1981.4767177}
}

@article{patton2012efficient,
  title={Efficient design of radar waveforms for optimised detection in coloured noise},
  author={Patton, LK and Frost, SW and Rigling, BD},
  journal={IET Radar, Sonar \& Navigation},
  volume={6},
  number={1},
  pages={21--29},
  year={2012},
  publisher={IET},
  doi = {10.1049/iet-rsn.2011.0071}
}

@article{jha2015topology,
  title={Topology optimisation for energy management in underwater sensor networks},
  author={Jha, Devesh K and Wettergren, Thomas A and Ray, Asok and Mukherjee, Kushal},
  journal={International Journal of Control},
  volume={88},
  number={9},
  pages={1775--1788},
  year={2015},
  publisher={Taylor \& Francis},
  doi = {10.1080/00207179.2015.1017006}
}

@inproceedings{tahmasbi2020active,
  author={Tahmasbi, Mehrdad and Bloch, Matthieu R.},
  booktitle={2020 IEEE International Symposium on Information Theory (ISIT)}, 
  title={Active Covert Sensing}, 
  year={2020},
  volume={},
  number={},
  pages={840-845},
  doi={10.1109/ISIT44484.2020.9174220}
}

@inproceedings{tahmasbi2021signaling,
   author={Tahmasbi, Mehrdad and Bash, Boulat A. and Guha, Saikat and Bloch, Matthieu},
  booktitle={2021 IEEE International Symposium on Information Theory (ISIT)}, 
  title={Signaling for Covert Quantum Sensing}, 
  year={2021},
  volume={},
  number={},
  pages={1041-1045},
  doi={10.1109/ISIT45174.2021.9517722}
}

@article{hao2022demonstration,
  title = {Demonstration of Entanglement-Enhanced Covert Sensing},
  author = {Hao, Shuhong and Shi, Haowei and Gagatsos, Christos N. and Mishra, Mayank and Bash, Boulat and Djordjevic, Ivan and Guha, Saikat and Zhuang, Quntao and Zhang, Zheshen},
  journal = {Phys. Rev. Lett.},
  volume = {129},
  issue = {1},
  pages = {010501},
  numpages = {7},
  year = {2022},
  month = {Jun},
  publisher = {American Physical Society},
  doi = {10.1103/PhysRevLett.129.010501},
  url = {https://link.aps.org/doi/10.1103/PhysRevLett.129.010501}
}

@inproceedings{bash2017fundamental,
  author={Bash, Boulat A. and Gagatsos, Christos N. and Datta, Animesh and Guha, Saikat},
  booktitle={2017 IEEE International Symposium on Information Theory (ISIT)}, 
  title={Fundamental limits of quantum-secure covert optical sensing}, 
  year={2017},
  volume={},
  number={},
  pages={3210-3214},
  doi={10.1109/ISIT.2017.8007122}
}

@article{helstrom1969quantum,
  title={Quantum detection and estimation theory},
  author={Helstrom, Carl W},
  journal={Journal of Statistical Physics},
  volume={1},
  pages={231--252},
  year={1969},
  publisher={Springer},
  doi = {10.1007/BF01007479}
}

@article{bhattacharyya1946measure,
ISSN = {00364452},
 URL = {http://www.jstor.org/stable/25047882},
 author = {A. Bhattacharyya},
 journal = {Sankhyā: The Indian Journal of Statistics (1933-1960)},
 number = {4},
 pages = {401--406},
 publisher = {Springer},
 title = {On a Measure of Divergence between Two Multinomial Populations},
 urldate = {2024-09-17},
 volume = {7},
 year = {1946}
}

@article{jozsa1994fidelity,
author = {Richard Jozsa},
title = {Fidelity for Mixed Quantum States},
journal = {Journal of Modern Optics},
volume = {41},
number = {12},
pages = {2315--2323},
year = {1994},
publisher = {Taylor \& Francis},
doi = {10.1080/09500349414552171}
}

@article{raginsky2001fidelity,
title = {A fidelity measure for quantum channels},
journal = {Physics Letters A},
volume = {290},
number = {1},
pages = {11-18},
year = {2001},
issn = {0375-9601},
doi = {10.1016/S0375-9601(01)00640-5},
url = {https://www.sciencedirect.com/science/article/pii/S0375960101006405},
author = {Maxim Raginsky}
}

@article{marian2012uhlmann,
 title = {Uhlmann fidelity between two-mode Gaussian states},
  author = {Marian, Paulina and Marian, Tudor A.},
  journal = {Phys. Rev. A},
  volume = {86},
  issue = {2},
  pages = {022340},
  numpages = {6},
  year = {2012},
  month = {Aug},
  publisher = {American Physical Society},
  doi = {10.1103/PhysRevA.86.022340},
  url = {https://link.aps.org/doi/10.1103/PhysRevA.86.022340}
}

@book{watrous2018theory, 
place={Cambridge}, 
title={The Theory of Quantum Information}, 
publisher={Cambridge University Press}, 
author={Watrous, John}, 
year={2018}
}

@book{nielsen2010quantum,
  place={Cambridge}, 
title={Quantum Computation and Quantum Information: 10th Anniversary Edition}, 
publisher={Cambridge University Press},
 author={Nielsen, Michael A. and Chuang, Isaac L.}, 
year={2010}
}

@article{fuchs1999cryptographic,
  author={Fuchs, C.A. and van de Graaf, J.},
  journal={IEEE Transactions on Information Theory}, 
  title={Cryptographic distinguishability measures for quantum-mechanical states}, 
  year={1999},
  volume={45},
  number={4},
  pages={1216-1227},
  doi={10.1109/18.761271}
}

@article{audenaert2007discriminating,
  title = {Discriminating States: The Quantum Chernoff Bound},
  author = {Audenaert, K. M. R. and Calsamiglia, J. and Mu\~noz-Tapia, R. and Bagan, E. and Masanes, Ll. and Acin, A. and Verstraete, F.},
  journal = {Phys. Rev. Lett.},
  volume = {98},
  issue = {16},
  pages = {160501},
  numpages = {4},
  year = {2007},
  month = {Apr},
  publisher = {American Physical Society},
  doi = {10.1103/PhysRevLett.98.160501},
  url = {https://link.aps.org/doi/10.1103/PhysRevLett.98.160501}
}

@article{calsamiglia2008quantum,
 title = {Quantum Chernoff bound as a measure of distinguishability between density matrices: Application to qubit and Gaussian states},
  author = {Calsamiglia, J. and Mu\~noz-Tapia, R. and Masanes, Ll. and Acin, A. and Bagan, E.},
  journal = {Phys. Rev. A},
  volume = {77},
  issue = {3},
  pages = {032311},
  numpages = {15},
  year = {2008},
  month = {Mar},
  publisher = {American Physical Society},
  doi = {10.1103/PhysRevA.77.032311},
  url = {https://link.aps.org/doi/10.1103/PhysRevA.77.032311}
}

@article{nussbaum2009chernoff,
  ISSN = {00905364, 21688966},
 URL = {http://www.jstor.org/stable/30243657},
 author = {Michael Nussbaum and Arleta Szkoła},
 journal = {The Annals of Statistics},
 number = {2},
 pages = {1040--1057},
 publisher = {Institute of Mathematical Statistics},
 title = {The Chernoff Lower Bound for Symmetric Quantum Hypothesis Testing},
 urldate = {2024-09-17},
 volume = {37},
 year = {2009}
}

@article{kargin2005chernoff,
 ISSN = {00905364},
 URL = {http://www.jstor.org/stable/3448612},
 author = {Vladislav Kargin},
 journal = {The Annals of Statistics},
 number = {2},
 pages = {959--976},
 publisher = {Institute of Mathematical Statistics},
 title = {On the Chernoff Bound for Efficiency of Quantum Hypothesis Testing},
 urldate = {2024-09-17},
 volume = {33},
 year = {2005}
}

@article{audenaert2008asymptotic,
  title={Asymptotic error rates in quantum hypothesis testing},
  author={Audenaert, Koenraad MR and Nussbaum, Michael and Szko{\l}a, Arleta and Verstraete, Frank},
  journal={Communications in Mathematical Physics},
  volume={279},
  pages={251--283},
  year={2008},
  publisher={Springer},
  doi = {10.1007/s00220-008-0417-5}
}

@article{weedbrook2012gaussian,
  title = {Gaussian quantum information},
  author = {Weedbrook, Christian and Pirandola, Stefano and Garc\'{\i}a-Patr\'on, Ra\'ul and Cerf, Nicolas J. and Ralph, Timothy C. and Shapiro, Jeffrey H. and Lloyd, Seth},
  journal = {Rev. Mod. Phys.},
  volume = {84},
  issue = {2},
  pages = {621--669},
  numpages = {0},
  year = {2012},
  month = {May},
  publisher = {American Physical Society},
  doi = {10.1103/RevModPhys.84.621},
  url = {https://link.aps.org/doi/10.1103/RevModPhys.84.621}
}

@book{sage1971estimation,
    author={Sage, Andrew P. and Melsa, James L. and Steinway, W. J.},
  journal={IEEE Transactions on Systems, Man, and Cybernetics}, 
  title={Estimation Theory with Applications to Communication and Control}, 
  year={1971},
  volume={SMC-1},
  number={4},
  pages={405-405},
  doi={10.1109/TSMC.1971.4308330}
}

@book{kay1993fundamentals,
  title={Fundamentals of statistical signal processing: estimation theory},
  author={Kay, Steven M},
  year={1993},
  publisher={Prentice-Hall, Inc.}
}

@article{ly2017tutorial,
 title = {A Tutorial on Fisher information},
journal = {Journal of Mathematical Psychology},
volume = {80},
pages = {40-55},
year = {2017},
issn = {0022-2496},
doi = {10.1016/j.jmp.2017.05.006},
url = {https://www.sciencedirect.com/science/article/pii/S0022249617301396},
author = {Alexander Ly and Maarten Marsman and Josine Verhagen and Raoul P.P.P. Grasman and Eric-Jan Wagenmakers}
}

@article{paris2009quantum,
author = {PARIS, MATTEO G. A.},
title = {QUANTUM ESTIMATION FOR QUANTUM TECHNOLOGY},
journal = {International Journal of Quantum Information},
volume = {07},
number = {supp01},
pages = {125-137},
year = {2009},
doi = {10.1142/S0219749909004839}
}

@article{braunstein1994statistical,
  title = {Statistical distance and the geometry of quantum states},
  author = {Braunstein, Samuel L. and Caves, Carlton M.},
  journal = {Phys. Rev. Lett.},
  volume = {72},
  issue = {22},
  pages = {3439--3443},
  numpages = {0},
  year = {1994},
  month = {May},
  publisher = {American Physical Society},
  doi = {10.1103/PhysRevLett.72.3439},
  url = {https://link.aps.org/doi/10.1103/PhysRevLett.72.3439}
}

@book{devore1995probability,
  title={Probability and Statistics for Engineering and the Sciences},
  author={Devore, Jay L and others},
  volume={5},
  year={1995},
  publisher={Duxbury Press Belmont}
}

@article{pirandola2018advances,
  title={Advances in photonic quantum sensing},
  author={Pirandola, Stefano and Bardhan, B Roy and Gehring, Tobias and Weedbrook, Christian and Lloyd, Seth},
  journal={Nature Photonics},
  volume={12},
  number={12},
  pages={724--733},
  year={2018},
  publisher={Nature Publishing Group UK London},
  doi = {10.1038/s41566-018-0301-6}
}

@article{polino2020photonic,
 author = {Polino, Emanuele and Valeri, Mauro and Spagnolo, Nicolò and Sciarrino, Fabio},
    title = {Photonic quantum metrology},
    journal = {AVS Quantum Science},
    volume = {2},
    number = {2},
    pages = {024703},
    year = {2020},
    month = {06},
    issn = {2639-0213},
    doi = {10.1116/5.0007577}
}

@book{holevo2011probabilistic,
  title="{Probabilistic and statistical aspects of quantum theory}",
  author={Holevo, Alexander S},
  volume={1},
  year={2011},
  publisher={Springer Science \& Business Media},
  doi = {10.1007/978-88-7642-378-9}
}

@article{adesso2014continuous,
  author = {Adesso, Gerardo and Ragy, Sammy and Lee, Antony R.},
title = {Continuous Variable Quantum Information: Gaussian States and Beyond},
journal = {Open Systems \& Information Dynamics},
volume = {21},
number = {01n02},
pages = {1440001},
year = {2014},
doi = {10.1142/S1230161214400010}
}

@article{braunstein2005quantum,
  title = {Quantum information with continuous variables},
  author = {Braunstein, Samuel L. and van Loock, Peter},
  journal = {Rev. Mod. Phys.},
  volume = {77},
  issue = {2},
  pages = {513--577},
  numpages = {0},
  year = {2005},
  month = {Jun},
  publisher = {American Physical Society},
  doi = {10.1103/RevModPhys.77.513},
  url = {https://link.aps.org/doi/10.1103/RevModPhys.77.513}
}

@article{case2008wigner,
author = {Case, William B.},
    title = "{Wigner functions and Weyl transforms for pedestrians}",
    journal = {American Journal of Physics},
    volume = {76},
    number = {10},
    pages = {937-946},
    year = {2008},
    month = {10},
    issn = {0002-9505},
    doi = {10.1119/1.2957889}
}

@book{weyl1950theory,
  title={The theory of groups and quantum mechanics},
  author={Weyl, Hermann},
  year={1950},
  publisher={Courier Corporation}
}

@article{banchi2015quantum,
  title = {Quantum Fidelity for Arbitrary Gaussian States},
  author = {Banchi, Leonardo and Braunstein, Samuel L. and Pirandola, Stefano},
  journal = {Phys. Rev. Lett.},
  volume = {115},
  issue = {26},
  pages = {260501},
  numpages = {6},
  year = {2015},
  month = {Dec},
  publisher = {American Physical Society},
  doi = {10.1103/PhysRevLett.115.260501},
  url = {https://link.aps.org/doi/10.1103/PhysRevLett.115.260501}
}

@article{pirandola2008computable,
  title = {Computable bounds for the discrimination of Gaussian states},
  author = {Pirandola, Stefano and Lloyd, Seth},
  journal = {Phys. Rev. A},
  volume = {78},
  issue = {1},
  pages = {012331},
  numpages = {8},
  year = {2008},
  month = {Jul},
  publisher = {American Physical Society},
  doi = {10.1103/PhysRevA.78.012331},
  url = {https://link.aps.org/doi/10.1103/PhysRevA.78.012331}
}

@article{nair2018quantum,
 title = {Quantum-Limited Loss Sensing: Multiparameter Estimation and Bures Distance between Loss Channels},
  author = {Nair, Ranjith},
  journal = {Phys. Rev. Lett.},
  volume = {121},
  issue = {23},
  pages = {230801},
  numpages = {6},
  year = {2018},
  month = {Dec},
  publisher = {American Physical Society},
  doi = {10.1103/PhysRevLett.121.230801},
  url = {https://link.aps.org/doi/10.1103/PhysRevLett.121.230801}
}

@article{caves1982quantum,
  title = {Quantum limits on noise in linear amplifiers},
  author = {Caves, Carlton M.},
  journal = {Phys. Rev. D},
  volume = {26},
  issue = {8},
  pages = {1817--1839},
  numpages = {0},
  year = {1982},
  month = {Oct},
  publisher = {American Physical Society},
  doi = {10.1103/PhysRevD.26.1817},
  url = {https://link.aps.org/doi/10.1103/PhysRevD.26.1817}
}

@book{gerry2023introductory,
  place={Cambridge}, 
title={Introductory Quantum Optics}, 
publisher={Cambridge University Press}, 
author={Gerry, Christopher and Knight, Peter}, 
year={2004}
}

@article{nair2020fundamental,
author = {Ranjith Nair and Mile Gu},
journal = {Optica},
number = {7},
pages = {771--774},
publisher = {Optica Publishing Group},
title = {Fundamental limits of quantum illumination},
volume = {7},
month = {Jul},
year = {2020},
url = {https://opg.optica.org/optica/abstract.cfm?URI=optica-7-7-771},
doi = {10.1364/OPTICA.391335}
}

@book{haus2000electromagnetic,
  title={Electromagnetic noise and quantum optical measurements},
  author={Haus, Hermann A},
  year={2000},
  publisher={Springer Science \& Business Media},
  doi = {10.1007/978-3-662-04190-1}
}

@article{pirandola2011quantum,
    title = {Quantum Reading of a Classical Digital Memory},
  author = {Pirandola, Stefano},
  journal = {Phys. Rev. Lett.},
  volume = {106},
  issue = {9},
  pages = {090504},
  numpages = {4},
  year = {2011},
  month = {Mar},
  publisher = {American Physical Society},
  doi = {10.1103/PhysRevLett.106.090504},
  url = {https://link.aps.org/doi/10.1103/PhysRevLett.106.090504}
}

@article{bash2015quantum,
  title={Quantum-secure covert communication on bosonic channels},
  author={Bash, Boulat A and Gheorghe, Andrei H and Patel, Monika and Habif, Jonathan L and Goeckel, Dennis and Towsley, Don and Guha, Saikat},
  journal={Nature Communications},
  volume={6},
  number={1},
  pages={8626},
  year={2015},
  publisher={Nature Publishing Group UK London},
  doi = {10.1038/ncomms9626}
}

@article{bullock2020fundamental,
 author={Bullock, Michael S. and Gagatsos, Christos N. and Guha, Saikat and Bash, Boulat A.},
  journal={IEEE Journal on Selected Areas in Communications}, 
  title={Fundamental Limits of Quantum-Secure Covert Communication Over Bosonic Channels}, 
  year={2020},
  volume={38},
  number={3},
  pages={471-482},
  doi={10.1109/JSAC.2020.2968995}
}

@article{gagatsos2019covert,
    title = {Covert sensing using floodlight illumination},
  author = {Gagatsos, Christos N. and Bash, Boulat A. and Datta, Animesh and Zhang, Zheshen and Guha, Saikat},
  journal = {Phys. Rev. A},
  volume = {99},
  issue = {6},
  pages = {062321},
  numpages = {15},
  year = {2019},
  month = {Jun},
  publisher = {American Physical Society},
  doi = {10.1103/PhysRevA.99.062321},
  url = {https://link.aps.org/doi/10.1103/PhysRevA.99.062321}
}

@article{shapiro2019quantum,
author = {Jeffrey H. Shapiro and Don M. Boroson and P. Ben Dixon and Matthew E. Grein and Scott A. Hamilton},
journal = {J. Opt. Soc. Am. B},
number = {3},
pages = {B41--B50},
publisher = {Optica Publishing Group},
title = {Quantum low probability of intercept},
volume = {36},
month = {Mar},
year = {2019},
url = {https://opg.optica.org/josab/abstract.cfm?URI=josab-36-3-B41},
doi = {10.1364/JOSAB.36.000B41}
}

@article{tan2008quantum,
  title = {Quantum Illumination with Gaussian States},
  author = {Tan, Si-Hui and Erkmen, Baris I. and Giovannetti, Vittorio and Guha, Saikat and Lloyd, Seth and Maccone, Lorenzo and Pirandola, Stefano and Shapiro, Jeffrey H.},
  journal = {Phys. Rev. Lett.},
  volume = {101},
  issue = {25},
  pages = {253601},
  numpages = {4},
  year = {2008},
  month = {Dec},
  publisher = {American Physical Society},
  doi = {10.1103/PhysRevLett.101.253601},
  url = {https://link.aps.org/doi/10.1103/PhysRevLett.101.253601}
}

@article{nair2022optimal,
  title = {Optimal Gain Sensing of Quantum-Limited Phase-Insensitive Amplifiers},
  author = {Nair, Ranjith and Tham, Guo Yao and Gu, Mile},
  journal = {Phys. Rev. Lett.},
  volume = {128},
  issue = {18},
  pages = {180506},
  numpages = {7},
  year = {2022},
  month = {May},
  publisher = {American Physical Society},
  doi = {10.1103/PhysRevLett.128.180506},
  url = {https://link.aps.org/doi/10.1103/PhysRevLett.128.180506}
}

@article{jonsson2022gaussian,
doi = {10.1088/1751-8121/ac83fa},
url = {https://dx.doi.org/10.1088/1751-8121/ac83fa},
year = {2022},
month = {aug},
publisher = {IOP Publishing},
volume = {55},
number = {38},
pages = {385301},
author = {Robert Jonsson and Roberto Di Candia},
title = {Gaussian quantum estimation of the loss parameter in a thermal environment},
journal = {Journal of Physics A: Mathematical and Theoretical}
}

@article{shi2023ultimate,
  title={Ultimate precision limit of noise sensing and dark matter search},
  author={Shi, Haowei and Zhuang, Quntao},
  journal={npj Quantum Information},
  volume={9},
  number={1},
  pages={27},
  year={2023},
  publisher={Nature Publishing Group UK London},
  doi = {10.1038/s41534-023-00693-w}
}

@article{nair2023quantum,
title={Quantum sensing of phase-covariant optical channels}, 
      author={Ranjith Nair and Mile Gu},
      year={2023},
      eprint={2306.15256},
      archivePrefix={arXiv},
      primaryClass={quant-ph},
      url={https://arxiv.org/abs/2306.15256}, 
      doi = {10.48550/arXiv.2306.15256}
}

@book{serafini2017quantum,
  title={Quantum continuous variables: a primer of theoretical methods},
  author={Serafini, Alessio},
  year={2017},
  publisher={CRC Press},
  doi = {10.1201/9781315118727}
}

@article{banchi2020quantum,
  title = {Quantum-Enhanced Barcode Decoding and Pattern Recognition},
  author = {Banchi, Leonardo and Zhuang, Quntao and Pirandola, Stefano},
  journal = {Phys. Rev. Appl.},
  volume = {14},
  issue = {6},
  pages = {064026},
  numpages = {15},
  year = {2020},
  month = {Dec},
  publisher = {American Physical Society},
  doi = {10.1103/PhysRevApplied.14.064026},
  url = {https://link.aps.org/doi/10.1103/PhysRevApplied.14.064026}
}

@article{zhuang2020entanglement,
  title={Entanglement-enhanced testing of multiple quantum hypotheses},
  author={Zhuang, Quntao and Pirandola, Stefano},
  journal={Communications Physics},
  volume={3},
  number={1},
  pages={103},
  year={2020},
  publisher={Nature Publishing Group UK London},
  doi = {10.1038/s42005-020-0369-4}
}

@article{pereira2021idler,
  title = {Idler-free channel position finding},
  author = {Pereira, Jason L. and Banchi, Leonardo and Zhuang, Quntao and Pirandola, Stefano},
  journal = {Phys. Rev. A},
  volume = {103},
  issue = {4},
  pages = {042614},
  numpages = {12},
  year = {2021},
  month = {Apr},
  publisher = {American Physical Society},
  doi = {10.1103/PhysRevA.103.042614},
  url = {https://link.aps.org/doi/10.1103/PhysRevA.103.042614}
}

@book{ramaswami2009optical,
  title={Optical networks: a practical perspective},
  author={Ramaswami, Rajiv and Sivarajan, Kumar and Sasaki, Galen},
  year={2009},
  publisher={Morgan Kaufmann},
  doi = {10.1016/C2009-0-17339-7}
}

@article{clerk2010introduction,
  title = {Introduction to quantum noise, measurement, and amplification},
  author = {Clerk, A. A. and Devoret, M. H. and Girvin, S. M. and Marquardt, Florian and Schoelkopf, R. J.},
  journal = {Rev. Mod. Phys.},
  volume = {82},
  issue = {2},
  pages = {1155--1208},
  numpages = {0},
  year = {2010},
  month = {Apr},
  publisher = {American Physical Society},
  doi = {10.1103/RevModPhys.82.1155},
  url = {https://link.aps.org/doi/10.1103/RevModPhys.82.1155}
}

@article{caves2012quantum,
  title = {Quantum limits on phase-preserving linear amplifiers},
  author = {Caves, Carlton M. and Combes, Joshua and Jiang, Zhang and Pandey, Shashank},
  journal = {Phys. Rev. A},
  volume = {86},
  issue = {6},
  pages = {063802},
  numpages = {21},
  year = {2012},
  month = {Dec},
  publisher = {American Physical Society},
  doi = {10.1103/PhysRevA.86.063802},
  url = {https://link.aps.org/doi/10.1103/PhysRevA.86.063802}
}

@article{lahteenmaki2014advanced,
  title={Advanced concepts in Josephson junction reflection amplifiers},
  author={L{\"a}hteenm{\"a}ki, Pasi and Vesterinen, Visa and Hassel, Juha and Paraoanu, GS and Sepp{\"a}, Heikki and Hakonen, Pertti},
  journal={Journal of Low Temperature Physics},
  volume={175},
  pages={868--876},
  year={2014},
  publisher={Springer},
  doi = {10.1007/s10909-014-1170-0}
}

@article{chia2020phase,
  title = {Phase-Preserving Linear Amplifiers Not Simulable by the Parametric Amplifier},
  author = {Chia, A. and Hajdu$\check{s}$ek, M. and Nair, R. and Fazio, R. and Kwek, L. C. and Vedral, V.},
  journal = {Phys. Rev. Lett.},
  volume = {125},
  issue = {16},
  pages = {163603},
  numpages = {6},
  year = {2020},
  month = {Oct},
  publisher = {American Physical Society},
  doi = {10.1103/PhysRevLett.125.163603},
  url = {https://link.aps.org/doi/10.1103/PhysRevLett.125.163603}
}

@article{haus1962quantum,
  title = {Quantum Noise in Linear Amplifiers},
  author = {Haus, H. A. and Mullen, J. A.},
  journal = {Phys. Rev.},
  volume = {128},
  issue = {5},
  pages = {2407--2413},
  numpages = {0},
  year = {1962},
  month = {Dec},
  publisher = {American Physical Society},
  doi = {10.1103/PhysRev.128.2407},
  url = {https://link.aps.org/doi/10.1103/PhysRev.128.2407}
}

@book{agarwal2012quantum,
place={Cambridge}, 
title={Quantum Optics}, 
publisher={Cambridge University Press}, 
author={Agarwal, Girish S.}, 
year={2012}
}

@article{caruso2006one,
doi = {10.1088/1367-2630/8/12/310},
url = {https://dx.doi.org/10.1088/1367-2630/8/12/310},
year = {2006},
month = {dec},
publisher = {IOP Publishing},
volume = {8},
number = {12},
pages = {310},
author = {F Caruso and V Giovannetti and A S Holevo},
title = {One-mode bosonic Gaussian channels: a full weak-degradability classification},
journal = {New Journal of Physics}
}

@article{braun2018quantum,
  title = {Quantum-enhanced measurements without entanglement},
  author = {Braun, Daniel and Adesso, Gerardo and Benatti, Fabio and Floreanini, Roberto and Marzolino, Ugo and Mitchell, Morgan W. and Pirandola, Stefano},
  journal = {Rev. Mod. Phys.},
  volume = {90},
  issue = {3},
  pages = {035006},
  numpages = {47},
  year = {2018},
  month = {Sep},
  publisher = {American Physical Society},
  doi = {10.1103/RevModPhys.90.035006},
  url = {https://link.aps.org/doi/10.1103/RevModPhys.90.035006}
}

@article{aspachs2010optimal,
  title = {Optimal Quantum Estimation of the Unruh-Hawking Effect},
  author = {Aspachs, Mariona and Adesso, Gerardo and Fuentes, Ivette},
  journal = {Phys. Rev. Lett.},
  volume = {105},
  issue = {15},
  pages = {151301},
  numpages = {4},
  year = {2010},
  month = {Oct},
  publisher = {American Physical Society},
  doi = {10.1103/PhysRevLett.105.151301},
  url = {https://link.aps.org/doi/10.1103/PhysRevLett.105.151301}
}

@article{gaiba2009squeezed,
title = {Squeezed vacuum as a universal quantum probe},
journal = {Physics Letters A},
volume = {373},
number = {10},
pages = {934-939},
year = {2009},
issn = {0375-9601},
doi = {10.1016/j.physleta.2009.01.026},
url = {https://www.sciencedirect.com/science/article/pii/S0375960109000802},
author = {Roberto Gaiba and Matteo G.A. Paris}
}

@article{bradshaw2021optimal,
  title = {Optimal probes for continuous-variable quantum illumination},
  author = {Bradshaw, Mark and Conlon, Lorc\'an O. and Tserkis, Spyros and Gu, Mile and Lam, Ping Koy and Assad, Syed M.},
  journal = {Phys. Rev. A},
  volume = {103},
  issue = {6},
  pages = {062413},
  numpages = {7},
  year = {2021},
  month = {Jun},
  publisher = {American Physical Society},
  doi = {10.1103/PhysRevA.103.062413},
  url = {https://link.aps.org/doi/10.1103/PhysRevA.103.062413}
}

@article{gregory2020imaging,
author = {T. Gregory  and P.-A. Moreau  and E. Toninelli  and M. J. Padgett },
title = {Imaging through noise with quantum illumination},
journal = {Science Advances},
volume = {6},
number = {6},
pages = {eaay2652},
year = {2020},
doi = {10.1126/sciadv.aay2652},
URL = {https://www.science.org/doi/abs/10.1126/sciadv.aay2652}
}

@article{ortolano2021experimental,
  author = {Giuseppe Ortolano  and Elena Losero  and Stefano Pirandola  and Marco Genovese  and Ivano Ruo-Berchera },
title = {Experimental quantum reading with photon counting},
journal = {Science Advances},
volume = {7},
number = {4},
pages = {eabc7796},
year = {2021},
doi = {10.1126/sciadv.abc7796}
}

@article{monras2011measurement,
  title = {Measurement of damping and temperature: Precision bounds in Gaussian dissipative channels},
  author = {Monras, Alex and Illuminati, Fabrizio},
  journal = {Phys. Rev. A},
  volume = {83},
  issue = {1},
  pages = {012315},
  numpages = {12},
  year = {2011},
  month = {Jan},
  publisher = {American Physical Society},
  doi = {10.1103/PhysRevA.83.012315},
  url = {https://link.aps.org/doi/10.1103/PhysRevA.83.012315}
}

@article{sharma2018bounding,
  title={Bounding the energy-constrained quantum and private capacities of phase-insensitive bosonic Gaussian channels},
  author={Sharma, Kunal and Wilde, Mark M and Adhikari, Sushovit and Takeoka, Masahiro},
  journal={New Journal of Physics},
  volume={20},
  number={6},
  pages={063025},
  year={2018},
  publisher={IOP Publishing},
  doi = {10.1088/1367-2630/aac11a}
}

@article{sharma2022optimal,
  title = {Optimal tests for continuous-variable quantum teleportation and photodetectors},
  author = {Sharma, Kunal and Sanders, Barry C. and Wilde, Mark M.},
  journal = {Phys. Rev. Res.},
  volume = {4},
  issue = {2},
  pages = {023066},
  numpages = {15},
  year = {2022},
  month = {Apr},
  publisher = {American Physical Society},
  doi = {10.1103/PhysRevResearch.4.023066},
  url = {https://link.aps.org/doi/10.1103/PhysRevResearch.4.023066}
}

@article{wang2020quantum,
  title = {Quantum sensing of open systems: Estimation of damping constants and temperature},
  author = {Wang, J. and Davidovich, L. and Agarwal, G. S.},
  journal = {Phys. Rev. Res.},
  volume = {2},
  issue = {3},
  pages = {033389},
  numpages = {8},
  year = {2020},
  month = {Sep},
  publisher = {American Physical Society},
  doi = {10.1103/PhysRevResearch.2.033389},
  url = {https://link.aps.org/doi/10.1103/PhysRevResearch.2.033389}
}

@article{harney2021ultimate,
  title = {Ultimate limits of thermal pattern recognition},
  author = {Harney, Cillian and Banchi, Leonardo and Pirandola, Stefano},
  journal = {Phys. Rev. A},
  volume = {103},
  issue = {5},
  pages = {052406},
  numpages = {14},
  year = {2021},
  month = {May},
  publisher = {American Physical Society},
  doi = {10.1103/PhysRevA.103.052406},
  url = {https://link.aps.org/doi/10.1103/PhysRevA.103.052406}
}

@article{zhuang2021quantum,
  title = {Quantum Ranging with Gaussian Entanglement},
  author = {Zhuang, Quntao},
  journal = {Phys. Rev. Lett.},
  volume = {126},
  issue = {24},
  pages = {240501},
  numpages = {7},
  year = {2021},
  month = {Jun},
  publisher = {American Physical Society},
  doi = {10.1103/PhysRevLett.126.240501},
  url = {https://link.aps.org/doi/10.1103/PhysRevLett.126.240501}
}

@book{barnett2002methods,
author = {Barnett, Stephen M. and Radmore, Paul M.},
    title = "{Methods in Theoretical Quantum Optics}",
    publisher = {Oxford University Press},
    year = {2002},
    month = {11},
    isbn = {9780198563617},
    doi = {10.1093/acprof:oso/9780198563617.001.0001},
    url = {https://doi.org/10.1093/acprof:oso/9780198563617.001.0001}
}

@book{rohatgi2015introduction,
  title={An introduction to probability and statistics},
  author={Rohatgi, Vijay K and Saleh, AK Md Ehsanes},
  year={2015},
  month = {08},
  publisher={John Wiley \& Sons},
  isbn = {9781118799642},
  doi = {10.1002/9781118799635}
}

@article{nair2011optimal,
  title = {Optimal Quantum States for Image Sensing in Loss},
  author = {Nair, Ranjith and Yen, Brent J.},
  journal = {Phys. Rev. Lett.},
  volume = {107},
  issue = {19},
  pages = {193602},
  numpages = {5},
  year = {2011},
  month = {Oct},
  publisher = {American Physical Society},
  doi = {10.1103/PhysRevLett.107.193602},
  url = {https://link.aps.org/doi/10.1103/PhysRevLett.107.193602}

}

@article{fujiwara2001quantum,
  title = {Quantum channel identification problem},
  author = {Fujiwara, Akio},
  journal = {Phys. Rev. A},
  volume = {63},
  issue = {4},
  pages = {042304},
  numpages = {4},
  year = {2001},
  month = {Mar},
  publisher = {American Physical Society},
  doi = {10.1103/PhysRevA.63.042304},
  url = {https://link.aps.org/doi/10.1103/PhysRevA.63.042304}

}

@article{meyer2020single,
author = {Meyer-Scott, Evan and Silberhorn, Christine and Migdall, Alan},
    title = "{Single-photon sources: Approaching the ideal through multiplexing}",
    journal = {Review of Scientific Instruments},
    volume = {91},
    number = {4},
    pages = {041101},
    year = {2020},
    month = {04},
    issn = {0034-6748},
    doi = {10.1063/5.0003320}
}

@article{shirokov2018uniform,
  title={Uniform continuity bounds for information characteristics of quantum channels depending on input dimension and on input energy},
  author={Shirokov, Maksim E},
  journal={Journal of Physics A: Mathematical and Theoretical},
  volume={52},
  number={1},
  pages={014001},
  year={2018},
  publisher={IOP Publishing},
  doi = {10.1088/1751-8121/aaebac}
}

@article{pirandola2017ultimate,
  title = {Ultimate Precision of Adaptive Noise Estimation},
  author = {Pirandola, Stefano and Lupo, Cosmo},
  journal = {Phys. Rev. Lett.},
  volume = {118},
  issue = {10},
  pages = {100502},
  numpages = {6},
  year = {2017},
  month = {Mar},
  publisher = {American Physical Society},
  doi = {10.1103/PhysRevLett.118.100502},
  url = {https://link.aps.org/doi/10.1103/PhysRevLett.118.100502}
}

@article{winter2017energy,
  title={Energy-constrained diamond norm with applications to the uniform continuity of continuous variable channel capacities},
  author={Winter, Andreas},
  journal={arXiv preprint arXiv:1712.10267},
  year={2017},
  doi = {10.48550/arXiv.1712.10267}
}

@article{shirokov2018energy,
  title={On the energy-constrained diamond norm and its application in quantum information theory},
  author={Shirokov, Maksim E},
  journal={Problems of Information Transmission},
  volume={54},
  pages={20--33},
  year={2018},
  publisher={Springer},
  doi = {10.1134/S0032946018010027}
}

@article{shirokov2019energy,
  title={Energy-Constrained Diamond Norms and Quantum Dynamical Semigroups},
  author={Shirokov, ME and Holevo, AS},
  journal={Lobachevskii Journal of Mathematics},
  volume={40},
  pages={1569--1586},
  year={2019},
  publisher={Springer},
  doi = {10.1134/S199508021910024X}
}

@article{becker2020convergence,
  title={Convergence rates for quantum evolution and entropic continuity bounds in infinite dimensions},
  author={Becker, Simon and Datta, Nilanjana},
  journal={Communications in Mathematical Physics},
  volume={374},
  pages={823--871},
  year={2020},
  publisher={Springer},
  doi = {10.1007/s00220-019-03594-2}
}

@article{becker2021energy,
  title = {Energy-Constrained Discrimination of Unitaries, Quantum Speed Limits, and a Gaussian Solovay-Kitaev Theorem},
  author = {Becker, Simon and Datta, Nilanjana and Lami, Ludovico and Rouz\'e, Cambyse},
  journal = {Phys. Rev. Lett.},
  volume = {126},
  issue = {19},
  pages = {190504},
  numpages = {7},
  year = {2021},
  month = {May},
  publisher = {American Physical Society},
  doi = {10.1103/PhysRevLett.126.190504},
  url = {https://link.aps.org/doi/10.1103/PhysRevLett.126.190504}
}

@article{takeoka2016optimal,
  title={Optimal estimation and discrimination of excess noise in thermal and amplifier channels},
  author={Takeoka, Masahiro and Wilde, Mark M},
  journal={arXiv preprint arXiv:1611.09165},
  year={2016},
  doi = {10.48550/arXiv.1611.09165}
}

@article{audenaert2012comparisons,
author = {Audenaert, Koenraad M. R.},
title = {Comparisons between quantum state distinguishability measures},
year = {2014},
issue_date = {January 2014},
publisher = {Rinton Press, Incorporated},
address = {Paramus, NJ},
volume = {14},
number = {1–2},
issn = {1533-7146},
journal = {Quantum Info. Comput.},
month = {jan},
pages = {31–38},
numpages = {8},
doi = {10.5555/2600498.2600500}
}

@article{liu2020quantum,
  title={Quantum Fisher information matrix and multiparameter estimation},
  author={Liu, Jing and Yuan, Haidong and Lu, Xiao-Ming and Wang, Xiaoguang},
  journal={Journal of Physics A: Mathematical and Theoretical},
  volume={53},
  number={2},
  pages={023001},
  year={2020},
  publisher={IOP Publishing},
  doi = {10.1088/1751-8121/ab5d4d}
}

@article{guo2020distributed,
  title={Distributed quantum sensing in a continuous-variable entangled network},
  author={Guo, Xueshi and Breum, Casper R and Borregaard, Johannes and Izumi, Shuro and Larsen, Mikkel V and Gehring, Tobias and Christandl, Matthias and Neergaard-Nielsen, Jonas S and Andersen, Ulrik L},
  journal={Nature Physics},
  volume={16},
  number={3},
  pages={281--284},
  year={2020},
  publisher={Nature Publishing Group UK London},
  doi = {10.1038/s41567-019-0743-x}
}

@article{martin2011using,
  title = {Using Berry's Phase to Detect the Unruh Effect at Lower Accelerations},
  author = {Mart\'{\i}n-Mart\'{\i}nez, Eduardo and Fuentes, Ivette and Mann, Robert B.},
  journal = {Phys. Rev. Lett.},
  volume = {107},
  issue = {13},
  pages = {131301},
  numpages = {5},
  year = {2011},
  month = {Sep},
  publisher = {American Physical Society},
  doi = {10.1103/PhysRevLett.107.131301},
  url = {https://link.aps.org/doi/10.1103/PhysRevLett.107.131301}
}

@article{ahmadi2014quantum,
  title = {Quantum metrology for relativistic quantum fields},
  author = {Ahmadi, Mehdi and Bruschi, David Edward and Fuentes, Ivette},
  journal = {Phys. Rev. D},
  volume = {89},
  issue = {6},
  pages = {065028},
  numpages = {10},
  year = {2014},
  month = {Mar},
  publisher = {American Physical Society},
  doi = {10.1103/PhysRevD.89.065028},
  url = {https://link.aps.org/doi/10.1103/PhysRevD.89.065028}
}

@article{ahmadi2014relativistic,
  title={Relativistic quantum metrology: Exploiting relativity to improve quantum measurement technologies},
  author={Ahmadi, Mehdi and Bruschi, David Edward and Sab{\'\i}n, Carlos and Adesso, Gerardo and Fuentes, Ivette},
  journal={Scientific reports},
  volume={4},
  number={1},
  pages={4996},
  year={2014},
  publisher={Nature Publishing Group UK London},
  doi = {10.1038/srep04996}
}

@article{braunstein1998teleportation,
  title = {Teleportation of Continuous Quantum Variables},
  author = {Braunstein, Samuel L. and Kimble, H. J.},
  journal = {Phys. Rev. Lett.},
  volume = {80},
  issue = {4},
  pages = {869--872},
  numpages = {0},
  year = {1998},
  month = {Jan},
  publisher = {American Physical Society},
  doi = {10.1103/PhysRevLett.80.869},
  url = {https://link.aps.org/doi/10.1103/PhysRevLett.80.869}
}

@article{hosler2013parameter,
  title = {Parameter estimation using NOON states over a relativistic quantum channel},
  author = {Hosler, Dominic and Kok, Pieter},
  journal = {Phys. Rev. A},
  volume = {88},
  issue = {5},
  pages = {052112},
  numpages = {5},
  year = {2013},
  month = {Nov},
  publisher = {American Physical Society},
  doi = {10.1103/PhysRevA.88.052112},
  url = {https://link.aps.org/doi/10.1103/PhysRevA.88.052112}
}

@article{joo2011quantum,
  title = {Quantum Metrology with Entangled Coherent States},
  author = {Joo, Jaewoo and Munro, William J. and Spiller, Timothy P.},
  journal = {Phys. Rev. Lett.},
  volume = {107},
  issue = {8},
  pages = {083601},
  numpages = {4},
  year = {2011},
  month = {Aug},
  publisher = {American Physical Society},
  doi = {10.1103/PhysRevLett.107.083601},
  url = {https://link.aps.org/doi/10.1103/PhysRevLett.107.083601}
}

@article{slussarenko2017unconditional,
  title={Unconditional violation of the shot-noise limit in photonic quantum metrology},
  author={Slussarenko, Sergei and Weston, Morgan M and Chrzanowski, Helen M and Shalm, Lynden K and Verma, Varun B and Nam, Sae Woo and Pryde, Geoff J},
  journal={Nature Photonics},
  volume={11},
  number={11},
  pages={700--703},
  year={2017},
  publisher={Nature Publishing Group UK London},
  doi = {10.1038/s41566-017-0011-5}
}

@article{mitchell2004super,
  title={Super-resolving phase measurements with a multiphoton entangled state},
  author={Mitchell, Morgan W and Lundeen, Jeff S and Steinberg, Aephraem M},
  journal={Nature},
  volume={429},
  number={6988},
  pages={161--164},
  year={2004},
  publisher={Nature Publishing Group UK London},
  doi = {10.1038/nature02493}
}

@article{lee2021quantum,
  title = {Quantum illumination via quantum-enhanced sensing},
  author = {Lee, Su-Yong and Ihn, Yong Sup and Kim, Zaeill},
  journal = {Phys. Rev. A},
  volume = {103},
  issue = {1},
  pages = {012411},
  numpages = {6},
  year = {2021},
  month = {Jan},
  publisher = {American Physical Society},
  doi = {10.1103/PhysRevA.103.012411},
  url = {https://link.aps.org/doi/10.1103/PhysRevA.103.012411}
}

@article{guha2009gaussian,
  title = {Gaussian-state quantum-illumination receivers for target detection},
  author = {Guha, Saikat and Erkmen, Baris I.},
  journal = {Phys. Rev. A},
  volume = {80},
  issue = {5},
  pages = {052310},
  numpages = {4},
  year = {2009},
  month = {Nov},
  publisher = {American Physical Society},
  doi = {10.1103/PhysRevA.80.052310},
  url = {https://link.aps.org/doi/10.1103/PhysRevA.80.052310}
}

@article{sanders1989quantum,
  title = {Quantum dynamics of the nonlinear rotator and the effects of continual spin measurement},
  author = {Sanders, Barry C.},
  journal = {Phys. Rev. A},
  volume = {40},
  issue = {5},
  pages = {2417--2427},
  numpages = {0},
  year = {1989},
  month = {Sep},
  publisher = {American Physical Society},
  doi = {10.1103/PhysRevA.40.2417},
  url = {https://link.aps.org/doi/10.1103/PhysRevA.40.2417}
}

@article{boto2000quantum,
  title = {Quantum Interferometric Optical Lithography: Exploiting Entanglement to Beat the Diffraction Limit},
  author = {Boto, Agedi N. and Kok, Pieter and Abrams, Daniel S. and Braunstein, Samuel L. and Williams, Colin P. and Dowling, Jonathan P.},
  journal = {Phys. Rev. Lett.},
  volume = {85},
  issue = {13},
  pages = {2733--2736},
  numpages = {0},
  year = {2000},
  month = {Sep},
  publisher = {American Physical Society},
  doi = {10.1103/PhysRevLett.85.2733},
  url = {https://link.aps.org/doi/10.1103/PhysRevLett.85.2733}
}

@article{lee2002quantum,
author = {Hwang Lee, Pieter Kok and Jonathan P. Dowling},
title = {A quantum Rosetta stone for interferometry},
journal = {Journal of Modern Optics},
volume = {49},
number = {14-15},
pages = {2325--2338},
year = {2002},
publisher = {Taylor \& Francis},
doi = {10.1080/0950034021000011536}
}

@article{dowling2008quantum,
author = {Jonathan P. Dowling},
title = {Quantum optical metrology – the lowdown on high-N00N states},
journal = {Contemporary Physics},
volume = {49},
number = {2},
pages = {125--143},
year = {2008},
publisher = {Taylor \& Francis},
doi = {10.1080/00107510802091298}
}

@article{tham2023quantum,
  title = {Quantum Limits of Covert Target Detection},
  author = {Tham, Guo Yao and Nair, Ranjith and Gu, Mile},
  journal = {Phys. Rev. Lett.},
  volume = {133},
  issue = {11},
  pages = {110801},
  numpages = {7},
  year = {2024},
  month = {Sep},
  publisher = {American Physical Society},
  doi = {10.1103/PhysRevLett.133.110801},
  url = {https://link.aps.org/doi/10.1103/PhysRevLett.133.110801}
}

@article{sacchi2005optimal,
  title = {Optimal discrimination of quantum operations},
  author = {Sacchi, Massimiliano F.},
  journal = {Phys. Rev. A},
  volume = {71},
  issue = {6},
  pages = {062340},
  numpages = {4},
  year = {2005},
  month = {Jun},
  publisher = {American Physical Society},
  doi = {10.1103/PhysRevA.71.062340},
  url = {https://link.aps.org/doi/10.1103/PhysRevA.71.062340}
}

@article{sacchi2005entanglement,
  title = {Entanglement can enhance the distinguishability of entanglement-breaking channels},
  author = {Sacchi, Massimiliano F.},
  journal = {Phys. Rev. A},
  volume = {72},
  issue = {1},
  pages = {014305},
  numpages = {2},
  year = {2005},
  month = {Jul},
  publisher = {American Physical Society},
  doi = {10.1103/PhysRevA.72.014305},
  url = {https://link.aps.org/doi/10.1103/PhysRevA.72.014305}
}

@article{shapiro2020quantum,
  author={Shapiro, Jeffrey H.},
  journal={IEEE Aerospace and Electronic Systems Magazine}, 
  title={The Quantum Illumination Story}, 
  year={2020},
  volume={35},
  number={4},
  pages={8-20},
  doi={10.1109/MAES.2019.2957870}
}

@article{shapiro2009quantum,
  author={Shapiro, Jeffrey H.},
  journal={IEEE Journal of Selected Topics in Quantum Electronics}, 
  title={The Quantum Theory of Optical Communications}, 
  year={2009},
  volume={15},
  number={6},
  pages={1547-1569},
  doi={10.1109/JSTQE.2009.2024959}
}

@article{shapiro2012physics,
  title={The physics of ghost imaging},
  author={Shapiro, Jeffrey H and Boyd, Robert W},
  journal={Quantum Information Processing},
  volume={11},
  number={4},
  pages={949--993},
  year={2012},
  publisher={Springer},
  doi = {10.1007/s11128-011-0356-5}
}

@article{shapiro2009quantumcoherent,
  title={Quantum illumination versus coherent-state target detection},
  author={Shapiro, Jeffrey H and Lloyd, Seth},
  journal={New Journal of Physics},
  volume={11},
  number={6},
  pages={063045},
  year={2009},
  publisher={IOP Publishing},
  doi = {10.1088/1367-2630/11/6/063045}
}

@article{glauber1963coherent,
  title = {Coherent and Incoherent States of the Radiation Field},
  author = {Glauber, Roy J.},
  journal = {Phys. Rev.},
  volume = {131},
  issue = {6},
  pages = {2766--2788},
  numpages = {0},
  year = {1963},
  month = {Sep},
  publisher = {American Physical Society},
  doi = {10.1103/PhysRev.131.2766},
  url = {https://link.aps.org/doi/10.1103/PhysRev.131.2766}
}

@article{lee2023bound,
author = {Su-Yong Lee and Dong Hwan Kim and Yonggi Jo and Taek Jeong and Zaeill Kim and Duk Y. Kim},
journal = {Opt. Express},
number = {23},
pages = {38977--38988},
publisher = {Optica Publishing Group},
title = {Bound for Gaussian-state quantum illumination using a direct photon measurement},
volume = {31},
month = {Nov},
year = {2023},
url = {https://opg.optica.org/oe/abstract.cfm?URI=oe-31-23-38977},
doi = {10.1364/OE.505405}
}

@article{giovannetti2014nature,
  title={Advances in quantum metrology},
  author={Giovannetti, Vittorio and Lloyd, Seth and Maccone, Lorenzo},
  journal={Nature Photonics},
  volume={5},
  number={4},
  pages={222--229},
  year={2011},
  publisher={Nature Publishing Group UK London},
  doi = {10.1038/nphoton.2011.35}
}

@article{caves1981quantum,
  title = {Quantum-mechanical noise in an interferometer},
  author = {Caves, Carlton M.},
  journal = {Phys. Rev. D},
  volume = {23},
  issue = {8},
  pages = {1693--1708},
  numpages = {0},
  year = {1981},
  month = {Apr},
  publisher = {American Physical Society},
  doi = {10.1103/PhysRevD.23.1693},
  url = {https://link.aps.org/doi/10.1103/PhysRevD.23.1693}
}

@article{demkowicz2015quantum,
title = {Chapter Four - Quantum Limits in Optical Interferometry},
editor = {E. Wolf},
series = {Progress in Optics},
publisher = {Elsevier},
volume = {60},
pages = {345-435},
year = {2015},
issn = {0079-6638},
doi = {https://doi.org/10.1016/bs.po.2015.02.003},
url = {https://www.sciencedirect.com/science/article/pii/S0079663815000049},
author = {Rafal Demkowicz-Dobrzański and Marcin Jarzyna and Jan Kołodyński}
}

@article{brida2010experimental,
  title={Experimental realization of sub-shot-noise quantum imaging},
  author={Brida, Giorgio and Genovese, Marco and Ruo Berchera, Ivano},
  journal={Nature Photonics},
  volume={4},
  number={4},
  pages={227--230},
  year={2010},
  publisher={Nature Publishing Group UK London},
  doi = {10.1038/nphoton.2010.29}
}

@article{whittaker2017absorption,
  title={Absorption spectroscopy at the ultimate quantum limit from single-photon states},
  author={Whittaker, Rebecca and Erven, Chris and Neville, Alex and Berry, Monica and O’Brien, JL and Cable, Hugo and Matthews, JCF},
  journal={New Journal of Physics},
  volume={19},
  number={2},
  pages={023013},
  year={2017},
  publisher={IOP Publishing},
  doi = {10.1088/1367-2630/aa5512}
}

@article{monras2007optimal,
  title = {Optimal Quantum Estimation of Loss in Bosonic Channels},
  author = {Monras, Alex and Paris, Matteo G. A.},
  journal = {Phys. Rev. Lett.},
  volume = {98},
  issue = {16},
  pages = {160401},
  numpages = {4},
  year = {2007},
  month = {Apr},
  publisher = {American Physical Society},
  doi = {10.1103/PhysRevLett.98.160401},
  url = {https://link.aps.org/doi/10.1103/PhysRevLett.98.160401}
}

@article{chang2011chernoff,
  author={Chang, Seok-Ho and Cosman, Pamela C. and Milstein, Laurence B.},
  journal={IEEE Transactions on Communications}, 
  title={Chernoff-Type Bounds for the Gaussian Error Function}, 
  year={2011},
  volume={59},
  number={11},
  pages={2939-2944},
  doi={10.1109/TCOMM.2011.072011.100049}
}

@article{skolnik1962introduction,
  title={Introduction to radar},
  author={Skolnik, Merrill I},
  journal={Radar Handbook},
  volume={2},
  pages={21},
  year={1962},
}

@book{richards2010principles,
  editor = {Mark A. Richards},
   affiliation = {Electrical and Computer Engineering Department, Georgia Insitute of Technology},
   editor = {James A. Scheer},
   affiliation = {Georgia Insitute of Technology},
   editor = {William A. Holm},
   affiliation = {Georgia Institute of Technology},
      title = {Principles of Modern Radar: Basic principles},
      publisher = {Institution of Engineering and Technology},
      year = {2010},
      series = {Radar, Sonar and Navigation},
      url = {https://digital-library.theiet.org/content/books/ra/sbra021e},
      doi = {10.1049/SBRA021E},
      isbn = {9781891121524}
}

@article{levanon1988radar,
  title={Radar principles},
  author={Levanon, Nadav},
  journal={New York},
  year={1988},
  isbn = {9780471858812}
}

@article{weedbrook2016discord,
  title={How discord underlies the noise resilience of quantum illumination},
  author={Weedbrook, Christian and Pirandola, Stefano and Thompson, Jayne and Vedral, Vlatko and Gu, Mile},
  journal={New Journal of Physics},
  volume={18},
  number={4},
  pages={043027},
  year={2016},
  publisher={IOP Publishing},
  doi = {10.1088/1367-2630/18/4/043027}
}

@article{bradshaw2017overarching,
  title = {Overarching framework between Gaussian quantum discord and Gaussian quantum illumination},
  author = {Bradshaw, Mark and Assad, Syed M. and Haw, Jing Yan and Tan, Si-Hui and Lam, Ping Koy and Gu, Mile},
  journal = {Phys. Rev. A},
  volume = {95},
  issue = {2},
  pages = {022333},
  numpages = {10},
  year = {2017},
  month = {Feb},
  publisher = {American Physical Society},
  doi = {10.1103/PhysRevA.95.022333},
  url = {https://link.aps.org/doi/10.1103/PhysRevA.95.022333}
}

@article{volkoff2024not,
  title={Not even 6 dB: Gaussian quantum illumination in thermal background},
  author={Volkoff, TJ},
  journal={Journal of Physics A: Mathematical and Theoretical},
  volume={57},
  number={6},
  pages={065301},
  year={2024},
  publisher={IOP Publishing},
  doi = {10.1088/1751-8121/ad1e18}
}

    \begin{appendices}
        \chapter{Characteristic function of thermal state for thermal loss channel}\label{appendixA}
In this section, we will give a detailed derivation of the characteristic function of the thermal state for thermal loss channel in Eq.~(\ref{eq:characteristicfunction}) from the main text. Remember that the characteristic function of the output state from thermal loss channel is evaluated as
\begin{align}
    \chi_{TL}^{out}=\chi^{in}(\xi)\expval{e^{-\xi^*\sqrt{1-\eta}\hat{a}_{th}+\xi\sqrt{1-\eta}\hat{a}_{th}^\dag}}.
\end{align}
Expressing thermal state in P-representation $\rho_{\text{th}}=\frac{1}{\pi N_B}\int d^2\alpha e^{-\abs{\alpha}^2/N_B}\ket{\alpha}\bra{\alpha}$ for $\ket{\alpha} = \ket{\alpha_r+i\alpha_i}$ being coherent state, hence
\begin{align}
    \expval{e^{-\xi^*\sqrt{1-\eta}\hat{a}_{th}+\xi\sqrt{1-\eta}\hat{a}_{th}^\dag}}&=\Tr\qty(\rho_{\text{th}}e^{-\xi^*\sqrt{1-\eta}\hat{a}_{th}+\xi\sqrt{1-\eta}\hat{a}_{th}^\dag})\nonumber\\
    &=\frac{1}{\pi N_B}\int d^2\alpha e^{-\abs{\alpha}^2/N_B}\bra{\alpha}e^{-\xi^*\sqrt{1-\eta}\hat{a}_{th}+\xi\sqrt{1-\eta}\hat{a}_{th}^\dag}\ket{\alpha}\nonumber\\
    &=\frac{e^{-\abs{\xi\sqrt{1-\eta}}^2/2}}{\pi N_B}\int d^2\alpha e^{-\abs{\alpha}^2/N_B}\bra{\alpha}e^{-\xi^*\sqrt{1-\eta}\hat{a}_{th}}e^{\xi\sqrt{1-\eta}\hat{a}_{th}^\dag}\ket{\alpha}\nonumber\\
    &=\frac{e^{-\abs{\xi\sqrt{1-\eta}}^2/2}}{\pi N_B}\int d^2\alpha e^{-\abs{\alpha}^2/N_B}e^{\sqrt{1-\eta}(\xi\alpha^* - \xi^*\alpha)},
\end{align}
where we invoked the Kermack-McCrae identity in the third line to re-express the displacement operator. Splitting the integral into the real part and imaginary part of $\alpha$ yields
\begin{align}
    \int d^2\alpha e^{-\abs{\alpha}^2/N_B}e^{\sqrt{1-\eta}(\xi\alpha^* - \xi^*\alpha)}&=\int d^2\alpha e^{-(\alpha_r^2+\alpha_i^2)/N_B}e^{\sqrt{1-\eta}(-2i\text{Re}[\xi]\alpha_i+2i\text{Im}[\xi]\alpha_r)}\nonumber\\
    &=\int d\alpha_r e^{-[\alpha_r^2/N_B - 2i\sqrt{1-\eta}\text{Im}[\xi]\alpha_r]}\int d\alpha_i e^{-[\alpha_i^2/N_B + 2i\sqrt{1-\eta}\text{Re}[\xi]\alpha_i]}\nonumber\\
    &=\qty[\sqrt{\frac{\pi}{1/N_B}}e^{\frac{\qty(2i\sqrt{1-\eta}\text{Im}[\xi])^2}{4\qty(1/N_B)}}]\qty[\sqrt{\frac{\pi}{1/N_B}}e^{\frac{\qty(-2i\sqrt{1-\eta}\text{Re}[\xi])^2}{4\qty(1/N_B)}}]\nonumber\\
    &=\pi N_B e^{-(1-\eta)N_B(\text{Im}[\xi]^2 + \text{Re}[\xi]^2)}\nonumber\\
    &=\pi N_B e^{-(1-\eta)N_B\abs{\xi}^2}.
\end{align}
Therefore, we arrive at the final expression
\begin{align}
    \expval{e^{-\xi^*\sqrt{1-\eta}\hat{a}_{th}+\xi\sqrt{1-\eta}\hat{a}_{th}^\dag}}&=\frac{e^{-\abs{\xi\sqrt{1-\eta}}^2/2}}{\pi N_B}\int d^2\alpha e^{-\abs{\alpha}^2/N_B}e^{\sqrt{1-\eta}(\xi\alpha^* - \xi^*\alpha)}\nonumber\\
    &=\frac{e^{-\abs{\xi\sqrt{1-\eta}}^2/2}}{\pi N_B}\pi N_B e^{-(1-\eta)N_B\abs{\xi}^2}\nonumber\\
    &=e^{-(1-\eta)\abs{\xi}^2(2N_B+1)/2}.
\end{align}

        \chapter{Lower error probability bound of Alice in distinguishing thermal loss channel}\label{appendixB}
This section provides a detailed derivation of the lower bound of Alice's error probability in distinguishing between the output states of any two thermal loss channel. For the pure state probes $\Psi = \ket{\psi}\bra{\psi}_{IS}$ prepared by Alice with
\begin{align}
    \ket{\psi}_{IS}=\sum_{\mathbf{n}}\sqrt{p_{\mathbf{n}}}\ket{\chi_{\mathbf{n}}}_I\ket{\mathbf{n}}_S,
\end{align}
number-diagonal-signal (NDS) states is a class of states satisfying the orthogonality of the idler mode $\bra{\chi_{\mathbf{n}}}\ket{\chi_{\mathbf{n'}}}=\delta_{\mathbf{n},\mathbf{n'}}$. From the phase covariance of thermal loss channels and the results from ref.~\cite{sharma2018bounding}, considering all probes with a given PMF $\{p_{\mathbf{n}}\}$, NDS transmitters results in the smallest error probability. As such, we can limit the choices of probe to NDS states for deriving the lower bound on Alice's error probability $P_e^A$. We consider the general case where Alice is tasked to distinguish between the output states from any two thermal loss channel with the given input $\Psi=\ket{\psi}\bra{\psi}_{IS}$,
\begin{align}
     &\text{H$_0$: }\rho_0 = \qty(\mathbb{I}_I\otimes \mathcal{L}_{\kappa_0,N_0}^{\otimes M})\Psi = \qty(\mathbb{I}_I\otimes\qty(\mathcal{A}_{G_0}\circ\mathcal{L}_{\Tilde{\kappa}_0})^{\otimes M})\Psi,\label{eq:H0_thermalloss}\\
    &\text{H$_1$: }\rho_1 = \qty(\mathbb{I}_I\otimes \mathcal{L}_{\kappa_1,N_1}^{\otimes M})\Psi = \qty(\mathbb{I}_I\otimes\qty(\mathcal{A}_{G_1}\circ\mathcal{L}_{\Tilde{\kappa}_1})^{\otimes M})\Psi\label{eq:H1_thermalloss},
\end{align}
where we have express the thermal loss channel as a concatenation of pure loss and quantum-limited amplifier channel for $b = \{0,1\}$ such that  
\begin{align}
    G_b &= (1-\kappa_b)N_B + 1,\\
    \Tilde{\kappa}_b &= \kappa_b/G_b.
\end{align}
The following proposition gives the result of the fidelity between the output states from two quantum-limited amplifier and is largely inspired from ref.~\cite{nair2022optimal}.
\begin{prop}\label{prop:amplifierfidelity}
    We consider two $M$-mode NDS state with $\ket{\mathbf{n}}=\ket{n_1}\cdots\ket{n_M}$ being the number state of the S mode
    \begin{align}
        \ket{\psi^{(0)}}_{IS} &= \sum_{\mathbf{n}}\sqrt{s^{(0)}_{\mathbf{n}}}\ket{\chi_{\mathbf{n}}}_I\ket{\mathbf{n}}_S,\\
         \ket{\psi^{(1)}}_{IS} &= \sum_{\mathbf{n}}\sqrt{s^{(1)}_{\mathbf{n}}}\ket{\chi_{\mathbf{n}}}_I\ket{\mathbf{n}}_S,
    \end{align}
where $\{s^{(0)}_{\mathbf{n}}\}$ and $\{s^{(1)}_{\mathbf{n}}\}$ are multimode photon ditributions for the respective states, and $\{\ket{\chi_{\mathbf{n}}}\}$ is a set of orthonormal states in mode $I$. Consider a quantum-limited amplifier $\mathcal{A}_G$ with unitary $\hat{U}_{\mathcal{A}}(\theta)$ acting on a number state input $\ket{\mathbf{n}}\ket{\mathbf{0}}$,
\begin{align}
    \hat{U}_{\mathcal{A}}(\theta)\ket{\mathbf{n}}\ket{\mathbf{0}} = \sum_{\mathbf{a\geq 0}}\prod_{m=1}^M\qty[\sech\theta]^{n_m+1}\sqrt{\binom{n_m+a_m}{a_m}}\qty[\tanh\theta]^{a_m}\ket{\mathbf{n}+\mathbf{a}}\ket{\mathbf{a}}\label{eq:gain_unitary},
\end{align}
where we have defined $\cosh^2\theta = G$. Suppose that now we apply the quantum-limited amplifier channel onto the two $M$-mode NDS state 
\begin{align}
   \tau_0 &= \qty(\mathbb{I}_I\otimes \mathcal{A}_{G_0}^{\otimes M})\qty(\ket{\psi^{(0)}}\bra{\psi^{(0)}}_{IS}) = \sum_{\mathbf{a\geq 0}}\varket{\varphi^{(0)}_{\mathbf{a}}}\varbra{\varphi^{(0)}_{\mathbf{a}}}\label{eq:H0_QLA},\\
    \tau_1 &= \qty(\mathbb{I}_I\otimes \mathcal{A}_{G_1}^{\otimes M})\qty(\ket{\psi^{(1)}}\bra{\psi^{(1)}}_{IS}) = \sum_{\mathbf{a\geq 0}}\varket{\varphi^{(1)}_{\mathbf{a}}}\varbra{\varphi^{(1)}_{\mathbf{a}}}\label{eq:H1_QLA}, 
\end{align}
with 
\begin{align}
    \varket{\varphi^{(0)}_{\mathbf{a}}} &= \qty[\tanh\theta^{(0)}]^{\Tr\mathbf{a}}\sum_{\mathbf{n\geq 0}}\sqrt{s^{(0)}_{\mathbf{n}}}\qty[\sech\theta^{(0)}]^{\Tr\mathbf{n} + M}\sqrt{\prod_{m=1}^M\binom{n_m+a_m}{a_m}}\ket{\chi_{\mathbf{n}}}_I\ket{\mathbf{n}}_S,\\
    \varket{\varphi^{(1)}_{\mathbf{a}}} &= \qty[\tanh\theta^{(1)}]^{\Tr\mathbf{a}}\sum_{\mathbf{n\geq 0}}\sqrt{s^{(1)}_{\mathbf{n}}}\qty[\sech\theta^{(1)}]^{\Tr\mathbf{n} + M}\sqrt{\prod_{m=1}^M\binom{n_m+a_m}{a_m}}\ket{\chi_{\mathbf{n}}}_I\ket{\mathbf{n}}_S,
\end{align}
being NDS states where we have define $\cosh^2\theta^{(0)}=G_0$ and $\cosh^2\theta^{(1)}=G_1$ respectively. Hence, the fidelity between the outputs of two quantum-limited amplifier is evaluated as
\begin{align}
    &\Tr\sqrt{\sqrt{\tau_0}\tau_1\sqrt{\tau_0}} \nonumber\\
    &\hspace{20pt}=\sum_{\mathbf{a\geq 0}}\llangle{\varphi^{(0)}_{\mathbf{a}}}|{\varphi^{(1)}_{\mathbf{a}}}\rrangle\nonumber\\
    & \hspace{20pt}= \sum_{\mathbf{n\geq 0}}\sqrt{s^{(0)}_{\mathbf{n}}s^{(1)}_{\mathbf{n}}}\qty[\sech\theta^{(0)}\sech\theta^{(1)}]^{n+M}\sum_{\mathbf{a\geq 0}}\qty[\prod_{m=1}^M\binom{n_m+a_m}{a_m}\qty[\tanh\theta^{(0)}\tanh\theta^{(1)}]^{a_m}],
\end{align}
where we have defined $\Tr\mathbf{n} \coloneqq n$. Using \textbf{Lemma B.1} from ref.~\cite{nair2022optimal}, we can further evaluate the fidelity 
\begin{align}
    &\Tr\sqrt{\sqrt{\tau_0}\tau_1\sqrt{\tau_0}}\nonumber\\
    &=\sum_{\mathbf{n\geq 0}}\sqrt{s^{(0)}_{\mathbf{n}}s^{(1)}_{\mathbf{n}}}\qty[\sech\theta^{(0)}\sech\theta^{(1)}]^{n+M}\sum_{a=0}^\infty\sum_{a\geq\mathbf{0}:\Tr\mathbf{a} = a}\qty[\prod_{m=1}^M\binom{n_m+a_m}{a_m}]\qty[\tanh\theta^{(0)}\tanh\theta^{(1)}]^a\nonumber\\
    &=\sum_{\mathbf{n\geq 0}}\sqrt{s^{(0)}_{\mathbf{n}}s^{(1)}_{\mathbf{n}}}\qty[\sech\theta^{(0)}\sech\theta^{(1)}]^{n+M}\sum_{a=0}^\infty\binom{n+M-1+a}{a}\qty[\tanh\theta^{(0)}\tanh\theta^{(1)}]^a\nonumber\\
    &=\sum_{\mathbf{n\geq 0}}\sqrt{s^{(0)}_{\mathbf{n}}s^{(1)}_{\mathbf{n}}}\qty[\sech\theta^{(0)}\sech\theta^{(1)}]^{n+M}\qty[1 - \tanh\theta^{(0)}\tanh\theta^{(1)}]^{-(n+M)}\nonumber\\
    &=\sum_{\mathbf{n\geq 0}}\sqrt{s^{(0)}_{\mathbf{n}}s^{(1)}_{\mathbf{n}}}\qty[\cosh\theta^{(0)}\cosh\theta^{(1)}-\sinh\theta^{(0)}\sinh\theta^{(0)}]^{-(n+M)}\nonumber\\
    &=\sum_{\mathbf{n\geq 0}}\sqrt{s^{(0)}_{\mathbf{n}}s^{(1)}_{\mathbf{n}}}\qty[\sech\qty(\theta^{(1)} - \theta^{(0)})]^{n+M}\nonumber\\
    &=\sum_{\mathbf{n\geq 0}}\sqrt{s^{(0)}_{\mathbf{n}}s^{(1)}_{\mathbf{n}}}\nu^{n+M},
\end{align}
where $\nu = \sech\qty(\theta^{(1)} - \theta^{(0)}) = \qty(\sqrt{G_0G_1} - \sqrt{(G_0-1)(G_1-1)})^{-1}\in(0,1]$.
\end{prop}\noindent
Consider a pure loss channel $\mathcal{L}_{\kappa}$ with a given number state input $\ket{\mathbf{n}}\ket{\mathbf{0}}$ where $\ket{\mathbf{n}}=\ket{n_1}\cdots\ket{n_m}$, the corresponding output will be
\begin{align}
    \hat{U}_{\mathcal{L}}(\phi)\ket{\mathbf{n}}\ket{\mathbf{0}} = \sum_{\mathbf{l=0}}^{\mathbf{n}}\prod_{m=1}^M\qty[\cos\phi]^{(n_m-l_m)}\qty[\sin\phi]^{l_m}\sqrt{\binom{n_m}{l_m}}\ket{\mathbf{n}-\mathbf{l}}\ket{\mathbf{l}}\label{eq:loss_unitary},
\end{align}
where we have defined $\kappa = \cos^2\phi$. Hence, for the pure loss channels acting on the state $\Psi$ in Eq.~(\ref{eq:H0_thermalloss}) and Eq.~(\ref{eq:H1_thermalloss}), we obtain the following states
\begin{align}
    \varsigma_0 &= \qty(\mathbb{I}_I\otimes \mathcal{L}_{\Tilde{\kappa}_0}^{\otimes M})\Psi = \sum_{\mathbf{l\geq 0}}\varket{\psi^{(0)}_{\mathbf{l}}}\varbra{\psi^{(0)}_{\mathbf{l}}}=\sum_{\mathbf{l\geq 0}}q^{(0)}_{\mathbf{l}}\ket{\psi^{(0)}_{\mathbf{l}}}\bra{\psi^{(0)}_{\mathbf{l}}} = \sum_{\mathbf{l\geq 0}}q^{(0)}_{\mathbf{l}}\Psi^{(0)}_{\mathbf{l}}\label{eq:H0_lossless},\\
    \varsigma_1 &= \qty(\mathbb{I}_I\otimes \mathcal{L}_{\Tilde{\kappa}_1}^{\otimes M})\Psi = \sum_{\mathbf{l\geq 0}}\varket{\psi^{(1)}_{\mathbf{l}}}\varbra{\psi^{(1)}_{\mathbf{l}}}=\sum_{\mathbf{l\geq 0}}q^{(1)}_{\mathbf{l}}\ket{\psi^{(1)}_{\mathbf{l}}}\bra{\psi^{(1)}_{\mathbf{l}}} = \sum_{\mathbf{l\geq 0}}q^{(1)}_{\mathbf{l}}\Psi^{(1)}_{\mathbf{l}}\label{eq:H1_lossless},
\end{align}
with
\begin{align}
    \varket{\psi^{(0)}_\mathbf{l}} &= \qty[\sin \phi^{(0)}]^{\Tr\mathbf{l}}\sum_{\mathbf{n\geq l}}\sqrt{p_{\mathbf{n}}}\qty[\cos\phi^{(0)}]^{\Tr\mathbf{n}-\Tr\mathbf{l}}\sqrt{\prod_{m=1}^M\binom{n_m}{l_m}}\ket{\chi_{\mathbf{n}}}_I\ket{\mathbf{n}-\mathbf{l}}_S\nonumber\\
    &= \qty[\sin \phi^{(0)}]^{\Tr\mathbf{l}}\sum_{\mathbf{n\geq 0}}\sqrt{p_{\mathbf{n+l}}}\qty[\cos\phi^{(0)}]^{\Tr\mathbf{n}}\sqrt{\prod_{m=1}^M\binom{n_m+l_m}{l_m}}\ket{\chi_{\mathbf{n+l}}}_I\ket{\mathbf{n}}_S\nonumber\\
    &=\sum_{\mathbf{n\geq 0}}\sqrt{r_{\mathbf{n}}^{(\mathbf{l})}}\ket{\zeta_{\mathbf{n}}^{(\mathbf{l})}}_I\ket{\mathbf{n}}_S\nonumber\\
    &=\sqrt{q^{(0)}_{\mathbf{l}}}\ket{\psi^{(0)}_{\mathbf{l}}},
\end{align}
being an NDS state with squared norm $q^{(0)}_\mathbf{l} =\bra{\psi^{(0)}_\mathbf{l}}\ket{\psi^{(0)}_\mathbf{l}}$. Similarly for $\varsigma_1$, we have
\begin{align}
    \varket{\psi^{(1)}_\mathbf{l}} &= \qty[\sin \phi^{(1)}]^{\Tr\mathbf{l}}\sum_{\mathbf{n\geq l}}\sqrt{p_{\mathbf{n}}}\qty[\cos\phi^{(1)}]^{\Tr\mathbf{n}-\Tr\mathbf{l}}\sqrt{\prod_{m=1}^M\binom{n_m}{l_m}}\ket{\chi_{\mathbf{n}}}_I\ket{\mathbf{n}-\mathbf{l}}_S\nonumber\\
    &= \qty[\sin \phi^{(1)}]^{\Tr\mathbf{l}}\sum_{\mathbf{n\geq 0}}\sqrt{p_{\mathbf{n+l}}}\qty[\cos\phi^{(1)}]^{\Tr\mathbf{n}}\sqrt{\prod_{m=1}^M\binom{n_m+l_m}{l_m}}\ket{\chi_{\mathbf{n+l}}}_I\ket{\mathbf{n}}_S\nonumber\\
    &=\sum_{\mathbf{n\geq 0}}\sqrt{r{'}_{\mathbf{n}}^{(\mathbf{l})}}\ket{\zeta_{\mathbf{n}}^{(\mathbf{l})}}_I\ket{\mathbf{n}}_S\nonumber\\
    &=\sqrt{q^{(1)}_{\mathbf{l}}}\ket{\psi^{(1)}_{\mathbf{l}}},
\end{align}
being an NDS state with squared norm $q^{(1)}_\mathbf{l}=\bra{\psi^{(1)}_{\mathbf{l}}}\ket{\psi^{(1)}_{\mathbf{l}}}$. In this case, we define $\Tilde{\kappa}_0 = \cos\phi^{(0)}$, $\Tilde{\kappa}_1 = \cos\phi^{(1)}$ and $\Tr x = \sum_{m=1}^M x_m$. From Eq.~(\ref{eq:H0_thermalloss}) and Eq.~(\ref{eq:H1_thermalloss}), we have 
\begin{align}
    \rho_0 &=\qty(\mathbb{I}_I\otimes\qty(\mathcal{A}_{G_0}\circ\mathcal{L}_{\Tilde{\kappa}_0})^{\otimes M})\Psi=\qty(\mathbb{I}_I\otimes\mathcal{A}_{G_0}^{\otimes M})\varsigma_0 = \sum_{\mathbf{l\geq 0}}q^{(0)}_{\mathbf{l}}\qty(\mathbb{I}_I\otimes\mathcal{A}_{G_0}^{\otimes M})\Psi^{(0)}_{\mathbf{l}},\\
    \rho_1 &=\qty(\mathbb{I}_I\otimes\qty(\mathcal{A}_{G_1}\circ\mathcal{L}_{\Tilde{\kappa}_1})^{\otimes M})\Psi=\qty(\mathbb{I}_I\otimes\mathcal{A}_{G_1}^{\otimes M})\varsigma_1= \sum_{\mathbf{l\geq 0}}q^{(1)}_{\mathbf{l}}\qty(\mathbb{I}_I\otimes\mathcal{A}_{G_1}^{\otimes M})\Psi^{(1)}_{\mathbf{l}},
\end{align}
which we can use the strong concavity property of fidelity and \textbf{Proposition~\ref{prop:amplifierfidelity}} to evaluate the upper bound of the fidelity between the output of the thermal loss channels
\begin{align}
    \mathcal{F}(\rho_0,\rho_1) & \geq \sum_{\mathbf{l\geq 0}}\sqrt{q^{(0)}_{\mathbf{l}}q^{(1)}_{\mathbf{l}}}\mathcal{F}\qty(\mathbb{I}_I\otimes\mathcal{A}_{G_0}^{\otimes M}\qty(\Psi^{(0)}_{\mathbf{l}}),\mathbb{I}_I\otimes\mathcal{A}_{G_1}^{\otimes M}\qty(\Psi^{(1)}_{\mathbf{l}}))\nonumber\\
    & = \sum_{\mathbf{l\geq 0}}\sum_{\mathbf{n\geq 0}}\sqrt{r^{(\mathbf{l})}_{\mathbf{n}}r{'}^{(\mathbf{l})}_{\mathbf{n}}}\nu^{n+M}\nonumber\\
    &=\nu^M\sum_{\mathbf{l\geq 0}}\sum_{\mathbf{n\geq 0}}p_{\mathbf{n+l}}\qty[\sin\phi^{(0)}\sin\phi^{(1)}]^{\Tr\mathbf{l}}\qty[\nu\cos\phi^{(0)}\cos\phi^{(1)}]^{\Tr\mathbf{n}}\qty[\prod_{m=1}^M\binom{n_m+l_m}{l_m}].
\end{align}
By shifting the index of summation $\mathbf{k = n+l}$, we rewrite the expression as
\begin{align}
    \mathcal{F}(\rho_0,\rho_1) &\geq \nu^M\sum_{\mathbf{k\geq 0}}p_{\mathbf{k}}\sum_{\mathbf{l\leq k}}\qty[\nu\cos\phi^{(0)}\cos\phi^{(1)}]^{\Tr[\mathbf{k}-\mathbf{l}]}\qty[\sin\phi^{(0)}\sin\phi^{(1)}]^{\Tr\mathbf{l}}\qty[\prod_{m=1}^M\binom{k_m}{l_m}]\nonumber\\
    &=\nu^M\sum_{\mathbf{k\geq 0}}p_{\mathbf{k}}\qty[\nu\cos\phi^{(0)}\cos\phi^{(1)}]^{\Tr\mathbf{k}}\sum_{\mathbf{l\leq k}}\qty[\nu^{-1}\tan\phi^{(0)}\tan\phi^{(1)}]^{\Tr\mathbf{l}}\qty[\prod_{m=1}^M\binom{k_m}{l_m}]\nonumber\\
    &=\nu^M\sum_{\mathbf{k\geq 0}}p_{\mathbf{k}}\qty[\nu\cos\phi^{(0)}\cos\phi^{(1)}]^{\Tr\mathbf{k}}\qty[1+\nu^{-1}\tan\phi^{(0)}\tan\phi^{(1)}]^{\Tr\mathbf{k}}\nonumber\\
    &=\nu^M\sum_{\mathbf{k\geq 0}}p_{\mathbf{k}}\qty[\nu\cos\phi^{(0)}\cos\phi^{(1)} + \sin\phi^{(0)}\sin\phi^{(1)}]^{\Tr\mathbf{k}}\nonumber\\
    &=\nu^M\sum_{n=0}^\infty p_n\qty[\nu\sqrt{\Tilde{\kappa}_0\Tilde{\kappa}_1} + \sqrt{(1-\Tilde{\kappa}_0)(1-\Tilde{\kappa}_1)}]^n,
\end{align}
which is the result of Eq.~(\ref{eq:fidelity_thermalloss}) from the main text. For the scenario of Alice's target detection in passive signature (PS) model, we define $\Tilde{\kappa}_0 = 0$, $G_0 = N_B+1$, $\Tilde{\kappa}_1 = \eta/G_1$, and $G_1 = (1-\eta)N_B+1$ to obtain Eq.~(\ref{eq:fidelitybound_Alice}), which is the fidelity lower bound of Alice's target detection in PS model. 
\\Applying Jensen's inequality, a lower bound dependent only as a function of the total probe energy $\mathcal{N}_S = \sum_{n=0}^\infty np_n$ can be obtained. From Eq.~(\ref{eq:fidelitybound_Alice}), we take the logarithm of the summation such that
\begin{align}
    \ln\qty[\sum_{n=0}^\infty p_n\qty(1-\frac{\eta}{(1-\eta)N_B + 1})^{n/2}] & \geq \sum_{n=0}^\infty p_n\ln\qty(1-\frac{\eta}{(1-\eta)N_B+1})^{n/2}\nonumber\\
    &=\frac{1}{2}\ln\qty(1-\frac{\eta}{(1-\eta)N_B+1})\sum_{n=0}^\infty n p_n\nonumber\\
    &=\ln\qty(1-\frac{\eta}{(1-\eta)N_B+1})^{\mathcal{N}_S/2},
\end{align}
such that we have the lower bound of the fidelity as
\begin{align}
    \Rightarrow \mathcal{F}(\rho_0,\rho_1)\geq\nu^M\sum_{n=0}^\infty p_n\qty(1-\frac{\eta}{(1-\eta)N_B + 1})^{n/2} &\geq \nu^M\qty(1 - \frac{\eta}{(1-\eta)N_B+1})^{\mathcal{N}_S/2}.
\end{align}
Applying the above bound to the lower bound of Eq.~(\ref{eq:QFB}),
\begin{align}
    P_e^A \geq \frac{1}{2}- \frac{1}{2}\sqrt{1-\mathcal{F}(\rho_0,\rho_1)},
\end{align}
we arrive at Eq.~(\ref{eq:errorprobabilityJensen_Alice}) of the main text, which characterises the error probability lower bound of Alice's target detection.
        \chapter{Analytical quantum Bhattacharyya exponent for GCS probe}\label{appendixC}
A detailed guide is provided in the derivation of the analytical result for Alice's target detection quantum Bhattacharyya exponent when using GCS probe from Eq.~(\ref{eq:perfcov_analytic_GCS}). Note that these methods are also used in deriving the result when TMSV probe is used, however not explicitly included in this thesis.\\
Consider a single mode coherent state $\ket{\alpha} = \ket{\alpha_r + i\alpha_i}$ transmitted by Alice to probe a target region. The mean vectors and the covariance matrices of the returned states for the two hypotheses are computed in the main text as follows
\begin{align}
     &\text{H}_0: \;q_0 = \begin{bmatrix}[0.8]
        0\\0
    \end{bmatrix};\quad V_0 = \begin{bmatrix}[0.8]
        N_B+\frac{1}{2} & 0\\ 0 & N_B+\frac{1}{2}
    \end{bmatrix},\\
    &\text{H}_1: \;q_1 = \begin{bmatrix}[0.8]
        \sqrt{2\eta\alpha_r}\\ \sqrt{2\eta\alpha_i}
    \end{bmatrix};\quad V_1 = \begin{bmatrix}[0.8]
        (1-\eta)N_B + \frac{1}{2} & 0 \\0 & (1-\eta)N_B+\frac{1}{2}
    \end{bmatrix},
\end{align}
whereby the corresponding symplectic matrices $S_b$ and the symplectic eigenvalues $\upsilon_b$ for $b\in\{0,1\}$ of the respective covariance matrices are derived,
\begin{align}
    V_0 &= S_0(\upsilon_0\mathbb{I}_2)S_0^T = \qty(\mathbb{I}_2)(N_B + 1/2)\mathbb{I}_2\qty(\mathbb{I}_2),\\
    V_1 &= S_1(\upsilon_1\mathbb{I}_2)S_1^T = \qty(\mathbb{I}_2)\qty[(1-\eta)N_B + 1/2)]\mathbb{I}_2\qty(\mathbb{I}_2),
\end{align}
where $\mathbb{I}_2$ is a $2$ by $2$ identity matrix. From Section~\ref{section:soverlap}, the above results are used to calculate the s-overlap of the returned states,
\begin{align}
   C_{1/2}[\alpha] &= \frac{\Pi_{1/2}}{\sqrt{\det \Sigma_{1/2}}}\exp\qty[-\frac{\delta_u^T\Sigma_{1/2}^{-1}\delta_u}{2}]\nonumber\\
   &=\mathcal{K}\exp\qty[-2\eta\Lambda^{-1}\abs{\alpha}^2],
\end{align}
for 
\begin{align}
    \mathcal{K} &= \frac{1}{N_B\sqrt{1-\eta} + \sqrt{N_B+1}\sqrt{(1-\eta)N_B+1}},\\
    \Lambda &= \frac{\sqrt{N_B+1} + \sqrt{N_B}}{\sqrt{N_B+1} - \sqrt{N_B}} + \frac{\sqrt{(1-\eta)N_B+1} + \sqrt{(1-\eta)N_B}}{\sqrt{(1-\eta)N_B+1} - \sqrt{(1-\eta)N_B}},
\end{align}
where we substituted $s=1/2$ to obtain the s-overlap for the subsequent calculation of the Bhattacharyya error exponent. Using this $C_{1/2}$, we evaluated Eq.~(\ref{eq:Perfect_GCS}) to obtain Alice's target detection Bhattacharyya exponent when using GCS probe
\begin{align}
    \chi_{GCS} &\simeq - \ln\int_\mathbb{C}d^2\alpha\;P(\alpha)\qty(C_{1/2}[\alpha])\nonumber\\
    &=-\ln\iint d\alpha_r\;d\alpha_i\; \frac{e^{-\abs{\alpha}^2/N_B}}{\pi N_B}\mathcal{K}\exp[-2\eta\Lambda^{-1}\abs{\alpha}^2]\nonumber\\
    &= -\ln\qty[\frac{\mathcal{K}}{\pi N_B}\iint d\abs{\alpha}\;d\theta\;\abs{\alpha} e^{-\abs{\alpha}^2\qty(1/N_B + 2\eta\Lambda^{-1})}]\nonumber\\
    &=\ln\qty[\frac{\mathcal{K}}{\pi N_B}\frac{2\pi}{2[1/N_B + 2\eta\Lambda^{-1}]}]\nonumber\\
    &=-\ln\qty[\frac{\mathcal{K}}{1 + 2\eta\Lambda^{-1}N_B}]\label{eq:xGCS}.
\end{align}
By assuming the target is a weakly reflecting target with $\eta\ll 1$, we perform Puiseux series expansion on Eq.~(\ref{eq:xGCS}) around $\eta = 0$
\begin{align}
    \chi_{GCS} \simeq -\ln\qty[1 - 2\eta N_B\qty(N_B - \sqrt{N_B(N_B+1)} + \frac{1}{2}) + \mathcal{O}(\eta^2)],
\end{align}
to obtain Eq~{\ref{eq:perfcov_analytic_GCS}} of the main text.
        \chapter{KKT condition on the range of allowable energies for an \texorpdfstring{$\epsilon$}{epsilon}-covert probe}\label{appendixD}

From the main text, we have discussed that the upper and lower limits on Alice's allowed signal energy beyond which $\epsilon$-covertness is violated can be obtained by solving the following optimisation problem:
\begin{align}
&\textbf{Extremise }f(q_1,q_2,\cdots)=\sum_{n=0}^\infty nq_n\nonumber\\
&\textbf{subject to }\nonumber\\
&\quad g(q_1,q_2,\cdots)=1-2\epsilon-\sum_{n=0}^\infty\sqrt{\binom{n+M-1}{n}\frac{N_B^n}{(N_B+1)^{n+M}}q_n}\leq 0\nonumber\\
&\quad h(q_1,q_2,\cdots)=\sum_{n=0}^\infty q_n-1=0\nonumber,
\end{align}
where the aim is to extremise the average energy of $\sum_{n=0}^\infty nq_n$ of the state received by Willie for his alternate hypothesis $\sigma_1$. The problem is subjected to two constraints: (1) the $\epsilon$-covert constraint following Eq.~(\ref{eq:ecovert_constraint}), (2) normalisation constraint on the probability mass function of Willie's total photon number $\{q_n\}$ under alternate hypothesis $\textsc{H}_1'$. This problem is considered to be non-linear programming as the $\epsilon$-covertness constraint $g(q_1,q_2,\cdots)$ follows an inequality. The Karush–Kuhn–Tucker (KKT) necessary conditions to this non-linear programming are thus as follows:
\\
\textbf{Stationarity}\\
To find the point $\vec{q}^*$ that minimises $f(q_1,q_2,\cdots)$ across all discrete probability distributions $\vec{q}$, the required condition is
\begin{align}
\frac{\partial}{\partial\vec{q}}\left.\qty[f+\lambda_1(g)+\lambda_2(h)]\right|_{\vec{q}=\vec{q}^*}=0,
\end{align}
which gives the system of equations for $k=1,2,\cdots$ such that 
\begin{align}
&\frac{\partial}{\partial q_k}\bigg[-\sum_{n=0}^\infty nq_n+\lambda_2\qty(\sum_{n=0}^\infty q_n-1)\nonumber\\
&\hspace{80pt}+\lambda_1\bigg(1-2\epsilon-\sum_{n=0}^\infty\sqrt{\binom{n+M-1}{n}\frac{N_B^n}{(N_B+1)^{n+M}}q_n}\bigg)\left.\bigg]\right|_{q_k=q_k^*}=0\nonumber\\
&-k-\lambda_1\sqrt{\binom{k+M-1}{k}\frac{N_B^k}{(N_B+1)^{k+M}}}\qty(\frac{1}{2\sqrt{q_k^*}})+\lambda_2=0\nonumber\\
&q_k^*=\frac{\lambda_1^2}{4(k-\lambda_2)^2}\qty[\binom{k+M-1}{k}\frac{N_B^k}{(N_B+1)^{k+M}}],\label{eq:lagrange1}
\end{align}
where $\lambda_1$ and $\lambda_2$ are the KKT multipliers.\\
\textbf{Primal feasibility}
\begin{align}
&\sum_{k=0}^\infty q_k^*-1=0\hspace{250pt}\\
&1 - 2\epsilon-\sum_{k=0}^\infty\sqrt{\binom{k+M-1}{k}\frac{N_B^k}{(N_B+1)^{k+M}}q_k^*}\leq 0
\end{align}
\textbf{Dual feasibility}
\begin{align}
&\lambda_1\geq 0\hspace{290pt}
\end{align}
\textbf{Complementary slackness}
\begin{align}
\lambda_1\qty(1-2\epsilon-\sum_{k=0}^\infty\sqrt{\binom{k+M-1}{k}\frac{N_B^k}{(N_B+1)^{k+M}}q_k^*})=0\hspace{190pt}
\end{align}
From the stationarity condition, it can be concluded that $\lambda_1\neq 0$ or the normalization condition for $\{q_n\}$ will be violated. Hence, by applying this condition into complementary slackness condition, it can be deduced that $\sum_{k=0}^\infty\sqrt{\binom{k+M-1}{k}\frac{N_B^k}{(N_B+1)^{k+M}}q_k^*}=1-2\epsilon$. In this case, the optimal solution is on the boundary of the constraint $g(q_1,q_2,\cdots)$. Substituting Eq. (\ref{eq:lagrange1}) into both constraints yield
\begin{align}
&\sum_{n=0}^{\infty}\frac{\lambda_1^2}{4(n-\lambda_2)^2}\qty[\binom{n+M-1}{n}\frac{N_B^n}{(N_B+1)^{n+M}}]=1,\nonumber\\
&\sum_{n=0}^{\infty} \qty[\binom{n+M-1}{n}\frac{N_B^n}{(N_B+1)^{n+M}}]\abs{\frac{\lambda_1}{2(n-\lambda_2)}}=1-2\epsilon.
\end{align}
Due to the infinite summation, we are unable to provide an analytical solution to the energy bounds. Hence to solve it numerically, we set the upper limit of summation to a finite number $d$,
\begin{align}
&\sum_{n=0}^d\mathcal{N}_1\frac{\lambda_1^2}{4(n-\lambda_2)^2}\qty[\binom{n+M-1}{n}\frac{N_B^n}{(N_B+1)^{n+M}}]=1,\nonumber\\
&\sum_{n=0}^d \mathcal{N}_2\qty[\binom{n+M-1}{n}\frac{N_B^n}{(N_B+1)^{n+M}}]\abs{\frac{\lambda_1}{2(n-\lambda_2)}}=1-2\epsilon,
\end{align}
where $\mathcal{N}_1$ and $\mathcal{N}_2$ are the re-normalisation factors for the finite summations. By selecting a sufficiently large $d$, and appropriate initial values of the KKT multipliers $\lambda_1$ and $\lambda_2$, the numerical solver of \textsc{Matlab} produces the numerical results for the set of $\{\lambda_1,\lambda_2\}$, which can be substituted into
\begin{align}
\sum_{n=0}^\infty nq_n=\sum_{n=0}^\infty n\frac{\lambda_1^2}{4(n-\lambda_2)^2}\qty[\binom{n+M-1}{n}\frac{N_B^n}{(N_B+1)^{n+M}}],
\end{align}
to obtain the upper limit and lower limit of the signal strength intercepted by Wille. Given the relationship between Alice's allowable signal energies and Willie's intercepted energies $\mathcal{N}_S = (\sum_n nq_n -\eta N_B)/(1-\eta)$, the upper and lower allowable probe energies  given $\epsilon$-covertness are obtained, as displayed in Fig.~\ref{fig:signalenergy} of the main text.
        \chapter{Tightness of Alice's fidelity bound under \texorpdfstring{$\epsilon$}{epsilon}-covert criteria}\label{appendixE}
In the main text, we have derived a lower bound for the fidelity between the states received by Alice under $\epsilon$-covertness criteria 
\begin{align}
    \mathcal{F}(\rho_0,\rho_1)\geq (1-2\epsilon)^2\nu^M\qty(N_B+1-\frac{N_B}{x})^M[\eta N_B(1-x)+1]^M,\label{eq:fidelitybound_ecovert}
\end{align}
using the relationship between the probability generating function (PGF) of the probe state and the PGF of the state received by Willie under alternate hypothesis $\sigma_1$. In order to check the tightness of this bound, we performed a numerical optimisation of Eq.~(\ref{eq:fidelitybound_Alice}) under the $\epsilon$-covertness constraint. Mathematically, we can rewrite the fidelity between the states received by Alice as a function of the PGF of the state $\sigma_1$,
\begin{align}
    \mathcal{F}(\rho_0,\rho_1) &\geq \nu^M\sum_{n=0}^\infty p_n\qty[1 - \frac{\eta}{(1-\eta)N_B+1}]^{n/2}\nonumber\\
    &=\nu^M\mathcal{P}_S(\xi)\nonumber\\
    &=\nu^M\qty[\frac{\eta_W G_W}{\xi(G_W - 1) - G_W(1 - \eta_W)+1}]^M\mathcal{P}_W\qty(1 - \frac{1-\xi}{G_W(\eta_W + \xi - 1) - \xi + 1})\nonumber\\
    &= \nu^M\qty[\frac{\eta_W G_W}{\xi(G_W - 1) - G_W(1 - \eta_W)+1}]^M\sum_{n=0}^\infty q_nx^n,
\end{align}
where $G_W = \eta  N_B + 1$, $\eta_W = (1-\eta)/G_W$, $\xi = \sqrt{1-\eta/[(1-\eta)N_B +1]}$ and $x = (1 - \xi)/[G_W(\eta_W + \xi - 1) - \xi + 1]$. The problem statement for the optimisation is hence
\begin{align}
\textbf{Minimise: }&f(q_1,q_2,\cdots)=\nu^M\qty[\frac{G_W\eta_W}{x(G_W - 1) - G_W(1 - \eta_W) + 1}]^M\sum_{n=0}^\infty q_n x^n\nonumber\\
\textbf{Subject to: }&g(q_1,q_2,\cdots)=\sum_{n=0}^\infty\sqrt{\binom{n+M-1}{n}\frac{N_B^n}{(N_B+1)^{n+M}}q_n}\geq 1-2\epsilon\nonumber\\
&h(q_1,q_2,\cdots)=\sum_{n=0}^\infty q_n = 1
\end{align}
where $g(q_1,q_2,\cdots)$ is the $\epsilon$-covert constraint following Eq.~(\ref{eq:ecovert_constraint}), and $h(q_1,q_2,\cdots)$ is the normalisation constraint on the probability mass function of Willie's total photon number $\{q_n\}$ under alternate hypothesis $\textsc{H}_1'$. The Karush-Kuhn-Tucker (KKT) necessary conditions for this non-linear programming problem are as follows:\\
\textbf{Stationarity}\\
To find the point $\Vec{q}^*$ that minimises $f(q_1,q_2,\cdots)$ across all discrete probability distributions $\Vec{q}$, the required condition is
\begin{align}
    \frac{\partial}{\partial q_k}\left.\qty[f + \lambda_1 g + \lambda_2 h]\right|_{q_k = q_k*}
\end{align}
which gives the system of equation, $k = 1,2,\cdots$ such that
\begin{align}
    q_k^* = \frac{\lambda_1^2 p_n}{4(\nu^M\mu^Mx^n + \lambda_2)^2},
\end{align}
where $\lambda_1$ and $\lambda_2$ are the KKT multipliers, $p_n = \binom{n+M-1}{n}\frac{N_B^n}{(N_B+1)^{n+M}}$, and $\mu = \frac{G_W\eta_W}{x(G_W - 1) - G_W(1 - \eta_W) + 1}$.\\
\textbf{Primal feasibility}
\begin{align}
  &1-2\epsilon - \sum_{n=0}^\infty\sqrt{\binom{n+M-1}{n}\frac{N_B^n}{(N_B+1)^{n+M}}q_n}\leq 0 \\
  &\sum_{n=0}^\infty q_n - 1 = 0
\end{align}
\textbf{Dual feasibility}
\begin{align}
    \lambda_1\geq 0
\end{align}
\textbf{Complementary slackness}
\begin{align}
    \lambda_1\qty(1-2\epsilon - \sum_{n=0}^\infty\sqrt{\binom{n+M-1}{n}\frac{N_B^n}{(N_B+1)^{n+M}}q_n}) = 0
\end{align}
From the stationarity condition, it can be concluded that $\lambda_1\neq 0$ or else the equality constraint is violated. Hence, by applying this condition into complementary slackness condition, it can be deduced that 
\begin{align} \nonumber
\qty(1-2\epsilon - \sum_{n=0}^\infty\sqrt{\binom{n+M-1}{n}\frac{N_B^n}{(N_B+1)^{n+M}}q_n}) = 0.
\end{align}
In this case, the optimal solution is on the boundary of the constraint $g(q_1,q_2,\cdots)$. Hence, expressing the two constraints as equality and confining the upper summation limit to a finite value,
\begin{align}
    &\sum_{n=0}^d\mathcal{N}_1\binom{n+M-1}{n}\frac{N_B^n}{(N_B+1)^{n+M}}\frac{\abs{\lambda_1}}{2\abs{\nu^M\mu^M x^n + \lambda_2}} = 1-2\epsilon,\\
    &\sum_{n=0}^d\mathcal{N}_2\binom{n+M-1}{n}\frac{N_B^n}{(N_B+1)^{n+M}}\frac{\lambda_1^2}{4(\nu^M\mu^M x^n + \lambda_2)^2}=1,
\end{align}
where $d$ is the upper summation limit, $\mathcal{N}_1$ and $\mathcal{N}_2$ are the re-normalisation factors for finite summations. Using numerical solver on \textsc{Matlab} with sufficiently large value of $d$, and appropriate initial values of $\lambda_1$ and $\lambda_2$, the above simultaneous equations can be solved and the optimal values of Alice's target detection fidelity are obtained. 
\begin{figure}
    \centering
    \begin{subfigure}
    \centering
        \includegraphics[width = 0.49\linewidth]{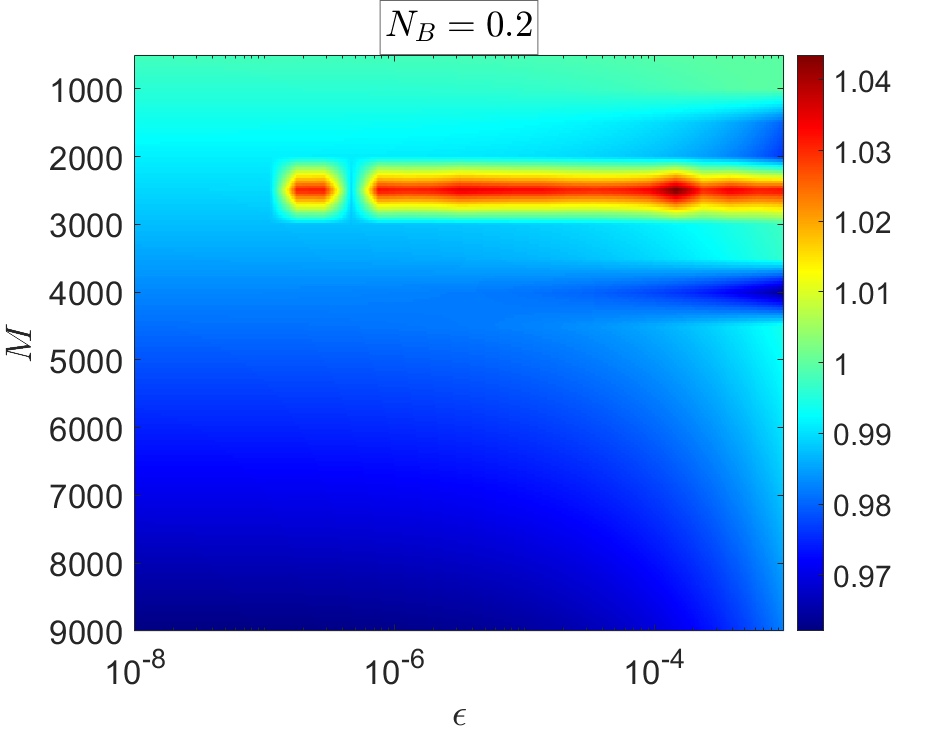}
    \end{subfigure}
        \begin{subfigure}
    \centering
        \includegraphics[width = 0.49\linewidth]{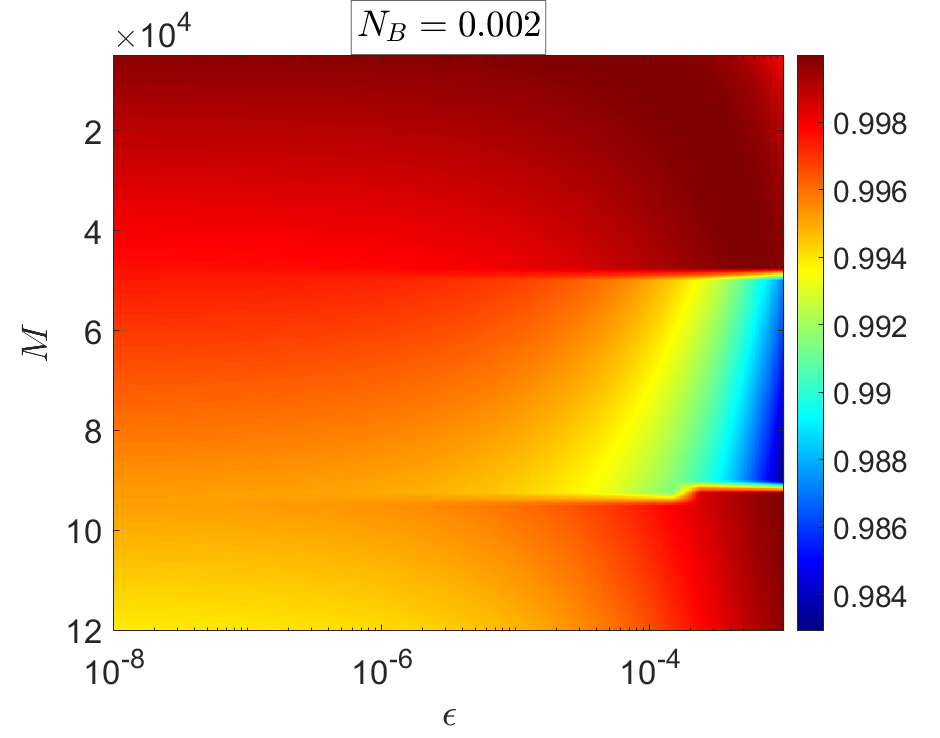}
    \end{subfigure}
    \captionsetup{width = \linewidth}
    \caption{Ratio of the analytical lower bound of fidelity derived in Eq.~(\ref{eq:Alice_fidelitybound_PGF}) to the numerically minimized fidelity under $\epsilon$-covertness.}
    \label{fig:AvsN}
\end{figure}
The heat maps of Fig.~{\ref{fig:AvsN}} compare the ratios of the numerically optimised fidelity and the analytical bound obtained in Eq.~(\ref{eq:Alice_fidelitybound_PGF}) across different values of $\epsilon$ and $M$ for two values of $N_B$. In our observations, we note that, aside from certain numerical artifacts that become evident for specific values of $\epsilon$ when $N_B=0.2$, the bound we have established closely aligns with the numerical minimum. This indicates a strong overall agreement between the analytical bound and the numerically determined minimum, with any discrepancies primarily manifesting as numerical artifacts rather than indicating a fundamental divergence between the two.
        \chapter{QFI of QLA gain sensing using coherent state probe}\label{appendixF}
We consider a single mode coherent state $\ket{\sqrt{N_m}}$ passing through a quantum-limited amplifier (QLA) $\mathcal{A}_{G}$ with gain parameter $G$. From the results of measurement, we are able to estimate a parameter $G'$ that is sufficiently close to the actual gain parameter such that the QLA is instead described as $\mathcal{A}_{G'}$. The output annihilation operator of a general QLA with gain $G$ is $\hat{a}_{out}=\sqrt{G}\hat{a}_{in}-\sqrt{G-1}\hat{e}_{in}^\dag$. Hence, the mean vectors and the covariance matrices of the output states from the two QLAs can be computed as
\begin{align}
    &\rho_{G}:\;q_{G} = \begin{bmatrix}[0.8]
        \sqrt{2N_mG}\\0
    \end{bmatrix};\qquad V_{G} = \begin{bmatrix}[0.8]
        (G-1)+\frac{1}{2} & 0\\0 & (G-1)+\frac{1}{2}
    \end{bmatrix},\\
     &\rho_{G'}:\;q_{G'} = \begin{bmatrix}[0.8]
        \sqrt{2N_mG'}\\0
    \end{bmatrix};\qquad V_{G'} = \begin{bmatrix}[0.8]
        (G'-1)+\frac{1}{2} & 0\\0 & (G'-1)+\frac{1}{2}
    \end{bmatrix}.
\end{align}
To calculate the quantum Fisher information, we can first compute the fidelity between the two output states. Since QLA operation is Gaussian-preserving, the output states remain Gaussian and we can thus use the method discussed in Section~\ref{section:fidelity_gaussian} to calculate the fidelity between the two states such that
\begin{align}
    \mathcal{F}(\rho_{G},\rho_{G'}) &= \Tr\sqrt{\sqrt{\rho_{G}}\rho_{G'}\sqrt{\rho_{G}}}\nonumber\\
    &=\mathcal{F}_0(V_{G},V_{G'})\exp\qty[-\frac{1}{4}\delta_u^T(V_{G} + V_{G'})^{-1}\delta_u],
\end{align}
where $\delta_u = q_{G'} - q_{G}$ is the difference in the mean vectors, and $\mathcal{F}_0(V_{G},V_{G'})$ is dependent only on the auxillary matrix defined as
\begin{align}
    V_{aux} &= \Omega^T(V_{G}+V_{G'})^{-1}\qty(\frac{\Omega}{4} + V_{G'}\Omega V_{G})\nonumber\\
    &=\frac{\frac{1}{4} + \qty(G - \frac{1}{2})\qty(G' - \frac{1}{2})}{G+G'-1}\mathbb{I}.
\end{align}
Hence, using Eq.~(\ref{eq:Ftot}), $\mathcal{F}_0(V_{G},V_{G'})$ can be computed to be
\begin{align}
    \mathcal{F}_0(V_{G},V_{G'}) &= \frac{F_{tot}}{\sqrt{G+G'-1}},
\end{align}
where we have the quantity
\begin{align}
    F_{tot} = \qty[\frac{\qty(2\frac{\sqrt{GG'(G-1)(G'-1)}}{G+G'-2GG'-1}+1)\qty(G + G' -2GG' - 1)}{G + G' - 1}]^{1/2}.
\end{align}
The fidelity between the output states of two QLA with distinct gain parameters for given coherent states input is thus computed to be
\begin{align}
    \mathcal{F}(\rho_{G},\rho_{G'}) &= \frac{F_{tot}}{\sqrt{G+G'-1}}\exp\qty[-\frac{\qty(\sqrt{G N_m} -\sqrt{G' N_m})^2}{2(G+G'-1)}].
\end{align}
With the quantum fidelity computed between the two output states of the QLAs, the quantum Fisher information can thus be derived to be
\begin{align}
    \mathcal{K}_{G}^{(\ket{\sqrt{N_m}})} &= -4\left.\frac{\partial^2\mathcal{F}(\rho_{G},\rho_{G'})}{\partial G'^2}\right|_{G'=G}\nonumber\\
    &=-4\qty[-\frac{1}{4G(G-1)} - \frac{N_m}{4G(2G-1)}]\nonumber\\
    &=\frac{1}{G(G-1)} + \frac{N_m}{G(2G-1)},
\end{align}
for a single-mode coherent probe state. In the main text, we used a $M$-mode coherent state probe $\ket{\sqrt{N_1}}\cdots\ket{\sqrt{N_M}}$, where each mode has a fraction of the total energy $N_m=N/M$. The resulting output state from the QLA is a product of single-mode Gaussian states. Hence, the quantum Fisher information for using the $M$-mode coherent state probe is simply
\begin{align}
    \mathcal{K}_{G}^{(\ket{\sqrt{N_1}}\cdots\ket{\sqrt{N_M}})} &=M\mathcal{K}_{G}^{(\ket{\sqrt{N_m}})}\nonumber\\
    &=\frac{M}{G(G-1)}+\frac{N}{G(2G-1)},
\end{align}
which gives the result in Eq.~(\ref{eq:QFI_gain_G_classical}) of the main text.
        \chapter{Nonunity detection efficiency on QLA gain sensing}\label{appendixG}
\renewcommand{\thesection}{\Roman{section}}
\let\oldaddcontentsline\addcontentsline
\renewcommand{\addcontentsline}[3]{}
For a non-ideal detection protocol, we simulate the loss in measurement by passing the output of the quantum-limited amplifier (QLA) through a pure loss channel as depicted in Fig.~\ref{fig:nonunity} of the main text. The annihilation operator of the measured mode is
\begin{align}
    \hat{b} &= \sqrt{\eta_d}\hat{a}_{out} + \sqrt{1-\eta_d}\hat{f}\nonumber\\
    &=\sqrt{\eta_d}\qty(\sqrt{G}\hat{a}_{in}+\sqrt{G-1}\hat{e}_{in}^\dag) + \sqrt{1-\eta_d}\hat{f}\label{eq:annihilation_loss_gain},
\end{align}
and the photon number operator of $\hat{b}$ becomes
\begin{align}
    \hat{N}_b &= \expval{\hat{b}^\dag\hat{b}}\nonumber\\
    &=\eta_d\hat{N}_{out} + \sqrt{\eta_d(1-\eta_d)}(\hat{a}_{out}^\dag\hat{f} + \hat{a}_{out}\hat{f}^\dag) + (1-\eta_d)\hat{f}^\dag\hat{f},
\end{align}
where $\eta_d$ is the transmittance of the loss channel, and $\hat{N}_{out} = \hat{a}_{out}^\dag\hat{a}_{out}$.

\section{Classical baseline}\label{appendixG_1}
In this section, we derive a theoretical quantum Fisher information (QFI) of the nonunity detection protocol for gain sensing using coherent state probe. Assuming the use of a single mode coherent state probe $\ket{\sqrt{N}}_S$ in Fig.~\ref{fig:nonunity} of the main text, but we allow for the possibility of performing any arbitrary measurements on the mode $\hat{b}$ downstream of the beamsplitter. This corresponds to the scenario where a system loss $\eta_d$ (including detection efficiency) is present in the system. The smallest mean squared error (MSE) (optimised over all quantum measurements) for estimating $G$ using classical probes is constrained by the quantum Cram\'er-Rao bound (QCRB)  associated with the state in the mode $\hat{b}$. Using Eq.~(\ref{eq:annihilation_loss_gain}), and noting that the modes $\hat{e}_{in}$ and $\hat{f}$ are in the vacuum state, the mean vector and the covariance matrix of the measured state in $\hat{b}$ mode are given by
\begin{align}
    \rho_G:\;q_G = \begin{bmatrix}[0.8]
        \sqrt{2\eta_d GN}\\0
    \end{bmatrix};\qquad V_G = \begin{bmatrix}[0.8]
        \eta_d(G-1) + \frac{1}{2} & 0\\0 &\eta_d(G-1) + \frac{1}{2}
    \end{bmatrix}.
\end{align}
Similarly, from the measurements of the output states, we made an assumption that the QLA has a parameter of $G'$. Since, both the QLA channel and pure loss channel are Gaussian-preserving, the resulting output state will be a Gaussian state (displaced thermal state), we can compute the fidelity and thus the QFI of parameters of the Gaussian states. Following closely to the method discussed in Appendix~\ref{appendixF}, we obtain the QFI of this setup to be
\begin{align}
    \mathcal{K}_G^{\text{coh}}=\frac{\eta_d N}{G[2\eta_d(G-1)+1]}+\frac{\eta_d}{(G-1)[\eta_d(G-1)+1]},
\end{align}
on $G$ in the lossy regime, which the result is plotted in Fig.~{\ref{fig:loss_sensing}} of the main text. Eq.~(\ref{eq:nonunity_coherent}) of the main text is the multimode coherent state probe generalisation of the above equation. As in the lossless case, it is unknown if the QFI can be achieved on a finite sample using a $G$-independent measurement.

\section{Quantum Fisher information for number-state probes}\label{appendixG_2}
We now consider the same scenario depicted in Fig.~\ref{fig:nonunity} of the main text, however with number-state probe $\ket{\psi}_S=\otimes_{m=1}^M\ket{n_m}_{S_m}$ instead. The fundamental bound on the MSE for estimating $G$ is set by the QCRB of the state family in the set of modes $\{\hat{b}_m\}_{m=1}^M$. Since the number-state probe are not Gaussian, we are unable to utilise the method of parameter estimation for Gaussian-state families as described in the previous section and Appendix~\ref{appendixF}. Instead, we first find the exact form of the measured state. From the action of the QLA on number-state inputs as described in Eq.~(\ref{eq:gain_unitary}), the output state of the amplifier for a single mode number state $\ket{N}_S$ can be evaluated to be
\begin{align}
    \rho_{\tau}&=\sech^{2(N+1)}\tau\sum_{a=0}^\infty\binom{N+a}{a}\tanh^{2a}\tau\ket{N+a}\bra{N+a}_S.
\end{align}
Passing this output state through a pure loss channel, the beam splitter acts on the number states according to Eq.~(\ref{eq:loss_unitary}) such that $\ket{n}\bra{n}_S\mapsto\sum_{k=0}^n\binom{n}{k}\eta_d^k(1-\eta_d)^{n-k}\ket{k}\bra{k}_S$, the state of the $\hat{b}$ mode becomes
\begin{align}
    \rho_\tau'&=\sech^{2(N+1)}\tau\sum_{a=0}^\infty\binom{N+a}{a}\tanh^{2a}\tau\qty[\sum_{k=0}^{N+a}\binom{N+a}{k}\eta_d^k(1-\eta_d)^{N+a-k}]\ket{k}\bra{k}_S\nonumber\\
    &=\sum_{k=0}^\infty\qty[\sech^{2(N+1)}\tau\sum_{a=\max(k-N,0)}^\infty\binom{N+a}{a}\binom{N+a}{k}\eta_d^k(1-\eta_d)^{N+a-k}\tanh^{2a}\tau]\ket{k}\bra{k}_S\nonumber\\
    &=\sum_{k=0}^\infty P_\tau(k)\ket{k}\bra{k}_S.
\end{align}
Using Eq.~(\ref{eq:fisherinformation_3}), we can find the classical Fisher information (FI) on $\tau$ and thus the lower bound the QFI following Eq.~(\ref{eq:QFI})
\begin{align}
    \mathcal{K}_\tau&\geq\mathcal{J}_{\tau}\nonumber\\
    &=-\sum_{k=0}^\infty P_\tau(k)\partial_\tau^2[\ln P_\tau(k)]\nonumber\\
    &=\sum_{k=0}^\infty \qty[P_\tau^{-1}(k)[\partial_\tau P_\tau(k)]^2-\partial_\tau^2 P_\tau(k)]\label{eq:QFI_loss}.
\end{align}
Since the state family $\{\rho_\tau'\}$ is diagonal in the number basis, the QFI on $\tau$ equals to the classical FI (saturation of the inequality in Eq.~(\ref{eq:QFI_loss})) of the family of photon number distributions $\{P_\tau(k);k=0,1,2,\cdots\}$ and is achieved by photodetection. Evaluating the derivatives give complex analytical expressions that cannot be further simplified to produce intuitive results:
\begin{align}
    \partial_\tau P_\tau(k) &=2\eta_d^k\sech^{2(N+2)}\tau\sum_{a=\max(k-N,0)}^\infty\binom{N+a}{a}\binom{N+a}{k}(1-\eta_d)^{N+a-k}\nonumber\\
    &\hspace{190pt}\times\qty[a-(N+1)\sinh^2\tau]\tanh^{2a-1}\tau,\\
     \partial^2_\tau P_\tau(k) &=2\eta_d^k\sech^{2(N+3)}\tau\sum_{a=\max(k-N,0)}^\infty\binom{N+a}{a}\binom{N+a}{k}(1-\eta_d)^{N+a-k}\nonumber\\
    &\hspace{10pt}\times\qty[2(N+1)^2\sinh^4\tau-(4aN+N+6a+1)\sinh^2\tau+2a^2-a]\tanh^{2a-1}\tau.
\end{align}
These expressions can be evaluated numerically to give the QFI $\mathcal{K}_\tau$. The QFI $\mathcal{K}_G$ on $G$ follows as $\mathcal{K}_G=\mathcal{K}_\tau/[4G(G-1)]$. The QFI for the multimode number-state probe is the sum of terms of the above form for each mode, and the result for multimode single-photon probes is displayed in Fig.~\ref{fig:loss_sensing}.

\section{Mean squared error of estimator \protect\(\check{G}\)}\label{appendixG_3}
In this section, we give a detailed calculation for the MSE achieved by the estimator for non-ideal detection protocol using $M$-mode number-state and coherent state probes with average total energy $N$,
\begin{align}
    \check{G} = \frac{\qty(\sum_{m=1}^M Y_m)/\eta_d + M}{N+M},
\end{align}
 where $\{Y_m\}$ denote the measured photocounts in the modes $\{\hat{b}_m\}$. Using the annihilation operator of the measured mode described in Eq.~(\ref{eq:annihilation_loss_gain}) for modes $\hat{e}$ and $\hat{f}$ in vacuum state, the first and second moments of the photocount in each mode satisfy (the mode subscript is omitted for simplification):
 \begin{align}
     \expval{Y} &=\expval{\hat{N}_b}=\eta_d\expval{\hat{N}_{out}}\label{eq:exp_Y},\\
     \expval{Y^2} &=\expval{\hat{N}_b^2}=\eta_d^2\expval{\hat{N}_{out}^2} + \eta_d(1-\eta_d)\expval{\hat{N}_{out}}\label{eq:exp_Ysq},
 \end{align}
where we have $\hat{N}_b =\hat{b}^\dag\hat{b}$, and $\expval{\hat{N}_{out}}=\hat{a}_{out}^\dag\hat{a}_{out}$. By using the Heisenberg-evolution equations for a QLA given in Eq.~(\ref{eq:annihilation_QLA}), the mean and second moment of $\hat{N}_{out}$ are evaluated to be:
\begin{align}
    \expval{\hat{N}_{out}} &= G\expval{\hat{N}_{in}} + G - 1\label{eq:exp_Nout},\\
    \expval{\hat{N}_{out}^2} &= G^2\expval{\hat{N}_{in}^2} + 3G(G-1)\expval{\hat{N}_{in}} + (G-1)(2G-1)\label{eq:exp_Noutsq},
\end{align}
where $\hat{N}_{in}=\hat{a}_{in}^\dag\hat{a}_{in}$. We can show that the estimator is unbiased by first evaluating the mean of $\sum_{m=0}^M Y_m$
\begin{align}
    \expval{\sum_{m=0}^M Y_m}&=\expval{\sum_{m=1}^M\hat{N}_b}\nonumber\\
    &=\sum_{m=1}^M \eta_d[GN_m + (G-1)]\nonumber\\
    &=\eta_d\qty[GN +M(G-1)],
\end{align}
where we have $\sum_{m=1}^M N_m = N$, and applied Eq.~(\ref{eq:exp_Y}) and Eq.~(\ref{eq:exp_Nout}). Hence, the mean of the estimator $\check{G}$ is calculated to be
\begin{align}
    \expval{\check{G}} &= \frac{1}{N+M}\qty[\frac{\eta_d[GN+M(G-1)]}{\eta_d}+M] =G,\nonumber\\
\end{align}
from which we can hence conclude that estimator is unbiased.\\
If a probe is of product form, in particular for a number-state probe $\otimes_{m=1}^M\ket{n_m}$ with $N=\sum_{m=1}^M n_m$, we find the variance of the operator $Y$ to be
\begin{align}
    \text{Var}\qty[\sum_{m=1}^M Y_m] &=\expval{\sum_{m=1}^M Y_m^2}-\expval{\sum_{m=1}^M Y_m}^2\nonumber\\
    &=\sum_{m=1}^M\qty[\expval{Y_m^2}-\expval{Y_m}^2]\nonumber\\
    &= G\eta_d^2[(G-1)(M+N)] +\eta_d(1-\eta_d)[G(M+N)-M],
\end{align}
where we have utilised the results from Eq.~(\ref{eq:exp_Ysq}) and Eq.~(\ref{eq:exp_Noutsq}), and $\expval{\hat{N}_{in}^2} = \bra{n_m}\hat{N}_{in}^2\ket{n_m}=n_m^2$. The MSE of estimator $\check{G}$ when using multimode number-state probe is hence
\begin{align}
    \text{MSE}^{\text{num}}[\check{G}] &= \text{Var}\qty[\frac{\qty(\sum_{m=1}^M Y_m)/\eta_d + M}{N+M}]=\frac{\text{Var}\qty[\sum_{m=1}^M Y_m]}{\eta_d^2(M+N)^2}\nonumber\\
    &=\frac{\eta_d^2G(G-1)(N+M) +\eta_d(1-\eta_d)[G(M+N)-M]}{\eta_d^2(M+N)^2}\nonumber\\
    &=\frac{G(G-1)}{N+M} + \frac{1-\eta_d}{\eta_d(N+M)}\qty[G-\frac{M}{N+M}],
\end{align}
which corresponds to Eq.~(\ref{eq:MSE_num}) of the main text. Similarly, for a coherent state probe $\otimes_{m=1}^M\ket{\sqrt{N_m}}$ with total signal energy of $N=\sum_{m=1}^M N_m$, its per-mode variance is:
\begin{align}  \expval{\hat{N}_{in}^2}&=\bra{\sqrt{N_m}}\hat{a}_{in}^\dag\hat{a}_{in}\hat{a}_{in}^\dag\hat{a}_{in}\ket{\sqrt{N_m}}\nonumber\\
&=\bra{\sqrt{N_m}}[\hat{a}_{in}^\dag\hat{a}_{in} + \hat{a}_{in}^\dag\hat{a}_{in}^\dag\hat{a}_{in}\hat{a}_{in}]\ket{\sqrt{N_m}}\nonumber\\
&=N_m + N_m^2.
\end{align}
Thus, the MSE of the estimator $\check{G}$ when using the multimode coherent state probe is found to be
\begin{align}
    \text{MSE}^{\text{coh}}&=\frac{\text{Var}\qty[\sum_{m=1}^M Y_m]}{\eta_d^2(M+N)^2}\nonumber\\
    &=\frac{\eta_d^2 G(G-1)(N+M) + G^2N\eta_d^2 + \eta_d(1-\eta_d)[G(M+N)-M]}{\eta_d^2(N+M)^2}\nonumber\\
    &=\frac{G(G-1}{N+M} + \frac{G^2 N}{(N+M)^2} + \frac{1-\eta_d}{\eta_d(N+M)}\qty[G - \frac{M}{N+M}],
\end{align}
which corresponds to Eq.~(\ref{eq:MSE_coh}) of the main text.
\let\addcontentsline\oldaddcontentsline
        \chapter{Returned States for Target Detection using SPES Probe}\label{appendixH}
This section will be dedicated for a step-by-step guide in deriving the returned state of target detection when using single-photon entangled state (SPES) as a probe. As independent and identically distributed (iid) SPES probe are used in this thesis, taking the $M$ times tensor product of the result for the single SPES result will yield the desired state. Hence the following derivation focuses a single SPES probe used. From Eq.~(\ref{eq:SPES_H1}), the thermal loss channel can be decomposed to
\begin{align}
    \rho_1 &= \qty(\mathbb{I}_I\otimes\mathcal{L}_{\eta,N_B/(1-\eta)})\Psi\nonumber\\
    &= \qty[\mathbb{I}_I\otimes\qty(\mathcal{A}_G\circ\mathcal{L}_{\tilde{\eta}})]\Psi,
\end{align}
where $\mathcal{A}_G$ is a quantum-limited amplifier channel with gain $G\geq 0$, $\mathcal{L}_{\tilde{\eta}}$ is a pure loss channel of transmittance $0\leq\tilde{\eta}\leq 1$, and $\Psi = \ket{\psi}\bra{\psi}_{SPES}$. For NPS model of target detection, we have $G=N_B+1$ and $\tilde{\eta} = \eta/G$. The interactions of quantum operations for the quantum-limited amplifier and pure loss channel on single-mode number states can be derived from the $M$-mode generalised unitary operators from Eq~(\ref{eq:gain_unitary}) and Eq.~(\ref{eq:loss_unitary}),
\begin{align}
    \text{Amplifier: }&\hat{U}_{\mathcal{A}}\ket{n}_S\ket{0}_E=[\sech\theta]^{n+1}\sum_{a=0}^{\infty}\sqrt{\binom{n+a}{a}}[\tanh\theta]^a\ket{n+a}\ket{a},\\
    \text{Loss: }&\hat{U}_{\mathcal{L}}\ket{n}_S\ket{0}_E = \sum_{l=0}^\infty[\cos\phi]^{n-l}[\sin\phi]^l\sqrt{\binom{n}{l}}\ket{n-l}\ket{l},
\end{align}
where $\binom{a}{b} = \frac{a!}{b!(a-b)!}$ is the binomial coefficient, and we have let $\cosh^2\theta = G$ and $\cos^2\phi = \tilde{\eta}$. Hence, passing SPES through the pure loss channel first yields 
\begin{align}
    \ket{\psi}_{ISE} &= \sqrt{N_S}\ket{0}\qty[\hat{U}_{\mathcal{L}}\ket{1}\ket{0}] + \sqrt{1-N_S}\ket{1}\qty[\hat{U}_{\mathcal{L}}\ket{0}\ket{0}]\nonumber\\
    &= \sum_{l=0}^1[\cos\phi]^{1-l}[\sin\phi]^{l}\sqrt{\binom{1}{l}}\sqrt{N_S}\ket{0}\ket{1-l}\ket{l} + \sqrt{1-N_S}\ket{1}\ket{0}\ket{0},
\end{align}
Tracing out the inaccessible mode $E$, the input to quantum-limited amplifier channel is
\begin{align}
    \rho_{IS}^{in}& = (1 - N_S)\ket{1}\bra{1}\otimes\ket{0}\bra{0}\nonumber\\
    &\hspace{10pt}+\sum_{l=0}^1[\cos\phi]^{2(1-l)}[\sin\phi]^{2l}\binom{1}{l}N_S\ket{0}\bra{0}\otimes\ket{1-l}\bra{1-l}\nonumber\\
    &\hspace{20pt} + [\cos\phi]\sqrt{N_S(1-N_S)}\qty[\ket{0}\bra{1}\otimes\ket{1}\bra{0}+\ket{1}\bra{0}\otimes\ket{0}\bra{1}].
\end{align}
Finally, applying the quantum-limited amplifier operations:
\begin{align}
    \rho_1 & = \Tr_E\qty[\rho^{in}_{ISE}]\nonumber\\
    &=\Tr_E\bigg[(1-N_S)\ket{1}\bra{1}\otimes\hat{U}_{\mathcal{A}}\ket{0}\ket{0}\bra{0}\bra{0}\hat{U}^{\dag}_{\mathcal{A}}\nonumber\\
    &\hspace{10pt} + [\cos\phi]\sqrt{N_S(1-N_S)}\ket{0}\bra{1}\otimes\hat{U}_{\mathcal{A}}\ket{1}\ket{0}\bra{0}\bra{0}\hat{U}^{\dag}_{\mathcal{A}}\nonumber\\
    &\hspace{10pt}+[\cos\phi]\sqrt{N_S(1-N_S)}\ket{1}\bra{0}\otimes\hat{U}_{\mathcal{A}}\ket{0}\ket{0}\bra{1}\bra{0}\hat{U}^{\dag}_{\mathcal{A}}\nonumber\\
    &\hspace{10pt}+\sum_{l=0}^1[\cos\phi]^{2(1-l)}[\sin\phi]^{2l}\binom{1}{l}N_S\ket{0}\bra{0}\otimes\hat{U}_{\mathcal{A}}\ket{1-l}\ket{0}\bra{1-l}\bra{0}\hat{U}^{\dag}_{\mathcal{A}}\bigg]\nonumber\\
    &=\Tr_E\bigg[\frac{(1-N_S)}{[\cosh\theta]^2}\sum_{a,a'\geq 0}[\tanh\theta]^{a+a'}\ket{1}\bra{1}\otimes\ket{a}\bra{a'}\otimes\ket{a}\bra{a'}\nonumber\\
    &\hspace{10pt}+\frac{[\cos\phi]}{[\cosh\theta]^3}\sqrt{N_S(1-N_S)}\sum_{a,a'\geq 0}[\tanh\theta]^{a+a'}\binom{a+1}{a}^{\frac{1}{2}}\ket{0}\bra{1}\otimes\ket{a+1}\bra{a'}\otimes\ket{a}\bra{a'}\nonumber\\
    &\hspace{10pt}+\frac{[\cos\phi]}{[\cosh\theta]^3}\sqrt{N_S(1-N_S)}\sum_{a,a'\geq 0}[\tanh\theta]^{a+a'}\binom{a'+1}{a'}^{\frac{1}{2}}\ket{1}\bra{0}\otimes\ket{a}\bra{a'+1}\otimes\ket{a}\bra{a'}\nonumber\\
    &\hspace{10pt}+\sum_{l=0}^1\frac{[\cos\phi]^{2(1-l)}[\sin\phi]^{2l}}{[\cosh\theta]^{2(2-l)}}\binom{1}{l}N_S\sum_{a,a'\geq 0}\binom{a-l+1}{a}^{\frac{1}{2}}\binom{a'-l+1}{a'}^{\frac{1}{2}}\nonumber\\
    &\hspace{100pt}\times[\tanh\theta]^{a+a'}\ket{0}\bra{0}\otimes\ket{a-l+1}\bra{a'-l+1}\otimes\ket{a}\bra{a'}\bigg].
\end{align}
By tracing the inaccessible $E$ mode and substituting $\cosh^2\theta = G$ and $\cos^2\phi = \tilde{\eta}$, we will arrive at the expression shown in Eq.~(\ref{eq:SPES_rho1}).
    \end{appendices}

\end{document}